\documentclass[amsmath,amssymb,aps,10pt,prd,letterpaper,nofootinbib,balancelastpage,notitlepage,superscriptaddress,twocolumn,floatfix]{revtex4-2}
\usepackage{graphicx}	
\usepackage{epstopdf}	
\usepackage{dcolumn}	
\usepackage{ragged2e}
\usepackage{enumitem}
\usepackage{bm}			
\usepackage[sort&compress]{natbib}	 	
	\setcitestyle{square,numbers,comma}	
\usepackage[colorlinks=true,urlcolor=blue,linkcolor=black,citecolor=blue]{hyperref}	
\usepackage{color}
   \definecolor{r}{RGB}{255,0,0}
   \definecolor{R}{RGB}{255,0,0}
   \definecolor{p}{RGB}{223,0,225}
   \definecolor{P}{RGB}{223,0,225}
   \definecolor{g}{RGB}{0,189,85}
     \definecolor{G}{RGB}{0,189,85}

\newcommand{\up}[1]{\textsuperscript{#1}}				
\newcommand{\tabref}[2][]{Tab{#1}.~\ref{tab:#2}}		
\newcommand{\figref}[2][]{Fig{#1}.~\ref{fig:#2}}		
\newcommand{\sectref}[2][]{Sec{#1}.~\ref{sect:#2}}		
\newcommand{\appref}[2][x]{Appendi{#1}~\ref{app:#2}}	
\renewcommand{\eqref}[2][]{Eq{#1}.~(\ref{eq:#2})}			
\newcommand{\citeR}[2][]{Ref{#1}.~\cite{#2}}			
\newcommand{\lb}{\ensuremath{\left}}					
\newcommand{\rb}{\ensuremath{\right}}					
\newcommand{\nl}{\nonumber \\ & \quad }					
\newcommand{\GeV}{\,\text{GeV}}
\newcommand{\Chand}{Chandrasekhar}

\newcommand{\Mpl}{M_{\text{Pl.}}}
\newcommand{\CX}{$($C$X)$}
\newcommand{\arcsinh}{\ensuremath\text{arcsinh}\,}

\begin{document}

\title{White Dwarf Bounds on CHAMPs}
\date{\today}
\author{Michael A.~Fedderke}
\email{mfedderke@stanford.edu}
\affiliation{Stanford Institute for Theoretical Physics, Department of Physics, Stanford University, Stanford, CA 94305, USA}
\author{Peter W.~Graham}	
\email{pwgraham@stanford.edu}
\affiliation{Stanford Institute for Theoretical Physics, Department of Physics, Stanford University, Stanford, CA 94305, USA}
\author{Surjeet Rajendran}
\email{srajend4@jhu.edu}
\affiliation{Department of Physics \& Astronomy, Johns Hopkins University, Baltimore, MD  21218, USA}

\begin{abstract}
White dwarfs (WD) effectively act as high-gain amplifiers for relatively small energy deposits within their volume via their supernova instability.
In this paper, we consider the ways a galactic abundance of $\mathcal{O}(1)$-charged massive relics (i.e., CHAMPs) could trigger this instability, thereby destroying old WD.
The dense central core structure formed inside the WD when heavy CHAMPs sink to its center can trigger a supernova via injection of energy during collapse phases, via direct density-enhanced (pycnonuclear) fusion processes of carbon nuclei dragged into the core by the CHAMPs, or via the formation of a black hole (BH) at the center of the WD.
In the latter scenario, Hawking radiation from the BH can ignite the star if the BH forms with a sufficiently small mass; if the BH instead forms at large enough mass, heating of carbon nuclei that accrete onto the BH as it grows in size may be able to achieve the same outcome (with the conservative alternative being simply that the WD is devoured by the BH).
The known existence of old WD that have not been destroyed by these mechanisms allows us to improve by many orders of magnitude on the existing CHAMP abundance constraints in the regime of large CHAMP mass, $m_X \sim 10^{11}$--$10^{18}\,$GeV.
Additionally, in certain regions of parameter space, we speculate that this setup could provide a trigger mechanism for the calcium-rich gap transients: a class of anomalous, sub-luminous supernova events that occur far outside of a host galaxy.
\end{abstract}

\maketitle

\tableofcontents

\section{Introduction}
\label{sect:introduction}

Charged Massive Particles (CHAMPs),%
\footnote{\label{ftnt:SMPs}%
		Also known more recently as charged Stable Massive Particles (SMPs).
	} %
defined here as massive early-universe relics with $\mathcal{O}(1)$ electrical charge, appear in many theories of physics beyond the Standard Model; for instance, the (N)LSP in $R$-parity conserving supersymmetric extensions of the SM can be electrically charged and (meta)stable (e.g., \citeR[s]{Ellis:1983ew,Byrne:2002ri}); theories of universal extra dimensions may have as their lightest KK-odd state a charged state (e.g., \citeR{Appelquist:2000nn,Byrne:2003sa}); and exotic stable composite bound states \cite{Wise:2014jva,Gresham:2017zqi,Grabowska:2018lnd,Bai:2018dxf} may have an $\mathcal{O}(1)$ net electrical charge without their stability being dramatically impacted.
Early interest in superheavy charged particles (e.g., \citeR{Cahn:1980ss}) long predated their proposal as an early dark matter (DM) candidate \cite{DeRujula:1989fe,Dimopoulos:1989hk} (since realized to be ruled out), and they have an extremely rich phenomenology (see, e.g., \citeR[s]{Perl:2001xi,Burdin:2014xma,2010JCAP...02..031S,Fairbairn:2006gg} for some recent reviews).

Across large regions of their mass range, a number of strong observational constraints limit the abundance of CHAMPs to at most a small fraction of the dark matter density: for instance, limits have been considered from the absence of anomalously heavy nuclei in bulk terrestrial samples (e.g., \citeR[s]{
Alvager:1967fgm,
Middleton:1979zz,
Smith:1979rz,
Smith:1982qu,
Turkevich:1984zz,
Dick:1984mk,
Nitz:1986gb,
Dick:1985wk,
Norman:1986ux,
Pichard:1987ub,
Norman:1988fd,
Polikanov:1990sf,
Hemmick:1989ns,
Verkerk:1991jf,
Yamagata:1993jq,
Javorsek:2001yu,
Mueller:2003ji,
PDGOtherSearches}, and see \citeR{Burdin:2014xma} for further references), from their impact on Big Bang Nucleosynthesis [BBN] (e.g., \citeR[s]{Cahn:1980ss,Pospelov:2006sc,Kohri:2006cn,Kaplinghat:2006qr,Bird:2007ge,Kawasaki:2007xb,Jedamzik:2007cp,Jedamzik:2007qk,Pospelov:2007js,Pospelov:2008ta,Kamimura:2008fx,PradlerThesis,Jedamzik:2009uy,Kusakabe:2010cb,Pospelov:2010hj,Kusakabe:2017brd}), from the absence of accelerated `CHAMP cosmic rays' (see, e.g., \citeR{Dunsky:2018mqs} for recent work), and from a variety of astrophysical observations (e.g., \citeR[s]{Gould:1989gw,SanchezSalcedo:2008zd,2010JCAP...02..031S,Burdin:2014xma,Perl:2001xi}).
For at least some ranges of CHAMP masses, terrestrial bounds have been subject to question owing to the complicated dynamics of charged particles in galactic magnetic fields \cite{Dimopoulos:1989hk,Chuzhoy:2008zy,Dunsky:2018mqs}.

Some of the more interesting and powerful constraints on the abundance of CHAMPs, $X$, at very large masses arise from the consideration of their impact on the survival of old neutron stars (NS) \cite{Gould:1989gw}.
The crucial idea advanced in \citeR{Gould:1989gw} is that a sufficient mass of CHAMPs accumulated inside a NS will collapse to form a black hole (BH) that can then accrete matter from inside the NS, destroying it.
Specifically, \citeR{Gould:1989gw} considered that halo CHAMPs will pass through and be captured by a protostellar cloud as it collapses to form the $M \sim 10$--$30M_{\odot}$ NS progenitor star.
This population of CHAMPs remains in the star throughout its lifetime, eventually contaminating the NS as it is formed.

Being extremely heavy, CHAMPs rapidly sink to the center of the NS after its formation,%
\footnote{\label{ftnt:presinking}%
		It is of course conservative to assume a uniform contamination of the CHAMPs exists throughout the NS precursor stellar material just prior to NS formation \cite{Gould:1989gw}; sinking of sufficiently heavy CHAMPs during the stellar evolution will almost certainly occur.
	} %
 and form an approximately isothermal, thermal-pressure-supported central structure%
 \footnote{\label{ftnt:otherChargedPlasmas}%
 		Although the properties of NS material near its center are not known with any degree of certainty, it is not entirely comprised of neutrons: at a minimum, a sizable population of protons and electrons co-exists in equilibrium with the neutrons \cite{Takibayev:2017xyz}.
		Although central NS temperatures and densities are much too high to allow the formation of electrostatic bound states of the CHAMP with any particle, the existence of these superposed charged plasmas permits screening of the CHAMP charges, allowing the formation of the thermal-pressure-supported CHAMP structure despite the mutually repulsive electrostatic forces existing between CHAMPs.
 	} %
that is stable so long as the CHAMPs do not locally come to dominate over the density of NS material. 
If however the CHAMPs do come to dominate the NS matter, the CHAMP structure becomes self-gravitating and unstable to collapse (the so-called `gravothermal catastrophe'; see, e.g., \citeR{1968MNRAS.138..495L}).
For a small enough CHAMP population, quantum degeneracy pressure would eventually re-stabilize the structure at a much smaller radius; however, if a super-\Chand\ mass of CHAMPs ($M_{\text{Chand.}}^X \ll M_\odot$ since $m_X \gg \text{GeV}$) is present, re-stabilization becomes impossible \cite{Chandrasekhar_1931}, and the collapse proceeds to BH formation on a short timescale.

If the BH thus formed has a mass $M_{\textsc{bh}} \gtrsim M_{\textsc{bh},\,\text{crit.,}}^{\text{NS}} \sim 4\times 10^{-20}M_{\odot}$, \citeR{Gould:1989gw} found that accretion of material from the NS onto the BH, even at Eddington-limited rates, exceeds mass loss from the BH due to the Hawking process \cite{Hawking:1974wb}, and the BH rapidly grows in size, consuming the entirety of the NS on an extremely short timescale.
The existence of observed old NSs can thus strongly constrain the abundance of CHAMPs.
On the other hand, if the BH has an initial mass $M_{\textsc{bh}} \lesssim M_{\textsc{bh},\,\text{crit.,}}^{\text{NS}}$, the mass loss by the BH due to Hawking radiation dominates the mass gain from accretion of NS matter, causing the BH to shrink in size and eventually evaporate.
This evaporation however inflicts no externally observable structural damage to the NS.
As a result, \citeR{Gould:1989gw} found that NS destruction bounds on the CHAMP abundance weaken considerably as the CHAMP mass increases above $m_X\gtrsim 10^{11}\,$GeV. 

Recently, there has been much interest in the literature in white dwarfs (WD) as unconventional, large spacetime volume, bolometric particle detectors \cite{Graham:2015apa,Bramante:2015cua,Graham:2018efk,Acevedo:2019gre,Janish:2019nkk}.
It has of course long been known that white dwarfs---``the biggest powder keg[s] in the Universe'' \cite{WoosleyPetschek:1990aa}---can be provoked into a thermal runaway leading to a type-Ia--like (SNIa) supernova explosion (visible at cosmological distances) by the concentrated local deposition of a sufficient large amount of energy inside the WD volume \cite{Timmes_1992}.
For a carbon--oxygen (CO) WD \cite{Koester_1990} in the mass range $M_{\textsc{wd}} \sim 0.8$--$1.35M_{\odot}$, an energy deposition of $(5\times10^{24})$--$(8\times10^{16})\,$GeV is sufficient to raise a region near the center of the WD with a physical size (`trigger length') of $\lambda_T \sim (6\times10^{-2})$--$(7\times 10^{-5})$\,cm to a critical temperature of around $0.5\,$MeV.
Such conditions result \cite{Timmes_1992,Graham:2015apa} in the birth of a stable propagating flame front that traverses the entire WD, burning approximately a solar mass of CO mixture to nuclear statistical equilibrium (NSE), and releasing $\sim 10^{54}$\,GeV of energy.

WDs thus behave as extremely high-gain natural amplifiers for sufficiently large local energy depositions occurring anywhere within a large fraction of their volume.
This observation has been used to place limits on the abundance of primordial black holes which transit thorough WD, inducing local heating by dynamical friction \cite{Graham:2015apa} (but see \citeR{Montero-Camacho:2019jte} for a recent reappraisal); on (asymmetric) dark matter which is captured in a WD, forms a core, and deposits energy in the WD material once the core becomes self-gravitating and collapses \cite{Bramante:2015cua,Acevedo:2019gre}; on dark matter (either transiting or captured) which transfers sufficient energy locally to the WD by scattering, decay, or annihilation \cite{Graham:2018efk}; and on dark matter that, once captured in a WD, forms a core that collapses to a black hole, which eventually triggers a supernova \cite{Acevedo:2019gre,Janish:2019nkk}.
It is this latter work on black-hole-induced SNIa-like supernovae that is of particular relevance to the present work.

The results of \citeR[s]{Acevedo:2019gre,Janish:2019nkk} indicate that, with some minor exceptions, the formation of a BH near the center of a sufficiently massive WD will either satisfy the heating requirements to trigger thermal runaway, or will simply consume the WD.
The former outcome can occur in one of two ways, depending on which of accretion of matter or Hawking radiation dominates the BH evolution.
If the accretion dominates, the BH formed will grow in mass and size.
Conservatively, this process would merely consume the entire WD.
However, an second possibility exists \cite{Janish:2019nkk}: the increasing mass of the BH of course leads to ever higher surface gravity, which will eventually lead to the sufficient acceleration of the carbon ions in the WD material in the immediate vicinity of the BH (or, more specifically, in the vicinity of the capture radius for accretion onto the BH) to enable local heating the WD above the critical threshold temperature, leading to thermal runway.
Alternatively, the increase in the local WD material density around the BH could allow pycnonuclear fusion to proceed even absent significant heating \cite{Janish:2019nkk,Kippenhahn:2012zqe}.%
\footnote{\label{ftnt:boomCollapseCaveat}%
		Note that this might also occur in the dense CHAMP-dominated BH-progenitor object that forms at the core of the WD just prior to collapse, although the rate estimates for this process are sufficiently uncertain that it is not clear that this can occur.
		See discussion in \sectref{TransChandMinus}.
	} %
On the other hand, if Hawking losses dominate the accretion, the BH will shrink, but here the outcome is dramatically different from the BH-in-a-NS case of \citeR{Gould:1989gw}.
As the BH evaporates, the Hawking temperatures rises and the energy loss rate increases.
Provided that certain minimal assumptions are met, the energy deposition rate from the Hawking process will eventually raise a sufficiently large volume of WD material above the critical temperature to trigger thermal runaway for WD in this mass range \cite{Acevedo:2019gre,Janish:2019nkk}.
The authors of \citeR{Janish:2019nkk} verify that this latter process completes within $\lesssim$\,Gyr of BH formation in most regions of parameter space, with only some minor exceptions possible.

\section{Executive summary}
\label{sect:executiveSummary}

In this paper, we consider the scenario in which CHAMPs have accumulated in a WD at first via a primordial contamination of the protostellar gas cloud of the WD progenitor star \`a la \citeR{Gould:1989gw}, and later in their evolution via direct, gravitationally boosted accretion from a population of halo CHAMPs over the first $\sim$\,Gyr of the WD lifetime. 
These CHAMPs will settle to the center of the WD, initially forming a thermal-pressure-supported structure. 

If a sufficient total mass of CHAMPs, is present in this structure, it will become self-gravitating and undergo gravothermal collapse.
While it is possible that the energy released by this collapse alone could trigger thermal runaway, it is conservative to assume that the WD survives this initial collapse phase.

Under that assumption, if a sub-\Chand\ mass of CHAMPs participates in the gravothermal collapse, the CHAMPs merely re-establish a hydrostatic equilibrium configuration at much smaller radius, supported by the degeneracy pressure of relativistic electrons in the WD core, communicated to the CHAMPs by electrostatic forces.

Later addition of further CHAMPs to this core, either because the entire primordial CHAMP abundance in the star did not form part of the thermal structure by the time the gravothermal collapse was initiated, or because further CHAMPs are accreted onto the star from the halo, will result in its central density increasing over time, until one of two possible instabilities (see below) occurs once the core mass reaches the trans- or super-\Chand\ mass regime.

Similarly, if a trans- or super-\Chand\ mass of CHAMPs participates in the initial gravothermal collapse, that collapse may initially be (momentarily) stalled by electron degeneracy pressure causing the reestablishment of a quasi-static core structure, but as the collapsing CHAMP cloud continues to add more CHAMPs to this core structure, its central density will also increase over time, again until one of the two aforementioned instabilities occurs.

For negatively charged CHAMPs, the first possible instability is that carbon ions drawn into the core along with the CHAMPs may begin to fuse immediately via density-enhanced pycnonuclear fusion mechanisms of the carbon ions once trans-\Chand\ masses are reached; this would would directly trigger thermal runaway.

Alternatively, if this does not occur, or if the CHAMPs are positively charged, then the growing degenerate core simply reaches and then eventually exceeds the maximum density that is supportable by degeneracy pressure as its mass reaches the trans-\Chand\ regime, leading inevitable to collapse to a black hole.
If this second alternative occurs, the BH-induced WD destruction mechanisms proposed in \citeR[s]{Acevedo:2019gre,Janish:2019nkk} become operational and can eventually trigger thermal runaway, unless the evolutionary timescales involved for this process are too long.
Except for the caveat about timescales, sufficiently CHAMP-contaminated, sufficiently old WDs are thus destroyed.

Using the known existence of very old, high mass WD of low surface magnetic field \cite{2017ASPC..509....3D}, we extend and improve upon the upper bounds placed on the CHAMP population in our Galaxy set by \citeR{Gould:1989gw}, finding that much stronger constraints can be set at large CHAMP masses ($m_X \sim 10^{11}$--$10^{18}$\,GeV); see \figref[s]{BoundsPlus} and \ref{fig:BoundsMinus} for our major results.

Additionally, making use of this supernova trigger mechanism at lower CHAMP masses ($m_X \sim 10^{7}$--$10^{11}$\,GeV; see \figref{CaRGT}), where the considerations advanced in, e.g., \citeR{Chuzhoy:2008zy} indicate that the majority of CHAMPs may possibly have been evacuated the central regions of galaxies and magnetically inhibited from re-entry, we advance a speculative explanation for a class of sub-luminous supernova events that occur preferentially displaced from the center of their host systems, the so-called calcium-rich gap transients (CaRGT).
CHAMPs, preferentially present in the outskirts of galaxies after having been ejected from the baryon-rich central regions by supernova shockwaves, can accrete onto WD found in such outer regions, thereby supplying the correct spatial morphology for these events, as supernovae of this type are then preferentially triggered in the outskirts of galaxies.
Moreover, the CHAMP-induced thermal runaway trigger mechanisms we discuss would be able to trigger those supernovae in the fairly abundant sub-\Chand\ population \cite{2016MNRAS.461.2100T,Kepler:2006ns} of WD with masses $M_{\textsc{wd}} \sim 0.8M_{\odot}$; recent modeling \cite{polin2019nebular} shows that such supernovae events may give roughly the correct luminosity and spectral characteristics for the CaGRT.

The remainder of this paper is structured as follows: 
in \sectref{ignition} we review generally the conditions for WD ignition by local energy deposition in the WD material in order to establish the physical conditions we will be seeking to attain in this work.
We then change tack, reviewing the `chemistry' of CHAMPs generally in \sectref{CHAMPchemistry}, before turning to a discussion of their behavior in galaxies in \sectref{galaxy}.
Following this, in \sectref{CHAMPsWD} we examine in detail the accumulation of CHAMPs in white dwarfs, and their behavior in such dense objects generally, with a particular focus on the dense central core structures they form in the WD, as well as the maximum supportable masses allowed before such cores undergo gravothermal collapse, or collapse to a BH.
We give estimates for relevant timescales in this discussion.
This discussion of the central core structure also includes an examination of a possible earlier trigger mechanism of the SNIa runaway of the WD by the compression of carbon nuclei in the core of the WD contaminated by $X^-$.
Assuming BH formation, and that WD survives until BH formation, in \sectref{BHinWD} we review the subsequent dynamical evolution of the BH showing that, even in this case, the WD is still destroyed in large regions of parameter space.
We turn these considerations into limits on the galactic abundance of CHAMPs in \sectref{limits} by virtue of the known existence of old, high mass, local WD with low surface magnetic field.
More speculatively, in \sectref{CaRGT}, we offer some observations that could allow this WD ignition mechanism to explain the calcium-rich gap transients.
Finally, we conclude in \sectref{conclusion}.
There are a number of appendices that offer extra information: \appref{CHAMPbinding} is a discussion of the binding energies of $X^-$ with positively charged nuclei; \appref{pycnonuclear} contains an estimate of the pycnonuclear fusion rate of carbons bound in $(\text{C}X)$ bound states in the central dense CHAMP-contaminated core of the WD; \appref{Accreting_Fraction} gives the full expressions we use to compute the CHAMP abundance accreting onto a WD over its lifetime; \appref{WDstructure} contains a detailed consideration both of canonical WD structure and CHAMP-contaminated WD structure; %
\appref[ces]{ElectronConduction} and \ref{app:freeFreeOpacity} contain the expressions we have used for the electron heat conduction and free-free opacity, respectively, that are used in computing the WD trigger conditions.

\section{White dwarf ignition}
\label{sect:ignition}
In order to understand the physical conditions required to initiate WD destruction, we begin our discussion by reviewing in detail the arguments advanced in a series of papers \cite{Timmes_1992,Graham:2015apa,Bramante:2015cua,Graham:2018efk,Montero-Camacho:2019jte,Acevedo:2019gre,Janish:2019nkk} which have examined the conditions required to initiate thermal runaway in a CO WD, and which most recently culminated in the proposal of \citeR[s]{Acevedo:2019gre,Janish:2019nkk} for a trigger mechanism for SNIa-like supernova events in WD due to the formation of a BH inside the WD (or, more conservatively, for destruction of the WD by accretion onto the BH).
\sectref{ignitionGeneral} contains a general discussion of ignition in a WD; we discuss the trigger criteria in detail in \sectref{WDignition}. 

While we independently re-estimate here various numerical values for important physical properties, the bulk of this section is based on the discussions in \citeR[s]{Timmes_1992,Graham:2015apa,Bramante:2015cua,Graham:2018efk,Montero-Camacho:2019jte,Acevedo:2019gre,Janish:2019nkk}.

\subsection{General discussion}
\label{sect:ignitionGeneral}

In ordinary stars, the pressure support necessary to maintain the local hydrostatic equilibrium conditions required for a stable stellar configuration is supplied by thermal motion (see generally \citeR{Kippenhahn:2012zqe}).
An abrupt local energy deposition occurring in the stellar material as a result of, e.g., a high-energy particle interaction within the stellar volume will result in a local temperature increase, causing an increase in the highly temperature-dependent rate of local nuclear burning, leading to further energy deposition and heating.
However, this process is self-regulating: the temperature increase causes an initial local over-pressure condition in the heated volume, which relaxes by causing the heated stellar material to expand quasi-adiabatically to a larger volume, lowering its temperature, and with it the rate of nuclear burning in the perturbed material.
Ordinary stars are thus highly stable against having their structure significantly disrupted by small local energy depositions.

In a WD however, the hydrostatic pressure support required for stability is supplied almost entirely by the quantum degeneracy pressure of the (possibly quite relativistic) electrons \cite{Koester_1990}, communicated to the heavier ionized nuclei by electrostatic forces. 
This pressure is sensitive to the electron density, but is highly insensitive to local temperature perturbations \cite{Landau:1980egq}.%
\footnote{\label{ftnt:largePerturbations}%
		So long as means $|\Delta T| \ll E_F$ where $E_F$ is the Fermi energy of the degenerate electrons; typically, $E_F \sim \mathcal{O}(\text{few MeV})$.
		A temperature increase $\Delta T \gg \text{MeV}$ would lift degeneracy and return the stellar material to a normal gas phase.
	} %
Local energy depositions in WD material resulting in a local heating of the ions are thus not subject to the self-regulating adiabatic expansion mechanism as ordinary stellar material.

This makes WD highly susceptible to a thermal runaway condition \cite{Koester_1990,WoosleyPetschek:1990aa}: some local energy deposit leads to local heating, which leads to a (highly) increased local nuclear burning and further local heating, and so on. 
The only temperature regulation mechanism available is thermal diffusion, either by conduction by degenerate electrons, or by radiative transport. 
If the rate of thermal energy flow by diffusion out of some (sufficiently large) perturbed volume is lower than the rate of energy injection from nuclear burning in that volume, the local temperature in the perturbed volume simply continues to rise as the nuclear burning rate increases with increasing temperature.
Moreover, the region of increased nuclear burning will propagate outward as a flame front,%
\footnote{\label{ftnt:deflagrationvsdetonation}%
		 Whether this occurs subsonically (a deflagration), supersonically (a detonation), or some combination (deflagration-to-detonation transition) remains an open problem \cite{doi:10.1146/annurev.astro.38.1.191}.
	} %
which eventually traverses and consumes the whole star \cite{WoosleyPetschek:1990aa,Kippenhahn:2012zqe}.
The total energy released by the nuclear burning of an $\mathcal{O}(1)$ fraction of the material in the WD to NSE, $\sim 10^{51}$\,erg, easily exceeds the energy required to lift the electron degeneracy, allowing the stellar material to finally violently expand as a result of the significant heating \cite{Kippenhahn:2012zqe}. 
This catastrophic energy release, being also in excess of the total gravitational binding energy of the star, $\sim \text{few}\times10^{50}$\,erg, leads to the total disruption of the WD, manifesting itself as a SNIa supernova \cite{Kippenhahn:2012zqe}, visible at cosmological distances (e.g., \citeR{Jones:2013dta}).%
\footnote{\label{ftnt:CanonicalTrigger}%
		The `canonical' trigger scenario for a SNIa is by mass accretion from a non-degenerate binary companion star, resulting in the WD mass increasing toward the Chandrasekhar limit \cite{Chandrasekhar_1931}, becoming ever more unstable to perturbation in the process, with ignition occurring shortly before the limit is reached owing to a combination of core heating and compression as a result of the mass accretion \cite{doi:10.1146/annurev.astro.38.1.191, Kippenhahn:2012zqe,Shapiro:2000abc}.
		Alternative scenarios such as double-degenerate collisions \cite{doi:10.1146/annurev.astro.38.1.191}, in which two degenerate, sub-Chandrasekhar WD collide and trigger the necessary runaway, have also been proposed.
	} %

\subsection{Trigger length}
\label{sect:WDignition}
The energy deposition necessary for the birth of a stable flame front in WD material is discussed in detail in \citeR[s]{Graham:2015apa,Bramante:2015cua,Graham:2018efk,Acevedo:2019gre,Janish:2019nkk,Montero-Camacho:2019jte}, following from the detailed numerical work of \citeR{Timmes_1992}.

As already stated, the basic criterion is that a volume of material of a certain minimum characteristic size $\lambda_T$, the \emph{trigger length}, should be heated to at least a minimum temperature $T_{\text{crit}}$, the \emph{trigger temperature}, such that the rate of energy loss by thermal diffusion from this volume to the surrounding medium, $-\dot{E}_{\text{diff.}}(T_{\text{crit}},\lambda_T)$, is smaller than the rate of energy deposition into that volume by nuclear fusion, $\dot{E}_{\text{nucl.}}(T_{\text{crit}})$ \cite{Timmes_1992}.
If this criterion is not satisfied, any localized initial temperature perturbation impressed into the material by an external source (e.g., a high-energy particle interaction occurring in the WD) may cause local nuclear burning to proceed all the way to NSE locally, but will ultimately simply diffuse away harmlessly into the bulk WD medium without triggering the formation of a self-sustaining, propagating nuclear flame front.

The criterion may also be alternatively phrased as requiring the deposition of a certain total minimum energy $E_T$, the \emph{trigger energy}, into the trigger volume within the characteristic time for heat diffusion from the trigger volume $\tau_{\text{diff.}}$; energy deposition into a volume larger than the trigger volume must be proportionally larger on a volumetric basis. 

We will be mostly concerned with cases where the temperature of the trigger volume, which we assume to be located near the center of a CO WD lying in the mass range $M_{\textsc{wd}} \sim 0.8$--$1.35M_{\odot}$, is raised at least as high as $T_{\text{crit.}}\sim 0.5$\,MeV \cite{Timmes_1992}.
The corresponding densities of the WD material are obtained by solving the Tolman--Oppenheimer--Volkhoff (TOV) equation \cite{Tolman:1939it,Oppenheimer:1939cs} to find the WD-mass--central-density relationship, $\rho_C(M_{\textsc{wd}})$, which for WDs with masses $M_{\textsc{wd}} \in [0.1,1.35]M_{\odot}$ can be given approximately by%
\footnote{\label{ftnt:EoSinToV}%
		We obtained this result assuming a fully degenerate electron equation of state (EoS) with electrons assumed to have their full dispersion relation (i.e., we did not use the non- or ultra-relativistic approximations) and exact local charge neutrality of the WD plasma; we did not however include Coulomb corrections to the EoS (see, e.g., \citeR{Salpeter:1961zz}).
		We then fitted a polynomial function to the numerical parameter $\alpha$ which completely dictates the WD structure.
		See \appref{WDstructure}.
	} %
\begin{align}
\rho_C &\sim 1.95 \times 10^6\, \text{g/cm}^3 \lb[ \alpha(M_{\textsc{wd}})^{-2} - 1 \rb]^{3/2} \label{eq:rhoCentral} \\
\alpha(M_{\textsc{wd}}) &\approx 1.0033 - 0.3087 x  -1.1652 x^2  + 2.0211 x^3 \nl - 2.0604 x^4 + 1.1687 x^5 - 0.2810 x^6 \nonumber \\
x &\equiv  M_{\textsc{wd}} / M_{\odot} \in [0.1,1.35];
\label{eq:alphaFit}
\end{align}
the central densities of WDs in the mass range $M_{\textsc{wd}} \sim 0.8-1.35M_{\odot}$ thus vary from $\sim 10^{7}\,\text{g/cm}^3$ to $\sim 10^{9}\,\text{g/cm}^3$.

To evaluate the trigger criterion, we must know the rate of diffusive heat transport.
Diffusive heat flow in presence of a temperature gradient is governed by Fourier's Law: $\bm{Q} = - k \bm{\nabla}u$, where $\bm{Q}$ is the (vector) heat flow, $u = u(t,\bm{x})$ is the temperature field, and $k$ is the thermal conductivity \cite{Kittel:1980abc}. 
The heat flow mechanisms in dense WD matter are dominantly radiative transport and electron conduction \cite{Timmes_1992,Kippenhahn:2012zqe,Graham:2015apa},%
\footnote{\label{ftnt:neutrinoLosses}%
		Losses by neutrino emission are subdominant \cite{Timmes_1992,Montero-Camacho:2019jte}, except during the final stages of thermal runaway \cite{Kippenhahn:2012zqe}.
	} %
so the thermal conductivity is given by \cite{Kippenhahn:2012zqe}
\begin{align}
k &= k_{\text{rad}} + k_{\text{cd}} = \frac{4\pi^2}{45} \frac{T^3}{\rho} \lb( \kappa_{\text{rad}}^{-1} + \kappa_{\text{cd}}^{-1} \rb),
\label{eq:thermalConductivity}
\end{align}
where $\kappa_{\text{rad,\,cd}}$ are, respectively, the radiative and conductive opacities; these in general depend on the density, temperature, and chemical composition of the material.
Electron thermal conductivities are given in \citeR{Potekhin:1999yv} and do not have a simple scaling at all densities, although for $\rho\gtrsim 8\times 10^8\,\text{g/cm}^3$ and at $T\sim 0.5\,$MeV, the scaling is approximately $\kappa_{\text{cd}} \propto \rho^{-1.4}$.
The radiative opacity can be taken to be given by the free-free Kramers opacity which scales as $\kappa_{\text{rad}} \propto \rho$, corrected by a suitable Gaunt factor~\cite{Cox:1968abc,Cox:1968def,Rybicki:2004abc,Kippenhahn:2012zqe} (although we ignore this correction here); see \appref[ces]{ElectronConduction} and \ref{app:freeFreeOpacity} for details.
For $T\sim 0.5\,$MeV, the electron conduction contribution to the thermal conductivity dominates at densities above $\rho \sim 8\times 10^8 \text{g/cm}^3$, while the radiative contribution dominates at lower densities; outside the range $\rho \sim 0.3$--$3\times 10^9\,\text{/cm}^3$ it is a good approximation that $k$ scales with density as implied by the dominant of the two individual contributions. 

The total rate of thermal energy loss from a spherical%
\footnote{\label{ftnt:AllConvexShapes}%
		This argument applies parametrically to any convex shape, with the radius of the sphere replaced by the smallest characteristic scale over with the heated region has a significant temperature gradient.
	} %
region of radius $\lambda_T$ by heat flow through its surface is $-\dot{E}_{\text{diff.}}(T_{\text{crit}},\lambda_T) = \lambda_T^2 \int d\Omega\, \bm{\hat{r}} \cdot\bm{Q}(T_{\text{crit}})$.
Assuming a radial temperature profile such that $\nabla u = \partial_r u \bm{\hat{r}}$ we have $-\dot{E}_{\text{diff.}}(T_{\text{crit}},\lambda_T) = + 4\pi k \lambda_T^2 (-\partial_r u)$.
We approximate $-\partial_r u \sim \Delta T / \lambda_T$ for a region of radius heated $\lambda_T$ heated to a temperature $T_{\text{crit}} = T_0 + \Delta T$ above the ambient temperature $T_0$, so that $-\dot{E}_{\text{diff.}}(T_{\text{crit}},\lambda_T) \sim 4\pi k \lambda_T \Delta T$.
For a region so heated under isobaric conditions, the excess energy is 
\begin{align}
\Delta E = M \int_{T_0}^{T_0 + \Delta T} dT' c_p(T') = (4\pi\lambda_T^3/3) \rho\mkern1mu \bar{c}_p(T_{\text{crit}})\Delta T,
\label{eq:Etrig}
\end{align}
where $c_p(T)$ is the constant-pressure specific (per-mass) heat capacity, and we have implicitly defined its average value in the relevant temperature range.
The diffusive cooling timescale is thus 
\begin{align}
\tau_{\text{diff.}} &\equiv \Delta E / |\dot{E}_{\text{diff.}} | \\
&= \frac{\rho \lambda_T^2}{3k} \lb[ \frac{1}{\Delta T} \int_{T_0}^{T_0 + \Delta T} dT' c_p(T') \rb]   \\
&\sim \rho \lambda_T^2 \bar{c}_p(T_{\text{crit}})/ (3 k).
\label{eq:taudiff}
\end{align}

The isobaric specific heat capacity can be approximated by as a sum of three terms, independently accounting for the ionic, electronic, and radiative contributions. 
We treat the ions as a free ideal gas, and keep only the leading term in the Sommerfeld expansion \cite{Landau:1980egq} for the electronic contribution, valid in the limit $T\ll E_F$.
That is, we take
\begin{align}
c_p(T,\rho) &\equiv  c^{\text{ions}}_p(T,\rho) + c^{\text{elec.}}_p(T,\rho)  + c^{\text{rad.}}_p(T,\rho) \label{eq:heatcap0} \\
c^{\text{ions}}_p(T,\rho) &\equiv \frac{5}{2\mu_a} \sum_i \frac{X_i}{A_i} \\
c^{\text{elec.}}_p(T,\rho) &\equiv \frac{\pi^2}{\mu_a\mu_e}\frac{T}{E_F}\lb[ 1-\lb(\frac{m_e}{E_F}\rb)^2\rb]^{-1} \\
c^{\text{rad.}}_p(T,\rho) &\equiv \frac{4\pi^4}{5\mu_a\mu_e} \lb( \frac{T}{E_F} \rb)^3 \lb[ 1- \lb( \frac{m_e}{E_F} \rb)^2 \rb]^{-3/2}
\label{eq:heatcap1}
\end{align}
where $\mu_a$ is the atomic mass unit, $E_F = [ 1 + ( 3\pi^2 n_e / m_e^3 )^{2/3} ]^{1/2}$ is the electron Fermi energy, $n_e = \rho / (\mu_a\mu_e)$ is the electron number density, $\mu_e \equiv \lb( \sum_i X_i Z_i/A_i \rb)^{-1}$ is the mean molecular mass per electron, and $X_i$, $Z_i$, and $A_i$ are, respectively, the mass fraction, charge, and atomic mass number of ion species $i$.
It follows that 
\begin{align}
\bar{c}_p(T_{\text{crit}}) &\equiv \frac{1}{\Delta T} \int_{T_0}^{T_0 + \Delta T} dT' c_p(T') \\
&\approx c^{\text{ions}}_p(T _{\text{crit}},\rho) + \frac{1}{2} c^{\text{elec.}}_p(T _{\text{crit}},\rho)  \nl 
	+ \frac{1}{4} c^{\text{rad.}}_p(T _{\text{crit}},\rho),
\label{eq:heatcap2}
\end{align}
assuming $T_{\text{crit}}\sim\Delta T \gg T_0$.

Assuming a specific (again, per-mass) nuclear energy generation rate of $\dot{S}_{\text{nucl.}}(T_{\text{crit}})$, the total nuclear energy generation rate in the same volume of material is $\dot{E}_{\text{nucl.}}(T_{\text{crit}}) = M \dot{S}_{\text{nucl.}} = (4\pi/3) \lambda_T^3 \rho \dot{S}_{\text{nucl.}}$, so that timescale for nuclear burning to double the excess thermal energy in a region of mass $M$ is 
\begin{align}
\tau_{\text{nucl.}} &\equiv \Delta E/ \dot{E}_{\text{nucl.}} \\
				&= \frac{1}{\dot{S}_{\text{nucl.}}} \int_{T_0}^{T_0 + \Delta T} dT' c_p(T')\\
				&\sim \bar{c}_p(T_{\text{crit.}}) \Delta T / \dot{S}_{\text{nucl.}}(T_{\text{crit.}}).
\end{align}

A parametric, order-of-magnitude estimate for the trigger condition can then be phrased as
\begin{align}
\dot{E}_{\text{nucl.}}(T_{\text{crit}}) &\sim -\dot{E}_{\text{diff.}}(T_{\text{crit}},\lambda_T) \Leftrightarrow \tau_{\text{nucl.}} =  \tau_{\text{diff.}} \\
\Rightarrow \dot{S}_{\text{nucl.}}(T_{\text{crit}}) &\sim 3 k \Delta T / (\rho \lambda_T^2) \sim 3 k T_{\text{crit}} / (\rho \lambda_T^2),
\label{eq:trigger}
\end{align}
where the final expression holds in the physically relevant limit $T_{\text{crit}} \gg T_0$.
Therefore, the trigger length estimate is given by 
\begin{align}
\lambda_T & \sim \sqrt{ 3 k T_{\text{crit}} / \rho\dot{S}_{\text{nucl.}}(T_{\text{crit}}) };
\label{eq:lambdaT}
\end{align}
see also \citeR[s]{Landau:1980egq,Timmes_1992}.

However, \eqref{lambdaT} should be employed with great care: the specific nuclear energy generation rates in the WD temperature range are extremely fast functions of temperature, and become also exponentially fast functions of density not too far above the unperturbed WD central densities for the WD we consider, owing to strong screening effects and pycnonuclear processes \cite{Gasques:2005ar,Yakovlev:2006fi,Kippenhahn:2012zqe}. 
Since the passage of a flame front necessarily involves rapid changes in both temperature and density, \eqref{lambdaT} does not necessarily give a very good numerical approximation to the actual trigger length \cite{Landau:1980egq,WoosleyPrivate}.
There is moreover an ambiguity as to the exact temperature at which the nuclear energy generation rate should be evaluated \cite{Timmes_1992,Landau:1980egq,WoosleyPrivate}: although we have indicated this temperature as $T_{\text{crit.}}$, we remind the reader that screened carbon fusion proceeds as a tunneling process for $T\lesssim \text{MeV}$, and the specific nuclear energy generation rates scale as fast as $d \ln\dot{S}_{\text{nucl.}} / d\ln T\sim 24$ in the temperature and density range relevant for WD flame propagation \cite{Gasques:2005ar,Yakovlev:2006fi}, making even $\mathcal{O}(1)$ errors in temperature highly relevant.

The correct procedure to determine the trigger length and trigger mass is to perform a numerical simulation of flame propagation in the style of \citeR{Timmes_1992}, using a large network of nuclear reactions to accurately evaluate $\dot{S}_{\text{nucl.}}$. 
The results of \citeR{Timmes_1992} for the trigger mass cover only a small range of densities $\rho \in [0.2,10]\times 10^{9}\,\text{g/cm}{}^3$ assuming $X_{\text{C}} = X_{\text{O}} = 1/2$, appropriate to the central densities of WD in the mass range $M_{\textsc{wd}} \in [1.25,1.41]M_{\odot}$, which does not cover the full (i.e., lower) WD mass range of interest to us.
We therefore follow the procedure of \citeR{Graham:2015apa} and analytically scale the results of \citeR{Timmes_1992} to densities lower than those numerically sampled in that reference by making use of the parametric scalings implied by \eqref{lambdaT} [and assuming%
\footnote{\label{ftnt:scalingViolation}%
		This assumption is already mildly violated at the upper end of this density range (for $T\sim0.5\,$MeV): owing to strong screening effects that increase the parametric scaling of the rate with increasing density, we have $d\ln\dot S_{\text{nucl.}}/d\ln \rho \sim 1.1$ by $\rho \sim 10^{8}\,\text{g/cm}^3$, and $d\ln\dot S_{\text{nucl.}}/d\ln \rho \sim 1.3$ for $\rho \sim 10^{9}\,\text{g/cm}^3$; this is actually indicative of the the onset of the pycnonuclear regime, in which the reaction rates become exponentially sensitive to density \cite{Gasques:2005ar,Yakovlev:2006fi,Kippenhahn:2012zqe}.
			} %
that $\dot{S}_{\text{nucl.}}(T_{\text{crit.}})\propto \rho$ for $\rho \lesssim10^9\,\text{g/cm}^3$]: $\lambda_T \propto \rho^{-2}$ for $\rho \ll 8\times 10^8 \text{g/cm}^3$ \cite{Graham:2015apa}.%
\footnote{\label{ftnt:Discrepancy}%
		Note that if we had to extrapolate above the numerically sampled range of \citeR{Timmes_1992}, we would find $\lambda_T \propto \rho^{-0.8}$ for $\rho \gg 8\times 10^8 \text{g/cm}^3$, in disagreement with the scaling $\lambda_T \propto \rho^{-0.5}$ used in \citeR{Graham:2015apa}.
This difference arises directly from the difference in the scaling we find here for $\kappa_{\text{cd}}$ from \citeR{Potekhin:1999yv}; we do not however need to resolve this discrepancy as extrapolation to higher densities is unnecessary given that the numerical results of \citeR{Timmes_1992} range up almost to the extremal central density of the \Chand-mass WD.
	} %
Explicitly, and although the density scaling that results has a discontinuous slope at the boundary of the following piecewise definition, we scale the results of \citeR{Timmes_1992} for $T\sim 0.5\,$MeV as follows:
\begin{widetext}
\begin{align}
\lambda_T^{0.5\,\text{MeV}} \approx \begin{cases} 1.3\times 10^{-4} \,\text{cm} \times \lb( \rho / \rho_1 \rb)^{-2}& \rho \leq \rho_1 \\[2ex] \lambda_1 \lb( \rho / \rho_1 \rb)^{ \ln (\lambda_2/\lambda_1) / \ln(\rho_2/\rho_1)} & \rho_1< \rho < \rho_2 \\[2ex] 2.5\times 10^{-5} \,\text{cm} & \rho = \rho_2
\end{cases},
\label{eq:TrigLength}
\end{align}
\end{widetext}
where in the range of numerical values sampled by \citeR{Timmes_1992} we have made a simple log-log linear interpolation,%
\footnote{\label{ftnt:LogLogLinearNote}%
		That is, a linear interpolation of $\log \lambda_T$ as a function of $\log \rho$.
	} %
owing to the graphical results of \citeR{Timmes_1992} being difficult to reliably read off in the intermediate regime. Here, $\lambda_{1,2} \equiv \lambda_T(\rho_{1,2})$, with $\rho_{1} = 2\times 10^{8}\,\text{g/cm}^3$ and $\rho_{2} = 1 \times 10^{10}\,\text{g/cm}^3$ being the endpoints of the numerically sampled values in \citeR{Timmes_1992}.

\begin{figure}[t]
\includegraphics[width=0.99\columnwidth]{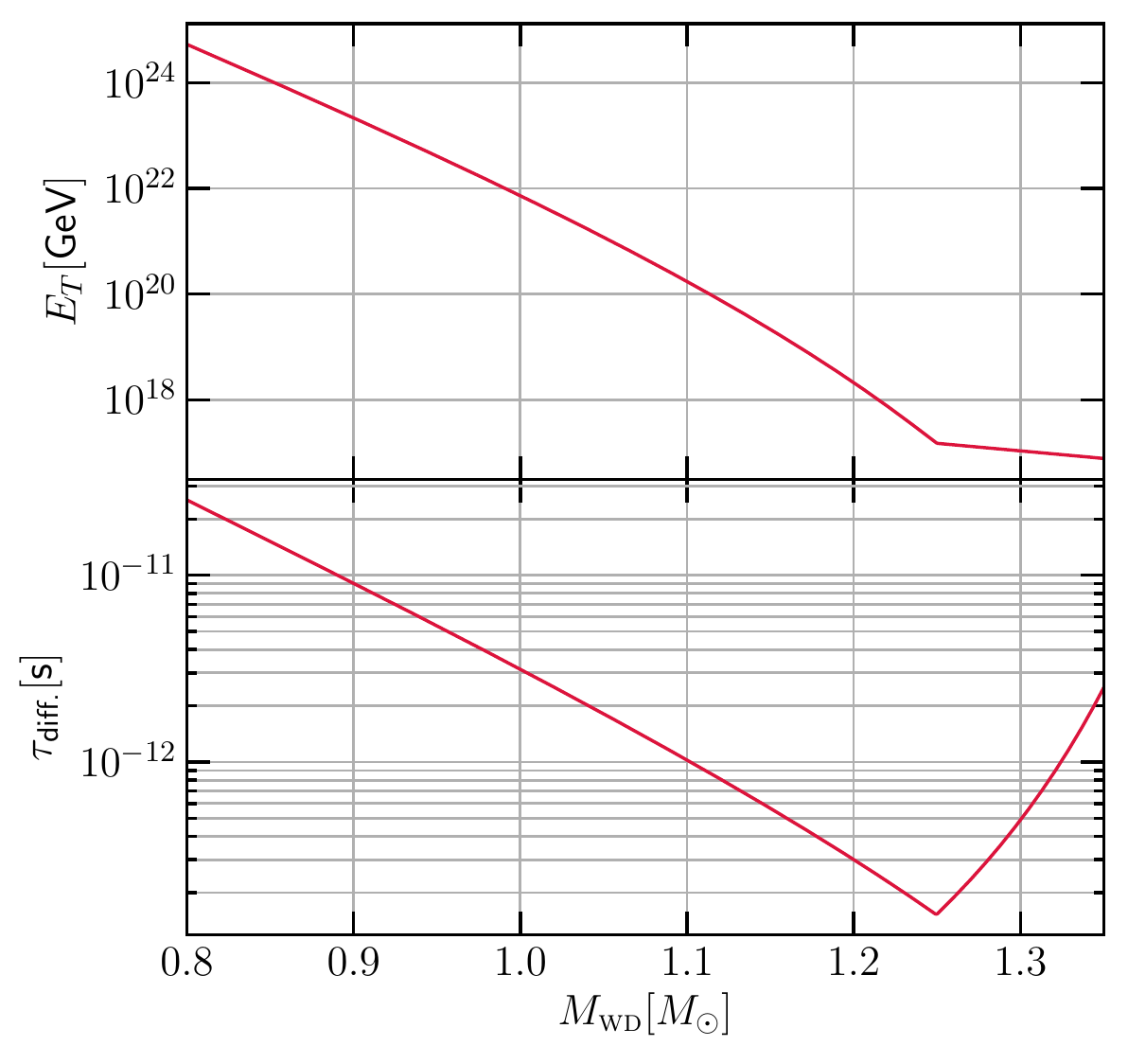}
\caption{\label{fig:ET}%
		\textsc{Upper Panel:} Energy deposition required with a trigger volume within a trigger time in order to initiate thermal runaway, as a function of the WD mass.
		We assume $T_{\text{crit.}}\sim 0.5\,$MeV and that the BH forms near the center of the WD. 
		\textsc{Lower Panel:} Diffusion timescale for the trigger volume as a function of the WD mass, under the same assumptions.}
\end{figure}

Armed thus with parameterizations for the thermal conductivity [\eqref{thermalConductivity}], isobaric specific heat capacity [\eqref[s]{heatcap0}--(\ref{eq:heatcap2})], trigger length assuming that $T_{\text{crit.}}=0.5\,$MeV (which is likely a conservatively high value) [\eqref{TrigLength}], WD-mass--central-density relationship [\eqref[s]{rhoCentral} and (\ref{eq:alphaFit})], trigger energy [\eqref{Etrig}], and diffusion timescale [\eqref{taudiff}], it is now possible to compute the trigger energy required to be injected to initiate thermal runaway. 
We show our results for the trigger energy and diffusion timescale in \figref{ET}, for WD in the mass range $M_{\textsc{wd}} \in [0.8,1.35]M_{\odot}$; the trigger energy results we find are more conservative than those in Fig.~1 of \citeR{Graham:2018efk} by up to 1.5 orders of magnitude.

We will be primarily interested in massive WD with $M_{\textsc{wd}} \sim 0.85M_{\odot}$ or $M_{\textsc{wd}} \sim 1.1M_{\odot}$.
We show representative physical characteristics for such WD (and one other) in \tabref{WDcharacteristics}.
For the remainder of this paper, we will sometimes quote results for critical masses and timescales both generally, and also specifically for WD of these two masses; in the latter cases, we will denote these specific values as `WD 0.85' and `WD 1.1', respectively.
\begin{table*}[t]
\begin{ruledtabular}
\caption{\label{tab:WDcharacteristics}%
		Physical characteristics for two representative classes of WD.
		Quoted are the mass in solar masses, radius in solar radii, central density in $\text{g/cm}^3$, sound speed $c_s$ at center of the WD as a fraction of the speed of light $c$, trigger energy in GeV, and diffusion time for a trigger volume in seconds.
		$R_{\textsc{wd}},\ \rho_{\text{c}}$, and $c_s$ are obtained using information from the solution of the TOV equation, assuming a fully degenerate electron EoS, without Coulomb corrections (see \appref{WDstructure}). 
	 }
\begin{tabular}{llllll}
$M_{\textsc{wd}}\ [M_{\odot}]$	&	$R_{\textsc{wd}}\ [R_{\odot}]$ 	&	$\rho_{\text{c}}\ [\text{g/cm}^3]$ 			& $c_s/c$		      &$E_T\ [\text{GeV}]$ 	&	$\tau_{\text{diff.}}\ [\text{s}]$ 		 \\ \hline
0.85				&	$1\times 10^{-2}$	&	$1.2\times 10^7$			      &	 $2\times 10^{-2}$  & $1\times 10^{24}$	& $1.5\times10^{-11}$	\\
1.1				&	$7\times 10^{-3}$	&	$5.5\times 10^7$		      &	 $2.8\times 10^{-2}$  & 	$2\times 10^{20}$	& $3\times 10^{-12}$ \\
1.2				&	$6\times 10^{-3}$	&	$1.3\times 10^8$		      &	 $3\times 10^{-2}$  & 	$2\times 10^{18}$	& $3\times 10^{-13}$ 
\end{tabular}
\end{ruledtabular}
\end{table*}

\section{CHAMP chemistry}
\label{sect:CHAMPchemistry}

In the previous section we reviewed the physical conditions required to bring about thermal runaway in a WD by localized energy deposition in the WD volume. 
As we will be arguing that such conditions can be brought about by CHAMPs contaminating a WD, we will now change tack and consider the behavior of CHAMPs in various relevant settings.
We begin that discussion in this section by recalling come basic properties of CHAMPs.

In this paper we consider CHAMPs to be either singly positively ($X^+$) or singly negatively ($X^-$) charged; many of our constraints in principle operate for CHAMPs with other charges, but such particles lie beyond our explicit consideration here.

The relative abundance of the population of the $X^\pm$ CHAMP states depends on details of the CHAMP production model, and whether $CP$ is violated in the production; as we are entirely agnostic to the production model, in this paper we consider any generic scenario in which the species are produced either symmetrically or asymmetrically (but always non-thermally, to avoid over-closure constraints \cite{Griest:1989wd}) in the early Universe, and survive until today.
Having $\mathcal{O}(1)$ electric charge, CHAMPs are of course subject to rich and complicated dynamics and `chemistry' after they are produced; see generally \citeR[s]{Cahn:1980ss,DeRujula:1989fe,Gould:1989gw,Chuzhoy:2008zy,Dunsky:2018mqs}.

Positively charged CHAMPs $X^+$ have quite straightforward chemistry, as the only stable negatively charged particles available for them to bind with are electrons; they thus have atomic-sized cross sections.

Negatively charged CHAMPs $X^-$ on the other hand have much more complicated behavior: they are able to form bound states [denoted $(NX)$] directly with nuclei [generically denoted $N$]%
\footnote{\label{ftnt:NisnotN}%
		We distinguish a generic nucleus $N$ from a nitrogen nucleus N by the use of italic and Roman fonts.
	} %
\cite{Cahn:1980ss,DeRujula:1989fe,Pospelov:2006sc}.
The $(NX)$ for $N$ heavier than He are in particular are very deeply bound, with typical ground-state binding energies of a few MeV, and average ground-state radii of order the nuclear size or smaller; see \tabref{EbindNX} and \appref{CHAMPbinding}.
These bound states, being net positively charged in all cases except $(pX)$, are thereafter free to bind with electrons, and in general have atomic-sized interaction cross sections with electrically charged Standard Model particles.

The case of $(pX)$ is unique, as the state is neutral, has a 25\,keV binding energy, and a ground-state Bohr radius of $r_B \sim \frac{2}{3} \langle r \rangle \sim 30\,\text{fm}$. 
As such, this state behaves almost as a heavy neutron, with a highly suppressed electromagnetic interaction cross section with matter, of order its geometrical size \cite{DeRujula:1989fe,Dimopoulos:1989hk}.
However, in dense environments, the $(pX)$ state is susceptible to disruption via the energetically favorable and classically allowed `exchange' reactions $(pX) + N \rightarrow p + (N X)$; combined with the fact that the $(pX)$ are the last states to recombine as the early Universe cools, this results in only a small fraction $\sim 10^{-4}$ of the total $X^-$ abundance being bound in $(pX)$ states primordially; see, e.g., \citeR[s]{Pospelov:2008ta,Kusakabe:2010cb,Pospelov:2010hj,Kusakabe:2017brd}.
An $\mathcal{O}(1)$ fraction of the CHAMPs become bound in $(\text{He}X$) primordially \cite{Pospelov:2008ta,Kusakabe:2010cb,Pospelov:2010hj,Kusakabe:2017brd}.%
\footnote{\label{ftnt:EarlierPXClaims}%
		Note that this is in contrast with claims in the earlier literature that a much larger fractional abundance of $(pX)$ exists primordially \cite{DeRujula:1989fe,Gould:1989gw}.
	} %

In all cases however, given their extremely large mass, CHAMPs have highly suppressed cross section-to-mass ratios, so their dynamics are vastly simpler than those of ordinary baryonic matter; they are more akin to those of dark matter, with some notable exceptions. 
For instance, heavy CHAMPs are prevented from collapsing into diffuse gas clouds in galaxies \cite{Gould:1989gw,Chuzhoy:2008zy,Dunsky:2018mqs} and can be present in galactic haloes, distributed much like dark matter (at least for $m_X \gtrsim 10^{11}$\,GeV).

Because the vast majority of $X^-$ are bound primordially to He, and such CHAMPs (if sufficiently heavy) do not become bound to diffuse collapsed structures in galaxies, the majority of $X^-$ that ever manage to enter a stellar environment (see discussion below) will do so for the first time as a $(\text{He}X)$ bound state from a halo distribution of CHAMPs.
For heavy $X^-$ $(m_X \gtrsim 10^{11}\,$GeV), we will thus make the assumption that the entire $X^-$ abundance that contaminates or accretes onto a star will initially enter that star as a $(\text{He}X)$ bound state.

For lower masses, $m_X \sim 10^5\,$GeV--$10^{11}$\,GeV, however, baryonic dynamics such as supernova shockwaves in galaxies can dramatically impact the spatial morphology and momentum distribution of CHAMPs (both $X^+$ and $X^-$) \cite{Chuzhoy:2008zy,Dunsky:2018mqs}, as we review in \sectref{galaxy}.
Even in this case, it is highly improbable that any processes occurring outside of a stellar environment could alter the `chemical' nature of any $X^-$ that exist in the form $(\text{He}X)$: processes than can disturb the $X^-$ chemical nature, e.g., $(\text{He}X) + N \rightarrow (NX) + \text{He}$ will necessarily involve a Coulomb barrier similar in size to that for fusion of a proton onto a heavy nucleus $N$, which is not a process that happens spontaneously at any appreciable rate outside dense stellar environments.%
\footnote{\label{ftnt:reducedmass}%
		Moreover, the reduced mass of the $(\text{He}X)$--$N$ system is larger than for $p$ fusion onto $N$, being $\mu\approx m_N$ rather than $\mu\approx m_p$; this further suppresses the tunneling rate through the Coulomb barrier, which is what controls the rate of escape of the $X^-$ to the heavier nucleus in such an exchange process.
	} %
Even these lighter $X^-$ thus likely remain in the form $(\text{He}X)$ until then enter a dense stellar environment for the first time.
For lower masses still, the CHAMPs in the form $(\text{He}X)$ may simply collapse into the baryonic disk \cite{Gould:1989gw,Chuzhoy:2008zy,Dunsky:2018mqs}; however, unless these CHAMPs again enter a dense stellar environment, they too likely survive as $(\text{He}X)$ for similar reasons.

On the other hand, CHAMPs that have been previously processed through a stellar environment and then re-released to the galaxy either during the late stages of stellar evolution, or in a previous galactic supernova event, will of course in general be bound up to heavier nuclei as $(NX)$: because nuclear processes with rates comparable to reactions such as $(\text{He}X) + N \rightarrow (NX) + \text{He}$ clearly do occur at appreciable rates in stars, so too do such exchange reactions; and the supernova event itself may involve many highly energetic processes capable of overcoming any Coulomb barriers and altering the chemical state of the CHAMPs.
Nevertheless, the fraction of the total number of CHAMPs in any galaxy so processed will be small, for the simple reason that the CHAMPs should be roughly homogeneously mixed with the hydrogen and helium gas primordially in the early Universe, and the total fraction of hydrogen and helium in a galaxy that gets processed through stars into heavier elements is small.
In any Population I or Population II star then, the total abundance of CHAMPs present that may be bound to heavy nuclei as a result of processes occurring \emph{before the birth of that particular star} will be no larger than a fraction approximately at the level of the metallicity of such a star (i.e., roughly at the percent level), and therefore negligible. 
Of course, CHAMPs that do enter a star are almost all subsequently processed in that star itself over its lifetime, but the above considerations imply that, as the starting point for considering the evolution of CHAMPs in a star, it is reasonable to assume that all CHAMPs initially enter it either as $X^+$ or bound as $(\text{He}X)$, with some exceptions that we discuss when they may be important.

\begin{table}[t]
\begin{ruledtabular}
\caption{\label{tab:EbindNX}%
		Numerically computed ground-state binding energies $E_B$ [MeV] and average radii [fm] of the $(NX)$ system, where the nucleus $N$ is modeled as a uniformly charged ball of radius $R=A^{1/3}r_0$ with $r_0 = 1.22\,\text{fm}$ (see, e.g., \citeR[s]{Cahn:1980ss,Tiburzi:2000rsq,Pospelov:2006sc}); see \appref{CHAMPbinding}.
		Also shown are the na\"ive estimated hydrogenic Coulomb binding energies that would be obtained ignoring the finite charge radius of the nucleus; these significantly overestimate the true binding energies for $N$ heavier than helium, owing to the bound state being localized within the nuclear volume.
	 }
\begin{tabular}{lllll}
$N$		&	$E_B$ [MeV]			&	$E_B^{\text{na\"ive}}$ [MeV] 	&	$\langle r \rangle$ [fm]		&	$R$ [fm]	\\ \hline \hline
$p$		&	0.025				&	0.025					&	43						&	---		\\
${}^4$He	&	0.35					&	0.40						&	6.1						&	1.9		\\
${}^8$Be	&	1.6					&	3.2						&	2.6						&	2.4		\\
${}^{12}$C&	2.9					&	11						&	2.1						&	2.8		\\
${}^{16}$O&	4.1					&	25						&	1.8						&	3.1		\\
${}^{24}$Mg&	6.1					&	86						&	1.7						&	3.5		\\
${}^{56}$Fe&	10.0					&	940						&	4.1						&	4.7
\end{tabular}
\end{ruledtabular}
\end{table}

\section{CHAMPs in the Galaxy}
\label{sect:galaxy}
As we have already mentioned, for sufficiently massive CHAMPs ($m_X \gtrsim 10^{11}\,$GeV \cite{Chuzhoy:2008zy,Dunsky:2018mqs}) the spatial morphology and momentum distribution of CHAMPs in the Galaxy is not significantly impacted by baryonic dynamics; such massive CHAMPs should thus have a halo distribution akin to that of the dark matter.
Lighter CHAMPs however can be significantly impacted by such baryonic dynamics \cite{Dimopoulos:1989hk,Dunsky:2018mqs,Chuzhoy:2008zy}; see also discussion in \citeR[s]{Gould:1989gw,SanchezSalcedo:2008zd,Foot:2010yz,McDermott:2010pa}.
Such considerations are themselves subject to significant uncertainty and there is some variation in the literature as to the fate of such CHAMPs.

\citeR{Chuzhoy:2008zy} argued that efficient evacuation from, e.g., the Milky Way (MW) disk, of CHAMPs that are not in the form ($pX$) (which are in any event a $\lesssim 10^{-4}$ fraction of CHAMPs \cite{Pospelov:2008ta,Kusakabe:2010cb,Pospelov:2010hj,Kusakabe:2017brd}) is possible by the combined action of supernova shockwaves, and the confining effects of galactic magnetic fields.
Their argument is that a $\mathcal{O}(1)$-charged CHAMP that is heavier than $m_X \sim 10^5 Q_X^2\,$GeV, where $Q_X$ is either the $X$ charge, or the charge of the state in which $X$ finds itself bound, will be accelerated by the periodic passage of a sufficiently intense supernova shockwave through the MW disk, and be unable to radiately dissipate the kinetic energy thus gained sufficiently quickly to avoid being accelerated above the disk escape speed on a timescale of $\mathcal{O}(100\,\text{Myr})$.
Moreover, the galactic disk magnetic fields, which are mostly in the plane of the disk, magnetically confine virialized CHAMPs in and around the disk to within approximately their gyroradius.
Given the MW CHAMP virial speed and typical $\sim 1$--$10\,\mu$G magnetic fields, virialized CHAMPs lighter than $m_X \sim 10^{11} Q_X$\,GeV have gyroradii smaller than the MW disk thickness.
\citeR{Chuzhoy:2008zy} thus argued that CHAMPs not in the form $(pX)$ which thus find themselves initially in the MW disk would be rapidly accelerated out of the disk by supernova shockwaves, and that virialized CHAMPs in the MW halo would be prevented from entering the disk by the confining action of the galactic disk magnetic fields, provided their mass is in the range $10^5Q_X^2\,\text{GeV} \lesssim m_X \lesssim 10^{11} Q_X\,\text{GeV}$.
Similar arguments presumably hold for other disk galaxies, and some cognate argument is also likely to hold for the central regions of elliptical galaxies.

We note again that since the majority of the CHAMPs present in non-collapsed structures in a galaxy will still be in the form $X^+$ or $($He$X)$, with or without an electron bound to them, it is appropriate to take $Q_X = +1$ for both cases $X^\pm$, when evaluating the boundaries of the region in which \citeR{Chuzhoy:2008zy} claims an effect.

The results of \citeR{Chuzhoy:2008zy} are however based on a greatly simplified picture of the dynamics of the diffuse CHAMP plasma under the combined action of complex shockwave dynamics, and the (turbulent) magnetic fields in galaxies.
More recent results, e.g., \citeR{Dunsky:2018mqs}, find that although $\mathcal{O}(1)$-charged CHAMPs in the mass range $m_X \sim 10^5$\,GeV--$10^{10}$\,GeV are sufficiently shock accelerated such that an $\mathcal{O}(1)$ fraction of those CHAMPs in the mass range indicated above may be evacuated, their diffusive re-entry into the disk is not completely inhibited by the magnetic fields; instead, an equilibrium between acceleration and diffusion is reached, and the CHAMP momentum spectrum is significantly altered. 
Their results also indicate that for $m_X \gtrsim 10^{10}Q_X$\,GeV, such CHAMPs form a DM-like halo that is not disturbed by baryonic dynamics, and that for $m_X \lesssim 10^{5}Q_X^2$\,GeV, such CHAMPs will collapse into the disk of the MW as it forms.
 
\citeR[s]{Dunsky:2018mqs} and \cite{Chuzhoy:2008zy} thus agree that sufficiently massive $\mathcal{O}(1)$-charged CHAMPs will form a DM-like halo, but they disagree on the details of how the CHAMP abundance, momentum distribution, and spatial morphology for lower CHAMP masses is impacted by baryonic dynamics.
Taken together, the conservative conclusion to draw from these results is that any CHAMP bounds that rely on knowing the galactic abundance or momentum distribution of CHAMPs for $m_X \sim 10^{5}$\,GeV--$10^{11}$\,GeV are subject to significant uncertainty.

We do however stress that those CHAMPs bound in the form $(pX)$ and which have not undergone some reaction that converts them to $(NX)$ [$N\neq p$] would not get significantly impacted by supernova shockwaves or magnetic field confinement, as they are not charged and hence can be neither efficiently entrained in a SN shockwave, nor deflected by a magnetic field.
Hence, for a CHAMP population which has any significant $X^-$ component, there is always a residual fraction of $X^-$ in the galaxy; this is however estimated to be no larger than $10^{-4}$ of the $X^-$ abundance \cite{Pospelov:2008ta,Kusakabe:2010cb,Pospelov:2010hj,Kusakabe:2017brd}. 

Finally, we note that annihilation of $X^+$ and $X^-$ in diffuse structures is inhibited: the $X^+$ and $(\text{He}X)$ states which are assumed to be the dominant forms of CHAMPs in diffuse structures at all times in the late Universe are both positively charged. 
Therefore, the rate for bringing an $X^+$ and a $(\text{He}X)$ to within $\sim$ fm of each other, in order to allow the $X^+$ wavefunction to significantly overlap with the highly localized $X^-$ wavefunction, is exponentially suppressed by a Coulomb barrier. 
The suppression is moreover exponentially more severe than the tunneling suppression factor for $pp$ fusion, since the tunneling exponential is suppressed by the reduced mass (see, e.g., \citeR{Clayton1983}), which is here $\mu \sim m_X/2 \gg m_p/2$. 
Combined with intrinsically low CHAMP densities in diffuse structures, any annihilation process that would tend to remove a symmetric component of the CHAMP abundance is simply much too slow to be relevant, and the $X^+$ and $X^-$ bound as $(\text{He}X)$ survive independently throughout the era when they are distributed diffusely.

\section{CHAMPs in white dwarfs}
\label{sect:CHAMPsWD}

Having discussed the chemistry, history, and behavior of CHAMPs outside stellar environments, we can now turn to the question of how CHAMPs come to contaminate WD, and their impact on WD dynamics.
Our discussion is guided in outline initially by that of \citeR{Gould:1989gw}.

In this section, we discuss first how CHAMPs come to be present in WDs and their evolution over time, considering in turn two mechanisms: (1) in \sectref{primordial}, we consider the same case as \citeR{Gould:1989gw}, in which CHAMPs accumulate in the protostellar cloud that collapses to form the stellar progenitor of the WD, leading to a contamination of the WD material as the WD is born. And (2), in \sectref{accumulated}, we consider the case where halo CHAMPs additionally accumulate onto the WD as during the first $\sim 1\,$Gyr of the WD existence (timescale to be discussed below).

Having considered the population of CHAMPs that can be present in a WD, we then estimate the timescale for that population to sink to the center of the WD [\sectref{timescales}], and the evolution of the initial structure formed by the CHAMPs at the center of the WD [\sectref{WDstructureThermal}].
We then consider in turn the fate of the CHAMPs in the cases of sub- or trans-\Chand\ [\sectref{WDstructure}], and super-\Chand\ [\sectref{superChandCHAMPs}] total CHAMP masses.
For the latter case, this discussion will naturally evolve into a discussion of black hole dynamics in a WD, which is the topic of the immediately following section [\sectref{BHinWD}].

\subsection{Primoridal CHAMP contamination}
\label{sect:primordial}
We consider first the case where CHAMPs accumulate in the protostellar cloud from which the WD progenitor forms.
We begin with a review of the arguments advanced in \citeR{Gould:1989gw} leading to the population estimate, and then turn to the question of how the CHAMPs behave as the star evolves.

\subsubsection{Population estimate}
\label{sect:PopnPrimordial}

The $X$, in whatever state they find themselves after recombination, will collapse into galaxies during early-Universe structure formation.
For both the $X^+$ and $(NX)$ [i.e., $(\text{He}X)$] forms of CHAMPs, $\sigma_{\textsc{sm-}X}/m_X$ will be small enough, if $m_X \gtrsim m_X^{\text{diffuse capture}}$, that even the multiple orbits executed through the MW galactic disk since galaxy formation would not be efficient in capturing the CHAMPs into diffuse disk gas clouds \cite{Gould:1989gw}.
Estimate for $m_X^{\text{diffuse capture}}$ vary from $m_X^{\text{diffuse capture}}\sim 10^5\GeV$ \cite{Dunsky:2018mqs} to $m_X^{\text{diffuse capture}}\sim 10^7\GeV$ \cite{Gould:1989gw}.
For lighter $X$, such trapping would be efficient, but magnetic heating effects could conceivably eject from the disk any such $X$ which are trapped in diffuse clouds \cite{Gould:1989gw}.
It is therefore unclear if a significant fraction of the $X$ in, e.g., the MW end up trapped in parts of the disk containing only diffuse gas if $m_X \lesssim m_X^{\text{diffuse capture}}$.
On the other hand, $X$ can be trapped by denser collapsing protostellar clouds; \citeR{Gould:1989gw} estimates that a protostellar cloud of molecular gas of mass $M_{\text{cloud}}$ would capture a total mass $M_{X^\pm}$ of the respective charge species of CHAMP, $X^\pm$, giving rise to star fractionally polluted at birth by each of $X^\pm$ by an amount $\eta_{\pm} \equiv M_{X^{\pm}} / M_{\text{cloud}}$ given by (see below for further discussion of the net contamination)~\cite{Gould:1989gw}
\begin{widetext}
\begin{align}
\eta_{\pm} \approx \begin{cases} 
\dfrac{192 \sqrt{2} \pi^{3/2}}{7e} \dfrac{\alpha^{1/2}\Mpl}{ v_{\text{rot.}}^3 m_p^2 m_e^{3/2}}  \dfrac{ f_\pm \rho_X M_{\text{cloud}} ^{1/4}}{m_X^{7/4}} g(y_{\text{crit.}}) & y_{\text{crit.}}\geq y_{\text{min.}} \\[3ex]
	\dfrac{3^{13/6}\, 2^{2/3} \pi^{1/3}}{7 e } \dfrac{ \alpha^4 \Mpl m_e^2 }{ v_{\text{rot.}}^3 m_p^{19/6} (n_H^{\text{min}})^{7/6} }  \dfrac{f_{\pm} \rho_X}{M_{\text{cloud}}^{1/3}} \lb[ 1 - \frac{7}{3} y_{\text{min.}}^{4} + \frac{4}{3} y_{\text{min.}}^{7}  \rb] & y_{\text{crit.}} < y_{\text{min.}}
\end{cases},
\label{eq:eta}
\end{align}
where 
\begin{align}
g(x) \equiv \lb[ \ln\lb(x^{-1}\rb) \rb]^{-7/4} \lb\{ 1 - \frac{7}{3} x^4 + \frac{4}{3} x^7 + 7 \int_{y_{\text{min}}/x}^1 dq \lb( \frac{x}{q} \rb)^4 \lb[ x^{-4q^4}-1 \rb] \rb\};
\end{align}
\end{widetext}
$y_{\text{crit}}$ is defined by the equation
\begin{align}
y_{\text{crit.}} \exp\lb[ \beta y_{\text{crit.}}^4 \rb] &= 1 \\
\beta  &= 4\alpha^2\Mpl^4 / ( m_p^2 m_X M_{\text{cloud}} ),
\end{align}
whose solution is given in terms of the Lambert $W$-function:%
\footnote{\label{ftnt:Wexpansion}%
		The asymptotic expansions of which are given by $x^{-1}W(x) \approx 1 - x + \tfrac{3}{2}x^2$ for $x\ll1$ and $W(x) \sim \ln(x) - \ln(\ln(x))$ for $x\gg 1$.
	} %
$y_{\text{crit.}} = \lb[ (4\beta)^{-1} W( 4 \beta ) \rb]^{1/4}$;
$y_{\text{min.}}$ dictates the onset of $X$-capture when the cloud reaches $n_H = n_H^{\text{min}} \sim 10^2\,\text{cm}^{-3}$, becoming sufficiently UV-shielded to be molecular:
\begin{align}
y_{\text{min.}} &\equiv \lb( \frac{4\pi}{3}\rb)^{1/6} \frac{ m_p^{2/3} (n_H^{\text{min}})^{1/6}  M_{\text{cloud}}^{1/3} }{\Mpl m_e^{1/2} \alpha } ;
\end{align}
$v_{\text{rot.}} \sim 220\,\text{km/s}$ is the local circular speed in the MW; $\rho_X$ is the total CHAMP mass density at the location of cloud collapse; and $f_\pm$ are the fractions of the total CHAMP abundance in the forms $X^\pm$, respectively.
In computing $f_-$, we include all $X^-$ bound in $(NX)$ states, \emph{except for} $(pX)$ \cite{Gould:1989gw}, for which this specific estimate does not apply because it is based on an electromagnetic interaction between the CHAMP and the diffuse gas which is absent for the neutral $(pX)$ state; however, they only constitute $\lesssim 10^{-4}$ of the total $X^-$ abundance \cite{Pospelov:2008ta,Kusakabe:2010cb,Pospelov:2010hj,Kusakabe:2017brd} and can thus be ignored.

The transition between the low-$m_X$ regime ($y_{\text{crit.}} < y_{\text{min.}}$) where $\eta$ is independent of $m_X$, and the high-$m_X$ regime ($y_{\text{crit.}} \geq y_{\text{min.}}$) where $\eta \propto m_X^{-7/4}$ occurs because the phase-space available for capture of more massive $X$ is suppressed relative to that for lighter $X$ until the cloud grows somewhat denser than $n_H^{\text{min}}$ and is hence physically smaller at the onset of efficient capture \cite{Gould:1989gw}.

Note that since the majority of the CHAMPs present in non-collapsed structures in a galaxy will still be in the form $X^+$ or $($He$X)$, it is appropriate to take $q_X = +1$ for both cases $X^\pm$, when evaluating $\eta_\pm$. 

For an $M_{\text{cloud}} = 4M_{\odot}$ protostellar cloud (appropriate for an $M_{\textsc{wd}}\sim 0.8M_{\odot}$ WD \cite{Catalan:2008tr}), the numerical estimate for $\eta_{\pm,\,4} \equiv \eta_\pm(M_{\text{cloud}}=4M_{\odot})$ is
\begin{widetext}
\begin{align}
\eta_{\pm,\,4} \approx \begin{cases} 
	2.2\times10^{-2} \dfrac{ f_\pm \rho_X }{\rho_{\text{halo}}} & m_X \lesssim 1.7 \times 10^{4}\,\text{GeV} \\[3ex]
	2.2\times10^{-2} \dfrac{ f_\pm \rho_X }{\rho_{\text{halo}}}  \lb( \dfrac{ m_X }{ 1.7 \times 10^{4}\, \text{GeV} } \rb)^{-7/4} \hat{g}(m_X) & m_X \gtrsim 1.7 \times 10^{4}\,\text{GeV}\end{cases},
\label{eq:eta10}
\end{align}
\end{widetext}
where $\rho_{\text{halo}} \sim 0.3\,$GeV/cm${}^3$ is the local MW DM halo density, and $\hat{g}(m_X) \equiv g[y_{\text{crit.}}(m_X)] / g[ y_{\text{crit.}}(1.7 \times 10^{4}\,\text{GeV})]$.
Given the asymptotic scalings of $y_{\text{crit.}}$, it follows that---for this size cloud---$\hat{g}(m_X)$ supplies only a log correction for $m_X \lesssim 10^{12}\,$GeV, but that $\hat{g}(m_X) \propto m_X^{3/4}$ for $m_X \gtrsim 10^{14}\,$GeV, which changes the power-law scaling of $\eta$ for large $m_X$.

CHAMPs that accumulate in protostellar clouds will inevitably be incorporated fairly uniformly into the stars formed by such clouds, as they are fairly well mixed; as a result, those stars are born with a baked-in CHAMP contamination fraction $\eta$ as estimated by \eqref{eta}  \cite{Gould:1989gw}.
To turn this estimate into an estimate for the total abundance of CHAMPs that are present in the WD at the end of the stellar evolution, we however need to make further assumptions. 
Since we assume that $m_X \gg m_p$, some gravitational sinking of the population of $X^+$ and $(NX)$ toward the center of the star will undoubtedly occur during the lifetime of the WD-progenitor star leading up to the formation of a WD-progenitor CO core, which would potentially boost the fractional mass contamination of that core relative to that of the initial protostellar cloud.
However, large-scale convective processes that can operate in certain regions, and during certain phases of the evolution, of stars that actively burn nuclear fuel make it challenging to give a quantitative estimate of this effect, and we follow \citeR{Gould:1989gw} in making the conservative assumption that the fractional mass pollution of the WD-progenitor CO core, and hence the WD itself upon formation, is just that of the initial protostellar cloud.
Under this assumption, the total abundance of CHAMPs that are expected to be present in the WD from this primordial contamination can be estimated as 
\begin{align}
M_{X^\pm}^{\text{prim.}} \approx \eta_\pm M_{\textsc{wd}}.
\label{eq:Mprim}
\end{align}
We assume further \cite{Gould:1989gw} that the $X$ contamination is approximately uniform throughout the WD core at the time of its formation.

In the next subsection, we examine further aspects of the behavior of the CHAMPs between the time of protostellar capture and WD birth that can impact the validity of the approximation \eqref{Mprim}.

\subsubsection{Behavior of CHAMPs after protostellar cloud contamination}
\label{sect:behaviorPrimordial}

$X^+$ will merely reside in the star without much change, whereas $(NX)$ [mostly beginning as $(\text{He}X)$] bound states can undergo significant processing in the core of the star, as the nuclear reaction $Q$-values for a large number of steps in a large number of different stellar burning reactions exceed the $(NX)$ binding energies for the nuclei involved (e.g., the modified CNO cycle reaction $p + ({}^{14}\text{N}X) \rightarrow {}^{15}\text{O} + X$ is possible with $Q\approx 3.8$\,MeV), which would enable ejection and subsequent recapture of the $X^-$ by a different nuclear species.
Catalyzed nuclear such as $({}^4\text{He}X) + {}^{14}\text{N} \rightarrow {}^{16}\text{O} + X$, or exchange reactions such as $({}^4\text{He}X) + {}^{14}\text{N} \rightarrow {}^{14}(\text{N}X) + {}^4\text{He}$ are also in principle energetically allowed (although they proceed as tunneling processes through a Coulomb barrier), and could result in the transfer the $X$ out of $({}^4\text{He}X)$ and into heavier bound states. 
However, note that even if the $X^-$ bound to helium do not exchange onto other heavier nuclei, the limits we will ultimately set will still conservative; see discussion in \sectref{WDstructureMinus}.

As it is most energetically favorable for an $X^-$ to bind with nuclei of higher charge (and hence greater mass) the natural end result of these processes would be the eventual accumulation in old, highly evolved stellar cores of $(N'X)$ bound states where $N'$ is the heaviest nucleus present in significant quantity.
The approach to the final distribution of $X$ in various heavy nuclei is in general complicated, and detailed study would require extensive numerical modeling utilizing a large reaction network populated with rates for all the reactions including the $X$ in addition to the usual nuclear reaction rates, as is done in ordinary solar modeling (e.g., \citeR[s]{Bahcall:2000nu,Serenelli:2011py,Vinyoles:2016djt}); such an effort is far beyond the intended scope of this work.
In particular, in the highly evolved $M\sim M_\odot$ WD-progenitor CO stellar core of a red giant with an initial mass in the appropriate range (i.e., $M_{\text{star}} \lesssim 9M_{\odot}$ \cite{Heger:2002by}), the $X^-$ will be present in the form $({}^{12}\text{C}X)$, $({}^{16}\text{O}X)$, or possibly bound to some heavier trace species (e.g., Ne, Mg, etc.) that was able to form in smaller quantity during the stellar evolution; all these $(NX)$ bound states have binding energies of at least $3\,$MeV.

Moreover, even if we were to assume that some significant population of $({}^4\text{He}X)$ were to survive unscathed during earlier burning phases in the stellar evolution, once such states become part of the dense He core of an old, massive red giant shortly before the helium flash that leads to the formation of the WD-progenitor core of CO material, the $X$ present in $($He$X)$ can catalyze fusion reactions such as $($He$X) + \text{He} \rightarrow ($Be$X) + \gamma$, which is allowed since $({}^{8}_{4}$Be$X)$ is stable $(E_B \approx 1.6\,$MeV; see \tabref{EbindNX}), in contrast to ${}^{8}_{4}$Be, which is famously unstable ($E_B \approx - 92\,$keV) \cite{TOI1996}.
Such stable $($Be$X)$ can then fuse with a further He nucleus giving a pathway to $($C$X)$ that is unsuppressed by the three-body nature of the triple-$\alpha$ process; it is thus highly likely that even $X$ still bound as $(\text{He}X)$ after the earlier stellar burning phases get into $($C$X)$ bound states quite early via this ersatz triple-$\alpha$ route, even before the full degenerate CO core itself forms as a result of the helium flash.
Such $X$ will be unlikely to escape to other nuclei until the WD thermal runaway since the binding energy of the $(\text{C}X)$ state, $\sim 3\,$MeV, is much larger than the temperature in the CO core ($T \lesssim 10\,$keV; reaching $T \lesssim 60\,$keV only in a near extremal CO WD just before thermal runaway \cite{Woosely:2003ng}).
Moreover, even though the degenerate electrons at the center of the near-extremal core WD do become relativistic as the Chandrasekhar mass is approached and thus have Fermi energies that can exceed the $(NX)$ binding energies, they are of course still Pauli blocked from transferring energy much larger than the temperature to (and thereby disrupting) the \CX\ bound states.

Although it would of course be energetically favorable for, e.g., $(\text{C}X) + \text{O} \rightarrow \text{C} + (\text{O}X)$ exchange reactions (or exchange reactions with heavier nuclei) to occur after $(\text{C}X)$ formation, such reactions are highly tunneling suppressed: the $C$ wavefunction of the $(\text{C}X)$ bound state (treating the $X$ is stationary as it is so heavy) is localized within $\sim$\ fm of the $X$ position, so the incoming O must also approach to within $\sim$\ fm of the $X$ to be captured. 
But since the O nucleus simply sees a $Q = +5$ object until it is within $\sim $\ fm of the $(\text{C}X)$ bound state, there is a Coulomb barrier of $\sim$ few MeV for this to occur.
Moreover, since the reduced mass of the $(\text{C}X) + \text{O}$ system is about twice that of a $\text{C} + \text{O}$ system, the tunneling suppression is much more pronounced that even for $\text{C} + \text{O}$ fusion (even though the charge is slightly lower, the reduced mass change more than compensates; see the discussion of Gamow energies in, e.g., \citeR{Clayton1983}), which is itself suppressed compared to the $\text{C} + \text{C}$ fusion probability.
Since the majority of C-ions in a non-extremal WD have not undergone a fusion reaction since CO core formation (else the core would have already burned), this implies that reactions like $(\text{C}X) + \text{O} \rightarrow \text{C} + (\text{O}X)$ are exceedingly unlikely to have occurred.

We do however mention that a CNO-like reaction such as $p + ({}^{12}\text{C}X) \rightarrow {}^{13}\text{N} + X$ could still lead to the ionization of the $X$; however, in highly evolved stellar cores (either in the He-rich environment where the ersatz triple-$\alpha$ processes occurs, or already in the degenerate CO core) the $p$ abundance has already been almost completely depleted by ordinary stellar burning \cite{Kippenhahn:2012zqe}, making this possibility rare.%
\footnote{\label{ftnt:CNOcycle}%
		Incidentally, we note that catalyzed fusion reactions such as $(pX) + {}^{14}\text{N} \rightarrow {}^{15}\text{C}+X$ could ostensibly dramatically speed up certain steps in the CNO solar energy generation cycle in our own Sun.
		This is because the Coulomb barrier for the $p$ to fuse onto the ${}^{14}\text{N}$ nucleus only becomes apparent at distances on the order of tens of fermi [the $(pX)$ bound state size], rather than at a couple of thousand fermi [the normal classical turning point under the thermodynamic conditions prevalent in the Sun].
		In principle this change could be expected to dramatically alter the relative neutrino yields from CNO vs. $pp$ cycles to be expected from the Sun.
		However, this fails for subtle reasons.
		Many of the steps in the CNO cycle have sufficient energy to directly eject the $X$ in the final state; such an ejected $X$ will find and bind with another nucleus, which is most likely a $p$ (on number-abundance grounds alone), which would allow a cycle of rapid fusions to proceed.
		However, there is a regulator: on number-abundance arguments alone, once every $\sim 10^4$ times the $X$ is ejected, it will find and bind with an ${}^{16}\text{O}$ nucleus (present in $\sim 10^{-4}$ number-abundance in the Sun), where it will remain until a proton fuses with the $(\text{O}X)$ nucleus in a side-branch of the CNO cycle, possibly ejecting the $X$ in the final state.
		Unfortunately, we estimate that the rate for that latter process is roughly as slow as the rate-limiting $p+{}^{14}\text{N}$ step in the main uncatalyzed CNO cycle.
		Therefore, given the small number of $X$ in the Sun and the fact that any one $X$ can only catalyze a limited number of reactions before being captured by a contaminant that keeps it bound up and unable to catalyze further fusions for about as long as the slowest CNO cycle step, we estimate that the overall impact on the CNO cycle neutrino output is negligible.
			} %
Thus, even if  such reactions were to occur in the He-rich core prior to the He flash, the $X$ would still likely be recaptured by another He nucleus, and the cycle would repeat until the $X$ got stuck in another heavy nucleus such as C or O.
If, on the other hand, such reactions occurred in the CO core after the He flash, the $X^-$ would again simply be recaptured by another heavy nucleus, likely C or O.
All of which is by way of saying that once a $X^-$ is bound to C or O, it is highly unlikely to be ionized again; or, if it is, it is likely to end up bound to another C or O nucleus.

In this paper, we will for simplicity therefore assume that the entirety of the $X^-$ population in the WD is bound to the heaviest nuclei that make up an $\mathcal{O}(1)$ fraction of the CO core of the WD-progenitor object that forms at the center of the red giant: that is, they will be bound to either ${}^{12}$C or ${}^{16}$O, forming $(\text{C}X)$ or $(\text{O}X)$, respectively.

We also note that \citeR{Gould:1989gw} explicitly considers only the net residual contamination of $X^+$ and $X^-$ present in the star after an assumed annihilation of the accumulated $X^+$ and $X^-$; that is, their result for the CHAMP contamination, which we claimed to have stated above as \eqref{eta}, is actually given only in terms of the net residual contamination $\eta \equiv |\eta_+ - \eta_-|$ (i.e., per \citeR{Gould:1989gw}, $\eta \equiv M_{X,\,\text{net}} / M_{\text{cloud}}$ is given by the same expression as at \eqref{eta}, but with $f_\pm \rightarrow |f_+ - f_-|$). 
Since \citeR{Gould:1989gw} is ultimately interested in CHAMPs accumulating in stars whose evolved cores later collapse to form neutron stars, in the cores of which the ambient temperatures are $T\sim \text{MeV}$ \cite{Baym:1979zs} and the ambient densities are nuclear, this is a justified assumption in their case as the $X^-$ are then unable to bind to nuclei, and are free to annihilate with $X^+$ when the CHAMPs collapse to an extremely dense state at the center of the star.

However, we will be concerned with CHAMP contamination of WDs, the interiors of which can only reach maximum temperatures (just prior to triggering thermal runaway in a near-extremal WD resulting in a full `ordinary' type-Ia SN) of $T \sim 7\times10^8\,\text{K} \approx 60\,\text{keV}$ (see \citeR{Woosely:2003ng} and references therein) and much lower densities.
This is much smaller than the $\gtrsim 3$\,MeV binding energies of typical [e.g., (C$X$), (O$X$)] states into which $X^-$ are bound.
As a result, the $(NX)$ bound states which are formed in the WD core will not be disrupted by the ambient conditions.
Therefore, similar to the argument already advanced for the diffuse galactic CHAMPs, the $X^-$ will be shielded (during the periods when they most densely accumulate) from annihilating against the $X^+$ by the presence of the large Coulomb barrier between the $Q=+1$ state $X^+$, and the $Q=+(Q_N-1) \sim 5$--$7$ state $(NX)$.
As such, we will not assume annihilation of the symmetric CHAMP component, and continue to track both the $X^+$ and $X^-$ contaminations independently; see further discussion in \sectref[s]{combinedCase} and \ref{sect:limitsCombined} on this point and how it impacts the final limits we are able to set, which are in principle stronger than those of \citeR{Gould:1989gw} in the $CP$-symmetric case.

An important caveat to the discussion in the preceding paragraph is that since individual nuclear processes that occur during the active nuclear burning phases of the stellar lifetime do have reaction $Q$-values large enough to potentially disrupt all the likely $(NX)$ bound states involved in various reactions, the $X^-$ will spend some portion of their lifetime during the active burning phases outside the protection of the positively charged nucleus, and would during such periods be susceptible to annihilation with $X^+$.
However, the dilute distribution of the $X^\pm$ in the ordinary stellar matter (assuming fairly uniform distribution in the WD-progenitor star) makes annihilation a much less likely scenario than, e.g., the $X^-$ just being immediately re-captured by another nearby nucleus. 
To judge the complete implausibility of any alternative outcome, consider a simple $\Gamma \sim n \sigma v$ argument: if the CHAMPs were to constitute 100\% of the local dark matter abundance (a deeply excluded possibility), the maximum mass of CHAMPs primordially present in the WD for a $m_X \sim 10^{11}$\,GeV CHAMP (which maximizes $\eta$ in the regime where we know that the CHAMP halo abundance is undisturbed by galactic dynamics) is of order $M_X \sim 10^{-13}M_{\odot}$, which yields an average number abundance in a $M_{\textsc{wd}}\sim 1.2M_{\odot}$ CO WD of $n_X \sim 10^{-23} n_{\text{ion}}$, where $n_{\text{ion}}$ is the number density of the C and O ions in the CO WD. 
Even charitably allowing $v_X$ as large as the WD escape speed $\sim 10^{-2} \sim 10 v_C^{\text{therm.}}$, it would still require an enormous ratio of the CHAMP annihilation cross section to the radiative capture cross section of the carbon ion on the CHAMP on the order of $10^{22}$ for the rate for an annihilation to be equal to the rate for radiative capture in any one instance when the $X^-$ is thus liberated in a nuclear reaction in the star.
Of course, the other way around this is if the $X^-$ participates in a truly enormous number of nuclear interactions in the lifetime of the star and therefore spends a large amount of time outside the protection of the positively charged ion nuclear charge cloud; this appears exceedingly unlikely on the grounds of the sheer number of nuclear reactions required for this to be an issue.
Once again, detailed evolutionary stellar modeling beyond the intended scope of this work would be required to fully resolve this issue; but on the arguments advanced here, we will simply state our results assuming that any $X^+$--$X^-$ annihilation which occurs during the active burning phases will be a small effect.

\subsection{Accumulated CHAMP contamination}
\label{sect:accumulated}
An additional mechanism exists to populate WD with CHAMPs: accumulation of CHAMPs from the halo which pass within the gravitational capture radius of the WD after its formation. 
In this subsection, we discuss the population estimate, estimate when this process is efficient, and discuss the subsequent behavior of the CHAMPs.

\subsubsection{Population estimate}
\label{sect:PopnAccumulated}
Owing to their unit charge, even very heavy CHAMPs can be stopped efficiently by stellar plasma, particularly the extremely dense CO plasma that exists in WD in the appropriate mass range.  
Given the total WD mass $M_{\textsc{wd}}$, radius $R_{\textsc{wd}}$, and fractional efficiency $\epsilon(v)$ in capturing a CHAMP with speed $v$ in the WD rest frame passing within the gravitational capture radius $R_{\text{capture}}(v) \sim R_{\text{star}} [ 1 + v_{\text{esc},\,{\textsc{wd}}}^2 / v^2 ]^{1/2}$ where $v_{\text{esc.}} = \sqrt{ 2 G M_{\text{star}} / R_{\text{star}} }$ is the WD escape velocity, the total accumulated $X^\pm$ CHAMP masses obtained within an accumulation time $\tau_{\text{accum.}}$ are given by
\begin{align}
&M^{\text{accum.}}_{X^\pm}(\tau_{\text{accum.}}) \nonumber \\
&= \int_{0}^{\tau_{\text{accum.}}} dt \int d^3v f(\bm{v}) \,  \epsilon(v) \, f_\pm \rho_{X}(t) \nl  \qquad\qquad \qquad\qquad\times \, \pi R^2_{\textsc{wd}} \lb[ 1 + \frac{v_{\text{esc},\,{\textsc{wd}}}^2}{v^2} \rb] v,
\label{eq:MXaccrete1}
\end{align}
where $f(\bm{v})$ is the CHAMP velocity distribution far from the star (see below), normalized to $\int_0^\infty  4\pi v^2 f(v) dv \equiv 1$, and the $\lb[\cdots\!\,\rb]$-factor captures the gravitational focusing enhancement to the geometrical capture cross section.

We again assume that the symmetric CHAMP abundance is prevented from annihilating away by the large Coulomb barrier between $(NX)$ and $X^+$, so we will still track the individual contamination abundances.
Note that $f_-$ here should count the $(pX)$ abundance, as the transfer mechanisms of $X^-$ from $(pX)$ to $(NX)$ are likely quite efficient given WD densities; this will be relevant if we discuss accumulation of CHAMPs for masses $m_X \lesssim 10^{11}\,$GeV where it is possible that CHAMPs other than $(pX)$ have been evacuated from the galaxy \cite{Chuzhoy:2008zy} (although, as we have already noted, this is subject to some controversy \cite{Dunsky:2018mqs}).
The contamination fraction of the WD from this accumulation of halo CHAMPs is given by $\eta_\pm = M_{X^\pm}/M_{\textsc{wd}}$.

We will assume that the star experiences a roughly constant CHAMP density throughout its lifetime so that $\rho_X(t) \approx \text{const.}$
Moreover, we will assume that CHAMPs are distributed in momentum space in the same way as fully virialized dark matter, which applies for $m_X \gtrsim 10^{11}$\,GeV \cite{Chuzhoy:2008zy,Dunsky:2018mqs}; it turns out (see \sectref{limits}) that this is the only mass range in which these additional accumulated CHAMPs are relevant for setting limits, so this is not a very strong additional assumption.
More specifically, we will take $f(v)$ to be given by a truncated Maxwellian speed distribution in the MW galactic rest frame \cite{Evans:2018bqy}:
\begin{align}
f(\bm{v}) &= \mathcal{N}^{-1} \exp\lb[ - \frac{|\bm{v}+\bm{v}_{\textsc{wd}}|^2}{v_0^2} \rb] \Theta\lb[ v_{\text{esc,\,\textsc{mw}}} - |\bm{v}+\bm{v}_{\textsc{wd}}| \rb]\label{eq:fCHAMPs}\\
\mathcal{N} &= \pi^{3/2} v_0^3 \Bigg[ \text{erf}\lb[ \frac{v_{\text{esc,\,\textsc{mw}}}}{v_0} \rb] \nl
	\qquad \qquad - \frac{2v_{\text{esc,\,\textsc{mw}}}}{\sqrt{\pi}v_0} \exp\lb[ - \frac{v_{\text{esc,\,\textsc{mw}}}^2}{v_0^2} \rb] \Bigg],
\label{eq:fCHAMPsnorm}
\end{align}
where $v_0 \approx 220\,\text{km/s}$ is the local circular speed in the MW; $v_{\text{esc,\,\textsc{mw}}} \approx 540\,\text{km/s}$ is the MW escape speed; and $\bm{v}_{\textsc{wd}}$ is the WD velocity in the galactic rest frame, which we will take to have a magnitude $v_{\textsc{wd}} \sim v_0$.

\begin{figure}[t]
\includegraphics[width=\columnwidth]{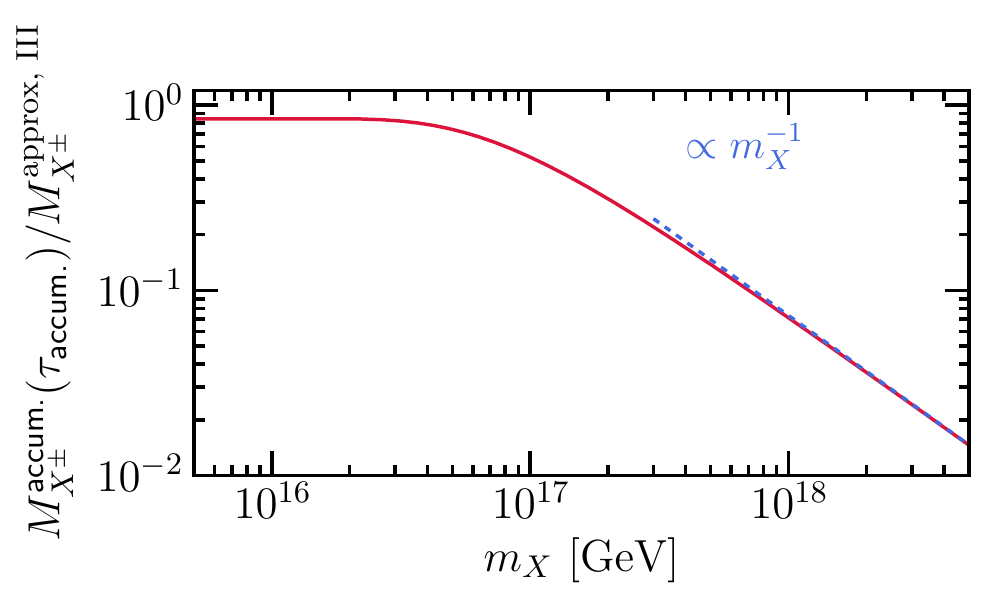}
\caption{ \label{fig:AccretingFraction} 
		The CHAMP abundance that becomes gravitationally bound to a WD, $M_{X^\pm}^{\text{accum.}}(\tau_{\text{accum.}})$ [\eqref[s]{CaseIFull}--(\ref{eq:CaseIIIFull})], compared to the na\"ive estimate $M^{\text{approx, III}}_{X^\pm}$ given at \eqref{MXaccrete2}, when the CHAMP speed distribution and maximum stopping speed are taken into account.
		This plot assumes that the CHAMP passes through a distance $\sim R_{\textsc{wd}}$ of WD material with a density equal to the average WD density for a $M_{\textsc{wd}} \sim 1M_{\odot}$ WD with $R_{\textsc{wd}} \sim 0.01R_{\odot}$, and that the CHAMP velocity distribution far from the WD is given by a truncated Maxwellian distribution, \eqref{fCHAMPs}.
		We assume that the WD moves with respect to the CHAMP distribution at the local circular speed in the MW, $v_{\textsc{wd}} \sim v_0 \sim 220$\,km/s, and that the MW escape speed is $v_{\text{esc,\,MW}} \sim 540$km/s.
		We further assume that $\sigma = 200$\,mb; see text.
		Note that the accumulated mass only begins to significantly deviate from its maximum value once $m_X \gtrsim 3 \times 10^{16}\,\text{GeV} \sim 2 m_X^{\text{max.}}$; because the estimate at \eqref{mXMax} was conservative in requiring \emph{all} CHAMPs to accrete, a significant fraction still accrete even somewhat about this mass. }
\end{figure}

We take 
\begin{align}
\epsilon(v) = \Theta( v_{\text{max}} - v ), 
\label{eq:epsilonV}
\end{align}
where we estimate the maximum speed (in the WD rest frame), $v_{\text{max}}$, of a CHAMP far from the WD that will become gravitationally bound to the WD after one passage through the WD as follows:
a CHAMP hitting the surface of the WD carries energy $E_i \sim \frac{1}{2} m_X \lb( v_{\text{esc},\,{\textsc{wd}}}^2 +v_{X,0}^2 \rb)$ where $v_{X,0} \lesssim 10^{-3}$ is the CHAMP speed far from the WD in the halo (in the WD rest frame).
This energy is lost to ions in the WD via elastic scattering events, each of which carries away a momentum of order $\Delta p \sim m_{\text{ion}} v_{\text{rel}}$ where $v_{\text{rel}}$ is the relative CHAMP-ion speed, and which can be approximated by $v_{\text{rel}} \sim \max\lb[ v_{X}, v_{\text{ion,\,therm.}} \rb]$.
In order to become gravitationally bound to the star, the CHAMP must lose enough energy that its energy drops below $E_f \sim \frac{1}{2}m_X v_{\text{esc},\,{\textsc{wd}}}^2$ after traversing an average distance of order $R_{\textsc{wd}}$ within the WD.
Approximating the WD as a uniform sphere of density $\bar{\rho} \sim M_{\textsc{wd}} / ( 4\pi R_{\textsc{wd}}^3 / 3 )$, and noting that $v_{\text{esc},\,{\textsc{wd}}}  = \sqrt{ 2 M_{\textsc{wd}} / M_{\text{Pl.}}^2 R_{\textsc{WD}}} \sim 2\times 10^{-2}$ while $v_{\text{halo}} \sim v_{\text{ion,\,therm.}} \sim 10^{-3}$ (assuming $T\sim \text{keV}$), we are in the regime where $v_{\text{rel}} \sim v_X$, and so the the CHAMP energy decreases exponentially with distance travelled through the WD: $E(x) \sim E_i \exp\lb( - 2 \bar{\rho}\sigma x / m_X \rb)$.
The largest initial speed the CHAMP can have far from the WD and still be captured after traversing an average distance $\sim R_{\textsc{wd}}$%
\footnote{\label{ftnt:EnergyLossRefinement}%
		This is of course approximate; a more refined computation would solve for the trajectory of the incoming CHAMPs (specified completely by their initial speed far from the WD, and their initial direction of motion relative to the WD velocity) from infinite distance up to the surface of the WD, and then from that surface onward as they pass through the WD losing energy, following a non-conservative trajectory which would likely need to be solved for numerically; such a computation could also make use of the true density profile of the WD to refine the estimate.
		We expect however that such a computation would yield results that are not dramatically different from the result we give here, and we have not performed such a refined computation.
	} %
inside the WD is thus
\begin{align}
v_{\text{max}} \sim v_{\text{esc},\,\textsc{wd}} \lb[ \exp\lb( \frac{2\bar{\rho}\sigma}{m_X} R_{\textsc{wd}} \rb) - 1 \rb]^{1/2},
\label{eq:vMax}
\end{align}
where $\sigma$ is the total cross section for momentum transfers of order $\Delta p \sim m_{\text{ion}} v_{\text{rel}}$, which we very conservatively%
\footnote{\label{ftnt:CoulombStopping}%
		While energy loss by charged particles in dense, degenerate WD material is in general a fairly complicated problem \cite{Graham:2018efk}, an alternative estimate for the linear stopping power of the WD material can be obtained from considering elastic electromagnetic scattering of the CHAMPs off the CO ions, represented by a Thomas--Fermi screened potential \cite{Graham:2018efk}. 
		For particles which are on the borderline between becoming bound to the WD and not becoming bound, we would have $v_X\sim v_{\text{esc,\,\textsc{mw}}}$ fairly constant within $\mathcal{O}(1)$ factors during the traverse of the WD since $v_{X,0} \sim 10^{-3}$ is smaller than $v_{\text{esc,\,\textsc{mw}}} \sim 2\times 10^{-2}$; as such, the Coulomb stopping power can be estimated as $dE/dx|_{\text{Coulomb}} \sim - \lb( 2\pi n_{\text{ion}} Z_{\text{ion}}^2 \alpha^2 \rb)/ \lb( m_{\text{ion}}  v_X^2 \rb) \times \Lambda$ \cite{Graham:2018efk} where $\Lambda$ is a Coulomb log of order $\Lambda \sim 10$--$15$ for the relevant parameters.  
		By comparison, the linear stopping power we have estimated with the geometric cross section is $dE/dx|_{\text{geometric}} \sim - n_{\text{ion}} m_{\text{ion}} \sigma v_X^2$.
		The ratio is $ \lb( dE/dx|_{\text{Coulomb}}  \rb) / \lb( dE/dx|_{\text{geometric}} \rb) \sim \lb( 2\pi Z_{\text{ion}}^2 \alpha^2 \rb)/ \lb( m_{\text{ion}}^2  v_X^4 \sigma \rb) \times \Lambda$.
		Consistent with the mean-ion approach, if we take $Z_{\text{ion}}\sim 7$, $m_{\text{ion}} \sim 7 \mu_a$, $v_X \sim 2\times 10^{-2}$, and $\Lambda \sim 15$, we find $\lb( dE/dx|_{\text{Coulomb}}  \rb) / \lb( dE/dx|_{\text{geometric}} \rb) \approx 70 \times ( \sigma / 200 \text{mb} )^{-1}$. 
		Our geometrical cross section estimate is thus perhaps too conservative by about a factor of 70.
		Taking the more aggressive cross section estimate would \emph{increase} the maximum stopping mass linearly by the same factor; see \eqref{mXMax}.
		The only change to the limits would be that the correction factor estimated in \figref{AccretingFraction} would remain $\sim 1$ until $m_X \sim 1\times 10^{18}\,\text{GeV}$, so the limits we present in \figref[s]{BoundsPlus} and \ref{fig:BoundsMinus} would be slightly stronger at the largest masses we consider.
	} %
estimate to be approximately the geometrical nuclear cross section $\sigma \sim \pi \lb( 1.22 \,\text{fm} \times A^{1/3} \rb)^2 \sim 170\,\text{mb} \sim 200\,\text{mb}$, with $A\sim 7$ consistent with the mean-ion approach we utilize throughout; see \citeR{Graham:2018efk} for a more detailed discussion of cross sections for stopping in WD material.
Demanding that $v_{\text{max}} \gtrsim v_{\text{esc,\,\textsc{mw}}}+v_0$ conservatively guarantees that the CHAMP will become bound after one passage through a distance $R_{\textsc{wd}}$ of WD material; a conservative estimate for the maximum CHAMP mass that is thus \emph{guaranteed} to become bound to the WD is given by
\begin{align}
m^{\text{max}}_X &\sim \frac{ 2\bar{\rho}\sigma R_{\textsc{wd}} }{ \ln\lb[ 1 + \lb(\frac{v_{\text{esc},\,\textsc{mw}}+v_{\circ}}{v_{\text{esc},\,\textsc{wd}}} \rb)^2 \rb] } \nonumber \\
& \sim 1.5\times 10^{16}\,\text{GeV} \times \frac{\sigma}{200\,\text{mb}},
\label{eq:mXMax}
\end{align}
where we again remind the reader that the cross section assumed is conservatively small given that the CHAMPs are electrically charged.
While \eqref{mXMax} is a good estimate for where these effects will begin to become relevant, an $\mathcal{O}(1)$ fraction of CHAMPs that traverse the WD will still be efficiently captured even for CHAMP masses somewhat above $m^{\text{max}}_X$; see \figref{AccretingFraction}.

For CHAMPs lighter than $m_X^{\text{max}}$, it is safe to assume $\epsilon = 1$ in \eqref{MXaccrete1} and to perform the integral without regard to the maximum stopping speed; this actually suffices for the majority of the mass range where we will be interested in the accumulation mechanism (see discussion below and in \sectref{limits}).
For heavier CHAMPs, we are required to take account of the maximum stopping speed, which imposes an upper bound $v\leq v_{\text{max.}}$ of the $dv$ integral in \eqref{MXaccrete1}, as implied by \eqref{epsilonV}.

The final result of a careful treatment of the integrals in \eqref{MXaccrete1} is a set of complicated functions for the accreted CHAMP abundance, which we show in \appref{Accreting_Fraction}; we use the full expressions \eqref[s]{CaseIFull}--(\ref{eq:CaseIIIFull}) for all quantitative results in this paper.
It is however useful to develop a simple approximate estimate for the accreted mass; this can easily be done directly from \eqref{MXaccrete1} in the case of $m_X \ll m_X^{\text{max.}}$.
Suppose we ignore the fact that the CHAMPs have a velocity distribution, and approximate $f(\bm{v}) = \delta( v - v_0 ) / 4\pi$, and also take $\epsilon = 1$; then we have that 
\begin{align}
M^{\text{accum.}}_{X^\pm}(\tau_{\text{accum.}})&\approx M^{\text{approx, III}}_{X^\pm} \nonumber \\
& \equiv \tau_{\text{accum.}} f_\pm \rho_{X} \pi R^2_{\textsc{wd}} \frac{v_{\text{esc,\,\textsc{wd}}}^2 }{v_0}
\label{eq:MXaccrete2}
\end{align}
where we taken $v_{\text{esc,\,\textsc{mw}}}^2 \gg v_0^2$.
Numerically, this estimate is a factor of roughly $1 / \text{erf}(1)\sim 1.2$ larger than the full result given at \eqref{CaseIIIFull} [which is well within the uncertainty on the estimate itself]; normalized to this approximate result $M^{\text{approx, III}}_{X^\pm}$, we plot the full result \eqref{CaseIIIFull} in \figref{AccretingFraction}.
The fractional suppression of the accumulating CHAMP abundance at large mass scales%
\footnote{\label{ftnt:Scaling}%
		The integral scales as $\int^{v_{\text{max}}}_0 dv \, v^2 \times v \times v^{-2} \propto v_{\text{max}}^2 \propto m_X^{-1}$, where the factor of $v^2$ comes from the integration measure, the factor of $v$ comes converting the CHAMP abundance to a flux, and the factor of $v^{-2}$ comes from the enhanced gravitational capture radius for slow-moving CHAMPs.
		The final scaling with $m_X$ is obtained by expanding the exponential in \eqref{vMax} for $m_X \gg 1.5\times 10^{16}$\,GeV [see \eqref{mXMax}].
	} %
roughly as $m_X^{-1}$.

Suppose we consider either a $M \sim 0.85 M_{\odot}$ WD or a $M \sim 1.1 M_{\odot}$ WD, take the accumulation duration to be the timescale for cooling of a WD to the point of crystallization (see discussion below) $\tau_{\text{accum.}} \sim 10^9\,$yr \cite{GarciaBerro:2007an}, and normalize to the local MW DM density $\rho_{\text{halo}} \sim 0.3 \,$GeV/cm${}^3$, then for $m_X \lesssim 1.5\times 10^{16}\,$GeV, we have
\begin{align}
\eta_{\pm,\,\text{accum.}}(0.85 M_{\odot} \ \text{WD}) \sim  1.9\times 10^{-13} \lb( \frac{ f_\pm \rho_{X} }{ \rho_{\text{halo}} } \rb)\\
\eta_{\pm,\,\text{accum.}}(1.1 M_{\odot} \ \text{WD}) \sim  1.3\times 10^{-13} \lb( \frac{ f_\pm \rho_{X} }{ \rho_{\text{halo}} } \rb),
\label{eq:etaAccum2}
\end{align}
which exceeds the CHAMP contamination fraction obtained via the protostellar channel (for a $M_{\text{cloud}}=7M_{\odot}$ protostellar cloud, as appropriate for a $M_{\textsc{wd}}\sim 1.1M_{\odot}$ WD \cite{Catalan:2008tr})%
\footnote{\label{ftnt:McloudVsMstar}%
		Note that while the mass of the WD progenitor star will be comparable to the collapsing protostellar cloud mass we input, significant shedding of the outer envelope of a highly evolved red giant occurs after the helium flash reduces the associated WD progenitor CO core mass to something $\sim M_{\odot}$ \cite{Heger:2002by}; see \citeR{Catalan:2008tr} for a discussion of this initial-mass--final-mass relationship.
		Our results are actually relatively insensitive to the exact assumed value of $M_{\text{cloud}}$, as $\eta \propto M_{\text{cloud}}^{-1/3}$ for large $m_X$ [see \eqref{eta} for $y_{\text{crit.}}<y_{\text{min}}$].
		} %
for $m_X \sim 1.6\times 10^{11}\,$GeV, assuming for the sake of argument the same $\rho_X$ (and ignoring any differences in $(pX)$ accumulation).

Note that this estimate will be subject to large Poisson uncertainties once the CHAMP mass becomes large enough that, given the CHAMP density, the number of CHAMPs that pass through the capture area in the stellar lifetime, $N^{\text{accum.}}_{X^{\pm}} = M_{X^{\pm}}^{\text{accum.}} / m_{X^\pm}$, becomes small.
Demanding conservatively that the fractional $1/\sqrt{N_{X^{\pm}}}$ uncertainty in the estimate is less than 10\% demands that $N_{X^{\pm}}\gtrsim 10^2$.
In order not to be subject this Poisson uncertainty, we must require that
\begin{align}
\lb( \frac{ f_\pm \rho_{X} }{ \rho_{\text{halo}} } \rb)_{\text{Poisson}}  \gtrsim \begin{cases} 
		6\times 10^{-38 }\frac{m_X}{10^5\,\text{GeV}} & m_X \lesssim 10^{17}\,\text{GeV} \\[2ex]
		1.2\times 10^{-22} \frac{m_X^2}{m_{\text{Pl.}}^2} & m_X \gtrsim 10^{18}\,\text{GeV}
		\end{cases},
\end{align}
with a small transition region in the scaling in the intermediate mass range.
See further discussion in \sectref{limits} where we consider the limits the accumulation of CHAMPs imposes on the galactic abundance of CHAMPs.

\subsubsection{Efficiency of accretion}
\label{sect:AccretionEfficiency}
We note that in order for CHAMPs accumulated by this mechanism to sink to the center of the WD and possibly feed a dense central core object, we should assume that the accretion is occurring while the interior of the WD is still in a liquid/non-solid phase, otherwise the sinking of the $X^\pm$ is likely significantly inhibited (see, e.g., \citeR{GarciaBerro:2007an}, in which the diffusion co-efficient for ${}^{22}$Ne contamination in a CO WD is set to zero after crystallization).
Since WD cooling models (e.g., \citeR{Renedo:2010vb}) indicate that that crystallization only occurs at ages $\tau \gtrsim 10^9$\,yr for CO WDs with masses around $M_{\text{WD}} \sim 0.85 M_{\odot}$, we restrict our attention to CHAMP accretion occurring within at most the first $\tau_{\text{accum.}} \sim \text{Gyr}$ after formation.
Note that it is not necessarily the case that sufficiently heavy CHAMPs would be prevented from sinking even after crystallization,%
\footnote{\label{ftnt:LatticeStability}%
		If the differential gravitational force acting on the CHAMP and a neighboring ion exceeds the electrostatic repulsive force exerted by the ion on the CHAMP at approximately the lattice spacing then, in a lattice configuration, the CHAMP would likely sink through the lattice.
		To get an idea of the mass scale involved, consider an $X^+$ located in a cold $M_{\textsc{wd}}\sim 1.2M_{\odot}$ WD, approximate the lattice spacing by $a \sim ( 3 m_{\text{ion}} / (4\pi \rho) )^{1/3}$, and approximate $\rho \sim 3 M_{\textsc{wd}} / (4\pi R_{\textsc{wd}}^3)$; taking $m_{\text{ion}} \sim 12 \mu_a$ and $Q_{\text{ion}} \sim 6$, we find that this criterion is reached for $m_X \gtrsim 10^{17}$\,GeV, so while the very heaviest $X^+$ CHAMPs could potentially sink through a solid lattice in a WD, lighter CHAMPs may be prevented from doing so. 
		The cognate mass estimate for $X^-$ bound as $(NX)$ is higher by the charge of the $(NX)$ state.
		Note however that this estimate is very rough, which is why we conservatively truncated our accumulation at a timescale such that the WD has not yet crystallized.
	} %
so this is conservative. 

Note that a short initial interval of CHAMP accumulation may be complicated by the persisting existence of an ionized planetary nebula blown off by the late-stage evolution of the WD progenitor star.
Given that planetary nebulae do not remain ionized for much more than $\sim \text{few}\times 10^{4}$ years (see, e.g., \citeR{Kwok_1994})---orders of magnitude shorter than the crystallization time for the WD---this will likely be a negligible effect; we have not attempted to account for it.

Relatedly, the results of \citeR{Dunsky:2018mqs} indicate accumulation of CHAMPs (here, onto the WD) in the presence of stellar winds and magnetic fields can be complicated by the entrainment of inwardly diffusing CHAMPs in magnetic field lines moving outward with the charged wind.
However, an absence of solar winds is expected for cooling WD with sufficiently high surface gravity $g\gtrsim 10^7\text{cm}/\text{s}^2$ and $T_{\text{eff}} \lesssim 5\times 10^4$\,K surface temperatures \cite{Unglaub:2008abc}; the surface gravity of a $M_{\textsc{wd}} \gtrsim 0.8 M_{\odot}$ WD is $g \gtrsim 2\times 10^8\,\text{cm}/\text{s}^2$ (and higher for more massive WD, which have smaller radii), so as long as the WD we are interested in have low enough surface temperatures, no winds should be expected.

On the other hand, certain WD are known to possess high magnetic fields \cite{2015SSRv..191..111F} (see also \citeR{2017ASPC..509....3D} and references therein).
Moreover, even if the $X^+$ or $(NX)$ bound state are neutral by virtue of having captured electrons, we estimate that the UV luminosity of even an old WD is sufficient to ionize at least one or more electrons from the $X^+$ or $(NX)$ state;%
\footnote{\label{ftnt:IonizeCHAMPs}%
		Consider a WD with a luminosity of $L_{\textsc{wd}} \sim 10^{-3} L_{\odot}$, which is typical for a $M_{\textsc{wd}} \sim M_{\odot}$ WD less than $\sim \text{Gyr}$ old \cite{Kippenhahn:2012zqe}, and conservative for the luminosity at an earlier age in the WD existence since WD cool over time.
		The surface temperature is then of order $T\sim \text{eV}$, and so the fraction of the number of photons emitted from the WD that have an energy sufficient to ionise the outer electron from the CHAMP state (which we take to be $\sim 10$\,eV) is, conservatively, $f_{>10\,\text{eV}} \sim 10^{-3}$.
		Approximating the hydrogenic photoionization cross section as $\sigma \sim 4\pi \alpha r_B^2$ where $r_B \sim 5\times 10^{-11}\,\text{m}$ is the Bohr radius, we find that at a distance of $R=10R_{\textsc{wd}}$ from the WD surface, the timescale for an ionization interaction to occur is 
		$\tau_{\text{ionize}}^{-1} \sim n_{\gamma} \sigma \sim ( f_{>10\,\text{eV}} L_{\textsc{wd}} / (4\pi R^2) / T ) ( 4\pi \alpha r_B^2 ) \sim 10^{-10} (L_{\odot}/\text{eV})  (r_B/R_{\textsc{wd}})^2 \sim 1/ ( 0.05 s )$.
		However, even moving at escape speed, a CHAMP would take a time $t \sim R/v_{\text{esc}} \sim R \sqrt{ R_{\textsc{wd}} M_{\text{Pl.}}^2 / 2M_{\textsc{wd}} } \sim 8\,\text{s} \gg \tau_{\text{ionize}}$ to traverse the remaining distance to the WD.
		Therefore, we estimate that the UV luminosity of the WD is sufficient to ionize electrons from the CHAMP state at least a distance $R \gtrsim 10R_{\textsc{wd}}$ from the surface of the WD.
	} %
 as a result, the WD magnetic fields could deflect incoming CHAMPs, making the accumulation estimate too aggressive.
However, this depends on the rigidity of the CHAMPs: we estimate that if a $X^+$ or singly-ionized $(NX)^+$ of mass $m_X \sim 10^{11}\,$GeV moving at a speed $v \sim \sqrt{2GM_{\textsc{wd}} / (2R_{\textsc{wd}})}$ were to experience a $\sim 7.5$\,MG magnetic field roughly one WD radius from the surface of a $M_{\textsc{wd}} \sim 1.1M_{\odot}$ WD, its Larmor radius $r_L \sim p/qB$ would be of order the WD radius, which would likely result in the particle being deflected enough to miss the WD surface.
Much weaker magnetic fields at this radius would not deflect the particle sufficiently to cause it to miss the surface; moreover, since any magnetic field drops off at large distance from the WD at least as rapidly as a dipole field $B\propto r^{-3}$, the Larmor radius at large distance grows much more rapidly than the distance from the WD, so it would likely not cause the particle to miss the WD owing to its effect further out from the WD.
Although it is not necessarily the case that fields much stronger than the one estimated here would cause the particle to miss the WD either (in a very strong field, the CHAMP would likely simply spiral down the magnetic field lines, still likely hitting the WD surface) this is a much more challenging situation to analyze.
We will thus wish to consider only WD with low magnetic field $\ll 7.5\,$MG when considering $m_X \gtrsim 10^{11}\,$GeV CHAMPs accumulating onto a $M_{\textsc{wd}} \sim 1.1M_{\odot}$ WD.
On the other hand, if we were to consider a $m_X \sim 3\times 10^7\,$GeV CHAMP (about as light a CHAMP as we will concerned with in the accumulating case) accumulating onto a $M_{\textsc{wd}} \sim 0.85M_{\odot}$ WD, a B field of 1.1kG at one WD radius above the surface would be problematic; very low magnetic field WD are thus required for bounds based on such accumulating WD to be robust.

Finally, note that if all the above caveats about this process being efficient are satisfied then, until such time as the WD crystalizes, this accumulation of CHAMPs is in addition to any primordial abundance of CHAMPs that may already have been present in the WD.

\subsubsection{Behavior of accumulated CHAMPs}
\label{sect:BehaviorAccumulated}
An $X^+$ incident on the WD is simply stopped, then diffusively sinks to the core of the WD (see the following sections).
The fate of an $X^-$ may be different: if it is incident as $(pX)$, it can first be stopped and then rapidly undergo an exchange reaction, becoming bound to a heaver nucleus $(NX)$, where $N$ is either He (in the WD atmosphere, if any) or is most likely C or O, which form the bulk of the WD.
Alternatively, the exchange reaction can occur during the stopping process.
In either case, after exchange, the heavier nucleus bound state will, after perhaps having undergone some further stopping, merely diffusively sink to the core of the WD.
If the incoming $X$ are in bound states $(NX)$ where $N$ is heavier than He, it is possible that the $X$ remains bound to that nucleus, and merely gets stopped and sinks.

\subsection{Sinking, stratification, and timescales}
\label{sect:timescales}
CHAMPs, whether $X^+$ or $X^-$ bound as $(NX)$, have a much smaller charge-to-mass ratio $Q_X/m_X$ than the CO material in which they are interspersed when the WD is born.
As such, they would be expected to sink (diffusively) in the WD toward the center of the star.
Moreover, once formed, a WD experiences no internal large-scale nuclear burning processes that would trigger convection until/unless thermal runaway is triggered, so convective mixing of the WD contents is unlikely to disrupt this sinking. 
In this subsection, we discuss the timescale for the sinking process.

A very rough argument for why $q/m$ is the relevant quantity \cite{1992A&A...257..534B,Bildsten:2001xb,Deloye:2002xb,GarciaBerro:2007an} when considering sinking is that the (non-degenerate) ions present at some radius in the WD would sink compared to the electrons absent a very small net positive charge overdensity interior to that radius which, in hydrostatic equilibrium, supplies exactly the correct electric field to balance the gravitational force on the ions at that radius.
However, since $F_E \sim q$ and $F_g \sim m$, such a bulk electrostatic force balance can only work for one value of $q/m$: particles with smaller than average $q/m$ at a given radius must sink, while those with a larger $q/m$ must rise. 
Indeed, the ${}^{12}_{\ 6}$C--${}^{16}_{\ 8}$O  ($q/m=0.5$) mixture in a WD does not stratify (while it remains liquid), but contaminants, e.g., ${}^{22}_{10}$Ne ($q/m = 0.45 < 0.5$) sink on cosmologically long timescales \cite{Bildsten:2001xb,Deloye:2002xb,GarciaBerro:2007an}.

Here we follow the discussion of \citeR{GarciaBerro:2007an,Deloye:2002xb} (see also \citeR{Burgers1969,Olson:1975ts,1992A&A...257..534B,Bildsten:2001xb}) to provide an estimate of the timescale for the sinking.
A first very approximate estimate for the timescale $\tau^{(1)}_{\text{sink}}$ for sinking an $\mathcal{O}(1)$ fraction of the WD radius is $\tau^{(1)}_{\text{sink}} \sim R_{\textsc{wd}} / w_X(R_{\textsc{wd}})$, where $w_X(R_{\textsc{wd}})$ is the CHAMP diffusion velocity at the surface of the WD.
Assuming that the CHAMPs are only a trace constituent in the WD background (easily satisfied during the initial sinking phase), and that the charge-to-mass ratio of the $X$ is very small, the (terminal) diffusion velocity can be estimated as%
\footnote{\label{ftnt:noQxDependence}%
		Note that there is no $Q_X$ dependence in $w_X$. 
		For $X$ a trace element of mass $m_X$ sinking through a background plasma with $q/m=2$, the full expressions of \citeR{GarciaBerro:2007an} give $|w_X| = | 2 Q_X - m_X/\mu_a | ( m_p g D / T )$, which yields \eqref{wSink} in the limit $m_X \gg \mu_a$ regardless of $Q_X$ (if we approximate $m_p \approx \mu_a$, which is allowed at the level of accuracy of these computations).
	} %
\begin{align}
|w_X(r)| \approx g D m_X / T,
\label{eq:wSink}
\end{align}
where $g = M(r)/(r^2M_{\text{Pl.}}^2)$ is the local acceleration due to gravity, and $D$ is a diffusion coefficient, which \citeR[s]{GarciaBerro:2007an,Deloye:2002xb} indicate can be estimated as the self-diffusion coefficient for the CO plasma:
\begin{align}
D &\approx 3 \omega_p a^2 \Gamma^{-4/3}
\end{align}
where
\begin{align}
\omega_p^2 &= 4\pi Z_{\text{ion}}^2 \rho \alpha / m^2_{\text{ion}} \\
a &\equiv ( 3 m_{\text{ion}} / 4\pi \rho )^{1/3} \\
\Gamma &= \alpha Z_{\text{ion}}^2 / ( aT ),\\
\Rightarrow D &\approx  \frac{9}{2\pi^{11/18}} \lb( \frac{3}{4} \rb)^{1/9} \frac{m_{\text{ion}}^{1/9} T^{4/3}}{Z_{\text{ion}}^{5/3} \alpha^{5/6} \rho^{11/18}},
\label{eq:SelfDiffusionCoefficient}
\end{align}
where `ion' here refers to an ion in the mixture through which the CHAMPs are sinking.
We will adopt the widely used `mean ion' approach and set $m_{\text{ion}} \sim 14\mu_a$ and $Z_{\text{ion}} \sim 7$, assuming roughly equal abundances of C and O.%
\footnote{\label{ftnt:Xisx}%
		At this level of approximation, it is irrelevant whether we assume the number or mass abundances are equal. 
		Technically, this assumes the number abundances are equal.
	} %

For the purposes of an initial rough estimate, we will take $T\sim10^7\,\text{K}\sim 1\,$keV (WD are approximately isothermal, and this is a typical WD core temperature), $M_{\textsc{wd}} \sim 1.1M_{\odot}$, $R_{\textsc{wd}} \sim 7\times 10^{-3} R_{\odot}$, and we will approximate the density with the average WD density: $\rho \approx \bar{\rho} \equiv 3 M_{\textsc{wd}} / (4\pi R_{\textsc{wd}}^3) \approx 4.5\times 10^6\,\text{g/cm}^3$.
The sinking timescale we then estimate is [clearly for $m_X \ll 10^{21}\,\text{GeV}$]%
\footnote{\label{ftnt:naiveIsDifferent}%
		We note that the terminal velocity estimate, and hence sinking time, are dramatically different from the na\"ive estimates obtained from setting the viscous drag force $F_{\text{drag}} \sim \rho_{\text{ion}} v_{\text{ion}} \sigma w_X$ equal to the gravitational force \cite{Gould:1989gw}, which would yield $w_X \sim g m_X / ( \rho_{\text{ion}} v_{\text{ion}} \sigma )$.
		If we take $v_{\text{ion}} \sim \sqrt{T / m_{\text{ion}}}$ and estimate $\sigma \sim 200\,\text{mb}$ as we have throughout, we would obtain $w_X \sim 15\,\text{cm/s} \times (m_X/10^5\,\text{GeV})$, assuming the same average WD density as in the main text.
		The difference is likely ascribable to the WD interior being a strongly coupled plasma which acts like a liquid, instead of a rarefied gas, so that the low $\sim$ nuclear cross section estimate we have used up to now (always in a way thus far that was conservative) potentially supplies a dramatic underestimate of the viscous drag force on the CHAMPs, and yields a much too aggressively short estimate of the sinking time.
		The extreme discrepancy between these estimates does however give some pause, and we note that the timescales we give here are possibly conservatively long.
	} %
\begin{align}
w_X(R_{\textsc{wd}}) &\sim 1 \times 10^{-7} \,\text{m/s} \times \lb( \frac{m_X}{10^5\,\text{GeV}} \rb) \\
&\approx 3\times 10^{-16} c \times \lb( \frac{m_X}{10^5\,\text{GeV}} \rb)\\
\tau_{\text{sink}}^{(1)} &\sim 1.6 \times 10^6\,\text{years} \times \lb( \frac{m_X}{10^5\,\text{GeV}} \rb)^{-1}.
\end{align}
For $m_X \gtrsim 10^{3}\,\text{GeV}$, this is less than the old WD lifetimes and/or crystallization times of $\sim\,\text{Gyr}$ we consider, but the estimate here is crude because we have taken $g$ and hence $w_X$ to be radially independent.

A better but still highly approximate analytically tractable estimate is obtained by taking into account the radial dependence of $w_X$, making the assumption that the WD is a sphere of uniform density (see, e.g., \citeR{Gould:1989gw}). 
Then, we have that $g(r) = M_{\text{enc.}}(r) / ( r^2 M^2_{\text{Pl.}}) = \lb[ M_{\textsc{wd}} (r/R_{\textsc{wd}})^3 \rb] / ( r^2 M^2_{\text{Pl.}}) =  g(R_{\textsc{wd}}) \times (r/R_{\textsc{wd}})$, implying that
\begin{align}
w_X(r) &= w_X(R_{\textsc{wd}}) \times \frac{r}{R_{\textsc{wd}}},
\end{align}
so that the timescale estimate to move from radius $r= R_i \sim R_{\textsc{wd}}$ to $r=R_f$ becomes
\begin{align}
\tau^{(2)}_{\text{sink}} \sim \int_{R_f}^{R_{\textsc{wd}}} \frac{dr}{w_X(r)} = \tau^{(1)}_{\text{sink}} \ln \lb( \frac{R_{\textsc{wd}}}{R_f} \rb),
\label{eq:tausink2}
\end{align}
which is longer than the previous estimate by a logarithmic factor that depends on the final radius.
In the next subsection, we estimate the initial radius of the CHAMP structure that forms at the core of the WD as a result of the sinking, which will show that this logarithmic factor is never large enough to cause the sinking timescale to become unacceptably long. 

Moreover, as we are also interested in the case of sinking where the star also contains an extremely compact core object (i.e., a BH) at the center of the WD, it is worth considering how the above estimate is modified in the case where $M_{\text{enc}} = M_{\textsc{wd}} (r/R_{\textsc{wd}})^3 + M_{\text{core}} \lb( 1 -(r/R_{\textsc{wd}})^3  \rb)$ for $M_{\text{core}} \ll M_{\textsc{wd}}$.
This implies that
\begin{align}
w_X(r) = w_X(R_{\textsc{wd}})  &\Bigg[  \frac{r}{R_\textsc{wd}} \lb( 1 - \frac{M_{\text{core}}}{M_{\textsc{wd}}} \rb) \nl 
+ \frac{M_{\text{core}}}{M_{\textsc{wd}}}  \lb(\frac{R_{\textsc{wd}}}{r}\rb)^2  \Bigg]
\label{eq:wXrefined}
\end{align}
If $M_{\text{core}} \ll M_{\textsc{wd}}$, then for $r \gtrsim r_{\text{cross}} \equiv R_{\textsc{wd}} \lb( M_{\text{core}} / M_{\textsc{wd}} \rb)^{1/3}$, the dynamics are still dominated by   the WD material enclosed at radius $r$, and the estimate at \eqref{tausink2} holds; however, for $r \lesssim r_{\text{cross}}$, the dynamics are dominated by the core, implying that $w_X \sim r^{-2}$, which regulates the logarithmic divergence in the total sinking time estimate at \eqref{tausink2}:
\begin{align}
\tau^{(3)}_{\text{sink}} &\sim \int_{R_f}^{R_{\textsc{wd}}} \frac{dr}{w_X(r)} \\
&\sim \tau^{(1)}_{\text{sink}}  \times \min \lb\{ \ln \lb( \frac{R_{\textsc{wd}}}{R_f} \rb) , \ln \lb[ \lb(\frac{M_{\textsc{wd}}}{M_{\text{core}}}\rb)^{1/3} \rb] \rb\}.
\label{eq:tausink3}
\end{align}
The maximum total time to sink to the center of the star in this case is thus 
\begin{align}
\max \tau^{(3)}_{\text{sink}} 
&\approx \tau^{(1)}_{\text{sink}}  \ln \lb[ \lb(\frac{M_{\textsc{wd}}}{M_{\text{core}}}\rb)^{1/3} \rb] \\
&\sim 2.5\times 10^7\,\text{years} \times \lb( \frac{m_X}{10^5\,\text{GeV}} \rb)^{-1} \nl
\times \lb\{ 1 + \frac{1}{15} \ln \lb[ \lb(\frac{10^{-20}M_{\odot}}{M_{\text{core}}}\rb)^{1/3} \rb] \rb\},
\label{eq:tausink3b}
\end{align}
which for $m_X \gtrsim 10^{3}\,\text{GeV}$ is less than WD lifetimes and/or crystallization times of $\sim\,\text{Gyr}$.

Moreover, \eqref[s]{tausink2} and (\ref{eq:tausink3b}) are still over-estimates of the sinking timescale because the density of the WD increases significantly above the average density as the core is approached, which makes $M_{\text{enc}}(r)$ and hence $w_X(r)$ (in the region $r>r_{\text{cross}}$) larger than that estimates assuming the uniform sphere.
Indeed, for a WD of mass $M_{\textsc{wd}}=0.8$--$1.2M_{\odot}$, the density exceeds the average density for $r/R_{\textsc{wd}} \lesssim 0.6$--$0.7$, with the central density eventually reaching a value roughly a factor of 9--16 larger than the average density.
Recomputing the timescale estimate with a realistic WD density profile, we find that for a $M_{\textsc{wd}} = 1.1M_{\odot}$ WD we have, within a factor of $\sim 2$ for $R_f \lesssim 10^{-2}R_{\textsc{wd}}$, that
\begin{align}
\tau^{(4), 1.1M_{\odot}}_{\text{sink}} &\sim \int_{R_f}^{R_{\textsc{wd}}} \frac{dr}{w_X(r)} \\
&\sim 2 \tau^{(1)}_{\text{sink}} \times \frac{ \bar{\rho} }{ \rho_{C}  } \nl 
 \times \min \lb\{ \ln \lb( \frac{R_{\textsc{wd}}}{R_f} \rb) , \ln \lb[ \lb(\frac{\rho_C}{\bar{\rho}}\frac{ M_{\textsc{wd}}}{M_{\text{core}}}\rb)^{1/3} \rb] \rb\} 
\label{eq:tausink4a}\\
 &\sim 4\times 10^6\,\text{years}  \times \lb( \frac{m_X}{10^5\,\text{GeV}} \rb)^{-1} \nl
\times \lb\{ 1 + \frac{1}{13} \ln \lb[ \lb(\frac{\rho_C/\bar{\rho}}{13} \frac{10^{-20}M_{\odot}}{M_{\text{core}}}\rb)^{1/3} \rb] \rb\}
\label{eq:tausink4}
\end{align}
assuming always that $R_f \ll R_{\textsc{wd}}$, and where we took $\rho_{C} / \bar{\rho} \approx 13$ for this WD.
Again, for $m_X \gtrsim 10^{3}\,\text{GeV}$ this is less than WD lifetimes and/or crystallization times of $\sim\,\text{Gyr}$.
Similar estimates hold for WD throughout the range of masses in which we are interested in this paper. 
Note that the logarithmic factor shown at \eqref{tausink4} differs from that at \eqref{tausink3} because the increased central density of WD material implies a smaller cross-over radius if a massive core is present.
Finally, note that the core mass assumed here is, within an order of magnitude or so, the \Chand\ mass for a $m_X \sim 10^{10}\,$GeV CHAMP, and that the logarithmic factor is clearly not very sensitive to the exact assumed core mass.

We must also consider whether it is actually possible to achieve the terminal diffusion velocity under the gravitational acceleration prevailing in the WD; if not, then the above estimates could be incorrect.
To see that this is easily possible, consider that the maximum velocity achievable by a particle free-falling through a uniform density sphere of mass $M$ and radius $R$ is given up to $\mathcal{O}(1)$ factors by $v_{\text{max.}} \sim \sqrt{ M / ( R M_{\text{{Pl.}}}^2 )}$.
For the $M_{\textsc{wd}} \sim M_{\odot}$ WD discussed above, this estimate is $v_{\text{max.}} \sim 10^{-2}c$, which is much greater than the terminal diffusion velocity for all $m_X \lesssim M_{\text{Pl.}}$; the terminal velocity is thus always reached.

Note that we will conservatively elect not to consider CHAMPs lighter than $m_X \lesssim 10^{3} \,\text{GeV}$, where the sinking time estimates given here approach WD lifetimes and/or crystallization times of $\sim \,\text{Gyr}$; see \figref[s]{BoundsPlus} and \ref{fig:BoundsMinus}.

\subsection{Thermally supported CHAMP-contaminated WD structure: self-gravitating collapse and timescales}
\label{sect:WDstructureThermal}

The $X^+$ and $(NX^-)$ present will sink diffusively to the center of the newly formed WD in a characteristic time $\tau_{\text{sink}}$ (see \sectref{timescales}), until (or if) they encounter sufficient pressure to halt this collapse/sinking process and establish a stable hydrostatic equilibrium (should such an equilibrium exist). 
In this subsection, we consider the initial formation of a CHAMP structure at the center of the WD; see also the discussions in \citeR[s]{Gould:1989gw,Janish:2019nkk,Acevedo:2019gre}.

Initially, so long as the WD material still dominates the central mass density of the WD, the CHAMPs will form an approximately isothermal thermal-pressure-supported structure%
\footnote{\label{ftnt:IdealGas}%
		There is a significant caveat to this discussion.
		For the picture of the thermal-pressure-supported structure that we advance in this section to be correct, the $X$ must contribute an ideal gas term to the pressure $P\supset n_X(r) T$ and this must be the only term in the pressure acting on the CHAMPs that varies significantly over the length scale $r_*$ (only pressure \emph{gradients} hydrostatically support structures against gravitational collapse).
		The $X^\pm$ are however electrically charged and immersed in strongly coupled, charged non-degenerate CO and degenerate electron plasmas, so this assumption is likely a gross approximation.
		It is therefore possible that the $X$ are not stalled in this thermal structure, but instead simply continue to diffusively sink toward the center of the star, directly forming the denser core structure at the center of the WD that we discuss in \sectref{WDstructure}, which has a maximum \Chand\ mass before it too must collapse.
		However, because we will set limits (see \sectref{limits}) requiring the presence of the larger of the self-gravitating mass or the \Chand\ mass of CHAMPs in the core structure, and because the self-gravitating mass exceeds the \Chand\ mass for large $m_X$ [\eqref{mXSG}], it is conservative to assume that this thermal structure must form and become self-gravitating before the CHAMPs can sink further.
	} %
at the center of the WD $\rho_X(r) \sim \rho_X(0) \exp[-(r/r_*)^2]$, with a characteristic scale height~\cite{Gould:1989gw}
\begin{align}
r_* &= \sqrt{ \frac{3TM_{\text{Pl.}}^2}{2\pi m_X \rho_{\textsc{wd}} } } \\
&\sim 350\,\text{m} \times \lb( \frac{T}{1\,\text{keV}} \rb)^{1/2} \times \lb( \frac{ m_X }{10^5\,\text{GeV}} \rb)^{-1/2} \nl
\times \lb( \frac{ \rho_{\textsc{wd}} }{ 5.5\times 10^{10}\,\text{g/cm}^3 } \rb)^{-1/2},
\label{eq:rStar}
\end{align}
and total mass
\begin{align}
M^X_{*} &= \pi^{3/2} \rho_X(0) r_*^3 \\
&\sim 6 \times 10^{-12}M_{\odot} \times \frac{ \rho_X(0) }{ \rho_{\textsc{wd}} } \times \lb( \frac{T}{1\,\text{keV}} \rb)^{3/2} \nl \times \lb( \frac{ m_X }{10^5\,\text{GeV}} \rb)^{-3/2} \times \lb( \frac{ \rho_{\textsc{wd}} }{ 5.5\times 10^{10}\,\text{g/cm}^3 } \rb)^{-1/2}
\label{eq:MthermCoreMass0}
\end{align}
where $\rho_{\textsc{wd}}$ is the central WD density, for which we have taken the fiducial value for a $M_{\textsc{wd}} \sim 1.1M_{\odot}$ WD (see \tabref{WDcharacteristics}).
$T$ is the central WD temperature, which we conservatively estimate as $T\sim 1\,\text{keV}$ (a higher temperature increases the self-gravitating mass, and would weaken our constraints; see \sectref{limits}): \citeR{1971IAUS...42...97V} indicates that a $M_{\textsc{wd}} \sim 1M_{\odot}$ WD with an age in the range $\tau_{\textsc{wd}}\sim10^8$--$10^9$\,yrs has a luminosity $L_{\textsc{wd}}/L_{\odot}\sim10^{-2.25}$--$10^{-3.5}$ (see their Fig.~1, `with neutrinos'), corresponding to a core temperature in the range $T\sim10^{7}$--$10^{6.3}$\,K, or $T\sim 0.8$--$0.2$\,keV.

Note that $r_*$ also supplies an alternative natural cutoff to the logarithmic divergence in \eqref{tausink2}: even for $m_X \sim 10^{18}\,\text{GeV}$ (as massive a CHAMP as we consider; see \sectref{limits}), $r_* \sim 10^{-4}\,\text{m} \sim 10^{-11}R_{\textsc{wd}}$ (for $R_{\textsc{wd}} \sim 7\times 10^{-3}R_{\odot}$), yielding a logarithmic enhancement of the sinking time by a factor of $\log(r_*/R_{\textsc{wd}}) \sim 25$; the sinking time estimate \eqref{tausink2} for a CHAMP of this mass thus becomes \ $\sim 10^2\,\text{s}$.
On the other hand, for $m_X \sim 10^5\,\text{GeV}$, we have $\log(r_*/R_{\textsc{wd}}) \sim 10$, so the sinking timescale \eqref{tausink2} is $\sim 10^7\,$years.
In both cases, these timescales are sufficiently short, and are within a factor of $\sim 5$--$10$ of the shorter estimate \eqref{tausink4}.

This thermal-pressure-supported structure is stable as long as $\rho_X(0) \lesssim \rho_{\textsc{wd}}$; at densities above this, the CHAMP structure will begin to dominate the mass density at the center of the WD, and the CHAMP structure becomes unstable to a collapse mode in which the CHAMP configuration loses total energy, heats up, and contracts \cite{Gould:1989gw,Janish:2019nkk}, also known as the `gravothermal catastrophe' \cite{1968MNRAS.138..495L}.%
\footnote{\label{ftnt:gravothermal}%
		A virial population of $N$ self-gravitating particles with average total kinetic energy $\langle E_K \rangle$, and temperature $T \sim \langle E_K \rangle/N$ obeys the virial theorem $\langle E_K \rangle \sim - (1/2) \langle U \rangle$ where $\langle U \rangle \sim - GM^2/R$ is the average total potential energy of the particles if their total mass is $M$ and the characteristic radius of the configuration is $R$.
		The average total energy of the system is thus $\langle E \rangle \sim \langle E_K \rangle+\langle U \rangle \sim (1/2) \langle U \rangle \sim - \langle E_K \rangle \sim - NT \sim - GM^2/(2R)$.
		The configuration thus has negative heat capacity $dE/dT \sim - N$, so an energy decrease causes a temperature increase and a decrease in $R$ for fixed $M$, implying a contraction.
	} %
This implies a maximum stable mass for the core,
\begin{align}
M_{{\text{s.g.,}\,X}} &= \pi^{3/2} \rho_{\textsc{wd}} r_*^3
= \lb( \frac{3TM_{\text{Pl.}}^2}{2 m_X \rho_{\textsc{wd}}^{1/3} } \rb)^{3/2} \\
&\sim 6 \times 10^{-12}M_{\odot} \times\lb( \frac{T}{1\,\text{keV}} \rb)^{3/2} \nl \times \lb( \frac{ m_X }{10^5\,\text{GeV}} \rb)^{-3/2}\nl \times \lb( \frac{ \rho_{\textsc{wd}} }{ 5.5\times 10^{10}\,\text{g/cm}^3 } \rb)^{-1/2}
\label{eq:MthermCoreMass}
\end{align}
[i.e., \eqref{MthermCoreMass0} with $\rho_X(0)=\rho_{\textsc{wd}}$]; core masses above this value will collapse spontaneously (see discussion in following sections).

The timescale for this collapse could be limited either by the timescale for the CHAMPs to lose energy to the WD material to allow the collapse to occur, or by the drift time for the CHAMPs to sink inwards from $r_*$.
The energy transfer timescale is extremely short: assuming that $v_X \ll v_{\text{ion}}$, our discussion of energy loss in \sectref{PopnAccumulated} is applicable, and the energy loss rate for a single CHAMP can be estimated as $dE/dt \sim - \rho_{\text{ion}} \sigma v_X^2 v_{\text{ion}}$.
The characteristic timescale for the CHAMP to lose $\mathcal{O}(1)$ of its kinetic energy is thus $\tau_{\text{energy}} \sim m_X v_X^2 / (dE/dt) \sim m_X / (\rho_{\text{ion}} \sigma v_{\text{ion}}) \sim 1.5 \times 10^{-16}\,\text{s} \times (m_X / 10^5\,\text{GeV}) \times ( \sigma / 200\,\text{mb})^{-1}$, where we used the same cross section estimate as in \sectref{PopnAccumulated}; see also \citeR{Gould:1989gw,Janish:2019nkk}.

On the other hand, if we track a particle collapsing inward with the collapsing distribution of CHAMPs of fixed total mass $M_{{\text{s.g.,}\,X}}$ from $r_*$ to smaller distances, we find that the timescale for infall assuming the estimate of the diffusive sinking speed given by \eqref{wSink}%
\footnote{\label{ftnt:StillOKToUseSinkingTime}%
		This estimate is still appropriate to use in this context, provided that $m_X \gg \text{GeV}$, since the CHAMPs can still dominate the mass density in the core while being a trace element at the onset of the sinking.
	} %
with $g(r) = M_{{\text{s.g.,}\,X}} / (r^2 M_{\text{Pl.}}^2 )$ is given by
\begin{align}
\tau^{\text{core}}_{\text{coll.}} \sim \int_{R_f}^{r_*} \frac{dr}{w_X(r)} &= \frac{T M^2_{\text{Pl.}} }{ 3 \pi^{3/2}  \rho_{\textsc{wd}} m_X D } \lb[ 1 - \lb( \frac{R_f}{r_*} \rb)^3 \rb] \\
&\approx \frac{T M^2_{\text{Pl.}} }{ 3 \pi^{3/2}  \rho_{\textsc{wd}} m_X D } \quad [R_f \ll r_*] \\
&\sim 1.5\times 10^5\,\text{yr} \times \lb( \frac{m_X}{10^5\,\text{GeV}} \rb)^{-1},
\label{eq:tausinkCore}
\end{align}
indicating that the diffusive sinking time is by some orders of magnitude the limiting timescale for all masses of interest to us ($m_X \lesssim 10^{18}\,\text{GeV}$; see \sectref{limits}).
Note also that this is much shorter (by a factor of $\sim 25$) than the total diffusive sinking time \eqref{tausink4} for CHAMPs distributed throughout the WD volume, so if $M_X^{\text{prim.}} \gg M_{{\text{s.g.,}\,X}}$, the core shrinking process completes more rapidly than any remaining CHAMPs distributed throughout the WD in excess of $M_{{\text{s.g.,}\,X}}$ sink to the core.

We note in passing that if the energy released in the self-gravitating collapse is sufficient to heat the star above the trigger criteria, this is an opportunity for early destruction of the WD.

The endpoint of the self-gravitating collapse phase depends on the total mass of CHAMPs and the CHAMP mass; we survey the possibilities in the following subsections.

\subsection{Degeneracy supported CHAMP-contaminated WD structure: sub-Chandrasekhar case}
\label{sect:WDstructure}
Assuming that the self-gravitating collapse discussed in the previous subsection proceeds, the endpoint of that collapse depends on the total mass of CHAMPs present in the star, and also the CHAMP mass (see \sectref{WDstructureComment} for a detailed discussion of when the discussion in this section is applicable). 

We begin in this subsection by considering the case where degeneracy pressure can re-stabilize the collapse of the thermal structure, examining the impact on the mechanical structure of a WD and explaining important evolutionary changes in the CHAMP chemistry that occur.

\subsubsection{Positively charged CHAMPs: $X^+$}
\label{sect:WDstructurePlus}

We discuss first the $X^+$ case. 
At typical (non-extremal) CO WD core temperatures of $T\sim 10^7\,\text{K}\sim1\,\text{keV}$, the $X^+$ have thermal de Broglie wavelengths $\lambda_{\text{th.\,dB}} \equiv \sqrt{ 2\pi /( m_X T ) } \sim 2\times10^{-15}\,\text{m}\sqrt{ 10^5 \, \text{GeV} / m_X }$ which are much smaller than the typical inter-ion spacing $\delta \equiv ( 3 / (4\pi n) )^{1/3} \sim 2\times 10^{-13}\,$m.
Here, we have used the particle number density at the core of an $M\sim M_{\odot}$ WD, $n_{\text{ion}} \sim 3\times 10^{31}\,\text{cm}^{-3}$, neglecting that the $X^+$ themselves will impact the mechanical structure of the WD, increasing the central number density and decreasing $\delta$.
Nevertheless, these estimates indicate that the central density would have to be increased by a factor of $\sim 10^6 \times ( m_X / 10^5 \, \text{GeV} )^{3/2}$ over that for a normal WD before $X^+$ degeneracy begins to be a concern.
Such density increases are however not observed to occur in numerical solutions of WD structure (see \appref{WDstructure}): assuming that the CHAMPs form a stratified core structure inside the WD, the maximum supportable number density in that core structure scales up from the ordinary WD density as $n \propto m_X$ as the core becomes more and more extremal, causing the inter-$X^+$ spacing to decrease as $\delta \sim m_X^{-1/3}$.
Assuming $n\sim 3\times 10^{31}\,\text{cm}^{-3} \times (10^5\,\text{GeV}/m_C) \sim 3\times 10^{35}\text{cm}^{-3}$ implies that $\delta \sim 10^{-14}\,\text{m}$, so the de Broglie wavelength computed above is still smaller than the inter-ion spacing at the center of the denser core for $m_X = 10^5$\,GeV.
Moreover, because $\lambda_{\text{th.\,dB}}\propto m_X^{-1/2}$, which is faster than the inter-ion spacing decreases under these circumstances, this hierarchy persists to higher $m_X$, and we can thus safely take the $X^+$ themselves to be non-degenerate in the WD at all times.

Being charged, the $X^+$ are however electromagnetically tightly coupled to the highly degenerate electron plasma (for a typical $M\sim M_{\odot}$ CO WD, $\lambda_{\text{th.\,dB}}^e \sim 40\delta_e$), which thus supplies electron degeneracy pressure support to the $X^+$, in exactly in the same fashion as it does for the ordinary positive C and O ions in a WD.
The electron degeneracy pressure scales as $P_{\text{deg.}} \propto n_e^{4/3}$ in the relativistic limit,%
\footnote{\label{ftnt:PdegScaling}%
		The degeneracy pressure scales as $n_e^{5/3}$ for non-relativistic electrons, which increases parametrically even faster with increasing $n_e$; the electrons at the center of even a $M\sim M_{\odot}$ WD already have $E_F = 2.5m_e$, which is already into a fairly relativistic regime, where $E$ scales approximately linearly with increasing $p$.
	} %
but is independent of the mass of the heavy ion to which this pressure support is communicated.
At the onset of self-gravitating collapse, when the central mass fraction of the CHAMPs and CO mixture are approximately equal, $X_X\sim X_{\text{C/O}}$, the $X^+$ constitute only a small number fraction of the central ions: $x_X \ll x_{\text{C/O}}$ because $m_X \gg m_{\text{C/O}}$.
As such, the mean molecular mass per free electron [\eqref{MeanMolecularMassPerFreeElectron}] is almost unperturbed by the presence of the CHAMPs, and the electron density and spatial distribution in the WD is similarly undisturbed.
However, as the self-gravitating collapse proceeds, the number fractions of the CHAMPs and C/O ions eventually become comparable at the center of the WD.
At this point, the electron number density and radial distribution begin to respond, increasing the pressure support to the $X^+$ and, provided the total $X^+$ CHAMP contamination $M_{X^+}$ is sufficiently small (to be quantified below), re-stabilizing the $X^+$ configuration at a much smaller radius.
We expect that the endpoint of the self-gravitating collapse in this scenario is the formation of a core at the center of the WD comprised of the two-fluid $X^+$/$e^-$ mixture, with an overburden of the multi-fluid ${}^{12}$C/${}^{16}$O/$e^-$ mixture of the canonical CO WD, with a transition region between these two stratified layers whose thickness is dictated by thermal effects.%
\footnote{\label{ftnt:InTheAlternative}%
		Note in the alternative that if the thermal-pressure-supported structure does not form for any reason (see, e.g., footnote \ref{ftnt:IdealGas}), then the CHAMPs sinking out of the full WD will simply accumulate in the core, and give rise to this stratified structure directly, once sufficiently many CHAMPs are present. 
	} %

\begin{figure*}[t]
\includegraphics[width=0.95\textwidth]{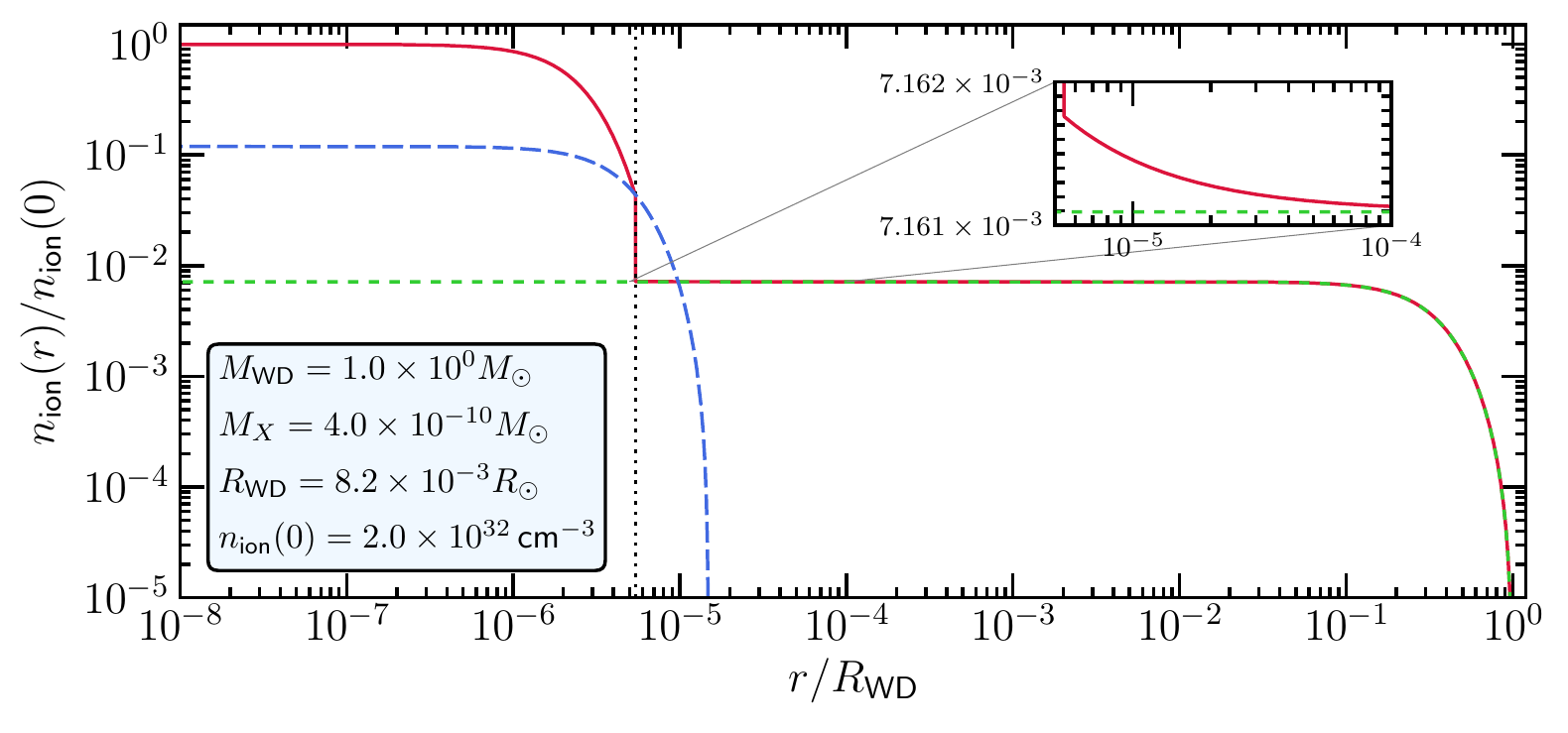}
\caption{\label{fig:XplusProfiles}%
		The number density of ions as a function of radius, normalized to the core number density ($n_{\text{ion}} \approx 2.0 \times 10^{32}\,$cm${}^{-3}$), of an $M = M_{\odot}$ pure-C WD with fractional $X^+$ contamination $\eta = 4\times 10^{-10}$, for an $m_X = 10^5\GeV$ CHAMP (red, solid curve).
		We assume as input that the $X^+$ have all settled to the core of the star (complete stratification), forming an inner core of pure-$X^+$ with mass $M_{X^+} \approx 0.8 M_{\text{Chand.},\, X}$, which is found to have radius $R_X \approx 5.4\times 10^{-6} R_{\text{WD}}$ (indicated by the vertical dotted black line); the whole star is found to have radius $R_{\text{WD}} = 8.2\times10^{-3}R_{\odot}$. 
		We ignore thermal effects, which would presumably mildly smooth out the discontinuity in ion number density at $r=R_X$ (see text).
		Also shown are (a) the ion number density profile for an equal-mass pure-C WD without $X^+$ contamination (green, short-dashed line), which is found to have an almost identical radius to the contaminated star, and (b) the ion number density for a pure $X^+$/$e^-$ WD of mass equal to the core mass (blue, long-dashed line).
		Note that, for ease of comparison, the latter two profiles are shown here normalized to the central density and radius of the $X^+$-contaminated star, and \emph{not} to their own central number densities.
		A comparison of these results indicates that, because $M_{X^+} \ll M_{\textsc{wd}}$, the $X^+$ have very little impact on the mechanical structure of the star outside the core (see inset).
		Moreover, because this particular core is not quite extremal, it does not well approximate the profile of the isolated, equal-mass, pure $X^+$/$e^-$ WD, as we expect a more extremal core would; see discussion in \appref{WDstructure}.
		}
\end{figure*}

We can roughly estimate the thickness $h$ of the transition region by setting the thermal kinetic energy of a CHAMP $E_{K}\sim (3/2)T$ equal to the gravitational potential energy gained upon rising a distance $h\ll R_{\text{core}}$ above the core boundary $\Delta E\sim h m_X M_{\text{core}} / (M_{\text{Pl}}^2 R^2_{\text{core}})$, leading to the estimate $h/R_{\text{core}} \sim (3/2) T R_{\text{core}}M_{\text{Pl.}}^2/\lb( m_XM_{\text{core}} \rb)$. 
For instance, with $m_X \sim 10^5$GeV, an $M_{\text{core}} \sim 4\times 10^{-10}M_{\odot}$ core (about 70\% of the \Chand\ mass for this CHAMP mass) with $R_{\text{core}} \sim 4.7\times 10^{-8}R_{\odot} \sim 30\,$m (see, e.g., \figref{XplusProfiles}) inside a $M_{\textsc{wd}} \sim M_{\odot}$ WD gives $h \sim 30\,\text{cm}\ll R_{\text{core}}$ for $T \sim 10$\,keV. 
If the core is near-extremal, $M_{\text{core}} \propto m_X^{-2}$, while $R_{\text{core}}$ decreases roughly as $m_X^{-4/3}$ (see \appref{WDstructure}), so it will always be the case that $h\ll R_{\text{core}}$ for near-extremal cores.
To give just one more explicitly computed set of values to verify this scaling, for $m_X \sim 10^{13}\,\text{GeV}$, an $M_{\text{core}} \sim 3.7\times 10^{-26}M_{\odot}$ core (about 70\% of the \Chand\ mass for this CHAMP mass) with $R_{\text{core}} \sim 4.6\times10^{-16} R_{\odot} \sim 3\times10^{-7}\,$m gives $h \sim 3\times10^{-9}\,\text{m}\ll R_{\text{core}}$ for $T \sim 10$\,keV. 
Since the transition layer is thin, we will ignore it and approximate the transition as immediate.

Note that the picture of near-complete stratification depends on (a) the very different charge-to-mass ratios \cite{1992A&A...257..534B} of the CO mixture and the $X^+$ (see \sectref{timescales}), and (b) the fact that the $X^+$ will `use up'  (for lack of better terminology) the ability of the electrons in the center of the WD to supply pressure support, making them unavailable to supply pressure support to the CO mixture, which must then be displaced upwards to maintain overall charge neutrality of the plasma (this is again in contrast to the neutron star case, where both the neutron star matter and the $X$ can co-exist without one displacing the other \cite{Gould:1989gw}).
Since the mean molecular weight per free electron $\mu_e$ [\eqref{MeanMolecularMassPerFreeElectron}]
of the $X^+$/$e^-$ fluid mixture is larger than for the CO mixture, this core will have a central ion number density higher than the central number density of the pure canonical CO WD (indeed, the extremal core [see below] central number density scales as $n_X \propto m_X$).
See \figref{XplusProfiles} for an example number density profile for such a cored WD, obtained by numerical solution of the Tolman--Oppenheimer--Volkoff (TOV) equation \cite{Tolman:1939it,Oppenheimer:1939cs} \emph{assuming as input} to the computation that complete stratification has occurred (see \appref{WDstructure}).
Note that the unphysical discontinuity in the ion number density at the core boundary shown in \figref{XplusProfiles} arises due to the continuity of the electron number density $n_e$ (and hence degeneracy pressure) at the interface, combined with the abrupt change of charge per ion from $+6$ for $r>R_X$ to $+1$ for $r<R_X$, which allows 6 times more ions to be present per unit volume just below the interface, compared to just above; this discontinuity would be smoothed by the thermal thickness of the transition region, which we have have approximated to be zero consistent with our estimates above.

While less extremal $X^+$/$e^-$ cores are `squashed' to some extent by the overburden of CO WD material (see, e.g., \figref[s]{XplusProfiles} and \ref{fig:massradius}), as the central core becomes more massive, it more and more closely approximates the physical structure of an isolated WD comprised of a pure $X^+$/$e^-$ mixture; see \appref{WDstructure} for a detailed discussion.
In particular, the stratified core has a maximum stable \Chand\ mass \cite{Chandrasekhar_1931}, which is equal to the \Chand\ mass for a isolated, pure $X^+$/$e^-$ WD: $M_{\text{Chand.},\, X^+} \approx 5.7 (\mu_{e}^{X^+})^{-2} M_{\odot} \ll M_{\odot}$, where $\mu_{e}^{X^+} = m_{X^+}/\mu_a \gg 1$; see again \appref{WDstructure} for a detailed technical argument as to why the \Chand\ mass scales in this fashion for $\mu_e \gg 2$.
Therefore, if $M_{X^+} > M_{\text{Chand.},\, X^+}$, the central core cannot be sufficiently pressure supported and, in the most conservative possible picture, will inevitably collapse to form a BH at the core of the WD.

\subsubsection{Negatively charged CHAMPs: $X^-$}
\label{sect:WDstructureMinus}
For sufficiently small total $X^-$ CHAMP contamination, $M_{X^-}$ (again to be quantified below), the $X^-$ case is na\"ively broadly similar to the $X^+$ case, but a detailed consideration of the putative stable structure that would form at the center of the WD in this case leads to another possible avenue for triggering WD runaway upon BH formation.

As we have already discussed, $X^-$ in the Universe take the form of $(NX)$, where inside a WD we can assume that $N$ is either ${}^{12}$C or ${}^{16}$O, in roughly equal fractions; see discussion in \sectref{behaviorPrimordial}.
For the first part of this subsection, we make the assumption that the $X^-$ remain immutably bound in such structures even when a CHAMP core forms inside a WD.
In the second part of this subsection, we will comment on the ions densities reached by such a configuration and whether modifications to this picture are required.

\paragraph{$X^-$ bound as $(NX)$}

Since the $(\text{C}X)$ and $(\text{O}X)$ bound states have charge-to-mass ratios that differ by only 40\%, it is not guaranteed that they will stratify within the lifetime of the WD in the same way as the CHAMP core stratifies from the CO overburden.
As such, we will assume that the $(\text{C}X)$ and $(\text{O}X)$ bound states form an approximately homogeneous mixture in the core of the WD.
This means that for the $(NX)$ case, we have the mean molecular mass per free electron [\eqref{MeanMolecularMassPerFreeElectron}]
\begin{align}
\frac{1}{\mu_{e}^{X^-}} = \frac{5X_{\text{C}}}{m_X/\mu_a} + \frac{7X_{\text{O}}}{m_X/\mu_a} \approx \frac{6\mu_a}{m_X} \approx \frac{6}{\mu_{e}^{X^+}},
\end{align}
where we assumed that the WD has composition $X_{\text{C}}\approx X_{\text{O}}\approx 0.5$.
This is approximately equivalent to assuming that the $X^-$ is bound to the `mean ion' with charge $\bar{Q} \sim 7$, implying $Q_{(NX)}=6$.
As such, the \Chand\ mass for the homogeneous $(NX)$ core at the center of the WD is $\approx 6^2=36$ times larger than that for an $X^+$ CHAMP of the same mass; essentially this can be understood by virtue of that fact that, on average, each electron in the core contributes pressure support for only $1/6$ of the mass of the CHAMP, as compared to the $X^+$ case, where the average electron supplies pressure support for the full CHAMP mass.

The mechanical structure of the $(NX)$ core is however largely similar to that shown in \figref{XplusProfiles}, except that the discontinuity in ion number density across the core boundary is reduced, since the average charge-per-ion in the CO mixture if 7, whereas that in the $(NX)$ mixture is 6.

We conclude that if $M_{X^-}$, the total mass of $X^-$ in the WD core exceeds the \Chand\ mass for the $(NX)$/$e^-$ fluid: $M_{\text{Chand.},\, X^-} \approx 5.7 (\mu_{e}^{X^-})^{-2} M_{\odot} \ll M_{\odot} $, where $\mu_{e}^{X^-} \approx m_{X^-}/(Q_{(NX)}\mu_a) \gg 1$ is the mean molecular weight per free electron for the $(NX)$/$e^-$ fluid and $Q_{(NX)} = 6$ is the `mean-ion charge', a central core of $(NX)$ cannot be sufficiently pressure supported and, in the most conservative possible picture, will inevitably collapse to form a black hole at the core of the WD; see also the discussion in \appref{WDstructure}.

We are also finally able to see why any limits we ultimately set would be conservative (see comment in \sectref{behaviorPrimordial}) even if we were to assume that the $X^-$ that enter the star bound to He were not to exchange onto a heavier nucleus: the \Chand\ mass for a $(\text{He}X)$ core is a factor of $Q_{(\text{He}X)}^2/Q_{(NX)}^2 = 6^2=36$ \emph{smaller} than the \Chand\ mass for an $(NX)$ core (assuming $Q_{(NX)}=6$ in the mean-ion approach), because (by the same argument as we have just made above for the $X^+$ state) the $(\text{He}X)$ state receives less pressure support than the $(NX)$ state while retaining approximately the same mass (in the limit $m_X \gg \mu_a$), all other things being equal.
As such, if we take as our limit criterion that a \Chand\ mass of CHAMPs are present (see \sectref{limits} for further discussion), then we are clearly conservative in assuming that the CHAMPs are all on the high-charge nuclei.
On the other hand, if%
\footnote{\label{ftnt:caveat7}%
		See footnote \ref{ftnt:IdealGas}.
	} %
the self-gravitating collapse is required to access this dense core structure per the discussion in \sectref{WDstructureThermal}, and the self-gravitating mass is larger than the \Chand\ mass [see discussion around \eqref{mXSG} below], then the charge of the ion in which the $X^-$ is bound is irrelevant anyway. 

\paragraph{High-sub-\Chand/trans-Chandrasekhar mass of $X^-$: the density of ions and pycnonuclear processes}
\label{sect:TransChandMinus}
The results of \appref{WDstructure} make clear that the central ion density of the fully stratified core of $(NX)$ material at the central of the $N$ star always exceeds both the central ion density of an equal-mass isolated pure $(NX)$ WD, and the central density of the uncontaminated pure $N$ star (see \figref{asymptote}); in particular, the central density of the core can increase to be some orders of magnitude above the central density of the CO star as the core becomes more and more extremal: $n_X^{\text{max}} \propto m_X$.
This however raises the possibility that, as the core mass becomes trans-\Chand, it is no longer self-consistent to compute the structure of the contaminated WD assuming the core is comprised immutably of $(NX)$ material.

Already in an ordinary CO WD, it is the case that three distinct outcomes can occur as the central density of the WD material increases \cite{Shapiro:2000abc}: (1) carbon-carbon fusion reactions can become so efficient as the core approaches the \Chand\ mass that the supernova instability is triggered before the \Chand\ mass is actually reached, (2) the core can exceed the \Chand\ mass and collapse (likely triggering the supernova instability as it collapses), or (3) the core can begin to neutronize, which modifies the equation of state.
In an ordinary CO WD, outcome (1) is believed to prevail at densities around $n_{\text{C}} \sim 5\times 10^{32} \,\text{cm}^{-3}$, although the central densities at which (2) and (3) occur---$n_{\text{C}}\sim 1.3\times 10^{33}\,\text{cm}^{-3}$ and $n_{\text{C}}\sim 2\times 10^{33}\,\text{cm}^{-3}$ \cite{Shapiro:2000abc}, respectively---are both within a factor of 3--4 of the central density at which (1) occurs \cite{Shapiro:2000abc}.

In our modified case, we have a similar concern: once the central ion density becomes too large, other processes may begin to occur that would disrupt the structure we have computed here assuming hydrostatic equilibrium and a fixed equation of state.
Similar to the ordinary CO WD, pycnonuclear fusion of carbon ions brought into close proximity by the increased density caused by the CHAMPs is the process most likely to disrupt the picture we have already outlined.
Obtaining an accurate rate for this process is challenging, as it occurs as a tunneling process between carbon nuclei, each individually bound to a CHAMP, within a larger ion lattice in which other nuclear species (e.g., O) are also present; moreover, electron screening enhancements of the fusion rate are also present.
In \appref{pycnonuclear} we develop a very simple estimate of when this process may become important, finding [c.f., \eqref{CXpyncoBoom}] that the density at which it becomes relevant is roughly $n_{[CX]}^{\text{boom}} \sim 0.2 \sqrt{ n_{\text{nucl}} n_{[C]}^{\text{boom}}}$, where $n_{\text{nucl}}\sim 2\times 10^{37}\,\text{cm}^{-3}$ is a nuclear number density, $n_{[C]}^{\text{boom}} \sim 5\times 10^{32}\,\text{cm}^{-3}$ is the number density at which pycnonuclear processes become important in ordinary CO WD material \cite{Shapiro:2000abc}, and $n_{[CX]}^{\text{boom}}\sim 10^{34}\,\text{cm}^{-3}$ is our estimate of when they become relevant in the $(NX)$ core of the WD. 

We thus see, in the trans-\Chand\ mass regime, the increasingly dense core may first undergo fusion processes that would trigger thermal runaway, destroying the WD before the core could even collapse to a BH.
However, as there is significant uncertainty in the estimate of the density at which these fusion processes first become relevant, we will in the following sections also proceed conservatively to discuss in detail the case in which the core \emph{will} proceed to collapse to a BH before it can be disrupted by fusion events (we must in any event discuss that case, as these pycnonuclear processes are inapplicable to the case of a $X^+$ core).
In this way, we will show that the inevitable outcome (with some minor exceptions in some regions of parameter space) is destruction of the WD, whether or not the pycnonuclear fusion process discussed here can trigger the runaway.

Note also that other processes would of course occur at higher central core densities still but, again, these would still lead to early destruction of the WD: for instance, by the time the ion density reaches nuclear density, fusion would be almost entirely unsuppressed, but there is additionally the consideration that, at such large densities, the $X^-$ would essentially no longer be individually associated with single positively charged nuclei, as the radii of the $(NX)$ bound states are of order the nuclear radius.
Instead, the nuclei $N$ may become delocalized in an $X^-$ lattice, akin to a metal structure.
In this case, if any $X^+$ are present, they would no longer be automatically prevented by a large Coulomb barrier from annihilating with an $X^-$, which could lead to an additional large energy deposition in the core, again increasing the likelihood of triggering the supernova runaway.

Finally, we comment that we have not included any Coulomb corrections to the EoS of the material in the WD, or in the CHAMP-contaminated core; at the extremely high charge densities implied by $n_X \propto m_X$, this may be a poor approximation, which adds additional uncertainty to any rate estimate we could give for any of these processes (it could not however modify the ultimate existence of a \Chand\ limit).

\subsubsection{Comment on applicability}
\label{sect:WDstructureComment}
We can now finally quantify our earlier caveats in \sectref[s]{WDstructurePlus} and \ref{sect:WDstructureMinus} that the discussion of the equilibrium structure in those preceding sections is applicable `[f]or sufficiently small total $X^{\pm}$ CHAMP contamination'.

We have now discovered that this condition is $M_{X^{\pm}} \leq M_{\text{Chand.,}\ X^\pm}$, where the \Chand\ mass here is the one applicable for the $X^\pm$, and $M_{X^\pm}$ is the total mass of $X^\pm$ in the stratified core at the time $t$ at which the structure is considered.
In particular, this discussion definitely holds if $M_{X^{\pm}}^{\text{prim.}} + M_{X^{\pm}}^{\text{accum.}}(\tau_{\text{accum.}}) < M_{\text{Chand.,}\ X^\pm}$.

To understand where else the discussion might hold, consider that there is a CHAMP mass $m_{X}^{\text{s.g.}}$ such that for
\begin{align}
m_X \gtrsim m_X^{\text{s.g.}} &\approx 0.85 \pi \frac{\rho_{\textsc{wd}}}{T^3} Q_X^4 \\
&\sim 6.4\times 10^8\,\text{GeV} \times Q_X^4 \times \lb( \frac{T}{1\,\text{keV}} \rb)^{-3} \nl
\times \frac{\rho_{\textsc{wd}}}{5.5\times 10^7\,\text{g/cm}^3}
\label{eq:mXSG}
\end{align}
(with the numerical coefficient found using the results of \appref{WDstructure}), the mass required for the initial thermal-pressure-supported structure%
\footnote{\label{ftnt:caveat1}%
		If it forms; see footnote \ref{ftnt:IdealGas}.
	} %
 to undergo self-gravitating collapse exceeds the \Chand\ mass for the same $m_X$: i.e., $M_{\text{s.g.},X} (m_X \gtrsim m_{X} ^{\text{s.g.}}) \geq M_{\text{Chand.},X^{\pm}} (m_X \gtrsim m_{X} ^{\text{s.g.}})$.
However, the discussion of the core structure in this subsection can still hold instantaneously for this case, as the timescale for the thermal-pressure-supported structure to collapse is reasonably long [\eqref{tausinkCore}] compared to the sound-crossing time of the core (which sets the dynamical timescale over which significant structural alterations can be compensated for), and a sub-Chandrasekhar core will still form at some point during this collapse, before the accretion of the additional mass from the thermal-pressure-supported structure onto the core drives the core mass over the \Chand\ limit; see \sectref{BHinWD} for further discussion of the subsequent dynamics.

Therefore, the discussion in \sectref[s]{WDstructurePlus} and \ref{sect:WDstructureMinus} also holds instantaneously for times $t<t_*$ where $t_*$ is the time at which the mass of $X^\pm$ in the stratified core equals the \Chand\ mass, $M^{\text{deg.}}_{X^{\pm}}(t_*)= M_{\text{Chand.,}\ X^\pm}$, under the following circumstances: (1) $m_X < m_X^{\text{s.g.}}$, and $M_{X^{\pm}}^{\text{prim.}} > M_{\text{Chand.,}\ X^\pm}$ [implying $t_* <  \tau_{\text{sink}}$]; and (2) $m_X < m_X^{\text{s.g.}}$, and $M_{X^{\pm}}^{\text{prim.}} < M_{\text{Chand.,}\ X^\pm}$, and $M_{X^{\pm}}^{\text{prim.}} + M_{X^{\pm}}^{\text{accum.}}(\tau_{\text{accum.}}) > M_{\text{Chand.,}\ X^\pm}$ [implying that $\tau_{\text{sink}} \lesssim t_* \lesssim \tau_{\text{accum.}}$].

It additionally holds for $t_{\text{s.g.}} \lesssim \tau_{\text{strat.}} < t<t_*$ where $t_{\text{s.g.}}$ is the time for the onset of self-gravitating collapse, $\tau_{\text{strat.}}$ is the time for the formation of a stratified core during the self-gravitating collapse, and $t_*$ is as above, under the following circumstances: (3) $m_X > m_X^{\text{s.g.}}$, and $M_{X^{\pm}}^{\text{prim.}} > M_{\text{s.g.,}\ X} > M_{\text{Chand.,}\ X^\pm}$ [implying $t_* <  t_{\text{s.g.}} + \tau_{\text{coll.}}^{\text{core}} \lesssim \tau_{\text{sink}}$]; and (4) $m_X > m_X^{\text{s.g.}}$, and $M_{X^{\pm}}^{\text{prim.}} < M_{\text{s.g.,}\ X}$ [implying $t_{\text{s.g.}} > \tau_{\text{sink}}$], and $M_{X^{\pm}}^{\text{prim.}} + M_{X^{\pm}}^{\text{accum.}}(\tau_{\text{accum.}}) > M_{\text{s.g.,}\ X} > M_{\text{Chand.,}\ X^\pm}$ [implying $t_* <  t_{\text{s.g.}} + \tau_{\text{coll.}}^{\text{core}} \lesssim \tau_{\text{accum.}}$].

\subsubsection{Combined $X^+$ and $X^-$ case}
\label{sect:combinedCase}
Absent a near-complete charge asymmetry between $X^+$ and $X^-$ (which could occur in some production mechanisms, or if all but a small residual charge asymmetric population of CHAMPs annihilates away), it is highly likely that the physical case will correspond to a WD that is contaminated by some abundance of both $X^+$, and $X^-$ bound as $(NX)$. 
As we have discussed in \sectref{CHAMPchemistry}, the large Coulomb barrier between $X^+$ and the positively charged $(NX)$ state will prevent the $X^+$ and $X^-$ annihilating, so both species can co-exist in the core.
This leads to two possible outcomes: (1) the core is a roughly homogenous mixture of $X^+$ and $(NX)$, or (2) the $X^+$ and $(NX)$ stratify, with the $X^+$ sinking to the core (smaller charge to mass ratio; see \sectref{timescales}).
To estimate which occurs, let us return to a timescale estimate for the sinking of $X^+$ through a $(NX)$ fluid similar to that we made in \sectref{timescales}.
Assuming $n_{X^+} \sim n_{(NX)} \sim 0.5$ as an approximation, $Q_{(NX)} = \bar{Q}_N - |Q_X|$ with $|Q_X| = 1$ and $\bar{Q}_N \sim 7$ in the mean-ion approach, and $m_X \gg \text{GeV}$, the sinking terminal velocity for the $X^+$ in the core is given by \cite{GarciaBerro:2007an,Deloye:2002xb}
\begin{align}
w_{X^+}^{\text{core}}(r) \sim \frac{g(r) D m_X}{T} \lb( 2 \frac{|Q_X|}{\bar{Q}_N} - 1 \rb) \sim - \frac{5}{7} \frac{g(r) D m_X}{T},
\label{eq:wXcore}
\end{align}
where $g(r) = M_{\text{enc}}(r)/(rM_{\text{Pl.}})^2$ with $M_{\text{enc}}(r) \approx M_{\text{core}} (r/R_{\text{core}})^3$ approximated as for a uniform density spherical core, and $D$ approximated by the self-diffusion coefficient \eqref{SelfDiffusionCoefficient}.
Note however that in evaluating $D$ we  must now take $m_{\text{ion}} \sim m_X$ and $Z_{\text{ion}} \sim 0.5 Q_{(NX)} + 0.5 |Q_{X}| \sim 3.5$ as appropriate for an initially homogeneous mixture in the core of $X^+$ and $(NX)$ with equal number densities through which the $X^+$ sink.
Assuming $M_{\text{core}}$ and $R_{\text{core}}$ to be the mass and radius of a near-extremal `mini-WD' degenerate core [see \eqref{MChand} and (\ref{eq:RChand})], we find that for $T\sim \text{keV}$ we have a sinking timescale of [$r_X(t) = r_X(0) \exp(-t/\tau_{X^+}^{\text{sink}})$]
\begin{align}
\tau_{X^+}^{\text{sink}} \sim 3\,\text{yr} \times \lb( \frac{m_X}{10^5\,\text{GeV} } \rb)^{-17/9} \times \lb( \frac{T}{\text{keV}} \rb)^{-1/3}.
\label{eq:tauXcore}
\end{align}
This estimate of the stratification timescale indicates that an initially mixed core should stratify almost immediately over most of our parameter space; in particular, this estimate is faster than the thermal core collapse time \eqref{tausinkCore} [relevant if $M_{\text{s.g.},\,X}>M_{\text{Chand.}}$] or the sinking time \eqref{tausink4}.
The equilibrium structure in this case is of course more complicated than either of the $X^+$ or $(NX)$ cores alone, consisting now of three nested spherical shells. 
However, since the \Chand\ mass for $X^+$ is a factor of $\sim 36$ lower than for $(NX)$, it will a reasonable approximation to estimate the maximum mass of the $X^+$ core to be that of the isolated $X^+$ core in most cases unless the total $(NX)$ mass exceeds $\sim 36$ times the $X^+$ mass.
Likewise, when the total $X^+$ mass is much below $\sim 1/36$ times the total $(NX)$ mass, it is likely a good approximation to ignore the central $X^+$ core and estimate the maximum allowed $(NX)$ mass as one would for the isolated $(NX)$ core.
In the intermediate regime where the total mass of $(NX)$ is roughly equal to 36 times the mass of the $X^+$, a more complicated situation will arise, but this is a tuned region of parameter space, and the allowed total mass will be at most an $\mathcal{O}(1)$ factor different from the maximum allowed $(NX)$ mass in this case.

\subsection{Super-\Chand\ mass of CHAMPs: BH formation, and timescales}
\label{sect:superChandCHAMPs}
As noted in \sectref{WDstructureComment}, our discussion thus far has been largely limited to the impact of a total mass of CHAMPs in the stratified core of the WD, $M_{X^\pm}^{\text{core}}$, that is sub-\Chand\ at all times (with the exception of our comments in \sectref{TransChandMinus}, which considered the case of $M_{X^\pm}^{\text{core}}$ in the high-sub-\Chand/trans-\Chand\ mass range; and our comments in \sectref{WDstructureComment}).
However, if a super-\Chand\ mass of CHAMPs is able to collect in the stratified core, the quiescent structure discussed in \sectref{WDstructure} will not be the equilibrium structure (or even the instantaneous quasi-static structure).
Instead, provided either that we consider $X^+$, or the considerations of \sectref{TransChandMinus} are inapplicable for $X^-$ in the high-sub-\Chand\ regime, the general relativistic instability to collapse of a sufficiently dense central core structure implies that once the sinking CHAMPs accumulate $\sim$ a Chandrasekhar mass in the stratified core, that core will collapse to a BH with an initial mass equal to the relevant \Chand\ mass.
Note however that in order to obtain such a massive stratified core, a larger total mass of CHAMPs than the \Chand\ mass may%
\footnote{\label{ftnt:caveat2}%
		See again footnote \ref{ftnt:IdealGas}.
	} %
be required to be in the star if $m_X \gtrsim m_X^{\text{s.g.}}$, so that self-gravitating collapse of the CHAMPs can occur and give rise to the stratified structure (see \sectref[s]{WDstructureThermal} and \ref{sect:WDstructureComment}). 
In either case, this abundance of CHAMPs could either be present primordially, or could accrete onto the WD before the crystallization time. 

The timescale for formation of the BH during the collapse of a trans-\Chand\ extremal core initially supported by degeneracy pressure is extremely short, of the order of the gravitational free-fall timescale (see, e.g., the discussion in \S36.3.1 of \citeR{Kippenhahn:2012zqe}): 
\begin{align}
\tau_{\text{collapse.}} &\sim \sqrt{ \frac{M_{\text{Pl.}}^2}{M_{\text{Chand.}}/R_{\text{Chand.}}^3} }\\
&\sim 1.5\, \mu\text{s} \times Q_X \times \frac{10^5\,\text{GeV}}{m_X}
\label{eq:tCollapse}
\end{align}
where $Q_X$ is the charge of the object in which $X$ appears in the WD [$Q_X = Q_{X^+}$ for $X^+$; $Q_X = Q_{(NX)}$ for $X^-$].

Note that this collapse provides another possible avenue to trigger thermal runaway: the gravitational binding energy released by the collapse can heat the WD.
Suppose we only track the collapse as far as when the core has collapsed to half its initial radius, at which point energy of order of its initial binding energy has been released by the collapsing CHAMPs. 
The energy can be estimated to be 
\begin{align}
E_B^{\text{init.}} &\sim \frac{M_{\text{Chand.}}^2 }{M_{\text{Pl.}}^2 R_{\text{Chand.}}} \\
&\sim  8 \times 10^{41}\,\text{GeV} \times Q_X^{8/3} \times \lb( \frac{10^5\,\text{GeV}}{m_X} \rb)^{8/3}
\label{eq:EBdinWDcore}
\end{align}
If a fraction of this energy could somehow converted to heating of the WD, thermal runaway may possibly be triggered.
Assuming that any such energy released would be deposited within a volume parametrically of size the trigger volume, a na\"ive criterion for the runaway to be initiated is that 
\begin{align}
E_{\text{deposited}} \sim \zeta \frac{ E_B^{\text{init.}} }{ \tau_{\text{collapse.}} } \min\lb[ \tau_{\text{diff.}} , \tau_{\text{collapse.}} \rb] \gtrsim E_T,
\label{eq:Edeposited}
\end{align}
where $\zeta \leq 1$ controls the fraction of available energy that is deposited in the WD as heating.
For a $M_{\textsc{wd}} \sim 1.1M_{\odot}$ WD, we find that for $m_X \gtrsim 5\times 10^{10} \, \text{GeV} \times Q_X$, we have $\tau_{\text{diff.}}\gtrsim \tau_{\text{collapse}}$, so that $\zeta \gtrsim 4\times 10^{-7} \times Q_X^{-8/3} \times \lb( {m_X} / {5\times 10^{10}\,\text{GeV}}\rb)^{8/3}$ is required, which is possible with $\zeta \lesssim 1$ for $m_X \lesssim 10^{13} \, \text{GeV} \times Q_X$.
For $m_X \lesssim 5\times 10^{10} \, \text{GeV} \times Q_X$, we have $\tau_{\text{diff.}}\lesssim \tau_{\text{collapse}}$, and we find that we need $\zeta \gtrsim 4\times 10^{-7} \times Q_X^{-5/3} \times \lb( {m_X} / {5\times 10^{10}\,\text{GeV}}\rb)^{5/3} < 4\times 10^{-7} \times Q_X^{-5/3}$.
We therefore see that, for $m_X \lesssim 10^{13} \, \text{GeV} \times Q_X$, there is a possibility of that some fraction $\zeta \leq 1$ (and possibly $\zeta \ll 1$) of this initial gravitational binding energy release being deposited in the WD and leading to heating would be sufficient to trigger thermal runaway.
Of course, our estimates here are schematic and approximate, although they are conservative in the sense that further energy is released as the core continues to collapse.
All we intend to argue here is that there is yet another plausible alternative WD destruction mechanism that could be operative for some range of CHAMP masses, even before the BH is reached; see also \citeR[s]{Acevedo:2019gre,Janish:2019nkk} for similar arguments in a different context.
Of course, if no such mechanism exists to cause a concomitant heating of the WD, the core simply collapses to a BH without possibly triggering thermal runaway.

Once the BH forms at the \Chand\ mass, its subsequent dynamics and ultimate fate depend on a number of different physical processes.
Moreover, the ultimate fate of the WD in which the BH forms also depends on which of the various BH dynamical mechanisms dominate at various points in the BH evolution.
In \sectref{BHinWD}, we undertake a detailed consideration of these points.

\section{Black holes in white dwarfs}
\label{sect:BHinWD}
In \sectref{superChandCHAMPs}, we noted that if a sufficient mass of CHAMPs is present primordially or can accrete onto the WD within the WD crystallization time $\tau_{\text{accum.}}$, a BH is born at the center of the WD. 
Here, a `sufficient mass' means the larger of the \Chand\ mass and the self-gravitating mass: gravothermal collapse is necessary to form a degenerate core that can later be pushed over the \Chand\ limit in order to trigger the gravitational collapse instability required to give birth to a BH.%
\footnote{\label{ftnt:caveat3}%
		Although see again the caveat at footnote \ref{ftnt:IdealGas}.
	} %
This holds true even in the case where the self-gravitating mass exceeds the \Chand\ mass, because the thermal structure discussed in \sectref{WDstructureThermal} takes some time to fully collapse, and a degenerate core must be born during that collapse, even if it is short lived.
We will assume throughout that the BH evolution begins with a BH born at the \Chand\ mass at time $t_{\textsc{bh}}$, but that various phases of evolution of the BH dynamics can follow.
In this section, we turn to an examination of the dynamics of the BH, and the implications for old WD. 

We begin in \sectref{BHdynamicalProcesses} by examining in detail the various dynamical processes that govern the BH mass evolution after BH formation, under various assumptions.
Thus armed with an understanding of the relevant dynamical processes, we turn in \sectref{BHevolution} to a consideration of the temporal evolution of the BH mass under the action of these dynamical contributions, mapping out how the BH mass evolves qualitatively (and, where necessary, quantitatively) in various regions of parameter space.
Finally, in \sectref{outcomes}, we turn to a consideration of the impact of the BH evolution on the fate of old WD.

Our discussion of the BH dynamics and WD outcomes is guided by \citeR[s]{Acevedo:2019gre,Janish:2019nkk}, which considered the trigger mechanism for SNIa-like supernova events in WD due to the formation of a BH inside the WD (or, more conservatively, for destruction of the WD by accretion onto the BH), while our discussion of WD outcomes draws also on a longer series of prior studies in this space \cite{Timmes_1992,Graham:2015apa,Bramante:2015cua,Graham:2018efk,Montero-Camacho:2019jte,Acevedo:2019gre,Janish:2019nkk}.
We note however that all numerical or quantitative estimates here are performed independently of past work to the extent this is possible.

Finally, we note that some of the accretion rate estimates we develop in this section are quite approximate models for the complex accretion dynamics; where relevant, we note where our physical conclusions depend on details of uncertain estimates.
Resolution of these uncertainties would likely require numerical modeling.

\subsection{Dynamical processes governing BH evolution in a WD}
\label{sect:BHdynamicalProcesses}
There are three possible dynamical processes that impact the evolution of the BH mass:
(1) Hawking radiation by the BH, assuming that this process occurs in the usual fashion it is believed to occur (see, e.g., \citeR{Kaplan:2018dqx} for one alternative viewpoint); 
(2) Bondi (or Eddington) accretion of CO WD material from the core of the WD; and 
(3) accretion of CHAMPs from the WD, which occurs at various rates throughout the BH evolution, depending on a variety of criteria.
In this subsection, we consider each contribution in turn.

\subsubsection{Hawking radiation from the BH}
\label{sect:Hawking}
Once the BH forms, it is widely believed that there will be a dynamical contribution to its evolution that will cause it to lose mass: the Hawking process \cite{Hawking:1974wb}.
This process causes the BH event horizon (which for a non-rotating BH lies at the Schwarzschild radius, $R_{\text{S}} = 2M_{\textsc{bh}}/M_{\text{Pl.}}^2$) to emit a thermal spectrum of particles at a temperature given by $T_H=M_{\text{Pl.}}^2/(8\pi M_{\textsc{bh}})$.
The Hawking mass loss rate is given approximately by
\begin{align}
\dot{M}_{\textsc{bh},\,\text{H}} &\sim - \frac{\pi^2}{60} \frac{g_{\text{eff.}}(T_H)}{2} T_H^4 (4\pi R_S^2) \\
&= - \frac{ (g_{\text{eff.}}(T_H)/2)M_{\text{Pl.}}^4}{15360\pi M_{\textsc{bh}}^2},
\label{eq:MdotHawking}
\end{align}
where $g_{\text{eff.}}(T)$ is the effective number of relativistic degrees of freedom at temperature $T$. 

If the Hawking radiation completely dominates the BH evolution, the BH mass will decrease, and the timescale to radiate away the entirety of the BH mass is finite; it is dominated by the time spent at the largest masses \cite{Janish:2019nkk}.
Making the conservative assumption that $g_{\text{eff}}(T_H)$ remains constant at some initial value $g_{\text{eff},\,0}$,%
\footnote{\label{ftnt:geffConservative}%
		 This is conservative because as the BH mass drops, the Hawking temperature increases, enabling the relativistic participation of more SM species, increasing $g_{\text{eff.}}$ and the rate of emission.
	} %
the timescale can be estimated by
\begin{align}
\tau_{\textsc{bh},\, \text{H}} \sim \frac{15360\pi }{3 (g_{\text{eff},\,0}/2) M_{\text{Pl.}}^4} M_{\textsc{bh},\, 0}^3,
\label{eq:tauHawking}
\end{align}
where $M_{\textsc{bh},\, 0}$ is the initial mass.
Numerically, we find
\begin{align}
\tau_{\textsc{bh},\, \text{H}} \sim 3.4\,\text{Gyr} \times \lb( \frac{M_{\textsc{bh},\, 0}}{10^{-19}M_{\odot}} \rb)^3,
\label{eq:tauHawkingNum}
\end{align}
where we took $g_{\text{eff},\,0}\approx 51/4$, as appropriate for temperatures $T_H\lesssim \text{MeV}$.

\subsubsection{Accretion of WD material onto the BH}
\label{sect:Bondi}

The BH will also accrete CO matter from the surrounding WD.
We assume that the accretion rate of WD matter on the BH is given (at least initially; see discussion below) approximately by the Bondi accretion rate \cite{Bondi_1952,Shapiro:2000abc}:
\begin{align}
\dot{M}_{\textsc{bh},\,\text{B}} \sim + 4\pi \lambda \lb( \frac{M_{\textsc{bh}} }{M_{\text{Pl.}}^2c_s^2} \rb)^2 \rho_{\textsc{wd}} c_s,
\label{eq:BondiAccretion}
\end{align}
where $\lambda \sim \mathcal{O}(1)$, $c_s$ is the WD sound speed, and $\rho_{\textsc{wd}}$ is the WD density; the latter two quantities are as evaluated at the center of the unperturbed WD (i.e., far from the BH event horizon, but still near the center of the WD).
Parametrically, this estimate follows, up to $\mathcal{O}(1)$ numerical factors, by assuming that particles crossing the sonic radius $R_{\text{sonic}} \sim R_{\textsc{s}}/c_s^2 \sim GM_{\textsc{bh}} / c_s^2$ at which the escape speed from the BH is equal to the sound speed in the (unperturbed) WD will continue to accrete onto the BH.

If Bondi accretion completely dominates the BH evolution, the BH grows, and the timescale for it to accrete up to the full WD mass (assuming nothing cuts this evolution off) is again finite; it is dominated by the time spent at the smallest masses \cite{Janish:2019nkk}.
The timescale for the BH to accrete from an initial mass $M_{\textsc{bh},\,0}$ to the full mass of the WD, assuming the Bondi accretion rate holds throughout, is given approximately by
\begin{align}
\tau_{\textsc{bh},\, \text{B}} \sim \frac{c_s^3M_{\text{Pl.}}^4}{4\pi\lambda \rho_{\textsc{wd}}} \lb[ M_{\textsc{bh},\, 0}^{-1} - M_{\textsc{wd}}^{-1} \rb],
\label{eq:tauBondi}
\end{align}
where $\lambda$ is the same $\mathcal{O}(1)$ constant as in \eqref{BondiAccretion}, $M_{\textsc{bh},\, 0} \ll M_{\odot}$ is the initial BH mass, and $M_{\textsc{wd}} \sim M_{\odot}$ is the WD mass.
Taking parameters appropriate for the center of a $M_{\textsc{wd}} \sim 1.1M_{\odot}$ WD (and treating $\rho_{\textsc{wd}}$ as constant, which is a reasonable approximation as the timescale is dominated by the time spent at the smallest masses for the BH, when the WD is mostly unperturbed), we find numerically that
\begin{align}
\tau_{\textsc{bh},\, \text{B}} \sim 3.1 \,\text{Gyr}  \times \lb( \frac{ M_{\textsc{bh},\, 0} }{ 10^{-18} M_{\odot} } \rb)^{-1}.
\label{eq:tauBondiNum}
\end{align}

It is also worth noting the mass at which the Bondi and Hawking rates balance (ignoring any CHAMP contribution; see below):
\begin{align}
M_{\textsc{bh}}^{\text{B}\sim\text{H}} &\sim \lb[ \frac{ (g_{\text{eff.},\,0}/2)M_{\text{Pl.}}^8 c_s^3 }{ 61440 \pi^2 \lambda \rho_{\textsc{wd}} } \rb]^{1/4}\\
&= 1.5\times 10^{38}\,\text{GeV} \times \lb( \frac{c_s}{2.8\times 10^{-2}} \rb)^{3/4} \nl \times \lb( \frac{g_{\text{eff.},\,0}}{51/4} \rb)^{1/4} \times  \lb( \frac{\lambda}{1} \rb)^{-1} \nl \times \lb( \frac{\rho_{\textsc{wd}}}{5.5\times 10^7\,\text{g/cm}^3}\rb)^{-2}
\end{align}
where we assumed fiducial values for a $M_{\textsc{wd}} \sim 1.1M_{\odot}$ WD (see \tabref{WDcharacteristics}).
Assuming those values, we have $M_{\textsc{bh}}^{\text{B}\sim\text{H}}  \sim 1.3\times 10^{-19}M_{\odot}$, which is interestingly \emph{and coincidentally} a region where the intrinsic growth and evaporation timescales for the BH happen to be about as long as current old WD ages, $\sim \text{few}\times\text{Gyr}$.

\paragraph{Eddington-limited accretion}
\label{sect:Eddington}
However, accretion can become (Eddington) limited if radiative backreaction on the in-falling matter in the region near the sound horizon supplies sufficient pressure support to re-establish hydrostatic equilibrium conditions near the sound horizon \cite{Eddington:1926abc,Kippenhahn:2012zqe,Shapiro:2000abc,Gould:1989gw,Janish:2019nkk}.
The Eddington luminosity is given by $L_{\text{edd}} = 4\pi M_{\textsc{bh}} /( \kappa_{\text{rad}}M_{\text{Pl.}}^2)$, where $\kappa_{\text{rad}} = \sigma / m_{\text{ion}}$ is the radiative opacity with $\sigma$ the photon scattering cross section with the stellar matter.
Following \citeR{Janish:2019nkk}, we will assume that $\sigma = \max\lb[ 1\,\text{mb} , 100\,\text{mb}\times (\omega/\text{MeV})^{-1} \rb]\gtrsim \sigma_* \equiv 200\,$mb for $\omega \lesssim 0.5$\,MeV ($\sim T_{\text{crit.}}$).

Assuming that a fraction $\epsilon$ of the in-falling matter is converted into outgoing radiation, the Eddington-limited mass accretion rate is 
\begin{align}
\dot{M}_{\textsc{bh},\,\text{E}} \sim 4\pi M_{\textsc{bh}} m_{\text{ion}} / (\epsilon \sigma M_{\text{Pl.}}^2).
\label{eq:MdotE}
\end{align}

Equating this to the Bondi accretion rate, we estimate that accretion becomes Eddington limited for
\begin{align}
M_{\textsc{bh}}^{\text{B}\sim\text{E}} &\sim \frac{c_s^3 m_{\text{ion}} M_{\text{Pl.}}^2}{\lambda \epsilon \sigma \rho_{\textsc{wd}}} \\
&\sim 3.5\times 10^{42} \,\text{GeV} \times \lb( \frac{c_s}{2.8\times 10^{-2}} \rb)^3 \times \lb( \frac{\epsilon}{0.1} \rb)^{-1} \nl \times  \lb( \frac{\lambda}{1} \rb)^{-1} \times \lb(  \frac{\sigma}{200\,\text{mb}} \rb)^{-1} \nl
\times \lb( \frac{\rho_{\textsc{wd}}}{5.5\times 10^7\,\text{g/cm}^3}\rb)^{-2},
\end{align}
where we assumed fiducial values for a $M_{\textsc{wd}} \sim 1.1M_{\odot}$ WD (see \tabref{WDcharacteristics}).
Note that, assuming the fiducial values, $M_{\textsc{bh}}^{\text{B}\sim\text{E}} \sim 3 \times 10^{-15} M_{\odot}$.

For $M_{\textsc{bh}} \lesssim M_{\textsc{bh}}^{\text{B}\sim\text{E}}$, the accretion of WD material is at the Bondi rate; for $M_{\textsc{bh}} \gtrsim M_{\textsc{bh}}^{\text{B}\sim\text{E}}$ the accretion is Eddington limited.
We take 
\begin{align}
\dot{M}_{\textsc{bh},\,\text{WD material}} = \min \lb[  \dot{M}_{\textsc{bh},\,\text{B}} , \dot{M}_{\textsc{bh},\,\text{E}} \rb]
\label{eq:MdotWDmat}
\end{align}

Note however that the Eddington-limited rates will be essentially irrelevant for most of our considerations, assuming fiducial values for a $M_{\textsc{wd}} \sim 1.1M_{\odot}$ WD. 
Even ignoring any (necessarily positive) CHAMP contribution to $\dot{M}_{\textsc{bh}}$, the mass accretion rate onto the BH assuming Bondi accretion of WD material already exceeds the Hawking rate in magnitude at a BH mass three orders of magnitude smaller than the mass at which the Bondi rate becomes Eddington limited.
The impact of the Eddington rate limitation is thus only felt deep into the regime where the BH is rapidly accreting matter from the WD: the timescale to accrete to the full WD mass from $M_{\textsc{bh}}^{\text{B}\sim\text{E}}$ is $\sim 3\,\text{Myr}$ assuming Bondi accretion, while the cognate estimate for the Eddington-limited accretion rate is 
\begin{align}
\tau_{\textsc{bh},\, \text{E}} &\sim \frac{ \epsilon \sigma M_{\text{Pl.}}^2}{4\pi m_{\text{ion}} } \ln\lb[ \frac{M_{\textsc{wd}}}{M_{\textsc{bh}}^{\text{B}\sim\text{Edd.}} } \rb] \\
&\sim 33\,\text{Myr}\times \lb( \frac{\epsilon}{0.1} \rb) \times \lb( \frac{\sigma}{200\,\text{mb}} \rb) \nl\times \lb( \frac{\rho_{\textsc{wd}}}{5.5\times 10^7\,\text{g/cm}^3}\rb).
\end{align}
While the Eddington accretion timescale here is $\mathcal{O}(10)$ times longer than the Bondi timescale, the absolute timescale is still extremely short, $\tau_{\textsc{bh},\, \text{E}} \ll \,\text{Gyr} \sim \tau_{\textsc{wd}}$, so this small additional amount of evolution time can be neglected completely in what follows.
These conclusions agree with those of \citeR{Janish:2019nkk}.

\subsubsection{CHAMP accretion onto the BH. Case I: $M_{\text{s.g.,}\,X} < M_{\text{Chand.,}\,X}$.}
\label{sect:CHAMPRainCase1}
It is our assumption throughout this paper that the initial CHAMP distribution in the WD upon formation is homogeneous (see later comments in \sectref{UniformComment}).
These CHAMPs will however begin to sink toward the center of the WD after WD formation, as discussed in \sectref{timescales}.
Our best estimate for the timescale for the sinking of CHAMPs from the outskirts of the WD to the core is given by \eqref{tausink4}, which we will approximate here as 
\begin{align}
\tau_{\text{sink}}\sim 4\times 10^6\,\text{yr} \times \frac{ 10^5\,\text{GeV}}{m_X},
\label{eq:tausinkapprox}
\end{align}
ignoring the small logarithmic correction.

In this subsection we will assume that we are in the regime where the self-gravitating mass of the initial thermal structure (if any) of CHAMPs formed at the core of the WD $M_{\text{s.g.,}\,X}$ has a mass smaller than the \Chand\ mass.
If this is the case, the mass contained in the self-gravitating core collapses to a stratified core object below the \Chand\ mass in the time \eqref{tausinkCore}, which is somewhat shorter than $\tau_{\text{sink}}$ or $\tau_{\text{accum.}}$.
We thus assume that this pre-BH evolutionary phase completes before the BH could form, so there is no diffuse overdense CHAMP structure at the center of the WD to feed the BH accretion at the time of BH formation.

Consider first the CHAMPs present in the WD primordially.
The rate at which these CHAMPs accrete onto the newly formed BH is somewhat uncertain, and we give two distinct estimates for it.

The first estimate is na\"ive: we simply assume that for $t \lesssim \tau_{\text{sink}}$, the primordial CHAMPs accrete onto the BH at a constant rate equal to the average accretion rate of all the primordial CHAMPs:
\begin{align}
\lb[\dot{M}_{\textsc{bh}}(t) \rb]^{(1)}_{\text{prim.,}\,X} \approx \tau_{\text{sink}}^{-1} M_{X^\pm}^{\text{prim.}} \Theta(\tau_{\text{sink}} - t),
\label{eq:MdotRain1}
\end{align}
where $\tau_{\text{sink}}$ is given approximately by \eqref{tausinkapprox}, $M_{X^\pm}^{\text{prim.}}$ is given by \eqref{Mprim}, and 
\begin{align}
\Theta(x) = \begin{cases} 1 & x>0 \\
					0 & x \leq 0 
		\end{cases}
\end{align}
is the Heaviside theta function.
Note that if $M_{X^\pm}^{\text{prim.}} < M_{\text{Chand.,}\,X}$, then this rate is zero since $\tau_{\text{sink}} < t_{\textsc{bh}}$, where  $t_{\textsc{bh}}$ is the BH formation time. 

The second estimate is more refined: assuming an initial uniform distribution of the CHAMPs, in the time interval $t\in[t_\star,t_\star+dt_\star]$, those CHAMPs present in the radial slice $r \in[r_\star,r_\star+dr_\star]$ will have sunk to the core of the WD, and will be accreted onto BH.
The values of $t_\star$ and $r_\star= r_\star(t_\star)$ are given by%
\footnote{\label{ftnt:notation}%
		Note that $r_\star$ as defined here should not be confused with $r_*$ refined by \eqref{rStar}.
	} %
\begin{align}
t_\star = \int_{R_f}^{r_\star} \frac{dr'}{w_X(r')},
\end{align}
where, to make the estimate tractable, we take $w_X$ to be given by the analytical estimate for the sinking speed of the CHAMPs at \eqref{wXrefined}.
It follows from the Leibniz rule that 
\begin{align}
dt_\star = \frac{dr_\star}{w_X(r_\star)}.
\end{align}
The mass accretion rate is thus given by
\begin{align}
\frac{dM(t_\star)}{dt} = \frac{4\pi r_\star^2 \rho_X dr_\star}{dr_\star / w_X(r_\star)} = 4\pi r_\star^2 \rho_X w_X(r_\star),
\end{align}
where $\rho_X$ is the average (assumed uniform) mass density of the primordial CHAMPs in the star: $\rho_X = 3  M_{X} / (4\pi R_{\textsc{wd}}^3)$.
Since the central object at time $t_\star$ consists of all CHAMPs accreted up to radius $r_\star$, we have 
\begin{align}
M_{\text{core}} \sim M(t_\star) \sim \frac{4\pi}{3} \rho_X r_\star^3 \sim M_X \frac{ r_\star^3 }{R^3} \sim \eta M_{\textsc{wd}} \frac{ r_\star^3 }{R_{\textsc{wd}}^3},
\end{align}
so that in \eqref{wXrefined} we can take
\begin{align}
& \lb[  \lb( 1 - \frac{M_{\text{core}}}{M_{\textsc{wd}}} \rb) + \frac{M_{\text{core}}}{M_{\textsc{wd}}}  \lb(\frac{R_{\textsc{wd}}}{r_\star}\rb)^3  \rb] \nonumber\\
&\sim \lb[  \lb( 1 - \frac{M_{\text{core}}}{M_{\textsc{wd}}} \rb) + \eta  \rb] \sim 1,
\end{align}
leading to 
\begin{align}
\frac{dM(t_\star)}{dt} \sim \frac{4\pi r_\star^3 \rho_X w_X(R_{\textsc{wd}})}{R} = 3 \frac{w_X(R_{\textsc{wd}})}{R_{\textsc{wd}}} M(t_\star).
\end{align}
Given that the central density of a WD is higher than the average density, our discussion in \sectref{timescales} leads us to conclude that this estimate is too small by, conservatively, a factor of $\sim 6.5$ [see, e.g., \eqref{tausink4a}] for the central density and core masses considered in arriving at \eqref{tausink4}; in the same approximation, we have $w_X(R_{\textsc{wd}})/R_{\textsc{wd}} \sim 2 / \tau_{\text{sink}}$ where $\tau_{\text{sink}}$ is as given by \eqref{tausinkapprox}.
Therefore, we can take the second estimate of the accretion rate to be 
\begin{align}
\dot{M}^{(2)}_{\text{prim.,}\,X}(t) \sim 40 \tau_{\text{sink}}^{-1} M(t),
\label{eq:MdotRain2}
\end{align}
where $\tau_{\text{sink}}$ is as given by \eqref{tausinkapprox} and where the numerical coefficient is a reasonable approximation when $M(t) \sim M_{\text{Chand.}, X}(m_X\sim 10^{10}\,\text{GeV}) \ll M_{\textsc{wd}}$, with only logarithmic dependence on this assumption so long as the approximations used in deriving this estimate remain satisfied.
This accretion rate is of course cut off when the entire primordial abundance of CHAMPs has accreted onto the central object.

Note that the second estimate [\eqref{MdotRain2}] is much smaller than the first [\eqref{MdotRain1}] when $M(t) \ll M^{\text{prim.}}_X$ [e.g., when $M(t)\sim M_{\text{Chand.,}\,X}$], and also that it depends explicitly on the mass of primordial CHAMPs already in the central object.
We comment below on the differing implications of the two rate estimates.

Provided that the WD is located where a significant galactic halo CHAMP abundance is present, there is a second CHAMP accretion mechanism that operates for a much longer time period.
As already discussed in \sectref{accumulated}, halo CHAMPs accrete onto the WD (provided that the WD has a sufficiently small magnetic field; see discussion in \sectref{AccretionEfficiency}).
If we assume that there is a quasi-steady-state period of WD evolution during which these CHAMPs simply sink through the star and are captured by the BH, then the mass accretion rate onto the BH during this quasi-steady-state period is simply equal to the CHAMP mass accretion rate onto the WD:
\begin{align}
\dot{M}_{\text{accum.,}\,X}(t) \approx &\ \tau_{\text{accum.}}^{-1} M^{\text{accum.}}_{X^\pm}(\tau_{\text{accum.}}) \nl \times  \Theta( \tau_{\text{accum.}} - t )\Theta( t- \tau_{\text{sink}} ),
\label{eq:MdotAccum}
\end{align}
where $M^{\text{accum.}}_{X^\pm}(\tau_{\text{accum.}})$ is given by the expressions in \sectref{PopnAccumulated} and \appref{Accreting_Fraction}, and $\tau_{\text{accum.}}$ is the timescale for WD crystallization (see the discussion in \sectref{PopnAccumulated}).

Note that we have set this rate to zero for times $t \lesssim \tau_{\text{sink}}$ because it will generally take a time on the order of $\tau_{\text{sink}}$ for the CHAMPs first accreted onto the WD to sink to the core and a steady-state flow to be established; none of our results depend sensitively on this assumption.

\subsubsection{CHAMP accretion onto the BH. Case II: $M_{\text{s.g.,}\,X} \gtrsim M_{\text{Chand.,}\,X}$.}
\label{sect:CHAMPRainCase2}
If the self-gravitating mass $M_{\text{s.g.,}\,X}$ exceeds the \Chand\ mass $M_{\text{Chand.,}\,X}$ there is a modification to the picture advanced in \sectref{CHAMPRainCase1}; although see footnote \ref{ftnt:IdealGas}.
In this case, when the BH is initially born at the \Chand\ mass, it is surrounded by a significant overdensity of CHAMPs in the collapsing self-gravitating cloud; this can temporarily boost the accretion rate of CHAMPs just as the BH is born, and until the self-gravitating cloud has been accreted in a time $\tau_{\text{coll.}}^{\text{core}}$ [\eqref{tausinkCore}].
We must therefore modify our estimates for the CHAMP mass accretion rates \eqref[s]{MdotRain1} or (\ref{eq:MdotRain2}), or \eqref{MdotAccum} [depending on which process is the prevailing rate at the time of self-gravitating collapse, $t_{\text{s.g.}}$, and which estimate we use for it] during this initial time.

We provide a simple estimate for this initial accretion rate, assuming just that the self-gravitating core is accreted at an average rate until it is completely depleted:
\begin{align}
\dot{M}_{\text{s.g.,}\,X}(t) \approx  (\tau_{\text{coll.}}^{\text{core}})^{-1} M_{\text{s.g.,}\,X^\pm} \Theta(  t_{\text{s.g.}} + \tau_{\text{coll.}}^{\text{core}} - t ),
\label{eq:MdotSG}
\end{align}
which holds during the time $t_{\text{s.g.}} \lesssim t_{\textsc{bh}} < t <  t_{\text{s.g.}} + \tau_{\text{coll.}}^{\text{core}}$.
This is a reasonable estimate because the mass in the self-gravitating cloud is initially concentrated in a Gaussian profile near the WD core, and not spread diffusely throughout the star.

Because $25 \tau_{\text{coll.}}^{\text{core}} \sim \tau_{\text{coll.}}^{\text{sink}}$ [cf., \eqref[s]{tausinkCore} and (\ref{eq:tausinkapprox})], the accretion rate estimate \eqref{MdotSG} is comparable to the estimate \eqref{MdotRain2} if $M(t) \sim M_{\text{s.g.,}\,X^\pm}$ is assumed in the latter rate.
Therefore, assuming that the full self-gravitating mass were to collapse into the BH (see discussion in \sectref{BHevolution} below), the matching between the accretion rates \eqref[s]{MdotSG} and (\ref{eq:MdotRain2}) would be roughly smooth [up to $\mathcal{O}(1)$ factors] at the end point of the self-gravitating collapse. 
Of course, for earlier times [e.g., supposing that $M(t) \sim M_{\text{Chand.,}\,X^\pm} \ll M_{\text{s.g.,}\,X^\pm}$], the rate at \eqref{MdotSG} exceeds that at \eqref{MdotRain2} [as expected], by roughly the ratio $M_{\text{s.g.,}\,X^\pm}/M_{\text{Chand.,}\,X^\pm}$.
Therefore, if $t_{\text{s.g.}} < \tau_{\text{sink}}$, our refined approach will be modified by assuming only the accretion rate \eqref{MdotSG} for the initial time period of the self-gravitating cloud collapse, then reverting to the rate \eqref{MdotRain2} until $t\sim \tau_{\text{sink}}$.%
\footnote{\label{ftnt:StillCorrect}%
		Roughly, the derivation of \eqref{MdotRain2} still holds after the initial phase where the CHAMPs are stalled in the thermal structure, provided we make the reasonable assumption that the dynamics of the CHAMPs sinking from radii large enough that they arrive in the vicinity of the core after the collapse of the self-gravitating cloud are not impacted by the earlier presence of that cloud.
	} %

However, \eqref{MdotSG} is smaller than the na\"ive accretion estimate \eqref{MdotRain1} when $M^{\text{prim.}}_{X^\pm} \gtrsim 25 M_{\text{s.g.,}\,X^\pm}$.
If the na\"ive approach is used, and $t_{\text{s.g.}} < \tau_{\text{sink}}$, we will instead assume that the accretion rate is the larger of \eqref{MdotSG} or \eqref{MdotRain1} during the initial self-gravitating cloud collapse, and if necessary then revert to the rate \eqref{MdotRain1} until $t\sim\tau_{\text{sink}}$.

In both cases, if $\tau_{\text{sink}}< t_{\text{s.g.}} < \tau_{\text{accum.}}$, we assume only the accretion rate \eqref{MdotSG} for the initial time period of the self-gravitating cloud collapse, and then revert instead to the rate estimate \eqref{MdotAccum} until $t\sim\tau_{\text{accum.}}$.

This piecewise approach to the accretion rate estimates is manifestly approximate, but it roughly captures the prevailing dynamics of the major epochs of CHAMP accretion under the various assumptions about the accretion dynamics.
The exact accretion behavior in the regimes where various CHAMP reservoirs are nearly exhausted (e.g., when the self-gravitating structure is nearly fully accreted) will of course be more complicated and our model for the accretion rate would not fully capture all the subtleties of the dynamics during these times.
We expect that the gross picture of the BH dynamics is captured by the approach we have outlined here, although details will be lost.

\subsection{Implications for BH evolution}
\label{sect:BHevolution}
Of the dynamical contributions to the BH mass evolution discussed in the \sectref{BHdynamicalProcesses}, the WD and CHAMP accretion processes tend to increase the BH mass, while the Hawking process tends to decrease it.
Moreover, the various accretion and emission processes have rates that are explicit functions of time, as well as functions of the BH mass in many cases.
In general, this leads to a rich and complicated dynamics of the BH, leading to a surprising complexity of possible BH evolutionary trajectories.

There are however only three distinct physical outcomes: (1) the BH will reach an evolutionary phase during which it undergoes runaway accretion with a rate sufficient in principle to accrete the entire WD mass within the WD lifetime, the countervailing effect of Hawking radiation notwithstanding; (2) the BH will reach a phase during which it undergoes Hawking evaporation at a rate sufficient in principle to radiate away its entire mass within the WD lifetime, the countervailing accretion of WD material and/or CHAMPs notwithstanding; or (3) the BH will follow an evolutionary trajectory along which it can eventually be either dominantly accreting or dominantly evaporating, but either owing to a balancing of rates of accretion and evaporation, or simply the intrinsic timescale for the dominant process being too long, the timescale for completion of the accretion or evaporation processes is longer than the observed age of the WD.
In scenarios (1) or (2), we will find that the accretion or evaporation likely does not actually need to proceed all the way to completion in order to trigger a back-reaction on the WD sufficient to trigger the supernova instability \cite{Janish:2019nkk,Acevedo:2019gre}, thereby terminating the BH evolution and giving a clear observable signal.
However, in case (3), there will (except for highly tuned regions of parameter space) be no observable impact on the WD in which the BH resides. 

Because case (3) is an important exception, we discuss the requirements to be in this region.
For concreteness, the qualitative discussion that follows is correct in detail (and has been checked quantitatively) for CHAMPs $X^+$ present in a WD with mass $M_{\textsc{wd}}\sim 1.1M_{\odot}$ that has characteristics as in \tabref{WDcharacteristics} and which has an assumed crystallization time of $\tau_{\text{accum.}} \sim \,\text{Gyr}$ and an assumed age/lifetime of $\tau_{\textsc{wd}} \sim 2\,$Gyr (see \sectref{limits} and \tabref{oldWD}).
Similar considerations, albeit with some modifications as to whether certain regions of behavior exist, are applicable for CHAMPs $X^-$ bound as $(NX)$, or for different WD parameters.

It turns out that, for $X^+$, the conditions under which outcome (3) can occur are such that the total mass of CHAMPs present \emph{primordially} in the WD exceeds the self-gravitating mass, which in turn exceeds the \Chand\ mass (as opposed to a situation in which the primordial mass of CHAMPs is too small to exceed the larger of the \Chand\ or self-gravitating masses, but the slow accretion of CHAMPs onto the WD triggers BH formation at a later time, but before the WD crystallization time).
To maintain concreteness, we discuss only this ordering of the self-gravitating and \Chand\ masses in what follows, assuming the existence of the thermal-pressure-supported phase (see footnote \ref{ftnt:IdealGas}).

As the CHAMPs in the star collect in the central thermal-pressure-supported structure, gravothermal collapse will eventually be triggered at some time $t_{\text{s.g.}} < \tau_{\text{sink}}$.
As the thermal cloud of CHAMPs collapses in a time $\tau_{\text{coll.}}^{\text{core}}$ thereafter, a stratified core will form, which will accrete CHAMPs from this initial thermal cloud structure. 
Once sufficiently many primordial CHAMPs have accreted onto this core, a BH will form at time $t_{\textsc{bh}}$ where $t_{\text{s.g.}} \lesssim t_{\textsc{bh}} \lesssim t_{\text{s.g.}} +\tau_{\text{coll.}}^{\text{core}}$, with a mass at the \Chand\ mass.

At this point, the remainder of the thermal cloud is still collapsing onto the newly formed BH, supplying a large positive contribution to the BH mass accretion rate.
Nevertheless, it is possible that the \Chand\ mass is so small that, including the effects of Hawking radiation from the newly formed BH, $\dot{M}_{\textsc{bh}}(M_{\textsc{bh}}=M_{\text{Chand.}};t_{\textsc{bh}}\lesssim t \lesssim t_{\text{s.g.}} +\tau_{\text{coll.}}^{\text{core}}) < 0$, and the BH would immediately begin to lose mass upon formation; outside of a highly tuned region, the timescale for this process is so fast that the BH would evaporate well within the WD lifetime. 
This avoids outcome (3).

If $\dot{M}_{\textsc{bh}}(M_{\textsc{bh}}=M_{\text{Chand.}};t_{\textsc{bh}}\lesssim t \lesssim t_{\text{s.g.}} +\tau_{\text{coll.}}^{\text{core}}) \approx 0$ (in a highly tuned region), the BH will neither grow nor shrink until such time as all the CHAMPs in the thermal cloud have been depleted, at which point the BH is still approximately at the \Chand\ mass.
However, once the CHAMPs initially in the thermal cloud are all accreted, $\dot{M}_{\textsc{bh}}(M_{\textsc{bh}}=M_{\text{Chand.}};t_{\text{s.g}} + \tau_{\text{coll.}}^{\text{core}}<t<\tau_{\text{sink}})$ may (depending on parameters and on which of the two estimates discussed in \sectref{CHAMPRainCase2} we give for the accretion rates) lose a positive contribution present in $\dot{M}_{\textsc{bh}}(M_{\textsc{bh}}=M_{\text{Chand.}};t_{\textsc{bh}}\lesssim t \lesssim t_{\text{s.g.}} +\tau_{\text{coll.}}^{\text{core}})$ and turn negative; this is particularly true if we use the refined approach of \sectref{CHAMPRainCase2} because $M_{\text{s.g.,}\,X} > M_{\text{Chand.},\,X}$. 
The intrinsic timescales involved for the Hawking process are sufficiently rapid that, outside a highly tuned region, the BH will then radiate away in a fraction of the remaining lifetime of the WD, and outcome (3) is avoided.
Alternatively, if we use the na\"ive approach detailed \sectref{CHAMPRainCase2} and \eqref{MdotRain1} is the accretion estimate throughout the thermal cloud collapse and sinking epochs, then no such positive contribution to $\dot{M}_{\textsc{bh}}$ is lost after $t\sim t_{\text{s.g}} + \tau_{\text{coll.}}^{\text{core}}$, and the BH will continue to sit at the \Chand\ mass until $t=\tau_{\text{sink}}$, at which time it definitely loses a large positive contribution to $\dot{M}_{\textsc{bh}}$, and will Hawking radiate away within the WD lifetime.
Outcome (3) is avoided in either alternative.
Note however that the boundary at which $\dot{M}_{\textsc{bh}}(M_{\textsc{bh}}=M_{\text{Chand.}};t_{\textsc{bh}}\lesssim t \lesssim t_{\text{s.g.}} +\tau_{\text{coll.}}^{\text{core}})=0$ depends on which of the two sets of rate estimates from \sectref{CHAMPRainCase2} we use, and this has some impact on the region in which outcome (3) is ultimately avoided overall; we will discuss this in \sectref{limits} below.

Finally, if $\dot{M}_{\textsc{bh}}(M_{\textsc{bh}}=M_{\text{Chand.}};t_{\textsc{bh}}\lesssim t \lesssim t_{\text{s.g.}} +\tau_{\text{coll.}}^{\text{core}}) > 0$ the initial BH will begin to increase in mass, eventually (outside of highly tuned regions where the Hawking process converts a significant portion of the sinking CHAMPs to radiation) reaching the full mass of the thermal cloud at a time $t = t_{\text{s.g.}} +\tau_{\text{coll.}}^{\text{core}}$.
It turns out that, for $X^+$, whenever $\dot{M}_{\textsc{bh}}(M_{\textsc{bh}}=M_{\text{Chand.}};t_{\textsc{bh}}\lesssim t \lesssim t_{\text{s.g.}} +\tau_{\text{coll.}}^{\text{core}}) > 0$, we also have $\dot{M}_{\textsc{bh}}(M_{\textsc{bh}}=M_{\text{s.g.}};t_{\text{s.g.}} +\tau_{\text{coll.}}^{\text{core}}<t<\tau_{\text{sink}}) > 0$, and the BH will always continue to accrete up in mass until it reaches the full primordial mass of CHAMPs present in the star around $t\sim\tau_{\text{sink}}$.
If we are in this region, we have thus far avoided BH evaporation on too fast a timescale, and we can still ultimately reach outcome (3).

To make further progress, we must consider the value of $\dot{M}_{\textsc{bh}}(M_{\textsc{bh}}=M_{X}^{\text{prim.}};\tau_{\text{sink}}<t<\tau_{\text{accum.}})$, which now has no further contribution from the primordial CHAMPs. 
If $\dot{M}_{\textsc{bh}}(M_{\textsc{bh}}=M_{X}^{\text{prim.}};\tau_{\text{sink}}<t<\tau_{\text{accum.}})< 0$, two outcomes are possible: in the first case, the intrinsic timescale for Hawking evaporation is fast enough that the BH radiates away from $M_{\textsc{bh}}=M_{X}^{\text{prim.}}$ to $M_{\textsc{bh}}=0$ before the WD crystallization time; this again avoids outcome (3).
Alternatively, in the second case, the timescale for the Hawking radiation of the BH can be too long [when $\dot{M}_{\textsc{bh}}(M_{\textsc{bh}}=M_{X}^{\text{prim.}};\tau_{\text{sink}}<t<\tau_{\text{accum.}})\sim 0$, but slightly negative], such that it survives until at least the WD crystallization time (likely without radiating much of its mass away, outside of tuned regions of parameter space, owing to the strong BH-mass dependence of the Hawking rate); such a BH can still end up with outcome (3).

On the other hand, if $\dot{M}_{\textsc{bh}}(M_{\textsc{bh}}=M_{X}^{\text{prim.}};\tau_{\text{sink}}<t<\tau_{\text{accum.}})> 0$, two outcomes are again possible: in the first case, the intrinsic timescale for accretion of CHAMPs and WD material is so fast that the BH will in principle be able to accrete up to the full WD mass within the WD crystallization time; this avoids outcome (3).
In the second case [when $\dot{M}_{\textsc{bh}}(M_{\textsc{bh}}=M_{X}^{\text{prim.}};\tau_{\text{sink}}<t<\tau_{\text{accum.}})\sim 0$, but slightly positive], the timescale for accretion of CHAMPs and WD material is too long to allow the BH to accrete up to the full WD mass before the WD crystallization time, but the BH can still accrete CHAMPs up to the full abundance that accumulate onto the WD before the crystallization time; such a BH can still end up with outcome (3).

Outside of tuned regions of parameter space, we thus have two possible BH that have survived to the crystallization time: those that had $\dot{M}_{\textsc{bh}}(M_{\textsc{bh}}=M_{X}^{\text{prim.}}; \tau_{\text{sink}}<t<\tau_{\text{accum.}})\sim 0$ but slightly negative, and arrive at the crystallization time with a mass still $M_{\textsc{bh}}\sim M_{X}^{\text{prim.}}$; and those that had $\dot{M}_{\textsc{bh}}(M_{\textsc{bh}}=M_{X}^{\text{prim.}};\tau_{\text{sink}}<t<\tau_{\text{accum.}})\sim 0$ but slightly positive, and arrive at the crystallization time with a mass $M_{\textsc{bh}}\sim M_{X}^{\text{prim.}} + M_{X}^{\text{accum.}}$ (ignoring any WD material accumulated, because there is almost no parameter space in which that contribution would be significant, but the WD is not also fully accreted by this point).

Once again, a positive contribution to $\dot{M}_{\textsc{bh}}$ from the slow CHAMP accumulation turns off at the crystallization time. 
For those BH that had $\dot{M}_{\textsc{bh}}(M_{\textsc{bh}}=M_{X}^{\text{prim.}};\tau_{\text{sink}}<t<\tau_{\text{accum.}})\sim 0$ but slightly negative before the crystallization time, $\dot{M}_{\textsc{bh}}(M_{\textsc{bh}}=M_{X}^{\text{prim.}};\tau_{\text{accum.}}<t<\tau_{\textsc{wd}})<0$ and there are two outcomes: firstly, the Hawking evaporation can be fast enough that the BH evaporates within the remaining WD lifetime; this avoids outcome (3).
Alternatively, the BH can take too long to evaporate in the remaining WD lifetime, resulting in outcome (3).
On the other hand, for BHs that had $\dot{M}_{\textsc{bh}}(M_{\textsc{bh}}=M_{X}^{\text{prim.}};\tau_{\text{sink}}<t<\tau_{\text{accum.}})\sim 0$ but slightly positive before the crystallization time and thus arrive at  this point with mass $M_{\textsc{bh}}\sim M_{X}^{\text{prim.}} + M_{X}^{\text{accum.}}$, there are three outcomes.
First, if $\dot{M}_{\textsc{bh}}(M_{\textsc{bh}}= M_{X}^{\text{prim.}} + M_{X}^{\text{accum.}};\tau_{\text{accum.}}<t<\tau_{\textsc{wd}})<0$, the BH begins to Hawking evaporate. 
It turns out, given the assumed parameters, that any such BH takes too long to Hawking evaporate within the remaining WD lifetime (this of course depends on the WD lifetime and could be violated with the WD lifetime was longer, say 4\,Gyr), resulting in outcome (3).
Second, $\dot{M}_{\textsc{bh}}(M_{\textsc{bh}} = M_{X}^{\text{prim.}} + M_{X}^{\text{accum.}};\tau_{\text{accum.}}<t<\tau_{\textsc{wd}})>0$ and sufficiently large that the BH accretes up to the full WD mass in the remaining WD lifetime; this avoids outcome (3).
Third, $\dot{M}_{\textsc{bh}}(M_{\textsc{bh}}= M_{X}^{\text{prim.}} + M_{X}^{\text{accum.}};\tau_{\text{accum.}}<t<\tau_{\textsc{wd}})>0$ but smaller, and the BH takes too long to accrete up to the full WD mass, resulting in outcome (3).

This detailed step-by-step analysis, although it is still crude and can miss some edge cases, provides a series of conditions that must be satisfied for the BH evolution to be result in outcome (3): i.e., be too slow to either Hawking evaporate away, or accrete up to the full WD mass, in the WD lifetime.
When we present limits below, we will exclude the region in which the following are all satisfied as, taken together, they approximately cover the region in which outcome (3) obtains%
\footnote{\label{ftnt:timescaleCaveat}%
		Evaporation timescales are estimated using only the intrinsic Hawking rate, neglecting any lengthening owing to partial cancellation against the positive contributions to $\dot{M}$, resulting in some small error in the boundary of the region.
		Similarly, accretion timescales are estimated using only the intrinsic Bondi (and, if applicable, slow CHAMP accretion rates), neglecting any lengthening owing to partial cancellation against the negative Hawking contribution to $\dot{M}$, resulting in some small error in the boundary of the region.
		We also neglect the Bondi/Eddington crossover here, per the discussion in \sectref{Eddington}.
	} %
\begin{enumerate}[label=(\alph*)]
\item $\dot{M}_{\textsc{bh}}(M_{\textsc{bh}} =M_{\text{Chand.}};t_{\textsc{bh}}\lesssim t \lesssim t_{\text{s.g.}} +\tau_{\text{coll.}}^{\text{core}}) > 0$; this is possibly two different conditions depending on which of the rate estimates discussed in \sectref{CHAMPRainCase2} is used, and we will show the boundaries of the regions of parameter space for which both of these conditions obtain in turn in \sectref{limits}.
\item $\dot{M}_{\textsc{bh}}(M_{\textsc{bh}}=M_{X}^{\text{prim.}};\tau_{\text{sink}}<t<\tau_{\text{accum.}})< 0$, and $\tau^{\text{evap.}}_{\text{H}}(M_{\textsc{bh}}=M_{X}^{\text{prim.}}\rightarrow 0) > \tau_{\text{accum.}}$, and $\tau^{\text{evap.}}_{\text{H}}(M_{\textsc{bh}}=M_{X}^{\text{prim.}}\rightarrow 0) > \tau_{\textsc{wd}} -  \tau_{\text{accum.}}$.
Note that is is slightly conservative to impose both the timescale conditions independently in this fashion, rather than as a single condition.
\item $\dot{M}_{\textsc{bh}}(M_{\textsc{bh}}=M_{X}^{\text{prim.}};\tau_{\text{sink}}<t<\tau_{\text{accum.}})> 0$, and $\tau^{\text{grow}}_{\text{B+accum.}}(M_{\textsc{bh}}=M_{X}^{\text{prim.}}\rightarrow M_{\textsc{wd}}) > \tau_{\text{accum.}}$ (timescale assuming Bondi accretion and slow CHAMP accretion), and $\tau_{\text{B}}(M_{\textsc{bh}}=M_{X}^{\text{prim.}}+M_X^{\text{accum.}}\rightarrow M_{\textsc{wd}}) > \tau_{\textsc{wd}} - \tau_{\text{accum.}}$.
\end{enumerate}
Note that, in principle, there is a region satisfying the following conditions which would not need to be excluded:
\begin{enumerate}
\item[(d)] $\dot{M}_{\textsc{bh}}(M_{\textsc{bh}}=M_{X}^{\text{prim.}};\tau_{\text{sink}}<t<\tau_{\text{accum.}})> 0$, $\dot{M}_{\textsc{bh}}(M_{\textsc{bh}}=M_{X}^{\text{prim.}}+M_X^{\text{accum.}};\tau_{\text{accum.}}<t<\tau_{\textsc{wd}})< 0$, and $\tau^{\text{evap.}}_{\text{H}}(M_{\textsc{bh}}=M_{X}^{\text{prim.}}+M_X^{\text{accum.}}\rightarrow M_{\textsc{wd}}) < \tau_{\textsc{wd}} - \tau_{\text{accum.}}$.
\end{enumerate}
However, with the lifetimes we have assumed in this discussion and for these WD parameters, region (d) does not exist.

\subsection{Implications of BH evolution for old WD}
\label{sect:outcomes}
Having discussed in detail the region in which the BH evolution does not have sufficient time to proceed to either devour the entire WD within the WD lifetime, or Hawking radiate away to zero mass within the same timeframe, we now switch focus to the case where either outcome could in principle happen, absent backreaction on the WD during the BH evolution.

In this subsection, we will discuss the backreaction on the WD as a result of these processes, and their implications of the survival of old WD.
Our discussion roughly follows that of \citeR[s]{Janish:2019nkk,Acevedo:2019gre}.

\subsubsection{Evaporating case}
\label{sect:outcomesEvaporating}
In this section we assume that the BH follows an evolutionary trajectory such that the final phase of the BH evolution is one in which the BH dynamics are dominated by evaporation rapid enough in principle to release the entire BH mass within the remaining WD lifetime [outcome (2) of \sectref{BHevolution}].
We see from the intrinsic Hawking radiation timescale estimate \eqref{tauHawkingNum} that a necessary (although, as the discussion at \sectref{BHevolution} makes clear, not sufficient) condition is for the BH to enter this phase with $M_{\textsc{bh},\,0} \lesssim 10^{38}\,\text{GeV}$, yielding an intrinsic evaporation timescale $\tau_{\text{H}} \lesssim \text{few Gyr}$, of order the WD lifetime.

A critical result of \citeR{Janish:2019nkk} is that the evaporation of such a BH inside a WD, assuming additionally only that $M_{\textsc{bh},\,0} > E_T(M_{\textsc{wd}})$ for its WD host, will always eventually deposit sufficient energy sufficiently rapidly to satisfy the trigger criteria for WD ignition discussed in \sectref{ignition}.
Indeed, assuming that the remaining BH lifetime is longer than the diffusion timescale for the WD trigger radius, then even for the largest trigger energies required (for the lightest WD possibly of interest to us), $E_T(M_{\textsc{wd}}\sim0.8M_{\odot})\sim 10^{25}\,$GeV (see \figref{ET}), we find conservatively that once the BH mass drops below $M^{\textsc{wd}\text{ trig.}}_{\textsc{bh}} \sim 10^{30}\,$GeV, the total energy Hawking radiated from the BH in one diffusion time exceeds $E_T$.%
\footnote{\label{ftnt:highTH}%
		$T_H$ is so high at this mass that all SM species are radiated ultra-relativistically; we took $g_{\text{eff.}}=101.5$ in \eqref{MdotHawking} to conservatively exclude gravitons and neutrinos.
	}\up{,}%
\footnote{\label{ftnt:JanishComparison}%
		\citeR{Janish:2019nkk} quotes a value for a $1.25M_{\odot}$ WD of $M^{\textsc{wd}\text{ trig., \cite{Janish:2019nkk}}}_{\textsc{bh}}(M_{\textsc{wd}}=1.25M_{\odot}) \sim 2\times 10^{35}\,$GeV.
		We would find a value of $M^{\textsc{wd}\text{ trig.}}_{\textsc{bh}}(M_{\textsc{wd}}=1.25M_{\odot}) \sim 10^{34}\,$GeV, largely because we have a slightly more conservative trigger condition.
		There is no qualitative difference here.
	} %
Since, as discussed at length in \citeR{Graham:2018efk}, thermalization of all particles emitted by the BH, except gravitons and possibly neutrinos, occurs within a volume not parametrically larger than the trigger volume, this would satisfy the trigger criterion, leading to thermal runaway.
For the two specific WD masses that will be of interest, the trigger condition is satisfied for $M^{\textsc{wd}\text{ 0.85 trig.}}_{\textsc{bh}} \sim 4\times10^{30}\,$GeV, and $M^{\textsc{wd}\text{ 1.1 trig.}}_{\textsc{bh}} \sim 1.8\times10^{32}\,$GeV, respectively.

Given that the minimum diffusion timescale for a WD in the mass range $M_{\textsc{bh}} \in [0.8,1.35]M_{\odot}$ is $\tau_{\text{diff.}} \sim 2\times 10^{-13}\,$s (see \figref{ET}), which is the lifetime for a BH of mass $M_{\textsc{bh}} \sim 2\times10^{28}\,\text{GeV}<M^{\textsc{wd}\text{ trig.}}_{\textsc{bh}}$, it is appropriate here to integrate the energy deposition over the full diffusion time.
Nevertheless, if for any reason the BH were either to drop below $M_{\textsc{bh}} \sim 2\times10^{28}\,\text{GeV}$ \emph{without} yet triggering thermal runaway, or if it simply forms below that mass, its (remaining) lifetime to radiate away its entire (remaining) mass is less than one diffusion time, and it is possible to draw a robust conclusion that, as long as the initial mass of the evaporating BH is above the trigger energy for the host WD (see \figref{ET} and \tabref{WDcharacteristics}), runaway will always be triggered.
Because a number of the evolutionary trajectories of BHs formed inside the WD result in the BH immediately Hawking radiating away upon formation (see \sectref{BHevolution}), so that $M_{\textsc{bh},\,0} \sim M_{\text{Chand.}}^{X}$, note that the smallest \Chand\ mass of any BH we consider occurs for $X^+$ for $m_X \sim 10^{18}\,\text{GeV}$, and is $M_{\text{Chand.}}^{X^+}(m_X=10^{18}\,\text{GeV}) \sim 5\times 10^{21}\,\text{GeV} > E_T(M_{\textsc{wd}}=1.1M_{\odot})$.

In summary, we have concluded in part that a sufficiently light BH which forms inside a WD of mass $M_{\textsc{wd}}\gtrsim 1.1 M_{\odot}$ would evaporate in a WD lifetime of $\sim 2\,\text{Gyr}$ in large regions of parameter space (see discussion in \sectref{BHevolution}) and, so long as its initial mass upon beginning to evaporate is above the trigger energy for the host WD (see \figref{ET} and \tabref{WDcharacteristics}), the evaporation will deposit sufficient energy to trigger thermal runaway in the WD, destroying it in a SNIa-like supernova event.
The existence of WD in this mass range older than $\sim 2\,$Gyr can thus place limits on this process having occurred.
On the other hand, one could also in principle search for the SNIa-like destruction events directly.
Apart from the caveats regarding the timescale and BH evolutionary trajectories, this agrees with the results of \citeR{Janish:2019nkk,Acevedo:2019gre}.

\subsubsection{Accreting case}
\label{sect:outcomesAccreting}
In this section we assume that the BH follows an evolutionary trajectory such that the final phase of the BH evolution is one in which the BH dynamics are dominated by accretion of WD material rapid enough in principle to devour the entire WD mass within the remaining WD lifetime [outcome (1) of \sectref{BHevolution}].
We see from the intrinsic Bondi accretion timescale estimate \eqref{tauBondiNum} (see also the discussion of the irrelevance of the Eddington limitation in \sectref{Eddington}) that a necessary (although, as the discussion at \sectref{BHevolution} makes clear, not sufficient) condition is for the BH to enter this phase with $M_{\textsc{bh},\,0} \gtrsim 10^{38}\,\text{GeV}$, yielding an intrinsic accretion timescale $\tau_{\text{H}} \lesssim \text{few Gyr}$, of order the WD lifetime.

If this is the way the WD ends, it is not with a bang but a whimper:%
\footnote{\label{ftnt:EliotApology}%
		With apologies to T.S.~Eliot.
	} %
the BH simply devours the whole WD, but there is no immediate observational signature of this process, other than the absence of WD older than $\sim \text{few}\,$Gyr.

The less conservative outcome proposed by \citeR{Janish:2019nkk} is that the accretion of WD matter onto the BH will be sufficiently violent that, in the vicinity of sonic radius from which Bondi accretion occurs, heating or compression of the WD material will trigger thermal runaway.
Since this must of course happen prior to the WD being completely devoured, the timescale for this to occur must necessarily be faster than the accretion timescale estimates given above.
Per \citeR{Janish:2019nkk}, for the direct heating of the material near around the sonic radius to have a chance to trigger thermal runaway, that radius would need to exceed the trigger length; we find that once $M_{\textsc{bh}} \gtrsim 10^{47}\,$GeV, we have $\lambda_T \gtrsim R_{\text{B}}$ for a $M_{\textsc{wd}} \sim 0.85 M_{\odot}$ WD (where we have estimated the trigger length using the unperturbed density, which is conservative).
Similarly, for a $M_{\textsc{wd}} \sim 1.1 M_{\odot}$ WD, $M_{\textsc{bh}} \gtrsim 10^{46}\,$GeV is required for $\lambda_T \gtrsim R_{\text{B}}$.
Since $c_s \sim 2$--$3\times10^{-2}$ at the sonic radius and $E_K = \frac{1}{2}m_{\text{ion}}c_s^2 \sim 2$--$5\,\text{MeV}\sim 10T_{\text{crit.}}$, the motion of the carbon ions assuming Bondi accretion would in principle be sufficiently fast that thermal runaway could be triggered in this ballpark (although, as discussed in \citeR{Janish:2019nkk}, some significant non-radial flow would be required to trigger the SN; this is however entirely plausible as carbon ions outside the sonic radius are not necessarily collapsing radially to the center of the star).

However, the above argument ignores the fact that for BH of these masses in WD of the relevant central densities, the BH mass at which $R_{\text{B}} \gtrsim \lambda_T$ occurs is already high enough that we estimate that the accretion is Eddington limited; see \sectref{Eddington}.
As a result, the flow of material onto the WD is choked by radiative back-reaction, and $c_s$ is an overestimate for the local speed of the ions.
As such, it is not clear that the trigger condition will be reached; a more detailed consideration of this point is beyond the scope of this work, and would likely require simulations of the accretion dynamics.

The other mechanism proposed by \citeR{Janish:2019nkk} is that the density increase of the WD material in the vicinity of the BH could trigger pycnonuclear fusion.
This is however much more speculative.
The density increase near the BH event horizon under Bondi accretion conditions is $\sim 1/c_s^2 \sim 10^3$, so for WD near $M_{\textsc{wd}}\sim 0.8M_{\odot}$ with $\rho\sim10^7\,\text{g/cm}^3$, the density could be boosted to near $\rho \sim 10^{10}\,\text{g/cm}^3$ near the BH event horizon, which could trigger pycnonuclear burning \cite{Janish:2019nkk,Gasques:2005ar,Yakovlev:2006fi,Kippenhahn:2012zqe}.
However, it is unclear that the heated region is either large enough, or that the flame front would propagate outward rapidly enough, to ignite the star \cite{Janish:2019nkk}.

We therefore reach conclusions similar to, although slightly more conservative than, \citeR{Janish:2019nkk}: if the BH trajectory leads it to a final evolutionary stage of accretion, a BH near formed near the center of a WD can destroy the WD within $\mathcal{O}(\text{Gyr})$, either by direct accretion into the BH, or possibly via the triggering of thermal runaway leading to a supernova explosion; a necessary condition is that the initial mass of the BH upon entering this evolutionary phase is $M_{\textsc{bh},\,0} \gtrsim 10^{38}\,\text{GeV}$.

\section{Limits from the existence of old white dwarfs}
\label{sect:limits}

\begin{table*}[!t]
\begin{ruledtabular}
\caption{\label{tab:oldWD}%
		Some representative old WD from the database in \citeR{2017ASPC..509....3D} (and references therein).
		We give some relevant physical characteristics: the mass in solar masses, surface magnetic field in MG (if known), cooling age in Gyr, effective surface temperature in Kelvin, surface gravitational acceleration in $\text{cm/s}^2$, and distance from Earth in parsec.
		Where multiple values for a specific physical parameter are shown in \citeR{2017ASPC..509....3D}, we quote the range of available values and refer the reader to the reference for more details.
		The WD labelled 1--4 are representative non-magnetic WD with masses similar to those of the fiducial WD we considered in \tabref{WDcharacteristics} and cooling ages $>$Gyr; the WD labelled 5--7 are high-mass magnetic WD with $B$ fields below those discussed in \sectref{AccretionEfficiency}, and cooling ages on the order of Gyr.
		With the exception of WD 5, which has one of the smallest measured $B$ fields of any WD in this mass range reported in \citeR{2017ASPC..509....3D}, we have selected these examples largely at random.
		}
\begin{tabular}{lllllllll}
\# & Name	&	$M_{\textsc{wd}}$ [$M_{\odot}$]	&	$B$ [MG]	& 	$t_{\text{cool}}$ [Gyr]	& $T_{\text{eff.}}$ [$10^{4}$K] & $\log_{10}\lb( g [\text{cm/s}^2] \rb)$ & $D$ [pc]	& Ref. \\	\hline
1	&	WDJ062144.86+753011.67	
			&	1.18--1.23	&	---	&	4.1	 &	0.6 	& 	9.0	&	67	&	\cite{WD1} 	\\
2	&	WDJ013839.12-254233.40
			&	1.17--1.22	&	---	&	4.2	&	0.7	&	9.0 	&	70	&	\cite{WD6}	\\
3	&	LP642-052	
			&	0.84		&	---	&	5.8	&	0.6	&	8.4	&	62	&	\cite{WD3}	\\ 
4	&	WDJ054706.58+753103.10
			&	0.76--0.87	&	---	&	4.0--5.2	&	0.6	&	8.3--8.4 &	75	&	\cite{WD7}	\\ \hline
5	&	WD 2051-208	
			&	1.21--1.24	&	0.20--0.26	&	0.62	&	1.9--2.1	& 9.0--9.1	&	31	&	\cite{WD2}	\\
6	&	WD 0903+083	
			&	1.16		&	6.0		&	1.24	&	1.3--4.9	&8.0--9.9	&	142	&\cite{WD4}	\\
7	&	WD 2202-000	
			&	1.08		&	1.0		&	2.19	&	1.0--2.2	& 8.0--9.0	&	152	&\cite{WD5}	\\
\end{tabular}
\end{ruledtabular}
\end{table*}

\begin{table*}[t]
\begin{ruledtabular}
\caption{\label{tab:ourParameters}%
		Parameters assumed in setting limits.
	}
\begin{tabular}{lll}
Parameter 			&	Symbol				&	 Value						\\	\hline
WD mass				&	$M_{\textsc{wd}}$		&	$1.1M_{\odot}$					\\
WD radius			&	$R_{\textsc{wd}}$		&	$7\times 10^{-3} R_{\odot}$		\\
WD central density		& 	$\rho_{\text{c}}$		&	$5.5 \times 10^7\,\text{g/cm}^3$	\\
WD crystallization time	&	$\tau_{\text{accum.}}$	&	1\,Gyr						\\
WD (cooling) age/lifetime&	$\tau_{\textsc{wd}}$		&	2\,Gyr						\\
WD central temperature	&	$T$					&	1\,keV						\\
WD surface effective temperature
					&	$T_{\text{eff}}$			&	$\ll 5\times 10^4\,K$				\\
WD surface gravitational acceleration
					&	$g$					&	$\gtrsim 10^7 \text{cm/s}^2$		\\
WD progenitor cloud mass&	$M_{\text{cloud}}$		&	$7\,M_{\odot}$	(range: $6$--$8\,M_{\odot}$) \cite{Catalan:2008tr}			\\
Speed of sound in WD center	
					&	$c_s$				&	$2.8\times 10^{-2}$				\\
Effective degrees of freedom for Hawking radiation [\eqref{MdotHawking}]
					& 	$g_{\text{eff}}$			&	51/4	(constant)			\\
Bondi accretion parameter [\eqref{BondiAccretion}]
					&	$\lambda$				&	1				\\
Location				&	---					&	In the disk, Earth vicinity			\\
Local DM density		&	$\rho_0$				&	$0.3\,\text{GeV/cm}^3$			\\
Local circular speed		&	$v_{\text{rot.}}$			&	220km/s						\\
WD speed relative to CHAMP halo
					&	$v_{\textsc{wd}}$		&	220km/s						\\
\end{tabular}
\end{ruledtabular}
\end{table*}

In this section, we present limits on the galactic CHAMP abundance by considering the regions of parameter space in which the WD destruction mechanisms outlined thus far would have destroyed a number of known old WD [see, e.g., \tabref{oldWD}] had such CHAMPs been present. 

In setting our bounds, we will conservatively consider a WD such as that labelled \#7 in \tabref{oldWD}: such a WD is fairly massive, quite old, has a measured magnetic field which is sufficiently small to avoid the concerns expressed in \sectref{AccretionEfficiency} about the accretion of halo CHAMPs onto the WD over its lifetime up to the crystallization time (at least for masses $m_X \gtrsim 10^{10}\,\text{GeV}$), and has low enough effective surface temperature and high enough surface gravity to avoid the concerns about stellar winds discussed in \sectref{AccretionEfficiency}.

For concreteness, the exact parameters we assume in setting our limits are shown in \tabref{ourParameters}.
Note that our parameter choices are conservative in a number of ways: (1) the WD mass (and hence central density) is not as high as that of some other old WD [e.g., \#1, 2, 5], (2) the age of the WD is not as large as that of other old WD [e.g., \#1--4], and (3) the magnetic field is large enough to be measurable: if the absence of magnetic field measurements for the older, more massive WD in \tabref{oldWD} is indicative of their magnetic fields being too small to measure, then those WD could present even more robust bounds.

The basic criterion for obtaining a bound on the existence of CHAMPs is that the WD at some point in its evolution contains at least \emph{the larger of} a \Chand\ mass worth of CHAMPs and a self-gravitating mass of CHAMPs (see \sectref{WDstructureThermal}),%
\footnote{\label{ftnt:caveat5}%
		But note also the caveat at footnote \ref{ftnt:IdealGas}; requiring the self-gravitating mass may not be necessary, and it is conservative to demand it.
	} %
either from the primordial contamination, or by accretion of the CHAMPs onto the WD.%
\footnote{\label{ftnt:}%
		We will present somewhat conservative limits assuming that $T\sim 1\,$keV is the WD central temperature; see discussion in \sectref{WDstructureThermal}. 
		In principle, the limits we present could thus be strengthened if very old, cold WD have core temperatures $T\ll 1\,$keV, but they can only be strengthened up to the limitation imposed by requiring at least a \Chand\ mass of CHAMPs.
		The maximum achievable improvement in the parameter range where we give limits occurs for $X^+$ at $m_X \sim 10^{18}\,$GeV, and is a factor of $\mathcal{O}( 3\times 10^{4} )$ improvement on the bound on $\rho_X/\rho_0$; see \figref{BoundsPlus}.
	} %
We first consider in turn the two cases where the WD is contaminated purely by $X^+$, and purely by $X^-$ in the form of $(NX)$, then we consider what happens if both $X^+$ and $(NX)$ are present.

Our limits will be presented graphically on the parameter space $(m_X,\,\rho_X / \rho_0)$, for either $f_{X^+} = 1$ or $f_{X^-} = 1$, and we make comments about how these limits can be applied to the mixed $X^+$/$X^-$ case.

\subsection{Limits on $X^+$}
\label{sect:limitsPlus}

\begin{figure*}[t]
\includegraphics[width=0.95\textwidth]{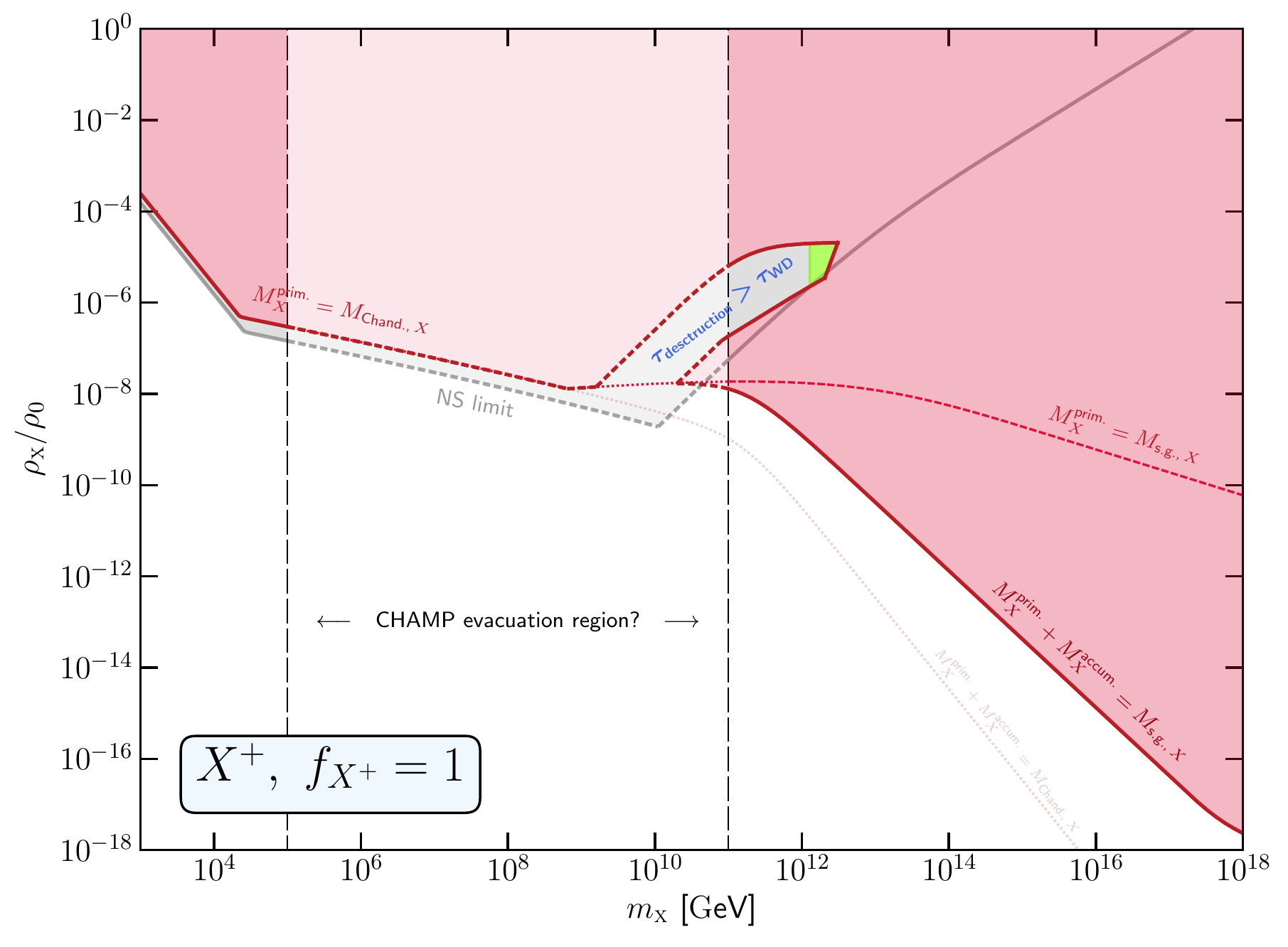}
\caption{\label{fig:BoundsPlus}%
		Constraints on the abundance of CHAMPs $X^+$, $\rho_X / \rho_0$, as a function of CHAMP mass $m_X$ in GeV.
		This plot is drawn assuming that $f_{X^+}=1$ (see discussion in text).
		Here, $\rho_X$ is the local CHAMP mass density, and $\rho_0$ is taken to be the local halo DM abundance $\rho_0 = 0.3\,\text{GeV}/\text{cm}^3$.
		The dark red shaded region bounded by the thick dark red lines (annotated as `$M_{X}^{\text{prim.}} + M_{X}^{\text{accum.}} = M_{\text{s.g.,}\,X}$' at large $m_X$, and annotated as `$M_{X}^{\text{prim.}} = M_{\text{Chand.,}\,X}$' at small $m_X$) denotes the region where (1) the sum of the primordial CHAMP abundance, $M_{X^+}^{\text{prim.}}$, and the CHAMP abundance accumulated onto the WD from the galactic CHAMP halo before the WD crystallization time, $M_{X^+}^{\text{accum.}}$, exceed the larger of the \Chand\ mass for $X^+$ or the self-gravitating mass \eqref{MthermCoreMass}, leading to BH formation; and (2) the subsequent BH evolutionary trajectory ends either with the BH accreting the entire WD mass (or triggering thermal runaway in the process of doing so) [see \sectref{outcomesAccreting}], or with the BH Hawking evaporating away, and triggering thermal runaway in the process [see \sectref{outcomesEvaporating}].
		This bound has a kink at $m^{\text{s.g.}}_X \sim 6.4\times 10^{8}\,$GeV owing to a cross-over between the \Chand\ and self-gravitating masses (with $M_{\text{s.g.,}\,X}>M_{\text{Chand.,}\,X}$ above this value of $m_X$, and smaller below).
		Nevertheless, we also continue to show the parameters for which the \Chand\ mass is exceeded at larger $m_X$ by $M_{X}^{\text{prim.}} + M_{X}^{\text{accum.}}$ with the faint red dotted line (annotated `$M_{X}^{\text{prim.}} + M_{X}^{\text{accum.}} = M_{\text{Chand.,}\,X}$' at large $m_X$), as this is the maximum extent to which the limits could improve if either (a) the conservatively high temperature we have assumed for the thermal structure ($T\sim \text{keV}$) is lowered [$M_{\text{s.g.,}\,X}\propto T^{3/2}$], or (b) the thermal-pressure-supported structure does not actually form; see footnote \ref{ftnt:IdealGas}.
		The shark-fin-shaped region around $m_X \sim 10^9\,$GeV--$10^{12}$GeV which is not shaded red and is annotated `$\tau_{\text{destruction}}>\tau_{\textsc{wd}}$' is the region in which a sufficient CHAMP abundance is present (primordially) in the WD to cause BH formation, but the evolutionary timescales for the BH are too long to destroy the WD within its assumed lifetime.
		The green shaded part of this shark-fin-shaped region would however evolve rapidly enough to destroy the BH (by Hawking evaporation) if the initial primordial CHAMP accretion rate is assumed to be given by the refined estimates in \sectref{CHAMPRainCase2}, instead of by the na\"ive estimates discussed there.
		The thin red dashed line (annotated as `$M_{X}^{\text{prim.}} = M_{\text{s.g.,}\,X}$' at large $m_X$) delineates the regions of parameter space where the primordial abundance alone is sufficient to exceed the self-gravitating mass (above the red dashed line), and where the primordial abundance must be augmented by the slow accretion of CHAMPs over the WD lifetime (until the crystallization time) in order to exceed a self-gravitating mass of CHAMPs in the star (below the red dashed line).
		These bounds all assume the WD parameters shown in \tabref{ourParameters}.
		Also shown are the NS-derived bounds reported by \citeR{Gould:1989gw}, with the region above the thick solid grey line excluded (annotated as `NS limit' and shaded grey except---for clarity---in the shaded red region).
		The region $10^5\,\text{GeV} \lesssim m_X \lesssim 10^{11}\,\text{GeV}$, where the results of \citeR{Chuzhoy:2008zy} call into question the presence of CHAMPs in the MW, is indicated by the thin long-dashed vertical black lines; in this region, we present all results as dotted lines (and the shading is lighter) to indicate the uncertainty as to their applicability; we do not claim robust bounds here (see discussion in text).
		CHAMPs lighter than $m_X \lesssim 10^{3}\,\text{GeV}$ have a sinking timescale in the WD estimated to approach or exceed the WD lifetime and/or crystallization time of $\sim \text{Gyr}$ and we conservatively do not present limits here; see \sectref{timescales}.
		}
\end{figure*}

For the case of pure $X^+$ contamination, $f_{X^+}=1$, we have argued that enough energy can in principle be released in the collapse of the \Chand\ mass core to the BH that the WD trigger criteria can be reached (possibly also true during the earlier self-gravitating cloud collapse); conservatively ignoring any early ignition scenarios though, and assuming that the BH successfully forms from the initial mini-WD-like core structure in the WD, we must demand that the evolutionary timescales for the BH to either accrete the WD mass (or until a trigger criterion is possibly reached), or radiate to effectively zero mass (or until the trigger criterion is reached), must be sufficiently short.

The shaded dark red region in \figref{BoundsPlus} indicates the region of parameter space where the larger of the \Chand\ mass and the self-gravitating mass of CHAMPs can accumulate in the WD, and where the dynamical timescales for the BH evolution are sufficiently short to guarantee WD destruction within the WD lifetime [the local DM density $\rho_0$ is used here purely as a convenient normalization for the CHAMP density].
The unshaded shark-fin-like region between $m_X \sim 10^9\,\text{GeV}$ and $m_X\sim10^{12}$\,GeV indicates the region in which a self-gravitating mass of CHAMPs (which exceeds the \Chand\ mass in this region) would be present in the WD (already primordially, as it turns out), but where the evolutionary timescales are too long to reliably conclude that the WD would be destroyed within its observed cooling age.
The tip of the shark-fin region is shaded green and indicates a region in which the evolutionary timescales are too long if the na\"ive initial CHAMP accretion estimates of \sectref{CHAMPRainCase2} are used; however, if the refined estimates discussed there are used instead, this region does evolve sufficiently rapidly to destroy the WD in its lifetime; see discussion in \sectref{BHevolution}.

The region below the thin dashed red line above $m_X \gtrsim 10^{11}\,$GeV is where the self-gravitating mass (which exceeds the \Chand\ mass in this region) is only exceeded owing to the slow accretion of CHAMPs onto the WD before the crystallization time; in the remainder of the region, the primordial abundance of CHAMPs suffices to meet the mass criterion.
Note in particular that this means that the bounds for $m_X \lesssim 10^{11}$\,GeV are in principle not open to question on the grounds of the WD magnetic field (see discussion in \sectref{AccretionEfficiency}) because the bounds in that region of parameter space do not rely on the slow accumulation of CHAMPs over the WD lifetime being efficient.
We also show as a faint red dotted line for $m_X \gtrsim 6.4\times 10^8\,\text{GeV}$ the region of parameter space in which the total accreted and primordial CHAMP abundance would exceed the \Chand\ mass, even though it is smaller there than the self-gravitating mass, both as an indication of the degree to which limits could be strengthened if a lower core temperature were assumed, and also because of the caveat expressed in footnote \ref{ftnt:IdealGas} about the necessity of considering the thermal-pressure-supported phase.

However, in the region $10^5\,\text{GeV} \lesssim m_X \lesssim 10^{11}\,\text{GeV}$ (delimited by the thin long-dashed vertical black lines in \figref{BoundsPlus}), we have indicated our bounds by dotted lines (and lighter shading) to indicate that, in this region, the results of \citeR{Chuzhoy:2008zy} (and to some extent \citeR{Dunsky:2018mqs}) call into question whether CHAMPs will be present in the region of the MW in which the WD-progenitor protostellar cloud is collapsing, and thus whether this bound is actually applicable in that region of parameter space; see discussion in \sectref{galaxy}.
In particular, we do not claim a robust bound in this region; see also the discussion in \sectref{CaRGT} below.

\figref{BoundsPlus} is drawn assuming that the primordial CHAMP contamination $\eta_\pm$ is obtained using a central value for the mass of the progenitor protostellar cloud for this WD of $M_{\text{cloud}} \sim 7M_{\odot}$ \cite{Catalan:2008tr} (implicitly assuming that the zero-age-main-sequence star has the same mass as the collapsing protostellar cloud used in the computation of $\eta_\pm$); however, owing to a fair degree of scatter in the initial-mass--final-mass relationship for WDs (see, e.g., Figures 1 and 2 of \citeR{Catalan:2008tr}), this value could easily vary in the range $M_{\text{cloud}} \sim 6$--$8M_{\odot}$. 
Such changes only move the boundaries of various excluded regions by $\mathcal{O}(1)$ numerical factors.

Note also that in \figref{BoundsPlus}, we have conservatively plotted all our bounds only up to $m_X \sim 10^{18}$\,GeV, at which point the limit on $\rho_{X^+} f_{X^+} / \rho_0$ is $(\rho_{X^+} f_{X^+} / \rho_0)_{\text{limit}}  \sim 3\times 10^{-18}$. 
This is still abundantly safe from the Poisson uncertainty on the accumulated number of CHAMPs that was discussed \sectref{PopnAccumulated}: a  $\sim 10\%$  Poisson uncertainty required only $(\rho_{X^+} f_{X^+} / \rho_0)_{\text{Poisson}} \gtrsim 10^{-24}$ at this mass.

Also shown in \figref{BoundsPlus} is the the region delineated by thick gray lines (and shaded gray where not otherwise already shaded) where \citeR{Gould:1989gw} claim a bound from the destruction of old NS owing to the formation of a BH inside the NS, assuming a neutron star mass of $M_{\text{NS}} = 2M_{\odot}$ with a progenitor protostellar cloud with $M_{\text{cloud}} = 10M_{\odot}$ (note that the same objection to the bounds of \citeR{Gould:1989gw} for $10^5\,\text{GeV} \lesssim m_X \lesssim 10^{11}\,\text{GeV}$ arises as for our WD bounds based on the arguments of \citeR{Chuzhoy:2008zy}; we indicate the bounds there as thick, dotted lines, and shade the region more lightly).
As already discussed in \sectref{introduction}, these bounds weaken sharply around $m_X \sim 10^{10}\,$GeV owing to the Hawking process causing a BH with initial mass $M_{\text{BH}}\lesssim 4\times10^{-20}M_{\odot}$ to evaporate before it can devour the whole NS, which in the NS case does not inflict externally observable damage on the NS (absent this consideration, their bound would continue to higher masses).

\subsection{Limits on $X^-$}
\label{sect:limitsMinus}

\begin{figure*}[t]
\includegraphics[width=0.95\textwidth]{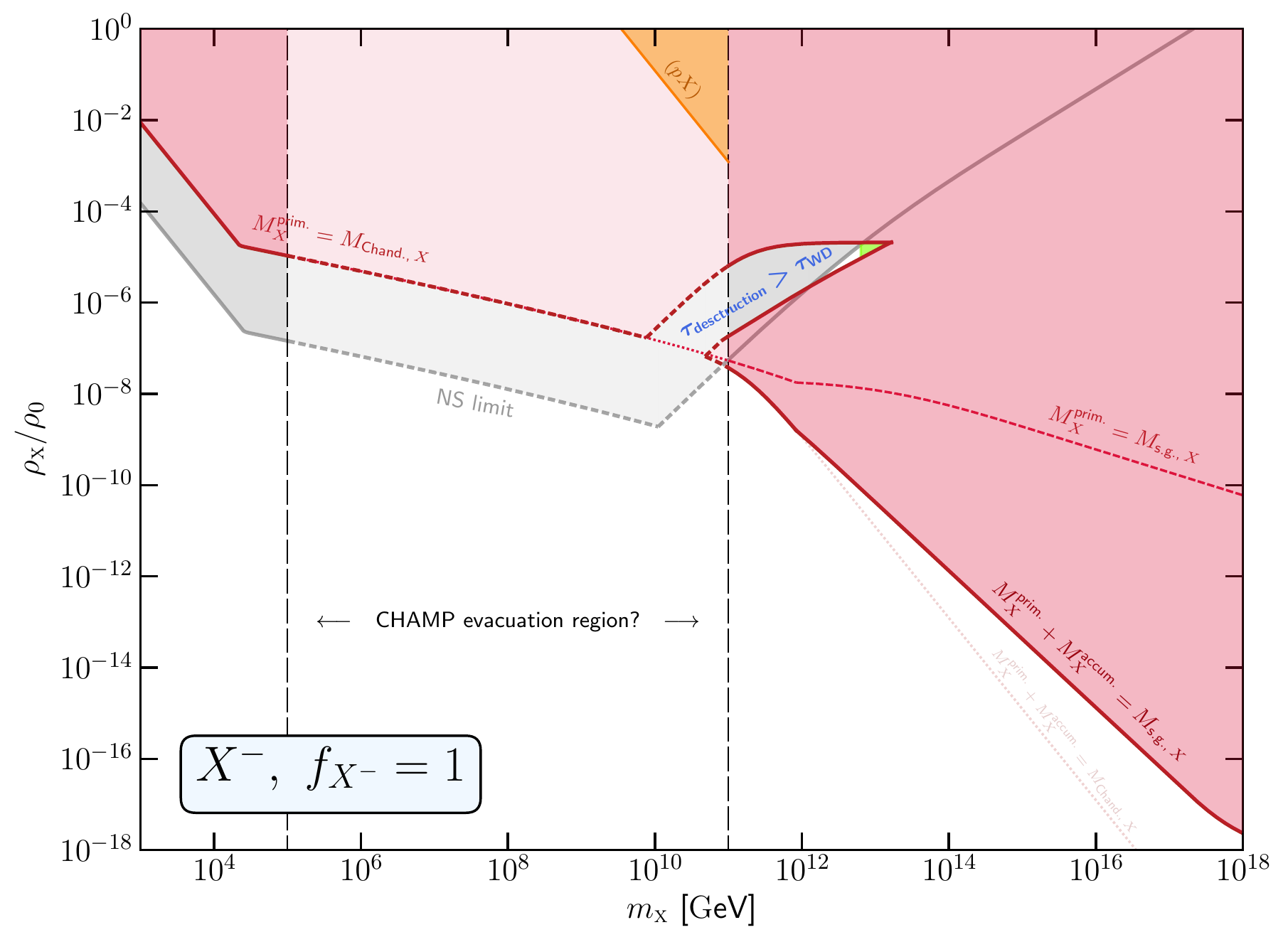}
\caption{\label{fig:BoundsMinus}%
		As for \figref{BoundsPlus}, but showing the constraints on the abundance of CHAMPs $X^-$ immutably bound as $(NX)$, assuming $f_{X^-}=1$ and $Q_{(NX)} = +6$.
		The kink in the bound at $m_X^{\text{s.g.}} \sim 8.3\times 10^{11}\,$GeV is where the \Chand\ and self-gravitating masses cross for $Q_{(NX)} = +6$.
		The additional orange shaded region [annotated as `$(pX)$'] as compared to \figref{BoundsPlus} indicates where, in the region $10^5\,\text{GeV} \lesssim m_X \lesssim 10^{11}\,\text{GeV}$, an abundance of $(pX)$ at the level of $10^{-4}$ of the total number abundance of $X^-$ would accumulate to a super-\Chand\ mass of CHAMPs (larger than the self-gravitating mass for these $m_X$) in the WD within the crystallization time, assuming that these neutral particles are not evacuated from the galactic disk by supernova shockwaves (and magnetically inhibited from re-entry) \cite{Chuzhoy:2008zy}, but are efficiently captured by dense WD material owing to exchange reactions such as $(pX) + N \rightarrow (NX) + p$ [see discussion in \sectref{accumulated}].
		Of course, the actual limit would not jump discontinuously from the red line to the orange region at $m_X \sim 10^{11}$\,GeV; there would be some smooth crossover region.
		CHAMPs lighter than $m_X \lesssim 10^{3}\,\text{GeV}$ have a sinking timescale in the WD estimated to approach or exceed the WD lifetime and/or crystallization time of $\sim \text{Gyr}$ and we conservatively do not present limits here; see \sectref{timescales}.
		}
\end{figure*}
For $(NX)$ we have argued that there are at least two points%
\footnote{\label{ftnt:thirdOption}%
		And possibly a third: during the collapse of the thermal-pressure-supported structure to a stratified core.
	} %
in the evolution prior to the BH formation where the WD thermal runaway could be triggered:  (1) by pycnonuclear processes in the trans-\Chand\ mass core of the WD, or (2) directly during collapse of the stratified core structure to a BH, if a fraction of the gravitational binding energy is transferred to the stellar matter.
We again conservatively ignore these early ignition mechanisms, and assume that the BH successfully forms from the initial stratified mini-WD-like core structure in the WD.
In order to destroy an old WD, we must again demand a sufficient mass of CHAMPs (as discussed in \sectref{limitsPlus}), and that the evolutionary timescales for the BH to either accrete the WD mass (or until a trigger criterion is possibly reached), or radiate to effectively zero mass (or until the trigger criterion is reached), must be sufficiently short.

Imposing these requirements, we find limits on $X^-$ bound as $(NX)$ that, for $f_{X^-}=1$, are broadly similar to those for $X^+$, but are weaker by a significant factor at low $m_X \lesssim m_X^{\text{s.g.}} \sim 8.3\times 10^{11}\,$GeV; see \figref{BoundsMinus}.
This is because the bound in this region is controlled by the \Chand\ mass, and the $(NX)$ \Chand\ mass is a factor of $\sim 6^2 \sim 36$ larger for the homogeneous $(NX)$ mixture as compared to the $X^+$ case, assuming the mean ion approach we have thus far taken.
For $m_{X} \gtrsim m_{X} ^{\text{s.g.}}$, the bounds for $(NX)$ and $X^+$ are identical, because the self-gravitating mass controls the bound, and the self-gravitating mass does not depend on the charge of the CHAMP (or CHAMP bound state); cf.~\eqref{MthermCoreMass}.
Note that boundaries of the shark-fin region where the evolutionary timescales are too long also shift slightly as compared to the $X^+$ case.

One important point to bear in mind however is that, in the region $10^5\,\text{GeV} \lesssim m_X \lesssim 10^{11}\,\text{GeV}$  where the results of \citeR{Chuzhoy:2008zy} call into question our bounds on the grounds that CHAMPs may get blown out of the galaxy by supernova shock waves, the $(pX)$ may not be so evacuated because, being neutral with binding energies $\sim 25\,$keV, they may not become efficiently entrained in supernova shock fronts \cite{Chuzhoy:2008zy}.
Moreover, the capture of $(pX)$ in the initial protostellar cloud is highly inefficient owing to their small cross sections (related to their charge neutrality); as such, $X^-$ bound as $(pX)$ are unlikely to primordially accumulate in the protostellar cloud (see discussion in \citeR{Gould:1989gw}).
However, we have argued that the presence of exchange reactions $(pX) + N \rightarrow (NX) + p$ combined with very high WD material densities make the stopping of $(pX)$ in a WD by the accumulation mechanism discussed in \sectref{accumulated} efficient.
Moreover, if they approach the WD still in the form of $(pX)$, they are not deflected by WD magnetic fields to nearly the same extent as charged particles, so even fairly light CHAMPs could accrete onto WD with large surface $B$ fields.
As such, we argue that a bound where the abundance of $(pX)$ accumulated onto the WD before the crystallization time exceeds the \Chand\ mass (or self-gravitating mass, whichever is larger) should still apply, but it is weaker than the cognate accumulation-only bound assuming a density $\rho_X$ of $X^-$ accumulate onto the WD by the ratio of abundance of $(pX)$ to other $(NX)$ states.
This fraction is of course difficult to estimate robustly, as it depends on the evolutionary history of the $(pX)$ states.
Nevertheless, to indicate the region where this bound would likely be applicable, we draw an orange shaded region in \figref{BoundsMinus} indicating where at least a \Chand\ mass of $X^-$ that are galactically present as $(pX)$ would accumulate onto the WD before the crystallization time, assuming that the $(pX)$ make up a fraction $10^{-4}$ of all $(NX)$ bound states, which is the early-Universe evolutionary estimate (see, e.g., \citeR[s]{Pospelov:2008ta,Kusakabe:2010cb,Pospelov:2010hj,Kusakabe:2017brd}).

\subsection{Limits when both $X^+$ and $X^-$ are present}
\label{sect:limitsCombined}
The case of most widely applicable physical relevance is some mixture of $X^+$ and $X^-$ [bound as $(NX)$] in the WD, with both species co-existing without annihilating by virtue of the large Coulomb barrier between the $X^+$ and the $X^-$ bound within the positively charged $N$ nuclear volume.
However, our discussion in \sectref{combinedCase} shows that the $X^+$ and $(NX)$ will rapidly stratify with the $X^+$ sinking toward the center of the core, and it would be conservative to demand that the total ${X^+}$ mass alone exceeds the $X^+$ \Chand\ mass.
Given that the $X^+$ bounds are, roughly speaking, a factor of $Q_{(NX)} \sim 6^2 = 36$ times stronger than the $(NX)$ bounds (by virtue of the $M_{\text{Chand.,}\,X}\propto Q_X^2$ scaling) at low $m_X \lesssim 6.4\times 10^8$\,GeV [where the $X^+$ \Chand\ and self-gravitating masses cross], so long as $f_{X^+} \gtrsim (Q_{X^+}/Q_{(NX)})^2 f_{X^-} \Rightarrow f_{X^+} \gtrsim 1/37$ (since $f_{X^+} + f_{X^-} = 1$), the limits on $\rho_X /\rho_0$ are to a good approximation given by
\begin{align}
\lb[ \frac{\rho_X}{\rho_0} \rb]_{\text{limit}} &\sim  \frac{1}{f_{X^+}} \times \lb[ \frac{\rho_X}{\rho_0} \rb]_{\text{limit},\, X^+}\nl 
	\text{for} \quad  Q_X^2/Q_{(NX)}^2 \lesssim f_{X^+} \leq 1 \nl
	\text{and} \quad m_X \lesssim 6.4\times 10^8\,\text{GeV}
\end{align}
where $\lb[ \rho_X / \rho_0 \rb]_{\text{limit}}^{X^+}$ is the limit assuming $f_{X^+}=1$, as shown in \figref{BoundsPlus}.

Likewise, once $f_{X^+} \ll 1/37$, the $X^-$ bounds on $\rho_X/\rho_0$ obtained by requiring a \Chand\ mass of $(NX)$ in the core would be stronger than those obtained from $X^+$; moreover, the total mass of $X^+$ is in this regime so small compared to the total mass of $(NX)$ that the impact of the small central $X^+$ core structure will likely not significantly impact the structure of the $(NX)$ layer above it.
Furthermore, the dense $X^+$ core would likely only aid to make the $(NX)$ core collapse earlier, so it is conservative in this case to fix the limit on $\rho_X/\rho_0$ by requiring a \Chand\ mass of $(NX)$ in the WD, which leads to the limit
\begin{align}
\lb[ \frac{\rho_X}{\rho_0} \rb]_{\text{limit}} &\sim  \frac{1}{f_{X^-}} \times \lb[ \frac{\rho_X}{\rho_0} \rb]_{\text{limit},\, X^- }\nl 
	\text{for} \quad 1-Q_X^2/Q_{(NX)}^2 \lesssim f_{X^-} \leq 1 \nl
	\text{and} \quad m_X \lesssim 6.4\times 10^8\,\text{GeV}.
\end{align}
In the tuned cross-over region where $f_{X^+} \sim (Q_{X^+}/Q_{(NX)})^2 f_{X^-} $ (and $m_X \lesssim 6.4\times 10^8\,\text{GeV}$), one would need to be more careful, but a limit correct within a factor that is $\mathcal{O}(1)$ would be set by requiring that an $X^-$ \Chand\ mass worth of CHAMPs is present in total in the star.

We propose the limit in the mixed $X^+$/$(NX)$ case can be given conservatively by 
\begin{align}
\lb[ \frac{\rho_X}{\rho_0} \rb]_{\text{limit}} \sim \min\Bigg[& \frac{1}{f_{X^+}}  \lb[ \frac{\rho_X}{\rho_0} \rb]_{\text{limit},\, X^+} ,  \nl 
\frac{1}{1-f_{X^+}} \lb[ \frac{\rho_X}{\rho_0} \rb]_{\text{limit},\, X^- } \Bigg]
\label{eq:combinedBound}
\end{align}
which will capture the true limit correct to within $\mathcal{O}(1)$ factors, for $m_X \lesssim 6.4\times 10^8\,\text{GeV}$.

In the region $m_X \gtrsim 8.3\times 10^{11}\,\text{GeV}$ [where the $(NX)$ \Chand\ and self-gravitating masses cross], the limits from $X^+$ and $(NX)$ become equal under our conservative assumptions, because the self-gravitating mass fixes the bound, and it does not depend on the charge of the CHAMP state; cf.~\eqref{MthermCoreMass}, but see footnote \ref{ftnt:IdealGas}.
In this case, it is again conservative to fix the bound to be given by \eqref{combinedBound}, which, since $\lb[ \rho_X / \rho_0 \rb]_{\text{limit},\, X^- }= \lb[ \rho_X / \rho_0 \rb]_{\text{limit},\, X^+}$ in this mass range, is at most a factor of 2 weaker than either individual bound: $\max_{f_{X^+}}\{ \min\lb[ 1/f_{X^+} , 1/(1-f_{X^+}) \rb] \} = 2$.

In the intermediate region, $ 6.4\times 10^{8}\,\text{GeV}\lesssim m_X\lesssim 8.3\times 10^{11}\,\text{GeV}$, \eqref{combinedBound} will still give a conservative estimate for the lower envelope of the possible bounded region; however, this is also roughly the range of parameter space in which the BH evolutionary timescales are too long to destroy the WD within its age, so some care is required to set a rigorous bound in this region, and obtain the correct shark-fin-shaped non-excludable region (see \figref[s]{BoundsPlus} and \ref{fig:BoundsMinus}).
Conservatively, one can of course choose to fix the lower edge of the bounded region by \eqref{combinedBound}, but avoid setting a bound in the union of the regions where the evolutionary timescales are too long assuming pure $X^+$ contamination, and where they are too long assuming $(NX)$ contamination.
While not exact, this will give an estimate of the actual bound that is likely no worse than the uncertainties in where the bound should lie owing to factors such as, e.g., the mass of the progenitor gas cloud, etc.

Finally, we note that the NS limits from \citeR{Gould:1989gw} bound only the \emph{asymmetry} $f \equiv |f_+ - f_-|$ because, in the NS, the $X^+$ and $X^-$ are not inhibited from annihilating by a Coulomb barrier as they are in our case (see discussion in \sectref{CHAMPchemistry}).
We place our overall bounds using the more restrictive of the bounds on the individual $X^+$ and $X^-$ contamination components, without assuming any annihilation. 
In particular, this means that for $f_{X^+} \sim f_{X^-}$ galactically, which is a highly motivated case that would occur with a $CP$ conserving production mechanism, our bounds are approximately 
\begin{align}
\lb[ \frac{\rho_X}{\rho_0} \rb]_{\text{limit}} \sim 2  \lb[ \frac{\rho_X}{\rho_0} \rb]_{\text{limit},\, X^+},
\end{align}
whereas the bounds from \citeR{Gould:1989gw} may possibly be significantly weakened (although they would not disappear entirely except maybe in a highly tuned region of parameter space, because of possible differences in accretion efficiency, etc. for the $X^\pm$ \cite{Gould:1989gw}).

\subsection{Comment on assumed uniform CHAMP distribution}
\label{sect:UniformComment}
We have assumed throughout this paper that the primordial CHAMP abundance is uniformly distributed in the WD at the time of WD formation.
However, some early sinking is to be expected, possibly even during the evolution of the WD-progenitor star lifetime.
Taken to the logical extreme, this could even cause the primordial abundance of CHAMPs to collapse to a BH earlier in the evolution of the WD progenitor star, before the CO core of this progenitor even becomes degenerate.
In this case, if the BH thus formed evaporates before the core is degenerate, it may not trigger a thermal runaway at all.

However, in almost the entire region where BH evaporation is responsible for destroying the WD per our canonical picture, the limits set by the additional CHAMPs accumulated onto the WD after formation are actually stronger than those coming from the primordial abundance alone.
Those limits would not be impacted by early CHAMP collapse and BH evaporation: even if the primordial abundance of CHAMPs collapsed early to form a BH which radiated away, a second BH could still form long after the WD is born (but before the crystallization time) owing to accretion of additional CHAMPs onto the WD.

Moreover, in the region of parameter space where limits from the primordial CHAMP abundance are much stronger than those from the additional accumulated CHAMPs, it turns out that the BH dynamics dictate that the BH would grow in size after formation to devour the whole star, instead of evaporating away.
In this case too, any early sinking is a non-issue, because the BH would still be present when the WD is born (if the progenitor star even survives that long).
Our bounds are thus quite robust to having the CHAMPs sink in the WD progenitor star prior to the formation of a degenerate core.

\section{Speculations: Ca-rich gap transients}
\label{sect:CaRGT}
In this section, we adopt a much more speculative attitude to make some comments on the so-called `Ca-rich Gap Transients' (hereinafter `CaRGT'). 

The CaRGTs are a class of approximately ten observed anomalous supernova-like events which are found to occur preferentially displaced from their most-likely host galaxies; see, e.g., \citeR[s]{Perets_2010,Kasliwal_2012,Lunnan:2016ake,De:2018wpe,shen2019progenitors} and references therein.
Events are included in this anomalous class based on specific criteria (see e.g., \citeR[s]{Kasliwal_2012,Lunnan:2016ake,De:2018wpe}) identifying them as rapidly evolving, calcium-rich, and faint (luminosities in the `gap' between novae and supernovae) transients; the statistically significant \cite{Lunnan:2016ake,Frohmaier:2018fed} preferentially large spatial offset from the most likely host is however \emph{not} one of the defining characteristics for inclusion, but is rather feature of the class to be explained.

The progenitors of these transient events are as-yet not known, although several `conventional astrophysics' explanations have been advanced (see \citeR{De:2018wpe} and references therein), such as low-mass He WDs which are detonated by some external perturbation (e.g., tidal deformation by a NS or BH in a close binary with the WD), and He shell detonation on a relatively low-mass CO WD core.

Here, we advance a possible unconventional explanation: that these events occur as a result of CHAMPs in the mass range $10^5\GeV < m_X < 10^{11}\GeV$ accumulating in sufficient quantity onto a sub-\Chand\ mass WD that they trigger the thermal runaway instability, destroying the WD in a supernova explosion. 
We suggest this CHAMP mass range as it is where the results of \citeR{Chuzhoy:2008zy} suggest that the CHAMP population in the center of galaxies (or in the disk) could be depleted [although see comments about $(pX)$ in \sectref[s]{galaxy} and \ref{sect:limitsMinus}].%
\footnote{\label{ftnt:RegionComment}%
		Of course, the results of \citeR{Chuzhoy:2008zy} were obtained specifically using parameters for the MW, so the boundaries of the CHAMP evacuation region may differ in a different system; as this section is in any event speculative, we simply adopt the MW results and show the allowed parameter space under that assumption.
		Any follow-up study that further examines our speculations would of course have to take this into account.
	} %

We imagine a WD allowed to form in the inner regions of a galaxy where the CHAMP density is low.
This WD can thus avoid being destroyed initially by protostellar or accreted CHAMP contamination.
If this WD then happens to be gravitationally ejected from the inner regions of the galaxy,%
\footnote{\label{ftnt:WDwanderingDistance}%
		Note that the long timeframe envisaged here provides ample time for the WD to move a significant distance in the galaxy: suppose we impose that the WD may take only 10\% of the crystallization time ($\sim 10^8$\,yrs) to move from the inner CHAMP-depleted region to the outer CHAMP-rich region, so that it still has 90\% of the crystallization time available to accrete CHAMPs in the CHAMP-rich region.
		Assuming that the WD moves with typical galactic speeds $v\sim10^{-3}$, it moves a distance of $\sim 30$\,kpc in this time, which is sufficient (see text).
	} %
it will enter a region in which the CHAMP density is not depleted, where it can begin to accrete CHAMPs from the remaining (unejected) virialized CHAMP density in the outer halo.
Alternatively, a WD could simply be born in the outer halo, but these regions of a galaxy are star-poor.
 
Should this WD come to accrete more than a \Chand\ mass (or a self-gravitating mass, whichever is larger) worth of CHAMPs within a crystallization time $\tau_{\text{accum.}}\sim10^9$\,yr of initial WD formation, a BH would form in the WD (unless the WD is destroyed earlier as we have discussed throughout), and in the interesting region of parameter, accrete up in mass over time.
A SN could then be triggered if the explosive mechanisms upon accretion of WD material \cite{Janish:2019nkk} discussed in \sectref{outcomesAccreting} are operative.
As this picture envisages CHAMP accretion that can only occur on the outskirts of galaxies, it could naturally explain the observed spatial distribution of the CaRGT, as well as their intermediate luminosities, as the SN is triggered in a sub-\Chand\ progenitor.
Although detailed modeling of the light curves and nucleosynthetic abundances of such an event are well beyond the intended scope of this paper, it is at least plausible that such a sub-\Chand\ progenitor could explain the high Ca yield, per recent modeling \cite{polin2019nebular,2011ApJ...734...38W,KasenPrivateCommunication}.
This picture may however be challenged by the some recent high estimates for the inferred rates for these events \cite{Frohmaier:2018fed,shen2019progenitors}, as it relies on some fairly rare events to be successful.

In order to roughly estimate the plausible region of parameter space in which this picture could operate, we consider a $M_{\text{WD}} = 0.85M_{\odot}$ WD with an extremely low magnetic field ($\lesssim 1\,$kG; see \sectref{AccretionEfficiency}) [see \tabref{WDcharacteristics} for other WD parameters], assuming that the progenitor of this WD formed in a region where the CHAMP abundance [except perhaps for $(pX)$] is zero owing to the mechanism of \citeR{Chuzhoy:2008zy}, so that the progenitor had zero initial CHAMP contamination [the $(pX)$ are not efficiently captured by the protostellar cloud \cite{Gould:1989gw}].
We assume that this WD then gets gravitationally ejected from the initial CHAMP-depleted region into an outer region of its parent galaxy (at galactocentric distance $\tilde{r}$),where the CHAMP abundance is non-zero.
For simplicity, we adopt a somewhat na\"ive model in which the WD, once ejected, experiences a constant, virialized CHAMP abundance $\rho_X(\tilde{r})$ for a period of $\tau \sim \tau_{\text{accum.}} \sim 10^9$\,yr, with the total attainable CHAMP mass in the star taken to be that given by the approximate expression \eqref{MXaccrete2}.
This is of course very schematic.
In particular, the CHAMP abundance will not just rapidly turn on as the WD moves in the galaxy, so by approximating the temporal integral in \eqref{MXaccrete1} by $\rho_X(\tilde{r}) \times \tau_{\text{accum.}}$, we make some error which is difficult to quantify given that we do not know the spatial profile of the evacuated CHAMPs (see below) and we do not specify the WD trajectory.

Moreover, the approximation \eqref{MXaccrete2} ignores the velocity distribution of the CHAMPs [which would in any event likely not be a Maxwellian distribution in the galactic rest frame if the CHAMPs have been significantly impacted by the expulsion dynamics (see, e.g., \citeR{Dunsky:2018mqs})]; however, the exact assumed velocity distribution only matters when considering much more massive CHAMPs ($m_X \gtrsim 10^{16}$\,GeV) where there is a question of whether the entire distribution can be efficiently captured onto the WD.
As our goal here is not to be exact, but rather to advance a plausibility argument that some region of parameter space could be available, we judge these rough approximations to be fit for purpose.

\begin{figure*}[t]
\includegraphics[width=0.45\textwidth]{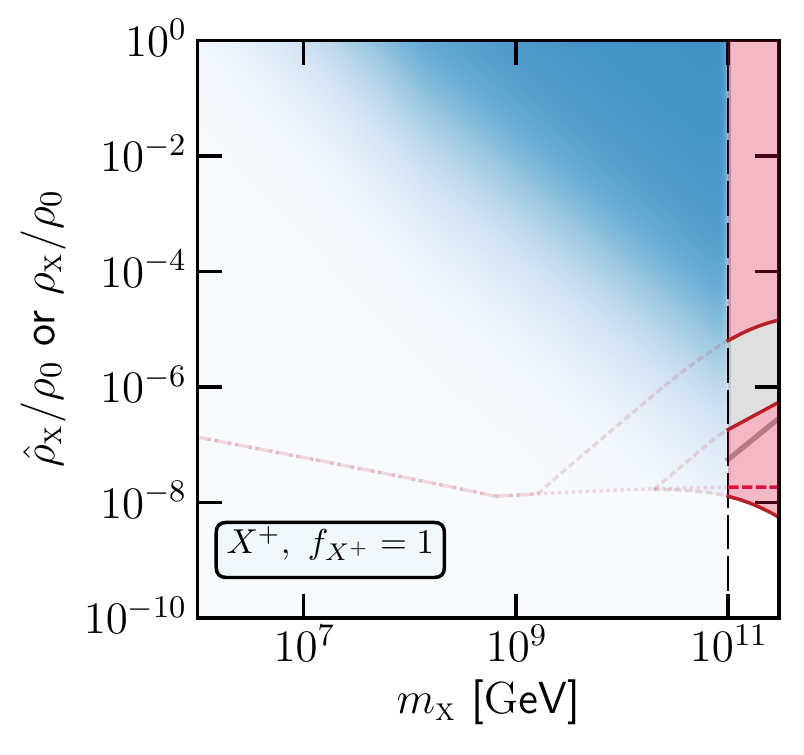}\hfill
\includegraphics[width=0.45\textwidth]{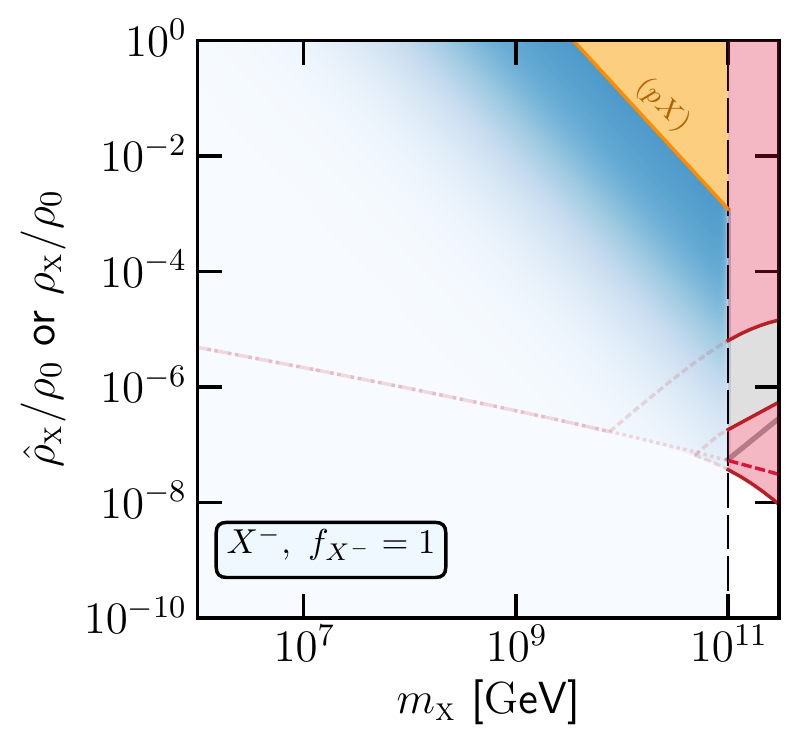}
\caption{\label{fig:CaRGT}%
	These plots indicate the regions (blue shaded) of $(m_X,\, \hat{\rho}_X/\rho_0)$ parameter space where our speculations about a possible trigger mechanism for the CaRGTs are applicable, both for $X^+$ (left panel) and $X^-$ bound as $(NX)$ (right panel).
	These shaded regions are shown with very fuzzy edges to emphasize the large uncertainties associated with our estimates, and their highly speculative nature.
	Also shown for $m_X \gtrsim 10^{11}\,$GeV are the bounds on $(m_X,\, \rho_X/\rho_0)$ from \figref[s]{BoundsPlus} and \ref{fig:BoundsMinus} [see captions there for explanations of the shading].
	The limits from \figref[s]{BoundsPlus} and \ref{fig:BoundsMinus} in the region $m_X \lesssim 10^{11}\,$GeV have been removed as, for our CaRGT picture to be applicable, the CHAMP expulsion mechanisms of \citeR{Chuzhoy:2008zy} must be operative, and this invalidates the limits from \figref[s]{BoundsPlus} and \ref{fig:BoundsMinus} in that mass range (but see footnote \ref{ftnt:RegionComment} for an important caveat about the mass range of interest); we have nevertheless included very light shaded lines that follow the outlines of those limits in order to guide the eye in a comparison of the regions of parameter space.
	Note that the CaRGT regions are shown for $\hat{\rho}_X/\rho_0$, while the limits from \figref[s]{BoundsPlus} and \ref{fig:BoundsMinus} are on $\rho_X/\rho_0$; in order to show both sets of results on the same plot and for the comparison to be meaningful, we have had to identify $\hat{\rho}_X$ and $\rho_X$ [see \sectref{CaRGT} and footnote \ref{ftnt:commentOnArbitraryRhoX}].
	The blue shaded regions assume $M_{\textsc{wd}} = 0.85M_{\odot}$ and $\tau_{\text{accum.}}\sim \text{Gyr}$; see \tabref{WDcharacteristics} for other parameters.
	}
\end{figure*}

In estimating $\rho_X(\tilde{r})$ we will make the approximation that although the inner region of the galaxy is actually evacuated of CHAMPs, the CHAMP density in the outer region still roughly tracks the spatial distribution of the DM halo at $r=\tilde{r} \sim 30\,$kpc, a representative median projected host--transient offset for the known CaRGTs (see, e.g., Figure 11 of \citeR{Lunnan:2016ake}), just with a different normalization to the DM; that is, we crudely approximate that
\begin{align}
\rho_X(r) \propto \rho_{\textsc{dm}}(r) \Theta[ r - r_{\text{evac.}} ],
\label{eq:rhoX1}
\end{align}
where $r_{\text{evac.}}$ is the radius within which we assume the CHAMPs would be evacuated by the SN shockwaves.
Here we are implicitly assuming either that (a) if the host galaxy is a disk galaxy, the evacuation of CHAMPs \cite{Chuzhoy:2008zy} (although see \citeR{Dunsky:2018mqs}) above and below the disk is still efficient even absent a large density of supernova shockwaves in these baryon-poor regions, or (b) the host galaxy is elliptical.
These assumptions are required in order that is plausible that a full 3D volume of CHAMPs have been evacuated (see below). 
However, in the latter case this does of course lead to a mismatch: the mechanism of \citeR{Chuzhoy:2008zy} on which we would need to rely to create this evacuated region specifically considered the properties of a MW-like disk galaxy.
Although SN shockwaves will certainly accelerate CHAMPs regardless of the galaxy type, there is a question as to whether there are appropriate and sufficiently large magnetic fields in elliptical galaxies to magnetically inhibit the entry of CHAMPs initially external to the evacuation volume into that volume.
We simply assume that this is the case and that the same range of $m_X$ are subject to the mechanism in the galaxies in which the CaRGT are observed as in the MW; if either of these assumptions is not the case, the picture presented here may break down, and/or require refinement.%
\footnote{\label{ftnt:OptimisticComment}%
		Alternatively, and optimistically, there may even be some discriminating power here: if different types of galaxies evacuate CHAMPs to differing degrees of efficiency, or in different spatial volumes, a high statistics sample of CaGRT events could even show distinct spatial morphologies of these events in different galaxy types if this is the trigger mechanism.
	} %
Note further that it is important for a full 3D region of CHAMPs to be evacuated, because if only a thin 2D slice (e.g., the MW disk) is evacuated then, once projection effects are accounted for, it is likely that the observed SNIa-like events triggered by the accumulation of CHAMPs would not exhibit the correct spatial morphology: a WD would only need to be ejected above or below the disk to experience a large CHAMP abundance, regardless if the radial distance from the center of the galaxy.

It is still necessary to settle on a specific parametrization convention for the DM abundance.
We will parametrize
\begin{align}
\rho_X(\tilde{r}) \equiv \frac{\hat{\rho}_X}{\rho_0}\,\frac{ \rho_{\textsc{dm}}(\tilde{r}) }{\rho_0}\, \rho_0,
\label{eq:rhoX2}
\end{align}
where $\hat{\rho}_X$ is an arbitrary constant density normalization,%
\footnote{\label{ftnt:commentOnArbitraryRhoX}%
		Although not completely arbitrary: if the CHAMP abundance thus parametrized where evaluated in the MW, and the CHAMPs were \emph{not} blown out of the MW disk, then $\rho_X(r_{\oplus}) = \hat{\rho}_X$.
	} %
and $\rho_0=0.3\,\text{GeV/cm}^3$ is the local DM density in the MW, which is simply used here as a convenient normalization.
Although none of the CaRGT events have been observed to occur in our galaxy, we will take MW-like parameters for $ \rho_{\textsc{dm}}(\tilde{r})$, assuming that has approximately the same radial dependence and normalization as the MW DM abundance (i.e., $\rho_{\textsc{dm}}(r_{\oplus}) \sim \rho_0$, where $r_{\oplus}$ is the distance from the Earth to the MW galactic center); this is of course a guess about the DM abundance in the CaRGT host galaxy, but the observed CaRGT events occur in galaxies of comparable sizes to the MW, so it is a well-motivated guess.
Note that under these assumptions, the value of $\rho_{\textsc{dm}}(\tilde{r})$ is insensitive at the $\sim$20\% level to the choice of NFWc($\gamma = 1,\ r_s = 20\,$kpc), NFWc($\gamma = 1.3,\ r_s = 20\,$kpc), Einasto($\alpha=0.17,\ r_s = 20\,$kpc), or isothermal ($r_s = 5\,$kpc) DM halo profile models (see, e.g., \citeR{Fedderke:2013pbc} and references therein for definitions), and can be approximated as 
\begin{align}
\frac{ \rho_{\textsc{dm}}(\tilde{r}) }{\rho_0} \sim 9\times 10^{-2}.
\label{eq:rhoDM}
\end{align}
Finally, in making our estimates, we assume that $v_{\textsc{wd}} \sim v_0(\tilde{r}) \sim \sqrt{ G M_{\text{galaxy}} / \tilde{r} } \sim 10^{-3}$, which must be correct within an order of magnitude unless the WD has been expelled from the inner region of the galaxy so fast that it is no longer gravitationally bound to the galaxy (any slower, and the WD would likely not move a sufficient distance in the galaxy to make our speculative mechanism operable, unless the WD being ignited are formed in the outskirts of the galaxy).

Under these assumptions, we compute the region of parameter space ($m_X$,\,$\hat{\rho}_X/\rho_0$), for the cases of pure $X^+$ and $X^-$ [bound as $(NX)$] contaminations, in which an initially CHAMP-free WD as described above would accrete at least the larger of a \Chand\ mass and a self-gravitating mass of CHAMPs before the crystallization time; see \figref{CaRGT}, where we show these regions (blue) with a fuzzy boundary to emphasize that there are large uncertainties in their exact location.
Note also that this region is drawn without regard to the region in which the evolutionary timescales for WD destruction are sufficient rapid [the region for $M_{\textsc{wd}} \sim 0.85\,M_{\odot}$ being somewhat different than that for $M_{\textsc{wd}} \sim 1.1M_{\odot}$].

We see that there is plausibly some region of parameter space in which this mechanism could clear the first hurdle and provide a trigger mechanism for a sub-\Chand\ WD in the spatial location at which CaRGT events occur; of course, this is not dispositive, both because of the large uncertainties on our estimates here, and because a careful estimate for the rate of this process would be required.
Nevertheless, taken at face-value, the fact that there appears to be some allowed region is interesting and allows us to speculate that this may be an explanation for these events that would somewhat naturally explain their spatial morphology.

\section{Conclusion}
\label{sect:conclusion}
In this paper we have considered the impact of an abundance of extremely massive, stable (or cosmologically long-lived), early-Universe relics with $\mathcal{O}(1)$ electrical charge on the late-stage evolution and survival of old white dwarfs (WD).

Such charged massive particles (CHAMPs) can come to contaminate old WD in at least two ways: by contaminating the protostellar cloud which collapses to form the main-sequence WD-progenitor star, thereby ending up in the WD at the end of the evolutionary phase of that massive progenitor, or---provided the WD has a small enough magnetic field to not deflect the incoming CHAMPs---by direct accretion on the WD over the course of its lifetime.
These CHAMPs, being extremely massive, sink in the WD, most probably forming first a thermally pressure-supported structure at the center, then later undergoing self-gravitating collapse to a miniature CHAMP-contaminated WD-like object at the center of the WD, provided a sufficient mass of CHAMPs is present.
If a sufficient mass of CHAMPs is present in the WD, either primordially upon WD formation, or accreted up over the WD lifetime, this central dense CHAMP-contaminated WD-like object eventually collapses to a black hole.
This black hole in general has complicated dynamics because it can accrete WD matter and CHAMPs to increase its mass, as well as Hawking radiate to reduce its mass. 

Our detailed study of the possible BH trajectories leads us to conclude that there are only three physical outcomes (see also \citeR[s]{Janish:2019nkk,Acevedo:2019gre}): (1) the BH survives for a length of time of order the WD age by virtue of a balancing between the mass accretion and mass loss rates and intrinsically long timescales for evolution, and nothing interesting happens observationally for an external observer; (2) the BH will evaporate within the WD lifetime if it ever forms, with the increasing high Hawking radiation power emitted by the BH serving to trigger the thermal runaway instability of the WD material, leading to a SNIa-like explosion; or (3) the BH will accrete enough matter within the WD lifetime to either conservatively devour the entire WD  (with no directly observable signature) or alternatively trigger the thermal runaway instability of the WD by heating of accreting carbon ions (or possibly other mechanisms), again resulting in a SNIa-like supernova explosion.

With the exception of the region of parameter space where the evolutionary timescales are too long, the WD destruction mechanisms outlined above impose severe bounds (see \figref[s]{BoundsPlus} and \ref{fig:BoundsMinus} for our main results) on the allowed galactic abundance of CHAMPs (either $X^+$ or $X^-$) that for $m_X \gtrsim 10^{11}\,$GeV are many orders of magnitude stronger than existing astrophysical bounds on such particles derived from their destruction of old neutron stars via a similar accretion mechanism (although without the possibility of a SNIa-like explosion).
We are able to place bounds on the abundances of $X^+$ and $X^-$ separately, and not on the residual asymmetry after their annihilation in the dense core of the WD, because the $X^-$ become deeply bound to nuclei in the WD, resulting in nuclear-sized bound states that are net positively charged and  which therefore prevent the $X^+$ and $X^-$ from approaching close enough to capture and annihilate.
This makes our bounds even stronger than existing constraints in the regime where the net charge asymmetry of the galactic CHAMP abundance is zero or small.

Some variations on the above picture are possible, but they too lead to WD destruction.
For instance, if the energy injection into the WD material at earlier phases of the evolution of the CHAMP core results in enough energy deposition into the WD material during any of the collapse phases (thermal-pressure-supported structure to mini-WD-like core structure, or mini-WD-like core structure to BH), the thermal runaway can be triggered.
Or, in the case of $X^-$, the $({}^{12}\text{C}X^-)$ bound states which form the bulk of the dense pre-BH mini-WD-like core structure inside the WD may become so dense that pycnonuclear (density-enhanced) fusion processes between carbon ions can occur at a sufficient rate to lead to thermal runaway; this can only happen just prior to the collapse of the core to the BH, if at all.

Additionally, we have speculated that in certain regions of parameter space, the possible WD trigger mechanism provided by an accreting BH could give a natural explanation for the so-called calcium-rich gap transient supernova events, because this mechanism can trigger sub-\Chand\ WD to go off as a supernova, which could naturally explain the Ca-rich spectra and sub-luminous nature of these events \cite{polin2019nebular}.
Moreover, if the supernova shocks are efficient at depleting the CHAMP abundance in baryon-rich regions of galaxies [with magnetic field inhibiting the (re-)entry of (expelled) CHAMPs] as has been argued in the literature \cite{Chuzhoy:2008zy}, the spatial morphology of these events, which are observed to occur preferentially far from the center of their host galaxies, could be naturally explained.
This mechanism is of course highly speculative, and other more conventional astrophysics explanations for these events may suffice.

In summary, our work improves astrophysical bounds on the allowed galactic CHAMP abundance by many orders of magnitude at the highest CHAMP masses $m_X \gtrsim 10^{11}\,$GeV, and advances a speculative explanation for a class of anomalous supernova events.

\acknowledgments
We thank David Dunsky, Ryan Janish, Daniel Kasen, Vijay Narayan, and Stanley Woosley for productive and informative conversations. 
M.A.F. would like to thank the Berkeley Center for Theoretical Physics at the University of California Berkeley and Lawrence Berkeley National Laboratory for their long-term hospitality during which much of this work was completed. 
M.A.F.~and P.W.G.~were supported by DOE Grant DE-SC0012012, by NSF Grant PHY-1720397, the Heising-Simons Foundation Grants 2015-037 and 2018-0765, DOE HEP QuantISED award \#100495, and the Gordon and Betty Moore Foundation Grant GBMF7946. 
S.R.~was supported in part by the NSF under grants PHY-1818899 and PHY-1638509, the Simons Foundation Award 378243 and the Heising-Simons Foundation grant 2015-038.
M.A.F.~acknowledges the Galileo Galilei Institute for Theoretical Physics in Florence, Italy for their support and hospitality, which enabled part of this work to be completed.

\appendix

\section{\texorpdfstring{$(NX)$}{(NX)} binding energies}
\label{app:CHAMPbinding}
In this Appendix, we outline the computation of the binding energy of a heavy, negatively charged CHAMP $X^-$ with a heavy nucleus $N$; see also \citeR[s]{Cahn:1980ss,Tiburzi:2000rsq,Pospelov:2006sc}.

We model the nucleus as a uniform charged ball of radius $R\approx R_0 A^{1/3}$ with $R_0=1.22\,\text{fm}$ \cite{Krane:1988abc}, and charge $Q_N$. 
Let $r$ be the relative co-ordinate separating the point-like CHAMP of charge $-|Q_X|$ and the center of the nucleus.
The electrostatic potential energy of this configuration is then 

\begin{align}
\Rightarrow V(r)  &= \begin{cases} - \dfrac{\hat{\alpha}}{r} & r > R \\[2ex]
							- \dfrac{\hat{\alpha}}{2R} \lb[ 3 - \dfrac{r^2}{R^2} \rb] & 0\leq r\leq R
				\end{cases},
\label{eq:VNXbinding}
\end{align}
where $\hat{\alpha} \equiv \alpha |Q_X| Q_N$.
The fact that the potential energy transitions from a $1/r$ potential for $r>R$ to a shifted harmonic oscillator potential for $0\leq r\leq R$ implies that the binding energies will be reduced from the na\"ive Bohr atom binding energies.
For reference, the na\"ive Bohr atom ground state would have Bohr radius and binding energy given respectively by \cite{Shankar:1994abc}
\begin{align}
a &= \frac{1}{\hat{\alpha} \mu} &
E_B^{\text{point}} &= \frac{1}{2} \hat{\alpha}^2 \mu,
\end{align}
where $\mu = m_X m_N / (m_X + m_N) \approx m_N$ is the reduced mass of the system.

Setting the wavefunction for the system to be $\Psi(t,\bm{r}) = e^{-iEt} \frac{u(r)}{r} Y_{l}^m(\theta,\phi)$, the 3D Schr\"odinger equation (SE) of course reads
\begin{align}
- \frac{1}{2\mu} u'' + \lb[ V(r) + \frac{l(l+1)}{2\mu r^2} \rb] u + E_B u &=0,,
\end{align}
where we set $E = - E_B$ with $E_B>0$ the binding energy of the system.

Consider then the 3D SE for $l=0$ with $u=u_1$ for $r>R$ and $u=u_2$ for $0\leq r\leq R$, with $C^1$-smoothness between $u_1$ and $u_2$ imposed at $r=R$; the boundary conditions (BCs) to be imposed on $u$ are $u_1(\infty) = 0$ and $u_2(0)=0$ (for discussion of the latter condition see, e.g., \S12.6 of \citeR{Shankar:1994abc}):
\begin{align}
- \frac{1}{2\mu} u_1''  - \frac{\hat{\alpha}}{r} u_1 + E_B u_1 &= 0\\
- \frac{1}{2\mu} u_2''  - \frac{\hat{\alpha}}{2R} \lb[ 3 - \frac{r^2}{R^2} \rb] u_2 + E_B u_2 &= 0.
\end{align}
Suppose we rescale $x \equiv r/R$, define $\beta \equiv R/a$ where $a$ is the Bohr radius as defined for the point charge setup, and define $\epsilon = 2\mu R^2 E_B \Rightarrow E_B = E_B^{\text{point}}\times(\epsilon/\beta^2)$, where $E_B^{\text{point}}$ is the binding energy of the point charge setup.
We then have
\begin{align}
(u_1)_{xx}  + \frac{2\beta}{ x} u_1 &= \epsilon u_1 \\
(u_2)_{xx}  - \beta x^2 u_2 &= ( \epsilon - 3\beta ) u_2 \\
u_1(\infty) = u_2(0) &= 0 \\
u_1(1) = u_2(1) &= 0 \\
u'_1(1) = u'_2(1) &= 0.
\end{align}
The exterior solution which obeys the BC is
\begin{align}
u_1(x) &= C_1 x e^{-x\sqrt{\epsilon}}\; U\!\lb(1-\frac{\beta}{\sqrt{\epsilon}},\; 2,\; 2x\sqrt{\epsilon} \rb),
\end{align}
where $U(a,b,z) \equiv \Gamma(a)^{-1} \int_0^\infty e^{-zt} t^{a-1} (1+t)^{b-a-1} dt$ is the confluent hypergeometric function.
The interior solution which obeys the BC is 
\begin{widetext}
\begin{align}
u_2(x) &= C_2 \lb\{\begin{array}{l}
	 D\lb[ \dfrac{1}{2} \lb( \dfrac{3\beta-\epsilon}{\sqrt{\beta}} - 1 \rb) ,\;  x \sqrt{2}\beta^{1/4} \rb] \\[3ex]
	 - \sqrt{\dfrac{2}{\pi}} \cos\lb[ \dfrac{\pi}{4} \lb( \dfrac{3\beta - \epsilon}{\sqrt{\beta}} - 1 \rb) \rb] \Gamma\lb[ \dfrac{1}{2} \lb( \dfrac{3\beta-\epsilon}{\sqrt{\beta}} + 1 \rb) \rb] D\lb[ \dfrac{1}{2} \lb( - \dfrac{3\beta-\epsilon}{\sqrt{\beta}} - 1 \rb) ,\;  i x \sqrt{2}\beta^{1/4} \rb]  
	 \end{array} \rb\},
\end{align}
where $D(\nu,z)$ is the parabolic cylinder function.

Continuity at $x=1$ demands that
\begin{align}
\frac{C_1}{C_2}e^{-\sqrt{\epsilon}} U\!\lb(1-\frac{\beta}{\sqrt{\epsilon}},\; 2,\; 2\sqrt{\epsilon} \rb) 
&= 
	 D\lb[ \dfrac{1}{2} \lb( \dfrac{3\beta-\epsilon}{\sqrt{\beta}} - 1 \rb) ,\;  \sqrt{2}\beta^{1/4} \rb] \nl
	 - \sqrt{\dfrac{2}{\pi}} \cos\lb[ \dfrac{\pi}{4} \lb( \dfrac{3\beta - \epsilon}{\sqrt{\beta}} - 1 \rb) \rb] \Gamma\lb[ \dfrac{1}{2} \lb( \dfrac{3\beta-\epsilon}{\sqrt{\beta}} + 1 \rb) \rb]  D\lb[ \dfrac{1}{2} \lb( - \dfrac{3\beta-\epsilon}{\sqrt{\beta}} - 1 \rb) ,\;  i \sqrt{2}\beta^{1/4} \rb] ,
\label{eq:Continuity}
\end{align}
while continuity of the derivative at $x=1$ imposes
\begin{align}
&\frac{C_1}{C_2} e^{-\sqrt{\epsilon}} U\!\lb(1-\frac{\beta}{\sqrt{\epsilon}},\; 2,\; 2\sqrt{\epsilon} \rb) \lb[ \lb( \sqrt{\epsilon}-1 \rb) +2(\sqrt{\epsilon}-\beta) \frac{ U\!\lb(2-\frac{\beta}{\sqrt{\epsilon}},\; 3,\; 2\sqrt{\epsilon} \rb)}{U\!\lb(1-\frac{\beta}{\sqrt{\epsilon}},\; 2,\; 2\sqrt{\epsilon} \rb)} \rb] \nonumber \\
&= \sqrt{\beta} D\lb[ \dfrac{1}{2} \lb( \dfrac{3\beta-\epsilon}{\sqrt{\beta}} - 1 \rb) ,\;  \sqrt{2}\beta^{1/4} \rb] - \sqrt{2}\beta^{1/4} D\lb[ \dfrac{1}{2} \lb( \dfrac{3\beta-\epsilon}{\sqrt{\beta}} + 1 \rb) ,\;  \sqrt{2}\beta^{1/4} \rb] \nl
+ \sqrt{\frac{2}{\pi}} \cos\lb[ \dfrac{\pi}{4} \lb( \dfrac{3\beta - \epsilon}{\sqrt{\beta}} - 1 \rb) \rb] \Gamma\lb[ \dfrac{1}{2} \lb( \dfrac{3\beta-\epsilon}{\sqrt{\beta}} + 1 \rb) \rb] \sqrt{\beta} D\lb[ \dfrac{1}{2} \lb( - \dfrac{3\beta-\epsilon}{\sqrt{\beta}} - 1 \rb) ,\;  i\sqrt{2}\beta^{1/4} \rb] \nl
+ \sqrt{\frac{2}{\pi}} \cos\lb[ \dfrac{\pi}{4} \lb( \dfrac{3\beta - \epsilon}{\sqrt{\beta}} - 1 \rb) \rb] \Gamma\lb[ \dfrac{1}{2} \lb( \dfrac{3\beta-\epsilon}{\sqrt{\beta}} + 1 \rb) \rb] i\sqrt{2}\beta^{1/4} D\lb[ \dfrac{1}{2} \lb( -\dfrac{3\beta-\epsilon}{\sqrt{\beta}} + 1 \rb) ,\;  i\sqrt{2}\beta^{1/4} \rb].
\label{eq:dContinuity}
\end{align}
\end{widetext}
Using \eqref{Continuity} to eliminate the $\lb[ C_1/C_2 e^{-\sqrt{\epsilon}} U(\,\cdots) \rb]$ expression that appears on the LHS of \eqref{dContinuity} in favor of the RHS of \eqref{Continuity} yields a single transcendental eigenvalue equation for $\epsilon$ as a function of $\beta$; this must be solved numerically to find the allowed ground state $\epsilon(\beta)$, with the binding energy constructed as 
\begin{align}
E_B &= 
\lb. \frac{\epsilon(\beta)}{2\mu R^2}\rb|_{\beta= |Q_X|Q_N \alpha \mu R}.
\label{eq:EBNX}
\end{align}

While the above procedure has the virtue of being an accurate solution to the problem of finding the binding energies of the system with the potential described by \eqref{VNXbinding}, it is numerically cumbersome.
We can develop a relatively accurate approximation by returning to \eqref{VNXbinding}, and considering only the part of the potential for $r<R$, which looks like the harmonic oscillator potential for a 3D oscillator with fundamental frequency $\omega_0 = \sqrt{\hat{\alpha}/(m_N R^3)}$, but with its energies offset by $-3\hat{\alpha} / (2R)$.
If we take this picture literally, the binding energies of the system should be the energies of the 3D oscillator
\begin{align}
(E_B^{\text{harmonic}})_n = -\frac{3}{2} \frac{\hat{\alpha}}{R} + \lb(n+\frac{3}{2}\rb)\sqrt{\frac{\hat{\alpha}}{m_N R^3}}.
\label{eq:EBharmonic}
\end{align}
However, the classical turning points for the ground state motion of this system are given by $x_* = \sqrt{3} ( R^3 / (\hat{\alpha}m_N) )^{1/4} \sim R \sqrt{3} (\hat{\alpha}m_N R)^{-1/4}$.
But $\hat{\alpha} m_N R \sim \alpha Q_N A \mu_a R_0 A^{1/3} \sim Q_NA^{4/3} / 24 \sim A^{7/3} / 48$, where we assumed $|Q_X|=1$ and at the last step we assumed $Q_N\sim A/2$; therefore, we have  $x_*/R \sim \sqrt{3} (\hat{\alpha} m_N R)^{-1/4} \sim 4.6 A^{-7/12} \sim 1.4 (A/8)^{-7/12} \sim 1.07 (A/12)^{-7/12}$.
While the classical turning points of the ground state motion for small $A$ therefore lie outside the region where the harmonic oscillator treatment is appropriate, we might still hope that the ground state binding energies are reasonable numerical approximations to the actual binding energies \eqref{EBNX} already for $A\sim 8$, with the accuracy of the approximation improving for larger $A$.
This turns out to be a correct conclusion; see \tabref{EbindNXapprox}.
We can therefore approximate the binding energies as 
\begin{align}
E_B &\approx (E_B^{\text{harmonic}})_0 \nonumber\\
&= \frac{3\alpha|Q_X|Q_N}{2R} \lb[ 1 - \lb( \alpha|Q_X|Q_N m_N R \rb)^{-1/2} \rb].
 \label{eq:EBNXapprox}
\end{align}

\begin{table}[t]
\begin{ruledtabular}
\caption{\label{tab:EbindNXapprox}%
		Ground-state binding energies $E_B$ [MeV] computed per \eqref{EBNX}, compared to the ground state energy level $(E_B^{\text{harmonic}})_{0}$ of the approximate harmonic oscillator treatment defined at \eqref{EBharmonic}.
		For $A\gtrsim 8$ and $Z\gtrsim4$,  it is a good approximation to take $E_B \approx (E_B^{\text{harmonic}})_{0}$; the approximation is poor for $(A,Z)=(4,2)$ as the true ground state is localized mostly outside the nuclear volume.
	 }
\begin{tabular}{lll}
$N$		&	$E_B$ [MeV]			&		$(E_B^{\text{harmonic}})_{0}$ [MeV]		\\ \hline \hline
${}^4$He	&	0.35					&		$<0$					\\
${}^8$Be	&	1.6					&		1.4					\\
${}^{12}$C&	2.9					&		2.9					\\
${}^{16}$O&	4.1					&		4.1					\\
${}^{20}$Ne&	5.2					&		5.2					\\
${}^{24}$Mg&	6.1					&		6.1					\\
${}^{56}$Fe&	\!\!\!10.0				&		9.9				
\end{tabular}
\end{ruledtabular}
\end{table}

\section{Pycnonuclear fusion rate estimate}
\label{app:pycnonuclear}
In order to estimate the ion number density in the $(NX)$ core at which pycnonuclear fusion may become relevant, we develop here an estimate of the tunneling suppression for this process; this estimate is similar to that developed in \citeR{Shapiro:2000abc}, which correctly captures the exponential suppression of the pycnonuclear fusion rate in the case of an ordinary CO WD (although it obtains the wrong prefactor).

We consider a simple one-dimensional tunneling problem in the following setup: let CHAMPs $X^-$ be located at $x=0,\pm L$, and assume that carbon ions C are bound to each CHAMP site.
We denote the carbon nuclear radius as $R$; each C wavefunction is thus localized within $\sim R$ of the locations $x=0,\pm L$.
Our approximation will treat the \CX\ bound states at $x=\pm L$ as immutable objects of charge $Q_{(\text{C}X)} = +5$, but we will track the $X$ at $x=0$ and the C initially localized around $x=0$ individually.
Because it is extremely massive, we will treat the $X$ at $x=0$ as stationary; additionally, because we actually imagine that the massive \CX\ bound states at $x=\pm L$ are actually localized in a quasi-periodic structure in which \CX\ are present at $x_k = kL$ for $k\in \mathbb{Z}$, we imagine also that the \CX\ bound states at $x=\pm L$ are stationary.
Within the context of a Wentzel--Kramers--Brillouin (WKB) approximation, we will ask for the probability that the carbon ion initially localized around $x=0$ is able to tunnel to a location $x=L-2R$, at which location it is within $\sim$ a nuclear diameter of the $C$ in the \CX\ bound state at $x=L$, and can undergo a fusion reaction.
This approximation is manifestly crude, but it should obtain the appropriate parametric scalings of the exponential tunneling suppression factor.

Let the $x$-coordinate of the dynamical carbon ion C that is initially localized around $x=0$ be $x$; accounting for the finite charge radius of the C ion in its binding with the $X^-$ located at $x=0$, the terms in the electrostatic potential that are of interest in this tunneling computation are (for $x\in[-L+2R,L-2R]$)
\begin{align}
V \supset V(x) &= - \frac{\alpha |Q_X| Q_C}{|x|} \beta(x) \nl
	 + \frac{\alpha Q_C Q_{(\text{C}X)} }{L-x} 	+ \frac{\alpha Q_C Q_{(\text{C}X)}} {L+x} \nl
	  - \frac{2\alpha Q_C Q_{(\text{C}X)}} {L} 
\label{eq:Vpycno}
\end{align}
where the first line is the interaction energy of the dynamical C ion with the $X^-$ at $x=0$ and
\begin{align}
\beta(x) &\equiv \begin{cases} 
				1 & |x| > R\\[2ex]
				\dfrac{1}{2} \dfrac{|x|}{R}	\lb[ 3 - \dfrac{x^2}{R^2} \rb] & |x|\leq R
			\end{cases};
\end{align}
the second line is the interaction energy of the dynamical C ion with the \CX\ states at $x=\pm L$, and the third is a constant offset in the potential energy conveniently chosen to zero out the potential energy contribution from the neighboring \CX\ ions at $x=0$.

Per the arguments advanced in \citeR{Shapiro:2000abc}, the probability per unit time per ion pair for a pycnonuclear fusion reaction to occur is given by 
\begin{align}
W \sim \frac{S(E)}{E} v_{\text{inc.}} | \psi_{\text{inc.}}|^2 \mathcal{T},
\end{align}
where $S(E)$ is the nuclear reaction $S$-factor which encodes all the nuclear physics, and is usually a slowly varying function of energy (absent resonances), $E$ is the (kinetic) energy of the ion which must tunnel though the Coulomb barrier to trigger the fusion, $v_{\text{inc}} \sim \sqrt{2 E/m}$ is the corresponding speed of that ion (in the usual symmetric tunneling case, this is technically the relative speed of the pair, so $m$ is replaced by the reduced mass $\mu$), $| \psi_{\text{inc.}}|^2$ is the ion wavefunction evaluated at the classical turning point for the motion of the ion, and $\mathcal{T}$ is the tunneling exponential.
Therefore, the dynamical C ion initially localized around $x=0$ will have a probability per unit time to fuse with the C ion bound to the $X^-$ at $x=L$ of 
\begin{align}
W \propto \sqrt{\frac{2}{E_K m_C}} |\psi_C(x_*)|^2 \mathcal{T},
\end{align}
where we have dropped the $S$ factor because it varies slowly with energy, and where $x_*$ is the classical turning point of the C ion motion in the potential \eqref{Vpycno} assuming that the ion has the ground state binding energy $E_B$ appropriate for the \CX\ bound state.
The tunneling exponential $\mathcal{T}$ is (note in connection with the sign under the square-root that the energy of the system  at the classical turning point is $E = -E_B<0$)%
\footnote{\label{ftnt:McinTunnelingExponent}%
		Note that since we are treating the other carbon ions as rigidly fixed to their respective massive CHAMPs in this approximation, the mass that appears in the tunneling exponential is the carbon mass, not the reduced mass of the carbon-carbon system.
	} %
\begin{align}
-\frac{1}{2} \ln \mathcal{T} = \sqrt{2 m_C} \int_{x_*}^{L-2R}dx'\sqrt{V(x')+E_B}.
\label{eq:TunnellingExponentialDefn}
\end{align}
Because the C ion in question is in a bound state with the $X^-$ located at $x=0$, we expect that $x_* \sim R$; see \appref{CHAMPbinding}.
To make this more precise, let us expand \eqref{Vpynco} for $|x|\lesssim R\ll L$ [the convenience of the the choice of the constant potential energy offset in \eqref{Vpycno} is now manifest]:
\begin{widetext}
\begin{align}
V(|x|\lesssim R\ll L) &\approx - \frac{\alpha |Q_X| Q_C}{R} \dfrac{1}{2} \lb[ 3 - \dfrac{x^2}{R^2} \rb] + \frac{2\alpha Q_C Q_{(\text{C}X)}}{L} \frac{x^2}{L^2}  \\
&= - \frac{3}{2} \frac{\alpha |Q_X| Q_C}{R} +  \frac{1}{2}  \dfrac{\alpha |Q_X| Q_C}{R^3} \lb[ 1 + 4 \frac{Q_{(\text{C}X)}  }{|Q_X|}\frac{R^3}{L^3} \rb] x^2;
\label{eq:Vpynco}
\end{align}
\end{widetext}
for $R\ll L$, the correction term in the $[\,\cdots]$ bracket on the second line above can be ignored, and the potential reduces to one of the same form as that we already considered at the end of \appref{CHAMPbinding} in developing the approximate treatment of the ground state binding energy of the $(NX)$ state, so $x_* \sim R$ is valid.

There is however a mismatch with the treatment in \appref{CHAMPbinding}, because we are treating the tunneling part of the problem here as a 1D problem, whereas we treated the problem using the 3D SE in \appref{CHAMPbinding}.
Following the treatment of the pycnonuclear rate estimate in \citeR{Shapiro:2000abc}, we will largely gloss over this mismatch and use a blend of the 1D and 3D results: (1) we estimate $|\psi_C(x_*)|^2 \sim (\sqrt{\pi} R)^{-3}$ as the value of the ground state wavefunction of the 3D harmonic oscillator evaluated around the classical turning point (taken for this part of the computation to be $x_*\sim R$ per \appref{CHAMPbinding}); (2) we estimate $m_C  E_K \sim (3/2) m_C \omega_0 \propto 1/R^2$ [this follows from the discussion around \eqref{EBharmonic}]; and (3) we will use the approximate ground state binding energy estimate \eqref{EBNXapprox} developed for the 3D problem to estimate $E_B$ in \eqref{TunnellingExponentialDefn}; but (4) we will otherwise continue to compute the tunneling exponential in the 1D approach.

For the tunneling part of the problem, we need $V(x)+E_B$ with $V(x)$ from \eqref{Vpycno} for $x\in [R,L-2R]$:
\begin{widetext}
\begin{align}
V(x\in[R,L-2R]) + E_B &= \frac{\alpha|Q_X|Q_C}{R} \Lambda - \frac{\alpha |Q_X| Q_C}{x} + \frac{2 \alpha Q_C Q_{(\text{C}X)}}{L} \lb[ \frac{ x^2 }{L^2-x^2}  \rb]  \\
&=\frac{\alpha|Q_X|Q_C}{R} \lb\{  \Lambda - \frac{R}{x} + 2 \frac{R}{L} \lb( 1 - q \rb) \lb[ \frac{ (x/L)^2 }{1-(x/L)^2}  \rb] \rb\},
\end{align}
\end{widetext}
where we have defined 
\begin{align}
\Lambda &\equiv \frac{3}{2} \lb[ 1 - \lb( \alpha |Q_X|Q_C m_C R \rb)^{-1/2} \rb] \sim 1,
\label{eq:LambdaDefn}
\end{align}
(the approximation being numerically satisfied for the nuclei of interest) and have used that $Q_{(\text{C}X)} = Q_C - |Q_X|$, and defined $q \equiv |Q_X|/Q_C$.
Therefore, if we parametrize $x_* \equiv \gamma R$ with $\gamma \sim 1$, and define $u \equiv x/L$ and $u_0 \equiv R/L < 1 $ (indeed, typically, $u_0\ll1$), then
\begin{widetext}
\begin{align}
-\frac{1}{2} \ln \mathcal{T} = \sqrt{2 \alpha|Q_X|Q_C m_C L^2/R }  \int_{\gamma u_0 }^{1-2u_0}du \lb\{ \Lambda - \frac{u_0}{u} + 2u_0 \lb( 1 - q \rb) \lb[ \frac{ u^2 }{1-u^2}  \rb] \rb\}^{1/2}.
\label{eq:TunnellingExponentialComp1}
\end{align}
\end{widetext}
Note that for this piece of the computation, we must set $\gamma$ to satisfy 
\begin{align}
\Lambda - \frac{1}{\gamma} + 2u_0^3 \lb( 1 - q \rb) \lb[ \frac{ \gamma^2 }{1-\gamma^2 u_0^2}  \rb] = 0;
\end{align}
since $u_0 \ll 1$ is assumed, the last term can be neglected, and this gives $\gamma \approx \Lambda$, which is again of $\mathcal{O}(1)$, approximately consistent with the estimate from the harmonic oscillator approximation (although the exact numerical values for $x_*$ derived in the two different approximations will differ).
Note also that the final term in the square root becomes maximum at the upper end of the integration range, where we can estimate its size as
\begin{align}
&\lb. 2u_0 \lb( 1 - q \rb) \lb[ \frac{ u^2 }{1-u^2}  \rb] \rb|_{u=1-2u_0} \\
& = \frac{1}{2} \lb( 1 - q \rb) \lb[ \frac{ 1 - 4 u_0 + 4u_0^2 }{ 1 - u_0 }  \rb] \approx \frac{1-q}{2} < \frac{1}{2}.
\label{eq:domination1}
\end{align}
Since (1) the integrand in \eqref{TunnellingExponentialComp1} vanishes at the lower integration limit owing to a cancellation of the first and second terms in the square-root (with the third term being negligible); (2) this cancellation persists only until $u \sim \text{few} \times u_0 \ll 1$, at which point the integrand becomes dominated by the first term in the square-root; and (3) this domination by the first term in the square-root persists at the $\mathcal{O}(1)$ level until the upper limit of the integration owing to \eqref[s]{LambdaDefn} and (\ref{eq:domination1}); we can estimate the integral in \eqref{TunnellingExponentialComp1} parametrically as $\int^{\cdots}_{\cdots} du \{\, \cdots \}^{1/2} \sim \sqrt{\Lambda}$, up to an $\mathcal{O}(1)$ factor.
Therefore, parametrically,
\begin{align}
\mathcal{T} \sim  \exp\lb[ -2 \sqrt{2 \alpha|Q_X|Q_C m_C L^2 \Lambda / R } \rb],
\label{eq:TunnExpEst}
\end{align}
where the exponent is correct up to an $\mathcal{O}(1)$ factor.

Putting together \eqref{TunnExpEst} with the the points (1) and (2) just below \eqref{Vpynco}, we expect parametrically that the rate per unit time per unit volume for this fusion process is 
\begin{align}
\Gamma_{\text{C}(\text{C}X)}/V &\propto n_{(\text{C}X)}^2 W_{\text{C}(\text{C}X)} \propto n_{(\text{C}X)}^2 R^{-2} \nl
\times \exp\lb[ -2 \sqrt{2 \alpha|Q_X|Q_C m_C L^2 \Lambda / R } \rb].
\end{align}
Let us trade out $R$ for the nuclear number density: $(4\pi / 3)n_{\text{nucl}} (2R)^3 \sim 1 \Rightarrow R \sim \lb[ 3 / (32\pi n_{\text{nucl}}) \rb]^{1/3}$; and $L$ for the $(\text{C}X)$ number density: $(4\pi / 3) n_{(\text{C}X)} L^3 \sim 1 \Rightarrow L \sim \lb[ 3/ (4\pi n_{(\text{C}X)}) \rb]^{1/3}$, yielding (we drop numerical prefactors)
\begin{align}
& \Gamma_{\text{C}(\text{C}X)}/V \nonumber \\
& \propto n_{(\text{C}X)}^2 n_{\text{nucl}}^{2/3} \nl
\times \exp\lb[ -4 \sqrt{\alpha|Q_X|Q_C m_C \Lambda n_{(\text{C}X)}^{-2/3} n_{\text{nucl}}^{1/3} \lb( 3 / (4\pi) \rb)^{1/3} } \rb].
\end{align}

On the other hand, the cognate estimate for the ordinary pycnonuclear C--C fusion process from \citeR{Shapiro:2000abc} yields
\begin{align}
\Gamma_{\text{CC}}/V \propto n_C^2 W_{\text{CC}} \propto n_C^2 \frac{L'}{(x'_0)^3} \times \exp\lb[ -2 (L')^2/(x'_0)^2 \rb],
\end{align}
where $L'$ is the average distance between carbon ions, and $x_0' \approx L'/\sqrt{2Q_C} \times \lb( \alpha m_C L' / 2 \rb)^{-1/4}$ is the classical turning point for the $C$ tunneling motion in the cognate computation, such that
\begin{align}
\Gamma_{\text{CC}}/V&\propto n_{C}^2 (L')^{-2} \lb( 2Q_C^2 \alpha  m_C L' \rb)^{3/4} \nl
\times \exp\lb[ - 2 \sqrt{ 2 Q_C^2 \alpha m_C L' } \rb];
\end{align}
if we similarly trade out $L' \sim \lb[ 3 / (4\pi n_{C}) \rb]^{1/3}$, we have (dropping numerical prefactors)
\begin{align}
\Gamma_{\text{CC}}/V&\propto n_{C}^{8/3} \lb( 2Q_C^2 \alpha  m_Cn_{C}^{-1/3} \lb( 3 / (4\pi) \rb)^{1/3} \rb)^{3/4} \nl
\times \exp\lb[ - 2 \sqrt{ 2 Q_C^2 \alpha m_C n_{C}^{-1/3} \lb( 3 / (4\pi) \rb)^{1/3} } \rb].
\end{align}

We would like to know where the volumetric rates are approximately equal: $\Gamma_{\text{CC}}/V \sim  \Gamma_{\text{C}(\text{C}X)}/V $.
This will allow us to estimate the number density at which the $X^-$ catalyzed pycnonuclear fusion process could trigger runaway in the collapsing core based on the number density at which this occurs in an ordinary CO WD.
We therefore seek
\begin{widetext}
\begin{align}
&n_{C}^{8/3} \lb( 2Q_C^2 \alpha  m_Cn_{C}^{-1/3} \lb( 3 / (4\pi) \rb)^{1/3} \rb)^{3/4} \exp\lb[ - 2 \sqrt{ 2 Q_C^2 \alpha m_C n_{C}^{-1/3} \lb( 3 / (4\pi) \rb)^{1/3} } \rb] \nonumber \\
&\sim n_{(\text{C}X)}^2 n_{\text{nucl}}^{2/3} \exp\lb[ -4 \sqrt{\alpha|Q_X|Q_C m_C \Lambda n_{(\text{C}X)}^{-2/3} n_{\text{nucl}}^{1/3} \lb( 3 / (4\pi) \rb)^{1/3} } \rb]\\
\Rightarrow 
&\ln\lb[ n_{C}^{8/3} n_{\text{nucl}}^{-2/3}n_{(\text{C}X)}^{-2} \lb( 2Q_C^2 \alpha  m_Cn_{C}^{-1/3} \lb( 3 / (4\pi) \rb)^{1/3} \rb)^{3/4} \rb] \nonumber \\
&\sim 2 \sqrt{ 2 Q_C^2 \alpha m_C n_{C}^{-1/3} \lb( 3 / (4\pi) \rb)^{1/3} }\lb[ 1 - \sqrt{2(|Q_X|/Q_C) \Lambda n_{(\text{C}X)}^{-2/3} n_{\text{nucl}}^{1/3}n_{C}^{1/3} }  \rb].
\label{eq:equalrateeqn}
\end{align}
\end{widetext}
To make progress, note that if we take the extreme carbon number densities shortly before pycnonuclear fusion is relevant \cite{Shapiro:2000abc} of around $\rho \sim 10^{10}\text{g/cm}^3 \Rightarrow n_C \sim 5\times 10^{32}\text{cm}^{-3}$, the factor $2 Q_C^2 \alpha m_C n_{C}^{-1/3} \lb( 3 / (4\pi) \rb)^{1/3} \sim 10^3$, while the largest that the logarithmic factor becomes assuming $n_C \lesssim n_{(\text{C}X)} \lesssim n_{\text{nucl}}$ and that $n_{\text{nucl}}\sim 10^{37}\text{cm}^{-3}$, is $\mathcal{O}(1)$.
Therefore, to good approximation, we can simply set the $[\,\cdots]$-bracket on the RHS of the final line of \eqref{equalrateeqn} to zero to find the condition for equal volumetric rates:
\begin{align}
n_{(\text{C}X)} &\sim \lb[ 2(|Q_X|/Q_C) \Lambda \rb]^{3/2} \sqrt{ n_{\text{nucl}} n_{C} }\\
& \sim 0.2 \sqrt{ n_{\text{nucl}} n_{C} } \\
&\sim 10^{34}\text{cm}^{-3},
\label{eq:CXpyncoBoom}
\end{align}
where we used $n_X \sim 5\times 10^{32}\text{cm}^{-3}$ as the usual carbon pycnonuclear fusion density \cite{Shapiro:2000abc}, and $n_{\text{nucl}}\sim 10^{37}\text{cm}^{-3}$.

\section{Full expressions for accreting CHAMPs}
\label{app:Accreting_Fraction}
We give here the full expressions for the mass of CHAMPs that accrete over the lifetime of the WD; see \sectref{PopnAccumulated}.

The form of the truncated Maxwellian distribution \eqref{fCHAMPs}, and more specifically the Heaviside theta function appearing therein, as well as the appearance of the additional Heaviside theta function in the expression assumed for $\epsilon(v)$ [\eqref{epsilonV}], dictate that caution must be exercised in the integration over velocity in \eqref{MXaccrete1}.
The integral over the azimuthal angle is always without bound and yields a factor of $2\pi$; however, the polar angle defined by $\bm{v}_{\textsc{wd}}\cdot{\bm{v}} = v v_{\textsc{wd}} \cos\theta$ may have a restriction, depending on the value of $v$.
There are three cases.
Case I: for $v_{\text{max}} \leq v_{\text{esc,\,\textsc{mw}}} - v_{\textsc{wd}}$, there is no angular restriction on $\theta$, and $\int d^3v \rightarrow 4\pi \int_0^{v_{\text{max}}} dv$ in \eqref{MXaccrete1}.
Case II: for $v_{\text{esc,\,\textsc{mw}}} - v_{\textsc{wd}} \leq v_{\text{max}} \leq v_{\text{esc,\,\textsc{mw}}} + v_{\textsc{wd}}$, the integral over $v$ must be broken into two domains.
For $v \leq v_{\text{esc,\,\textsc{mw}}} - v_{\textsc{wd}}$ there is no angular restriction, while for $v_{\text{esc,\,\textsc{mw}}} - v_{\textsc{wd}} < v 
\leq v_{\text{max}}$ there is a maximum allowed value of $\cos\theta \leq \cos\theta_*(v) \equiv \lb( v_{\text{esc,\,\textsc{mw}}}^2-v_{\textsc{wd}}^2-v^2 \rb)/(2v_{\textsc{wd}}v)$.
The integral in \eqref{MXaccrete1} must thus be performed as $\int d^3v \rightarrow 4\pi \int_0^{v_{\text{esc,\,\textsc{mw}}} - v_{\textsc{wd}}} dv + 2\pi \int_{v_{\text{esc,\,\textsc{mw}}} - v_{\textsc{wd}}}^{v_{\text{max}}} dv \int_{-1}^{\cos\theta_*(v)} d\cos\theta$.
Case III: for $v_{\text{max}} \geq v_{\text{esc,\,\textsc{mw}}} + v_{\textsc{wd}}$, the integral over $v$ would in principle need to be split into three domains, but one is identically zero.
For $v \leq v_{\text{esc,\,\textsc{mw}}} - v_{\textsc{wd}}$ there is no angular restriction, while for $v_{\text{esc,\,\textsc{mw}}} - v_{\textsc{wd}} < v 
\leq v_{\text{esc,\,\textsc{mw}}} + v_{\textsc{wd}}$ there is a maximum allowed value of $\cos\theta \leq \cos\theta_*(v) \equiv \lb( v_{\text{esc,\,\textsc{mw}}}^2-v_{\textsc{wd}}^2-v^2 \rb)/(2v_{\textsc{wd}}v)$.
The integral in \eqref{MXaccrete1} must thus be performed as $\int d^3v \rightarrow 4\pi \int_0^{v_{\text{esc,\,\textsc{mw}}} - v_{\textsc{wd}}} dv + 2\pi \int_{v_{\text{esc,\,\textsc{mw}}} - v_{\textsc{wd}}}^{v_{\text{esc,\,\textsc{mw}}} + v_{\textsc{wd}}} dv \int_{-1}^{\cos\theta_*(v)} d\cos\theta$.
The angular integral is identically zero for $v_{\text{esc,\,\textsc{mw}}} + v_{\textsc{wd}} < v \leq v_{\text{max}}$, and so the final result is independent of $v_{\text{max}}$.

Taking this into account, the accumulated mass of CHAMPs for Cases I, III, and II, are, respectively,
\begin{widetext}
\begin{align}
M_{X^{\pm}}^{\text{I}} &= M_{X^{\pm}}^{\text{approx,\,III}} \times  \lb\{\begin{array}{l}
							\dfrac{v_0}{v_{\textsc{wd}}} \lb[ \text{erf}\lb( \dfrac{v_{\textsc{wd}}}{v_0}\rb) + \dfrac{1}{2} \text{erf}\lb( \dfrac{v_{\text{max}}-v_{\textsc{wd}}}{v_0}\rb) - \dfrac{1}{2} \text{erf}\lb( \dfrac{v_{\text{max}}+v_{\textsc{wd}}}{v_0}\rb)  \rb] \times \lb( 1 + \dfrac{1}{2} \dfrac{v_0^2}{v_{\text{esc,\,\textsc{wd}}}^2}+ \dfrac{v_{\textsc{wd}}^2}{v_{\text{esc,\,\textsc{wd}}}^2}  \rb) \\[4ex]
							+ \dfrac{1}{\sqrt{\pi}} \dfrac{v_0^2}{v_{\text{esc,\,\textsc{wd}}}^2} \exp\lb[ - \dfrac{v_{\textsc{wd}}^2}{v_0^2} \rb] - \dfrac{1}{2\sqrt{\pi}} \dfrac{v_0^2}{v_{\text{esc,\,\textsc{wd}}}^2} \exp\lb[ - \dfrac{(v_{\text{max}}-v_{\textsc{wd}})^2}{v_0^2} \rb] \times \lb( 1 + \dfrac{v_{\text{max}}}{v_{\textsc{wd}}} \rb) \\[4ex]
							- \dfrac{1}{2\sqrt{\pi}} \dfrac{v_0^2}{v_{\text{esc,\,\textsc{wd}}}^2} \exp\lb[ - \dfrac{(v_{\text{max}}+v_{\textsc{wd}})^2}{v_0^2} \rb] \times \lb( 1 - \dfrac{v_{\text{max}}}{v_{\textsc{wd}}} \rb)
															  \end{array} \rb\} \nl  \qquad\qquad\qquad
											\times \lb\{ \text{erf}\lb[ \dfrac{v_{\text{esc,\,\textsc{mw}}}}{v_0} \rb] - \dfrac{2v_{\text{esc,\,\textsc{mw}}}}{\sqrt{\pi}v_0} \exp\lb[ - \dfrac{v_{\text{esc,\,\textsc{mw}}}^2}{v_0^2} \rb]  \rb\}^{-1}
\label{eq:CaseIFull} \\[5ex]
M_{X^{\pm}}^{\text{II}} &= M_{X^{\pm}}^{\text{approx,\,III}} \times  \lb\{\begin{array}{l}
							\dfrac{v_0}{v_{\textsc{wd}}} \lb[ \text{erf}\lb( \dfrac{v_{\textsc{wd}}}{v_0}\rb) + \dfrac{1}{2} \text{erf}\lb( \dfrac{v_{\text{max}}-v_{\textsc{wd}}}{v_0}\rb) - \dfrac{1}{2} \text{erf}\lb( \dfrac{v_{\text{esc,\,\textsc{mw}}}}{v_0}\rb)  \rb] \times \lb( 1 + \dfrac{1}{2} \dfrac{v_0^2}{v_{\text{esc,\,\textsc{wd}}}^2}+ \dfrac{v_{\textsc{wd}}^2}{v_{\text{esc,\,\textsc{wd}}}^2}  \rb) \\[4ex]
							+ \dfrac{1}{\sqrt{\pi}} \dfrac{v_0^2}{v_{\text{esc,\,\textsc{wd}}}^2} \exp\lb[ - \dfrac{v_{\textsc{wd}}^2}{v_0^2} \rb] - \dfrac{1}{2\sqrt{\pi}} \dfrac{v_0^2}{v_{\text{esc,\,\textsc{wd}}}^2} \exp\lb[ - \dfrac{(v_{\text{max}}-v_{\textsc{wd}})^2}{v_0^2} \rb] \times \lb( 1 + \dfrac{v_{\text{max}}}{v_{\textsc{wd}}} \rb) \\[4ex]
							-  \dfrac{1}{\sqrt{\pi}} \exp\lb[ - \dfrac{v_{\text{esc,\,\textsc{mw}}}^2}{v_0^2} \rb] \times \lb(
							\begin{array}{l}
							 1 + \dfrac{v_{\text{esc,\,\textsc{mw}}}^2}{v_{\text{esc,\,\textsc{wd}}}^2} + \dfrac{v_0^2}{v_{\text{esc,\,\textsc{wd}}}^2} \times \lb( 1 - \dfrac{1}{2} \dfrac{v_{\text{esc,\,\textsc{mw}}}}{v_{\textsc{wd}}} \rb) \\[4ex] 
							 - \dfrac{ v_{\text{esc,\,\textsc{mw}}}}{v_{\textsc{wd}}} \lb( 1 + \dfrac{1}{3} \dfrac{v_{\text{esc,\,\textsc{mw}}}^2}{v_{\text{esc,\,\textsc{wd}}}^2} \rb) + \dfrac{ v_{\text{max}}}{v_{\textsc{wd}}} \lb( 1 + \dfrac{1}{3} \dfrac{v_{\text{max}}^2}{v_{\text{esc,\,\textsc{wd}}}^2} \rb)  \\[4ex] 
							 - \dfrac{ v_{\text{esc,\,\textsc{mw}}} v_{\textsc{wd}} }{v_{\text{esc,\,\textsc{wd}}}^2} \lb( 1 - \dfrac{1}{3} \dfrac{v_{\textsc{wd}}}{v_{\text{esc,\,\textsc{mw}}}} \rb) 
							 \end{array} \rb)
															  \end{array} \rb\} \nl  \qquad\qquad\qquad
											\times \lb\{ \text{erf}\lb[ \dfrac{v_{\text{esc,\,\textsc{mw}}}}{v_0} \rb] - \dfrac{2v_{\text{esc,\,\textsc{mw}}}}{\sqrt{\pi}v_0} \exp\lb[ - \dfrac{v_{\text{esc,\,\textsc{mw}}}^2}{v_0^2} \rb]  \rb\}^{-1}
											\label{eq:CaseIIFull} \\[5ex]
M_{X^{\pm}}^{\text{III}} &= M_{X^{\pm}}^{\text{approx,\,III}} \times  \lb\{\begin{array}{l}
							 \dfrac{v_0}{v_{\textsc{wd}}} \text{erf}\lb( \dfrac{v_{\textsc{wd}}}{v_0}\rb) \times \lb[ 1 + \dfrac{1}{2}\dfrac{v_0^2}{v_{\text{esc,\,\textsc{wd}}}^2} + \dfrac{v_{\textsc{wd}}^2}{v_{\text{esc,\,\textsc{wd}}}^2} \rb] + \dfrac{1}{\sqrt{\pi}}  \dfrac{v_0^2}{v_{\text{esc,\,\textsc{wd}}}^2}\exp\lb[-\dfrac{v_{\textsc{wd}}^2}{v_0^2}\rb]\\[4ex]
							 - \dfrac{2}{\sqrt{\pi}} \exp\lb[- \dfrac{v_{\text{esc,\,\textsc{mw}}}^2}{v_0^2} \rb] \lb( 1 + \dfrac{v_{\text{esc,\,\textsc{mw}}}^2}{v_{\text{esc,\,\textsc{wd}}}^2}   + \dfrac{v_0^2}{v_{\text{esc,\,\textsc{wd}}}^2} +\dfrac{1}{3}  \dfrac{v_{\textsc{wd}}^2}{v_{\text{esc,\,\textsc{wd}}}^2} \rb) 
															  \end{array} \rb\} \nl  \qquad\qquad\qquad
											\times \lb\{ \text{erf}\lb[ \dfrac{v_{\text{esc,\,\textsc{mw}}}}{v_0} \rb] - \dfrac{2v_{\text{esc,\,\textsc{mw}}}}{\sqrt{\pi}v_0} \exp\lb[ - \dfrac{v_{\text{esc,\,\textsc{mw}}}^2}{v_0^2} \rb]  \rb\}^{-1} \label{eq:CaseIIIFull} 
\end{align}
\end{widetext}
where $M_{X^{\pm}}^{\text{approx,\,III}}$ is defined at \eqref{MXaccrete2}, and all other variables are as defined in \sectref{PopnAccumulated}.

\section{White dwarf structure}
\label{app:WDstructure}
In this Appendix, we review the computation of white dwarf mechanical structure, and the derivation of the \Chand\ limit.
We also discuss modifications introduced by CHAMP contamination.

\subsection{General problem statement}
\label{app:WDstructureGeneral}
The mechanical structure of a white dwarf of mass $M_{\textsc{wd}}$ and radius $R_{\textsc{wd}}$ is given by the stable, spherically symmetric equilibrium solutions to Einstein's equations (although in practice the GR corrections are small, except for a near-extremal, \Chand-mass WD) for a perfect fluid---$T^{\mu\nu} = (\rho+P)u^\mu u^\nu - Pg^\mu\nu$ with $u^2 = -1$---with the equation of state given to good approximation by the fully degenerate electron equation of state, with overall plasma neutrality additionally assumed to prevail everywhere throughout the star in order to guarantee the electrostatic communication of the electron degeneracy pressure to the much heavier ions that form the bulk of the mass density of the star; see, e.g., \citeR[s]{Chandrasekhar_1931,Kippenhahn:2012zqe}.

Working in a spherical coordinate system ($t,r,\theta,\phi$) with the line-element $ds^2 = e^{\nu(r)} dt^2 - \lb( 1 + h(r) \rb)^{-1} dr^2 - r^2 d\Omega^2$ leads to the two independent equations: the first being for the mass enclosed at radial coordinate $r$, and the second, the Tolmann--Oppenheimer--Volkhoff (TOV) equation \cite{Oppenheimer:1939cs,Tolman:1939it}, being the GR-corrected version of the Newtonian hydrostatic equilibrium equation:
\begin{align}
\frac{dM(r)}{dr} &= 4\pi r^2 \rho(r) \label{eq:MrEqn}\\
\frac{dP(r)}{dr} &= - \frac{1}{r^2\Mpl^2} \frac{\lb(\rho(r) + P(r) \rb) \lb(M(r)+4\pi r^3P(r)\rb)}{1-\frac{2M(r)}{r\Mpl^2}}. \label{eq:TOVeqn}
\end{align}
The boundary conditions are $M(0)=0$ and $P(0)=P_0$, where $P_0$ is a free parameter giving the central pressure; this defines a one-parameter class of solutions.
The radial-coordinate extent of the star and the Arnowitt--Deser--Misner (ADM) mass are then fixed in terms of the central pressure via the condition $P(R_{\textsc{wd}})=0$ and $M_{\textsc{wd}}=M(R_{\textsc{wd}})$; for $r\geq R_{\textsc{wd}}$, the solution matches onto the free-space Schwarzschild metric with mass $M_{\textsc{wd}}$.

Defining the radial-coordinate-dependent electron Fermi momentum to be $p_F(r) = m_e x_F(r)$, so that the Fermi energy is $E_F(r) \equiv m_e \sqrt{1+[x_F(r)]^2}$ and the electron density is $n_e(r) = (g_e m_e^3/6\pi^2)[x_F(r)]^3$ with $g_e=2$, the total pressure is given by the fully degenerate electron pressure (we ignore the ion thermal contribution, and the $T\neq 0$ corrections to the electron pressure)
\begin{align}
P(r) &= A g(x_F(r)) \\
g(x) &\equiv 3\,\arcsinh x + x(2x^2-3)\sqrt{1+x^2}\\
A&\equiv \frac{g_e m_e^4}{48\pi^2}.
\end{align}
The electron contribution to the total energy density is given by
\begin{align}
\rho_e(r) &= 3 A f(x_F(r))\\
f(x) &\equiv x(2x^2+1)\sqrt{1+x^2}-\arcsinh x.
\end{align}
Imposing plasma neutrality gives the much larger (for all densities relevant for the present work) ion energy density contribution:
\begin{align}
\rho_i &\equiv B \hat{\mu}_e(r) [x_F(r)]^3\\
B &\equiv \frac{g_e m_e^3 \mu_a \mu_e^R }{6\pi^2},
\end{align}
and we have defined the mean molecular weight per free electron $\mu_e(r) \equiv \mu^R_e \hat{\mu}_e(r)$ with $\mu_e^R\equiv \mu_e(R)$ by
\begin{align}
\frac{1}{\mu_e(r)} \equiv \sum_j \frac{X_j(r) Z_j}{A_j},
\label{eq:MeanMolecularMassPerFreeElectron}
\end{align}
where $Z_j,A_j$, and $X_j(r)$ are, respectively, the charge, mass number, and mass fraction at radius $r$ of ion species $j$; this definition assumes full ionization, but could be corrected for partial ionization if desired.
The total mass density is $\rho(r) = \rho_e(r) + \rho_i(r)$.
In this version of the computation, the chemical composition must be assumed as an input; for a pure CO WD, we will assume a completely homogeneous mixture, $X_j(r) = \text{const.}$, so that $\hat{\mu}_e(r)\equiv 1$.

\subsection{Non-dimensional governing equations}
\label{app:WDstructureNonDim}
In practice it is preferable to re-define variables $r\equiv r_0 \xi$, $M(r) \equiv M_0 \hat{M}(\xi)$, $\phi(\xi) \equiv \alpha \sqrt{ 1+ [x_F(r)]^2}$ with $\alpha \in(0,1]$ a free parameter, and where 
\begin{align}
r_0 &\equiv \frac{\alpha\Mpl}{B}\sqrt{\frac{2A}{\pi}}, \\
M_0&= \frac{4\pi r_0^3 B}{\alpha^3} = 8 \sqrt{\frac{2A}{\pi}} \frac{A}{B^2} \Mpl^3,
\end{align}
in terms of which the governing equations Eqs.~(\ref{eq:MrEqn}) and (\ref{eq:TOVeqn}) can be written as
\begin{widetext}
\begin{align}
\frac{d \phi(\xi)}{d\xi} &= - \frac{1}{\xi^2} \frac{ \lb( \hat{\mu}_e(\xi) + 8 C \phi(\xi)/\alpha \rb) \lb( \hat{M}(\xi) + C\alpha^3\xi^3\hat{g}[\phi(\xi)/\alpha] \rb)}{1- (16C/\alpha) (\hat{M}(\xi)/\xi) } \label{eq:WDEqn1}\\
\frac{d \hat{M}(\xi)}{d\xi} &= \alpha^3 \xi^2 \lb[ \hat{\mu}_e(\xi) \lb( \lb( \phi(\xi) / \alpha \rb)^2 - 1 \rb)^{3/2} + 3 C \hat{f}[\phi(\xi)/\alpha] \rb] \label{eq:WDEqn2}
\end{align}
\end{widetext}
where
\begin{align}
\hat{f}(y)&= f\lb[\sqrt{y^2-1}\,\rb] \nonumber \\
&= y(2y^2-1)\sqrt{y^2-1} - \arcsinh\sqrt{y^2-1},
\end{align}
and
\begin{align}
\hat{g}(y)&= g\lb[\sqrt{y^2-1}\,\rb] \nonumber \\
&= 3\,\arcsinh\sqrt{y^2-1} + y(2y^2-5)\sqrt{y^2-1},
\end{align}
and $C \equiv A/B = m_e / (8\mu_a \mu_e^R) \approx 3.4\times 10^{-5} \times (2/\mu_e^R) \ll 1$; the boundary conditions are fixed by $\phi(0)=1$ and $\hat{M}(0)=0$.
The one-parameter class is solution is now defined by the value of $\alpha$, with the WD radial-coordinate extent and ADM mass given by
\begin{align}
R_{\textsc{wd}} &= r_0 \xi_*\\
M_{\textsc{wd}} &= M_0 \hat{M}(\xi_*),
\end{align}
where $\phi(\xi_*) \equiv \alpha$ defines $\xi_*$.
The central density is given by [cf.,~\eqref{rhoCentral}]
\begin{align}
\rho_C \equiv \rho(0) &= B \lb[ \hat{\mu}_e(0) ( \alpha^{-2} - 1)^{3/2} + 3 C \hat{f}(1/\alpha) \rb] \\
& \approx B \hat{\mu}_e(0) ( \alpha^{-2} - 1)^{3/2}.
\label{eq:rhoC}
\end{align}

The governing equations Eqs.~(\ref{eq:WDEqn1}) and (\ref{eq:WDEqn2}) must still be solved numerically; see \eqref{alphaFit} for a numerical fit that gives $\alpha$ to an accuracy of better than 1\% for masses in the range $M_{\textsc{wd}} \in [0.1,1.35]M_{\odot}$, assuming $\hat{\mu}_e=1$ and $\mu_e^R=2$.

\subsection{Chandrasekhar mass}
\label{app:ChandrasekharMass}
It is of course well known that a WD has a maximum stable mass \cite{Chandrasekhar_1931}, the \Chand\ mass.
For $\mu_e=2$ (as in a CO WD), the maximum mass is found numerically to be $M_{\text{Chand.}} \approx 1.42\,M_{\odot}$, corresponding to $\alpha_\star \approx 4.4\times 10^{-2}$.

However, since $8C/\alpha \approx ( 6\pi^2 \rho_C/ g_e \mu_a^4 (\mu_e^R)^4 )^{1/3} \approx 1.0 \times 10^{-3} \times ( \rho_C / 10^8 \,\text{g\,cm}^{-3})^{1/3} \times (2/\mu_e^R)^{4/3} \ll 1$, the governing equations \eqref[s]{WDEqn1} and (\ref{eq:WDEqn2}) depend only weakly on the mean molecular mass per free electron, and the dependence of the mass of the physical solution on $\mu_e$ is dominated by the scaling factor $M_0$ that relates the non-dimensionalized solution to the physical one.
To see this, suppose we send $C/\alpha \rightarrow 0$ in the governing equations \eqref[s]{WDEqn1} and (\ref{eq:WDEqn2}) [but retain $A,B\neq0$ in the scaling factors $r_0,M_0$ relating the dimensional and non-dimensional solutions], and assume that the chemical composition is homogeneous.
Then the governing equations reduce to a single simplified equation:
\begin{align}
\frac{d}{d\xi} \lb( \xi^2 \frac{d\phi(\xi)}{d\xi} \rb) = - \xi^2 \lb( [\phi(\xi)]^2 - \alpha^2 \rb)^{3/2},
\label{eq:LaneEmden}
\end{align}
with the boundary conditions $\phi(0)=1$ and $d\phi(0)/d\xi = 0$, and with $\xi_*$ still defined by $\phi(\xi_*) = \alpha$.
Moreover, in this regime, $\hat{M}(\xi) = -\xi^2 d\phi(\xi)/d\xi$.
For $\alpha \ll 1$, as relevant for the extremal configurations of the star, $\phi(\xi) \sim 1 \gg \alpha$ for most $\xi < \xi_*$, so the approximate governing equation, \eqref{LaneEmden}, becomes independent of $\alpha$ for most $\xi$ (indeed, it reduces to the Lane--Emden equation with a polytropic index of 3 \cite{Kippenhahn:2012zqe}); it is only in the regime $\xi\approx \xi_*$, where $\phi(\xi) \sim \alpha$, that the dependence on $\alpha$ enters.
However, in this regime, the governing equation tells us that $d \hat{M}(\xi) / d\xi = (d/d\xi) \lb( -\xi^2 d\phi(\xi)/d\xi \rb) \sim 0$; therefore, in the only regime where the solution does depend on $\alpha \ll 1$, the value of $\hat{M}(\xi)$ is approximately stationary.
Taken together, these observations imply that $\hat{M}(\xi_*)$ is approximately independent of $\alpha$ when $\alpha \ll 1$, and takes the value $\hat{M}(\xi_*) \approx 1.96$.
As such, and because the scaling factor $M_0 \propto \mu_e^{-2}$ also does not depend on $\alpha$ explicitly, the limiting physical mass of a WD with $\mu_e \neq 2$ can be given to a very good approximation by
\begin{align}
M_{\text{Chand.}} &= \frac{1}{2}M_{\text{Pl.}} \sqrt{\frac{6\pi}{g_e}} \lb( \frac{M_{\text{Pl.}}}{\mu_a \mu_e^R} \rb)^2 \hat{M}(\xi_*)\\\
&  \approx 1.4\,M_{\odot} \times (2/\mu_e)^2,
\label{eq:MChand}
\end{align}
even for $\mu_e \gg 2$.
We have checked explicitly with the numerical solution that this scaling is obeyed well, even for masses as large as $m_X \sim 10^{19}\,\text{GeV}$. 

\begin{figure}[t]
\includegraphics[width=0.95\columnwidth]{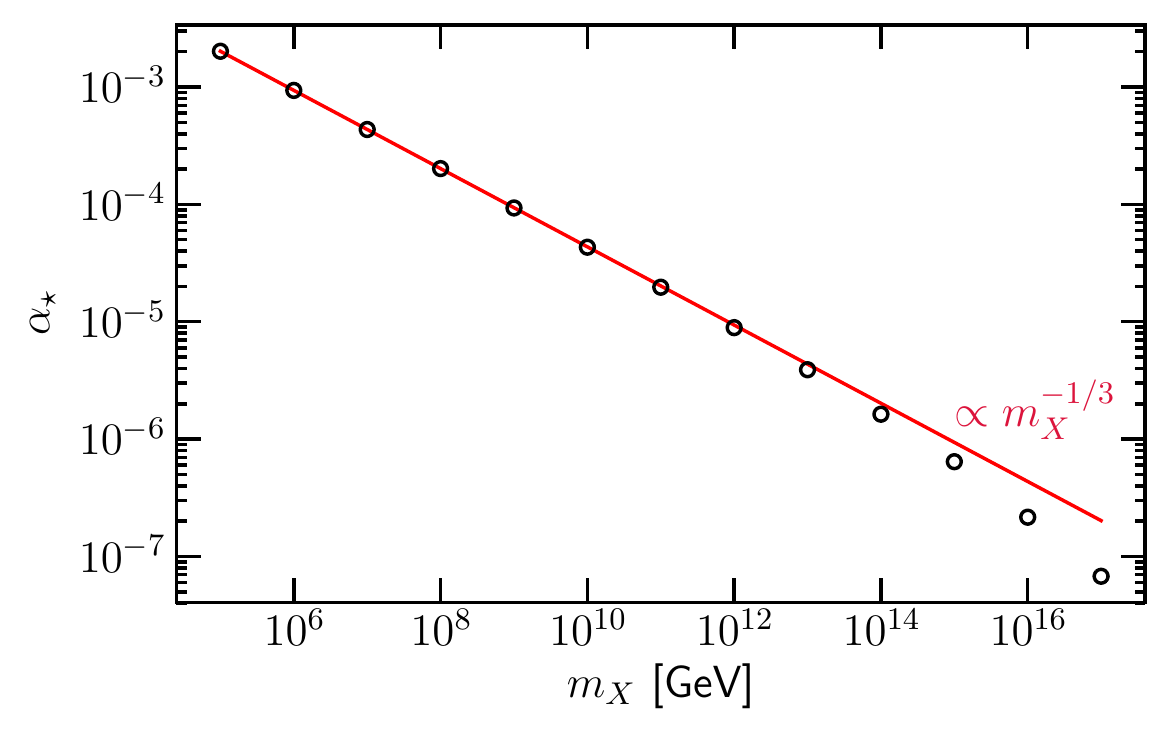}
\caption{ \label{fig:alpha_crits} The value of $\alpha_{\star}$ necessary to obtain the critical star assuming a pure $(NX)$ configuration with $Q_{(NX)}=+6$, as a function of the CHAMP mass $m_X$ (black circles). 
Over a wide range of masses, $\alpha_{\star}\propto m_X^{-1/3}$ (red line), as discussed in the text.}
\end{figure}

It is more difficult to develop an understanding of the scaling of the radius of the extremal star with $m_X$, as we need to know both how the value of $\alpha$ required to obtain the critical star scales, since $r_0 \propto \alpha / \mu_e$, and also how the value of $\xi_*$ depends on $\alpha$.
However, on the basis of the arguments advanced above about the independence of the solution to the Lane--Emden equation to the value of $\alpha$ for $\alpha \ll 1$ except in the region where $\xi\approx\xi_*$, when $\hat{M}$ is approximately stationary, we can argue that the value of $\xi_*$ is a fairly weak function of parameters when $\alpha \ll 1$. 
It remains to understand the scaling of $\alpha$ with $m_X$.

The critical stellar configuration is obtained when the EoS for the electrons at the stellar core comes too close to the extreme relativistic EoS, $P\propto\rho^{4/3}$ (see, e.g., \citeR{Shapiro:2000abc}); this in turn means that the condition is really one on the central electron number density of the star, and thus on the central mass density of the star (for fixed charge-to-mass ratio massive constituents).
Indeed, a GR fluid stability analysis (see, e.g., \citeR{Shapiro:2000abc}) shows that the relevant condition is that the central density of the extremal star scales as $\rho_C\propto \mu_e^2\propto m_X^2/Q_X^2$.
From \eqref{rhoC}, we see that for $\alpha \ll 1$, $\rho_C \propto \mu_e \alpha^{-3}$; we thus expect that $\alpha_\star \propto Q_X^{1/3} m_X^{-1/3}$ is a good approximation to the scaling of critical value of $\alpha$ with $m_X$, provided that $\alpha \ll 1$.
We verify numerically over a broad range of masses $m_X$ that the $m_X$ scaling is obeyed within an $\mathcal{O}(1)$ numerical factor; see \figref{alpha_crits}.
We also verify the $Q_X$ scaling numerically.
As a result, we expect that the radius of the extremal star obeys roughly $R_{\textsc{wd}}\propto Q_X^{4/3}m_X^{-4/3}$ to within an $\mathcal{O}(1)$ numerical factor, or 
\begin{align}
R_{\text{Chand.}} &\approx 1.5\times 10^{-3} \,R_{\odot} \times (2/\mu_e)^{4/3},
\label{eq:RChand}
\end{align}
where we used the \Chand\ radius for an extremal CO WD as the fiducial value.
Note also that this is self-consistent with the scaling of the central (technically, average) density $\rho_C \propto M_{\textsc{wd}}/R_{\textsc{wd}}^3 \propto Q_X^2 m_X^{-2} / (Q_X^{4/3} m_X^{-4/3})^3 \propto m_X^2/Q_X^2$.
Note that this implies that the central ion number density of the extremal star increases as $n_C \propto m_X/Q_X^2$.

We also that the electron mass does not enter \eqref{MChand} explicitly.

\subsection{CHAMP-modified WD structure}
\label{app:CHAMPmodWD}
The equilibrium configuration with a heavy CHAMP ($m_X \gg 20 \,\text{GeV}$, so that we can ignore the C and O ion masses relative to the $X$ mass whenever the ions and CHAMPs co-exist, or are bound) is a complete stratification of the $X$ material (for $Q_X>0$) or the homogeneously mixed $(N_jX)$ material (for $Q_X<0$),%
\footnote{\label{ftnt:stratification}%
		Strictly speaking, given sufficient time to settle, the homogeneous $(N_jX)$ material mixture will itself stratify into individual layers of pure $(N_jX)$ for each $j$ due to the small differences in the ion charges (and, with the mass being dominated by the CHAMP, the charge-to-mass ratios) of each species; we will ignore this additional complication in this work, as the timescale for a complete stratification of this type is likely extremely long (the charge differences are only on the order of 20\%), and we will assume instead that all the $(N_jX)$ species exist as one entirely homogeneous mixture at the center of the star.
	} %
and the ordinary homogeneous CO mixture; this occurs because of the relative buoyancy (more specifically, the much smaller charge-to-mass ratio) of the CHAMP-contaminated material as compared to the standard WD material; see discussion in \sectref{timescales}. 
We thus assume as input that the mean molecular mass per electron, $\hat{\mu}_e(\xi)$, undergoes a sharp transition (in reality, this would be smoothed by thermal effects; see \sectref{WDstructurePlus}) at a radius $\xi=\xi_X$ between its value for ordinary WD matter (for $\xi_X < \xi \leq \xi_*$), and its value for the CHAMP-contaminated matter (for $0\leq \xi \leq \xi_X$).
The value of $\xi_X$ is found consistently such that $M_X \equiv M(\xi_X)$ is the total CHAMP mass in the stratified core.

In the heavy CHAMP limit, we obtain the mean molecular mass per electron for the CHAMP-contaminated mixture using $A_X \equiv m_X / \mu_a$; $Z_X = Q_X$ (for $Q_X>0$) or $Z_{(N_jX)} = Q_j-|Q_X|$ (for $Q_X<0$); and $X_X = 1$ (for $Q_X>0$) or $X_{(N_j X)} = y_j$ (for $Q_X<0$), where $y_j$ is the fraction of negatively charged CHAMPs bound to ion species $j$ (satisfying $\sum_j y_j=1$):
\begin{align}
\mu_e(0\leq\xi<\xi_X) &\equiv 
\begin{cases}
	\dfrac{m_X}{\mu_a} Q_X^{-1} & Q_X>0 \\[3ex]
	\dfrac{m_X}{\mu_a} ( \bar{Q}_N -|Q_X| )^{-1} & Q_X<0,
\end{cases}
\end{align}
where $\bar{Q}_N \equiv \sum_j y_j Q_j$ is the mean charge of the ions to which the CHAMPs are bound (weighted by the fraction of CHAMPs bound to each ion species).
Given that we assume a CO mixture, $\bar{Q}_N \in [6,8]$; moreover, we assume a composition of equal mass-fraction abundances, so we will assume throughout this work as an approximation that $\bar{Q}_N \approx 7$.
This approximation of course ignores the small difference between the mass-fraction and number fraction of the ions, and also ignores the differential affinity of CHAMPs to bind to each ion species given their differing binding energies; however, since $\bar{Q}_N$ is in any event bounded in a small range, choosing the middle of that range introduces only an $\mathcal{O}(1)$-factor error in our results.

\begin{figure}[t]
\includegraphics[width=0.95\columnwidth]{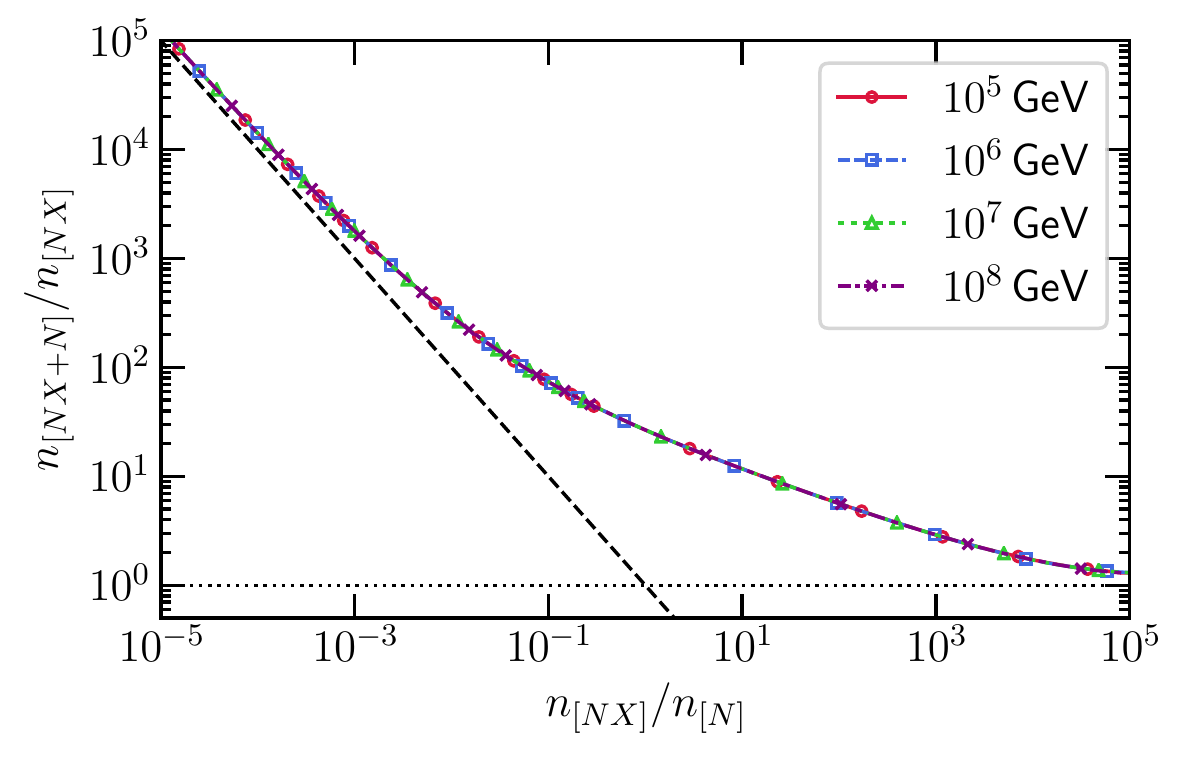}
\caption{ \label{fig:asymptote} 
	For an $X^-$ of varying mass as indicated in the legend, this plot shows a comparison of the central ion number densities in the central core of a fully stratified CO star containing some CHAMP contamination [denoted $n_{[NX+N]}$], with the central ion number densities in an isolated pure-($NX$) star of the same mass as the core [denoted $n_{[NX]}$], and the central ion number densities in the uncontaminated (CO) star [denoted $n_{[N]}$].
	In all cases, the WD structure is computed in the mean-ion approach assuming the `mean ion' in the unperturbed CO star has charge $+7$ (resulting in a core of heavy charge $+6$ ($NX$) objects).
	These results are obtained by fixing the total mass of the WD in the stratified and uncontaminated cases to be $M_{\textsc{wd}} = 1.2M_{\odot}$.
	It is clear that the central core object in the stratified star is always more dense than that of either the uncontaminated star or the isolated completely contaminated star, as a result of the overburden in the stratified case.
	However, as the central density in the isolated completely contaminated star becomes much larger [respectively, smaller] than the central ion density in the uncontaminated star, the central number density in the equal-mass core in the partially contaminated stratified case behaves more and more like that of the isolated completely contaminated star (the black dotted line shows $n_{[NX+N]}=n_{[NX]}$) [respectively, like that of the uncontaminated star (the black dashed line shows $n_{[NX+N]}=n_{[N]}$)].
	The value of the CHAMP mass $m_X$ appears to be irrelevant for this comparison, assuming it is large enough to guarantee complete stratification in the partially contaminated case case. 
}
\end{figure}

\begin{figure}[t]
\includegraphics[width=0.95\columnwidth]{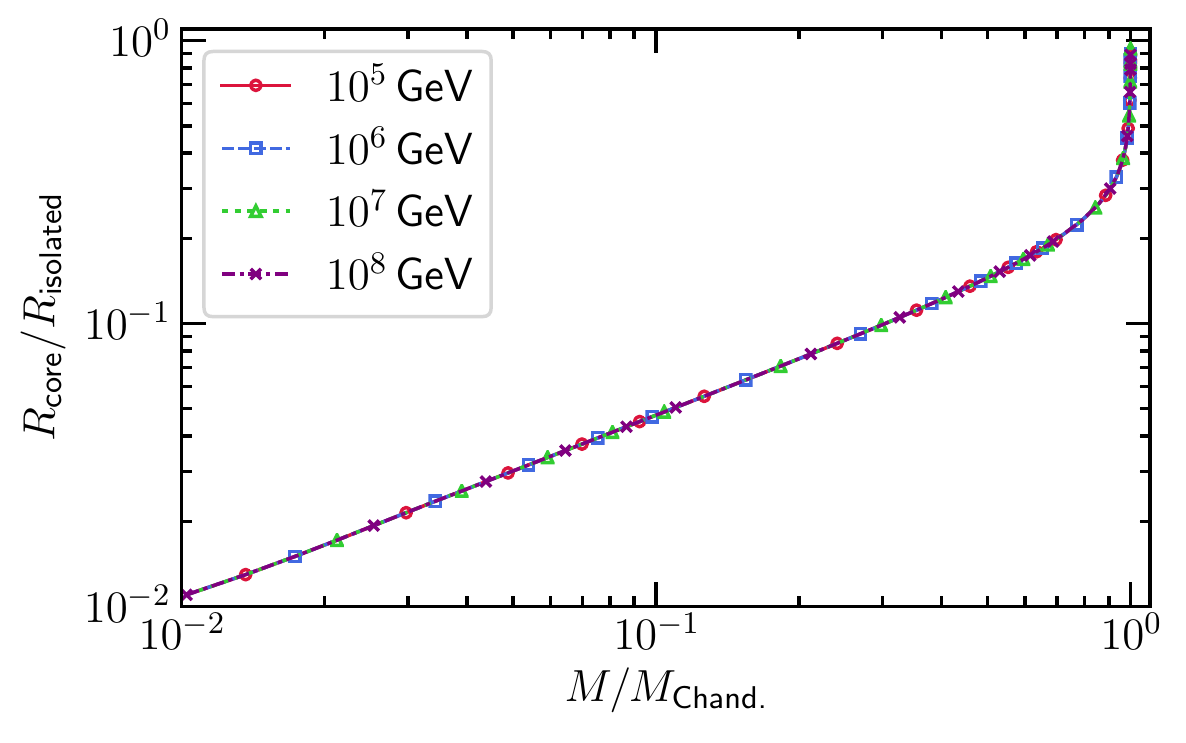}
\caption{ \label{fig:massradius} For an $X^-$ of varying mass $m_X$ as indicated in the legend, this plot shows a comparison of the radius of the central core of a fully stratified CO star containing some CHAMP contamination, with the radius of an equal-mass (denoted $M$) isolated pure-($NX$) star.
In both cases, the WD structure is computed in the mean-ion approach assuming the `mean ion' in the unperturbed CO star has charge $+7$ [resulting in a core of heavy charge $+6$ ($NX$) objects].
These results are obtained by fixing the total mass of the WD in the stratified case to be $M_{\textsc{wd}} = 1.2M_{\odot}$.
It is clear that the central core object in the stratified star is always more compact than the isolated completely contaminated star, as a result of the overburden in the stratified case `squashing' the central core; however, as the mass of the core in the stratified case approaches the \Chand\ mass of the isolated completely contaminated star, its radius and that of the isolated completely contaminated star converge.}
\end{figure}

For the case where the central pressure in the core vastly exceeds the ambient pressure in the CO material just outside the core (i.e., when the core itself is near-extremal), the solutions that are obtained via the above procedure have the \emph{approximate} appearance for $0\leq \xi \leq \xi_X$ of a small \emph{isolated} WD comprised of CHAMP-contaminated material (i.e., the CO overburden has little impact on the interior solution for near-extremal cores); see \figref[s]{XplusProfiles}, \ref{fig:asymptote}, and \ref{fig:massradius}.
This follows because the boundary condition for an isolated WD, $P(r=R)=0$, is more closely approximated as $P_{\text{ambient}}/P_{\text{core}} \rightarrow 0$.
Therefore, to maintain a stable hydrostatic equilibrium, the stratified inner CHAMP-contaminated material must obey a separate \Chand\ limit that can be given approximately as $M_X \lesssim 1.4 \, M_{\odot} \times ( 2Q' \mu_a / m_X )^2 \ll 1.4 \,M_\odot$, where $Q' = Q_X$ for $Q_X>0$ and $Q' =  \bar{Q}_N -|Q_X| \approx 7 - |Q_X|$ for $Q_X<0$.
Stratified cores of CHAMPs that violate this bound will thus collapse to a black hole in the center of the WD.
On the other hand, somewhat unsurprisingly, less extremal cores behave quite differently from an their equal-mass isolated WD brethren comprised of pure CHAMP-contaminated material: the CO overburden significantly reduces the radius of the core, and forces the central density of the core to be maintained at a much higher value than the isolated object; see again \figref[s]{XplusProfiles}, \ref{fig:asymptote}, and \ref{fig:massradius}.

For CHAMPs that are closer in mass to the ion species' masses, the picture will be more complicated; we do not treat this case.

\section{Electron heat conduction}
\label{app:ElectronConduction}
We take the electron heat conductivity $k_{\text{cd}}$ from \citeR{Potekhin:1999yv}. 
For completeness, we reproduce the relevant expressions from \citeR{Potekhin:1999yv} in full here; the formulae in this section hold in the natural unit system, $\hbar = c = k_B = 1$.
The thermal conductivity for electron conduction is given by
\begin{align}
k_{\text{cd}} &= \frac{\pi^2 T n_e}{3 E_F \nu_k},
\end{align}
where $n_e = (g_e m_e^3/6\pi^2) x_F^3$ is the electron density, $T$ is the temperature, $E_F = m_e \sqrt{1+x_F^2}$ is the electron Fermi energy, and the dominant contribution to the parameter $\nu_k = \nu_k^{ei} + \nu_k^{ee}$ is given by electron-ion scattering:
\begin{align}
\nu_k^{ei} &= \frac{4\pi \alpha^2 E_F}{p_F^3} \sum_j Z_j^2 n_j \Lambda_{k,j},
\end{align}
where $\alpha \approx 1/137$ is the fine structure constant, $p_F = m_e x_F$ is the electron Fermi momentum, and the sum runs over all ion species present in the multi-component plasma (MCP) [we adopt the approximate prescription detailed in the second-last paragraph in the left-hand column on page 6 of \citeR{Potekhin:1999yv} to handle the MCP case], with $Z_j$ and $n_j$ (respectively) the charge and number density of ion species $j$.
$\Lambda_{k,j}$ is the Coulomb logarithm for ion species $j$, defined by
\begin{align}
\Lambda_{k,j} &\equiv \lb[ \Lambda^1_{k}(s_j,w_j) + \lb( \frac{p_F}{E_F} \rb)^2 \Lambda^2_{k}(s_j,w_j) \rb] \nl \times G_{k,j}D_j,
\end{align}
where 
\begin{widetext}
\begin{align}
2 \Lambda^1_{k}(s,w) &\equiv s(1+s)^{-1}\lb( 1-e^{-w} \rb) + (1+sw)e^{sw}\Big[ \text{Ei}[-sw]-\text{Ei}[-(1+s)w] \Big] + \ln(1+1/s), \\
2 \Lambda^2_{k}(s,w) &\equiv \lb( w^{-1} + s^2(1+s)^{-1} \rb)\lb( 1 - e^{-w} \rb) -1 + s(2+sw)e^{sw} \Big[ \text{Ei}[-sw]-\text{Ei}[-(1+s)w] \Big] + 2s\ln(1+1/s),\\
G_{k,j} &\equiv \frac{\eta_j}{\sqrt{\eta_j^2+\lb[0.19 Z_j^{-1/6}\rb]^2}}\lb(1+0.122\,\beta_j^2\rb) + 0.0105( 1 - Z_j^{-1}) \lb[ 1 + (p_F/E_F)^3\beta_j \rb] \frac{\eta_j}{\lb(\eta_j^2+0.0081\rb)^{3/2}}, \\
D_j &\equiv \exp\lb[ - \alpha_{0,j} u_{-1} e^{-9.1\eta_j}/4 \rb], \qquad \alpha_{0,j} \equiv \frac{ 4 p_F^2 a_j^2 }{ 3 \Gamma_j \eta_j }
\end{align}
\end{widetext}
with $\text{Ei}(x) \equiv - \int_{-x}^\infty t^{-1} e^{-t} dt$ being the exponential integral function, and where
\begin{align}
s_j &\equiv (q_{s,j} / 2 p_F)^2, \\
w_j &\equiv u_{-2} \lb( 2p_F / q_{D,j} \rb)^2 \lb( 1 + \beta_j / 3 \rb),
\end{align}
with
\begin{align}
\eta_j &\equiv T / \omega_{p,j}, &
\beta_j&\equiv \pi\alpha Z_j p_F / E_F, \\
a_j &= ( 3 / 4\pi n_j )^{1/3}, &
\Gamma_j &\equiv \frac{ \alpha Z_j^{5/3} }{T} \lb( \frac{4\pi n_e}{3} \rb)^{1/3},
\end{align}
\begin{align}
q_{D,j} &= \sqrt{3\Gamma_j}/a_j ,&
k_{\text{TF}}^2 &=  \frac{4\alpha}{\pi} E_Fp_F,\\
q_{s,j} &= (q_{i,j}^2+k_{\text{TF}}^2)e^{-\beta_j} ,&
\omega_{p,j}^2 &\equiv 4\pi \alpha Z_j^2 n_j m_j^{-1} \\
q^2_{i,j} &= q^2_{D,j} ( 1 + 0.06 \Gamma_j ) e^{-\sqrt{\Gamma_j}}.
\end{align}
Finally, the constants $u_{-1} = 2.8$ and $u_{-2} = 13.0$.

We also include the sub-dominant contribution to the parameter $\nu_k$ from electron-electron scattering \cite{1997A&A...323..415P,Potekhin:1999yv}:
\begin{align}
\nu_k^{ee} &= \frac{12\alpha^2 T^2}{\pi^3 E_F} \lb(\frac{ p_F }{k_{\text{TF}}} \rb)^3 J(x_F,y),
\end{align}
where 
\begin{align}
y &= \sqrt{3} \,\omega_{p,e}  / T, & \omega_{p,e}^2 = 4\pi \alpha n_e / E_F,
\end{align}
and
\begin{align}
J(x,y) &\approx \lb( 1 + \frac{6}{5x^2} + \frac{2}{5x^4} \rb) \nl 
	\times \Bigg[  \frac{y^3}{3(1+0.07414y)^3} \ln\lb( 1 + \frac{2.81}{y} - \frac{0.81}{y} \frac{p_F^2}{E_F^2} \rb) \nl
	\qquad + \frac{\pi^5}{6} \frac{y^4}{(13.91+y)^4 } \Bigg].
\end{align}

\section{Free-free opacity}
\label{app:freeFreeOpacity}
The formulae in this section hold in the natural unit system, $\hbar = c = k_B = 1$.

The Rosseland mean free-free opacity is given by the Kramers opacity \cite{Kippenhahn:2012zqe}, suitably corrected \cite{Rybicki:2004abc,Timmes_1992} by a velocity averaged (i.e., thermally averaged) Gaunt factor $ \langle \bar{g}_{\text{ff}} \rangle(T) $ that is also appropriately averaged over frequency \cite{Rybicki:2004abc,Cox:1968abc}:
\begin{align}
\kappa_{\text{ff, rad}} &\approx \lb( \frac{2}{1+945\zeta(7)/\pi^6}\rb) \frac{2}{15} \sqrt{\frac{2\pi}{3}} \frac{\alpha^3}{m_e^{3/2}\mu_a^2} \langle \bar{g}_{\text{ff}} \rangle(T) \nl \times ( 1 + X ) ( X +  Y  + B)\rho T^{-7/2}  \label{eq:Kramers1} \\
&\approx 0.1175\, \text{cm}^2 \, \text{g}^{-1} \times \langle \bar{g}_{\text{ff}} \rangle(T) \times F(\rho,T) \nl \times ( 1 + X ) ( X +  Y  + B) \nl \times \lb( \frac{ \rho }{10^8\, \text{g\,cm}^{-3} } \rb) \times \lb( \frac{ T }{ 10^9\, \text{K} } \rb)^{-7/2},
\label{eq:radiativeOpacity}
\end{align}
where we have assumed that $Z/A = 1/2$ for all elements in the chemical composition, except hydrogen; $X$ and $Y$ are, respectively, the helium and hydrogen mass fractions (both zero in our case); $B = \sum_{i} Z_i^2 X_i / A_i $ is the heavy-element contribution (with the sum running over all elements heavier than helium), which takes the value $B=3.5$ for a $X({}^{12}\text{C})= X({}^{16}\text{O})=0.5$ mixture; $m_e$ is the electron mass; and $\mu_a$ is the atomic mass unit.
The fraction in the $(\,\cdots)$-bracket on the first line of \eqref{Kramers1} is numerically equal to $1.0044$.

The appropriate frequency-averaged value $\langle \bar{g}_{\text{ff}} \rangle(T)$ of the velocity-averaged free-free Gaunt factors $\bar{g}_{\text{ff}}(T,\nu)$  is given by
\begin{align}
\langle \bar{g}_{\text{ff}} \rangle(T) &\equiv \dfrac{ \int_0^\infty dx\, x^7 e^{2x} (e^x-1)^{-3} \bar{g}_{\text{ff}}(T,\nu = xT/2\pi)}{\int_0^\infty dx\, x^7 e^{2x} (e^x-1)^{-3}};
\label{eq:freqAveragedGaunt}
\end{align}
this definition ensures that the total free-free radiative opacity is the Rosseland mean \cite{Rybicki:2004abc,Kippenhahn:2012zqe}.
The thermally averaged Gaunt factors $\bar{g}_{\text{ff}}(T,\nu)$ are tabulated in \citeR{1990ApJS...74..291I}, which results indicate that this correction is usually $\mathcal{O}(1)$, or within an order of magnitude or two thereof; we omit this Gaunt factor correction.

\bibliographystyle{JHEP}
\bibliography{references.bib}

\providecommand{\href}[2]{#2}\begingroup\justify\begin{thebibliography}{100}

\bibitem{Ellis:1983ew}
J.~R. Ellis, J.~S. Hagelin, D.~V. Nanopoulos, K.~A. Olive and M.~Srednicki,
  \emph{{Supersymmetric Relics from the Big Bang}},
  \href{http://dx.doi.org/10.1016/0550-3213(84)90461-9}{\emph{Nucl. Phys.} {\bf
  B238} (1984) 453--476}.

\bibitem{Byrne:2002ri}
M.~Byrne, C.~F. Kolda and P.~Regan, \emph{{Bounds on charged, stable
  superpartners from cosmic ray production}},
  \href{http://dx.doi.org/10.1103/PhysRevD.66.075007}{\emph{Phys. Rev.} {\bf
  D66} (2002) 075007} [\href{https://arxiv.org/abs/hep-ph/0202252}{{\tt
  arXiv:hep-ph/0202252}}].

\bibitem{Appelquist:2000nn}
T.~Appelquist, H.-C. Cheng and B.~A. Dobrescu, \emph{{Bounds on universal extra
  dimensions}}, \href{http://dx.doi.org/10.1103/PhysRevD.64.035002}{\emph{Phys.
  Rev.} {\bf D64} (2001) 035002}
  [\href{https://arxiv.org/abs/hep-ph/0012100}{{\tt arXiv:hep-ph/0012100}}].

\bibitem{Byrne:2003sa}
M.~Byrne, \emph{{Universal extra dimensions and charged LKPs}},
  \href{http://dx.doi.org/10.1016/j.physletb.2003.12.064}{\emph{Phys. Lett.}
  {\bf B583} (2004) 309--314} [\href{https://arxiv.org/abs/hep-ph/0311160}{{\tt
  arXiv:hep-ph/0311160}}].

\bibitem{Wise:2014jva}
M.~B. Wise and Y.~Zhang, \emph{{Stable Bound States of Asymmetric Dark
  Matter}}, \href{http://dx.doi.org/10.1103/PhysRevD.90.055030,
  10.1103/PhysRevD.91.039907}{\emph{Phys. Rev.} {\bf D90} (2014) 055030}
  [\href{https://arxiv.org/abs/1407.4121}{{\tt arXiv:1407.4121}}].

\bibitem{Gresham:2017zqi}
M.~I. Gresham, H.~K. Lou and K.~M. Zurek, \emph{{Nuclear Structure of Bound
  States of Asymmetric Dark Matter}},
  \href{http://dx.doi.org/10.1103/PhysRevD.96.096012}{\emph{Phys. Rev.} {\bf
  D96} (2017) 096012} [\href{https://arxiv.org/abs/1707.02313}{{\tt
  arXiv:1707.02313}}].

\bibitem{Grabowska:2018lnd}
D.~M. Grabowska, T.~Melia and S.~Rajendran, \emph{{Detecting Dark Blobs}},
  \href{http://dx.doi.org/10.1103/PhysRevD.98.115020}{\emph{Phys. Rev.} {\bf
  D98} (2018) 115020} [\href{https://arxiv.org/abs/1807.03788}{{\tt
  arXiv:1807.03788}}].

\bibitem{Bai:2018dxf}
Y.~Bai, A.~J. Long and S.~Lu, \emph{{Dark Quark Nuggets}},
  \href{http://dx.doi.org/10.1103/PhysRevD.99.055047}{\emph{Phys. Rev.} {\bf
  D99} (2019) 055047} [\href{https://arxiv.org/abs/1810.04360}{{\tt
  arXiv:1810.04360}}].

\bibitem{Cahn:1980ss}
R.~N. Cahn and S.~L. Glashow, \emph{{Chemical Signatures for Superheavy
  Elementary Particles}},
  \href{http://dx.doi.org/10.1126/science.213.4508.607}{\emph{Science} {\bf
  213} (1981) 607--611}.

\bibitem{DeRujula:1989fe}
A.~De~Rujula, S.~L. Glashow and U.~Sarid, \emph{{Charged Dark Matter}},
  \href{http://dx.doi.org/10.1016/0550-3213(90)90227-5}{\emph{Nucl. Phys.} {\bf
  B333} (1990) 173--194}.

\bibitem{Dimopoulos:1989hk}
S.~Dimopoulos, D.~Eichler, R.~Esmailzadeh and G.~D. Starkman, \emph{{Getting a
  Charge Out of Dark Matter}},
  \href{http://dx.doi.org/10.1103/PhysRevD.41.2388}{\emph{Phys. Rev.} {\bf D41}
  (1990) 2388}.

\bibitem{Perl:2001xi}
M.~L. Perl, P.~C. Kim, V.~Halyo, E.~R. Lee, I.~T. Lee, D.~Loomba et~al.,
  \emph{{The Search for stable, massive, elementary particles}},
  \href{http://dx.doi.org/10.1142/S0217751X01003548}{\emph{Int. J. Mod. Phys.}
  {\bf A16} (2001) 2137--2164}
  [\href{https://arxiv.org/abs/hep-ex/0102033}{{\tt arXiv:hep-ex/0102033}}].

\bibitem{Burdin:2014xma}
S.~Burdin, M.~Fairbairn, P.~Mermod, D.~Milstead, J.~Pinfold, T.~Sloan et~al.,
  \emph{{Non-collider searches for stable massive particles}},
  \href{http://dx.doi.org/10.1016/j.physrep.2015.03.004}{\emph{Phys. Rept.}
  {\bf 582} (2015) 1--52} [\href{https://arxiv.org/abs/1410.1374}{{\tt
  arXiv:1410.1374}}].

\bibitem{2010JCAP...02..031S}
F.~J. {S{\'a}nchez-Salcedo}, E.~{Mart{\'\i}nez-G{\'o}mez} and J.~{Maga{\~n}a},
  \emph{{On the fraction of dark matter in charged massive particles
  (CHAMPs)}},
  \href{http://dx.doi.org/10.1088/1475-7516/2010/02/031}{\emph{JCAP} {\bf 2010}
  (2010) 031} [\href{https://arxiv.org/abs/1002.3145}{{\tt arXiv:1002.3145}}].

\bibitem{Fairbairn:2006gg}
M.~Fairbairn, A.~C. Kraan, D.~A. Milstead, T.~Sjostrand, P.~Z. Skands and
  T.~Sloan, \emph{{Stable Massive Particles at Colliders}},
  \href{http://dx.doi.org/10.1016/j.physrep.2006.10.002}{\emph{Phys. Rept.}
  {\bf 438} (2007) 1--63} [\href{https://arxiv.org/abs/hep-ph/0611040}{{\tt
  arXiv:hep-ph/0611040}}].

\bibitem{Alvager:1967fgm}
T.~Alv{\"a}ger and R.~A. Naumann, \emph{{Search for stable heavy massive
  particles of positive integral charge}},
  \href{http://dx.doi.org/10.1016/0370-2693(67)90367-X}{\emph{Phys. Lett.} {\bf
  24B} (1967) 647--648}.

\bibitem{Middleton:1979zz}
R.~Middleton, R.~W. Zurmuhle, J.~Klein and R.~V. Kollarits, \emph{{Search for
  an Anomalously Heavy Isotope of Oxygen}},
  \href{http://dx.doi.org/10.1103/PhysRevLett.43.429}{\emph{Phys. Rev. Lett.}
  {\bf 43} (1979) 429--431}.

\bibitem{Smith:1979rz}
P.~F. Smith and J.~R.~J. Bennett, \emph{{A Search for Heavy Stable Particles}},
  \href{http://dx.doi.org/10.1016/0550-3213(79)90006-3}{\emph{Nucl. Phys.} {\bf
  B149} (1979) 525--533}.

\bibitem{Smith:1982qu}
P.~F. Smith, J.~R.~J. Bennett, G.~J. Homer, J.~D. Lewin, H.~E. Walford and
  W.~A. Smith, \emph{{A Search for Anomalous Hydrogen in Enriched
  $\text{D}_2\text{O}$, using a Time-Of-Flight Spectrometer}},
  \href{http://dx.doi.org/10.1016/0550-3213(82)90271-1}{\emph{Nucl. Phys.} {\bf
  B206} (1982) 333--348}.

\bibitem{Turkevich:1984zz}
A.~Turkevich, K.~Wielgoz and T.~E. Economou, \emph{{Searching for supermassive
  Cahn-Glashow particles}},
  \href{http://dx.doi.org/10.1103/PhysRevD.30.1876}{\emph{Phys. Rev.} {\bf D30}
  (1984) 1876--1880}.

\bibitem{Dick:1984mk}
W.~J. Dick, G.~W. Greenlees and S.~L. Kaufman, \emph{{Search for Anomalous
  Isotopes of Sodium and Possible Superheavy Stable Particles}},
  \href{http://dx.doi.org/10.1103/PhysRevLett.53.431}{\emph{Phys. Rev. Lett.}
  {\bf 53} (1984) 431}.

\bibitem{Nitz:1986gb}
D.~Nitz, D.~Ciampa, T.~Hemmick, D.~Elmore, P.~W. Kubik, S.~L. Olsen et~al.,
  \emph{{A search for anomalously heavy isotopes of low Z nuclei}},
  \href{http://dx.doi.org/10.1063/1.36084}{\emph{AIP Conf. Proc.} {\bf 150}
  (1986) 1143--1146}.

\bibitem{Dick:1985wk}
W.~J. Dick, G.~W. Greenlees and S.~L. Kaufman, \emph{{Search for Possible
  Superheavy Particles in Sodium Nuclei}},
  \href{http://dx.doi.org/10.1103/PhysRevD.33.32}{\emph{Phys. Rev.} {\bf D33}
  (1986) 32}.

\bibitem{Norman:1986ux}
E.~B. Norman, S.~B. Gazes and D.~A. Bennett, \emph{{Searches for Supermassive
  $X^-$ Particles in Iron}},
  \href{http://dx.doi.org/10.1103/PhysRevLett.58.1403}{\emph{Phys. Rev. Lett.}
  {\bf 58} (1987) 1403--1406}.

\bibitem{Pichard:1987ub}
B.~Pichard, J.~Rich, M.~Spiro, F.~Biraben, G.~Grynberg, P.~Verkerk et~al.,
  \emph{{On The Possibility Of Detecting Superheavy Hydrogen Through
  Centrifugation And Atomic Spectroscopy}},
  \href{http://dx.doi.org/10.1016/0370-2693(87)91256-1}{\emph{Phys. Lett.} {\bf
  B193} (1987) 383--388}.

\bibitem{Norman:1988fd}
E.~B. Norman, R.~B. Chadwick, K.~T. Lesko, R.~M. Larimer and D.~C. Hoffman,
  \emph{{Search for Supermassive Cahn-Glashow Particles in Lead}},
  \href{http://dx.doi.org/10.1103/PhysRevD.39.2499}{\emph{Phys. Rev.} {\bf D39}
  (1989) 2499}.

\bibitem{Polikanov:1990sf}
S.~Polikanov, C.~S. Sastri, G.~Herrmann, K.~Lutzenkirchen, M.~Overbeck,
  N.~Trautmann et~al., \emph{{Search for supermassive nuclei in nature}},
  \href{http://dx.doi.org/10.1007/BF01288200}{\emph{Z. Phys.} {\bf A338} (1990)
  357--361}.

\bibitem{Hemmick:1989ns}
T.~K. Hemmick et~al., \emph{{A Search for Anomalously Heavy Isotopes of Low $Z$
  Nuclei}}, \href{http://dx.doi.org/10.1103/PhysRevD.41.2074}{\emph{Phys. Rev.}
  {\bf D41} (1990) 2074--2080}.

\bibitem{Verkerk:1991jf}
P.~Verkerk, G.~Grynberg, B.~Pichard, M.~Spiro, S.~Zylberajch, M.~E. Goldberg
  et~al., \emph{{Search for superheavy hydrogen in sea water}},
  \href{http://dx.doi.org/10.1103/PhysRevLett.68.1116}{\emph{Phys. Rev. Lett.}
  {\bf 68} (1992) 1116--1119}.

\bibitem{Yamagata:1993jq}
T.~Yamagata, Y.~Takamori and H.~Utsunomiya, \emph{{Search for anomalously heavy
  hydrogen in deep sea water at 4000-m}},
  \href{http://dx.doi.org/10.1103/PhysRevD.47.1231}{\emph{Phys. Rev.} {\bf D47}
  (1993) 1231--1234}.

\bibitem{Javorsek:2001yu}
D.~Javorsek, D.~Elmore, E.~Fischbach, T.~Miller, D.~Oliver and V.~Teplitz,
  \emph{{Experimental limits on the existence of strongly interacting massive
  particles bound to gold nuclei}},
  \href{http://dx.doi.org/10.1103/PhysRevD.64.012005}{\emph{Phys. Rev.} {\bf
  D64} (2001) 012005}.

\bibitem{Mueller:2003ji}
P.~Mueller, L.~B. Wang, R.~J. Holt, Z.~T. Lu, T.~P. O'Connor and J.~P.
  Schiffer, \emph{{Search for anomalously heavy isotopes of helium in the
  earth's atmosphere}},
  \href{http://dx.doi.org/10.1103/PhysRevLett.92.022501}{\emph{Phys. Rev.
  Lett.} {\bf 92} (2004) 022501}
  [\href{https://arxiv.org/abs/nucl-ex/0302025}{{\tt arXiv:nucl-ex/0302025}}].

\bibitem{PDGOtherSearches}
{\scshape Particle Data Group}, M.~Tanabashi et~al., \emph{Review of particle
  physics (other particle searches)}, {\emph{Phys. Rev.} {\bf D98} (2018)
  030001}.

\bibitem{Pospelov:2006sc}
M.~Pospelov, \emph{{Particle physics catalysis of thermal Big Bang
  Nucleosynthesis}},
  \href{http://dx.doi.org/10.1103/PhysRevLett.98.231301}{\emph{Phys. Rev.
  Lett.} {\bf 98} (2007) 231301}
  [\href{https://arxiv.org/abs/hep-ph/0605215}{{\tt arXiv:hep-ph/0605215}}].

\bibitem{Kohri:2006cn}
K.~Kohri and F.~Takayama, \emph{{Big bang nucleosynthesis with long lived
  charged massive particles}},
  \href{http://dx.doi.org/10.1103/PhysRevD.76.063507}{\emph{Phys. Rev.} {\bf
  D76} (2007) 063507} [\href{https://arxiv.org/abs/hep-ph/0605243}{{\tt
  arXiv:hep-ph/0605243}}].

\bibitem{Kaplinghat:2006qr}
M.~Kaplinghat and A.~Rajaraman, \emph{{Big Bang Nucleosynthesis with Bound
  States of Long-lived Charged Particles}},
  \href{http://dx.doi.org/10.1103/PhysRevD.74.103004}{\emph{Phys. Rev.} {\bf
  D74} (2006) 103004} [\href{https://arxiv.org/abs/astro-ph/0606209}{{\tt
  arXiv:astro-ph/0606209}}].

\bibitem{Bird:2007ge}
C.~Bird, K.~Koopmans and M.~Pospelov, \emph{{Primordial Lithium Abundance in
  Catalyzed Big Bang Nucleosynthesis}},
  \href{http://dx.doi.org/10.1103/PhysRevD.78.083010}{\emph{Phys. Rev.} {\bf
  D78} (2008) 083010} [\href{https://arxiv.org/abs/hep-ph/0703096}{{\tt
  arXiv:hep-ph/0703096}}].

\bibitem{Kawasaki:2007xb}
M.~Kawasaki, K.~Kohri and T.~Moroi, \emph{{Big-Bang Nucleosynthesis with
  Long-Lived Charged Slepton}},
  \href{http://dx.doi.org/10.1016/j.physletb.2007.03.063}{\emph{Phys. Lett.}
  {\bf B649} (2007) 436--439} [\href{https://arxiv.org/abs/hep-ph/0703122}{{\tt
  arXiv:hep-ph/0703122}}].

\bibitem{Jedamzik:2007cp}
K.~Jedamzik, \emph{{The cosmic Li-6 and Li-7 problems and BBN with long-lived
  charged massive particles}},
  \href{http://dx.doi.org/10.1103/PhysRevD.77.063524}{\emph{Phys. Rev.} {\bf
  D77} (2008) 063524} [\href{https://arxiv.org/abs/0707.2070}{{\tt
  arXiv:0707.2070}}].

\bibitem{Jedamzik:2007qk}
K.~Jedamzik, \emph{{Bounds on long-lived charged massive particles from Big
  Bang nucleosynthesis}},
  \href{http://dx.doi.org/10.1088/1475-7516/2008/03/008}{\emph{JCAP} {\bf 0803}
  (2008) 008} [\href{https://arxiv.org/abs/0710.5153}{{\tt arXiv:0710.5153}}].

\bibitem{Pospelov:2007js}
M.~Pospelov, \emph{{Bridging the primordial A=8 divide with catalyzed big bang
  nucleosynthesis}},  \href{https://arxiv.org/abs/0712.0647}{{\tt
  arXiv:0712.0647}}.

\bibitem{Pospelov:2008ta}
M.~Pospelov, J.~Pradler and F.~D. Steffen, \emph{{Constraints on Supersymmetric
  Models from Catalytic Primordial Nucleosynthesis of Beryllium}},
  \href{http://dx.doi.org/10.1088/1475-7516/2008/11/020}{\emph{JCAP} {\bf 0811}
  (2008) 020} [\href{https://arxiv.org/abs/0807.4287}{{\tt arXiv:0807.4287}}].

\bibitem{Kamimura:2008fx}
M.~Kamimura, Y.~Kino and E.~Hiyama, \emph{{Big-Bang Nucleosynthesis Reactions
  Catalyzed by a Long-Lived Negatively-Charged Leptonic Particle}},
  \href{http://dx.doi.org/10.1143/PTP.121.1059}{\emph{Prog. Theor. Phys.} {\bf
  121} (2009) 1059--1098} [\href{https://arxiv.org/abs/0809.4772}{{\tt
  arXiv:0809.4772}}].

\bibitem{PradlerThesis}
J.~Pradler, \emph{The long-lived stau as a thermal relic}.
\newblock PhD thesis, Technische Universit\"at M\"unchen, 2009.

\bibitem{Jedamzik:2009uy}
K.~Jedamzik and M.~Pospelov, \emph{{Big Bang Nucleosynthesis and Particle Dark
  Matter}}, \href{http://dx.doi.org/10.1088/1367-2630/11/10/105028}{\emph{New
  J. Phys.} {\bf 11} (2009) 105028}
  [\href{https://arxiv.org/abs/0906.2087}{{\tt arXiv:0906.2087}}].

\bibitem{Kusakabe:2010cb}
M.~Kusakabe, T.~Kajino, T.~Yoshida and G.~J. Mathews, \emph{{New results on
  catalyzed BBN with a long-lived negatively-charged massive particle}},
  \href{http://dx.doi.org/10.1103/PhysRevD.81.083521}{\emph{Phys. Rev.} {\bf
  D81} (2010) 083521} [\href{https://arxiv.org/abs/1001.1410}{{\tt
  arXiv:1001.1410}}].

\bibitem{Pospelov:2010hj}
M.~Pospelov and J.~Pradler, \emph{{Big Bang Nucleosynthesis as a Probe of New
  Physics}},
  \href{http://dx.doi.org/10.1146/annurev.nucl.012809.104521}{\emph{Ann. Rev.
  Nucl. Part. Sci.} {\bf 60} (2010) 539--568}
  [\href{https://arxiv.org/abs/1011.1054}{{\tt arXiv:1011.1054}}].

\bibitem{Kusakabe:2017brd}
M.~Kusakabe, G.~J. Mathews, T.~Kajino and M.-K. Cheoun, \emph{{Review on
  Effects of Long-lived Negatively Charged Massive Particles on Big Bang
  Nucleosynthesis}},
  \href{http://dx.doi.org/10.1142/S021830131741004X}{\emph{Int. J. Mod. Phys.}
  {\bf E26} (2017) 1741004} [\href{https://arxiv.org/abs/1706.03143}{{\tt
  arXiv:1706.03143}}].

\bibitem{Dunsky:2018mqs}
D.~Dunsky, L.~J. Hall and K.~Harigaya, \emph{{CHAMP Cosmic Rays}},
  \href{http://dx.doi.org/10.1088/1475-7516/2019/07/015}{\emph{JCAP} {\bf 1907}
  (2019) 015} [\href{https://arxiv.org/abs/1812.11116}{{\tt
  arXiv:1812.11116}}].

\bibitem{Gould:1989gw}
A.~Gould, B.~T. Draine, R.~W. Romani and S.~Nussinov, \emph{{Neutron Stars:
  Graveyard of Charged Dark Matter}},
  \href{http://dx.doi.org/10.1016/0370-2693(90)91745-W}{\emph{Phys. Lett.} {\bf
  B238} (1990) 337--343}.

\bibitem{SanchezSalcedo:2008zd}
F.~J. Sanchez-Salcedo and E.~Martinez-Gomez, \emph{{Galactic constraints on
  CHAMPs}},  \href{https://arxiv.org/abs/0812.0797}{{\tt arXiv:0812.0797}}.

\bibitem{Chuzhoy:2008zy}
L.~Chuzhoy and E.~W. Kolb, \emph{{Reopening the window on charged dark
  matter}}, \href{http://dx.doi.org/10.1088/1475-7516/2009/07/014}{\emph{JCAP}
  {\bf 0907} (2009) 014} [\href{https://arxiv.org/abs/0809.0436}{{\tt
  arXiv:0809.0436}}].

\bibitem{Takibayev:2017xyz}
N.~Takibayev and K.~Boshkayev, eds., \emph{Neutron Stars: Physics, Properties
  and Dynamics}.
\newblock Nova Science Publishers, New York, 2017.

\bibitem{1968MNRAS.138..495L}
D.~{Lynden-Bell} and R.~{Wood}, \emph{{The gravo-thermal catastrophe in
  isothermal spheres and the onset of red-giant structure for stellar
  systems}}, \href{http://dx.doi.org/10.1093/mnras/138.4.495}{\emph{Mon. Not.
  Roy. Astron. Soc.} {\bf 138} (1968) 495}.

\bibitem{Chandrasekhar_1931}
S.~Chandrasekhar, \emph{The maximum mass of ideal white dwarfs},
  \href{http://dx.doi.org/10.1086/143324}{\emph{Astrophys. J.} {\bf 74} (1931)
  81}.

\bibitem{Hawking:1974wb}
S.~W. Hawking, \emph{Black hole explosions?},
  \href{http://dx.doi.org/10.1038/248030a0}{\emph{Nature} {\bf 248} (1974)
  30--31}.

\bibitem{Graham:2015apa}
P.~W. Graham, S.~Rajendran and J.~Varela, \emph{{Dark Matter Triggers of
  Supernovae}}, \href{http://dx.doi.org/10.1103/PhysRevD.92.063007}{\emph{Phys.
  Rev.} {\bf D92} (2015) 063007} [\href{https://arxiv.org/abs/1505.04444}{{\tt
  arXiv:1505.04444}}].

\bibitem{Bramante:2015cua}
J.~Bramante, \emph{{Dark matter ignition of type Ia supernovae}},
  \href{http://dx.doi.org/10.1103/PhysRevLett.115.141301}{\emph{Phys. Rev.
  Lett.} {\bf 115} (2015) 141301} [\href{https://arxiv.org/abs/1505.07464}{{\tt
  arXiv:1505.07464}}].

\bibitem{Graham:2018efk}
P.~W. Graham, R.~Janish, V.~Narayan, S.~Rajendran and P.~Riggins, \emph{{White
  Dwarfs as Dark Matter Detectors}},
  \href{http://dx.doi.org/10.1103/PhysRevD.98.115027}{\emph{Phys. Rev.} {\bf
  D98} (2018) 115027} [\href{https://arxiv.org/abs/1805.07381}{{\tt
  arXiv:1805.07381}}].

\bibitem{Acevedo:2019gre}
J.~F. Acevedo and J.~Bramante, \emph{{Supernovae Sparked By Dark Matter in
  White Dwarfs}},
  \href{http://dx.doi.org/10.1103/PhysRevD.100.043020}{\emph{Phys. Rev.} {\bf
  D100} (2019) 043020} [\href{https://arxiv.org/abs/1904.11993}{{\tt
  arXiv:1904.11993}}].

\bibitem{Janish:2019nkk}
R.~Janish, V.~Narayan and P.~Riggins, \emph{{Type Ia supernovae from dark
  matter core collapse}},
  \href{http://dx.doi.org/10.1103/PhysRevD.100.035008}{\emph{Phys. Rev.} {\bf
  D100} (2019) 035008} [\href{https://arxiv.org/abs/1905.00395}{{\tt
  arXiv:1905.00395}}].

\bibitem{WoosleyPetschek:1990aa}
S.~E. Woosley, \emph{{T}ype {I} {S}upernovae: {C}arbon {D}eflagration and
  {D}etonation},  in \emph{Supernovae} (A.~G. Petschek, ed.).
\newblock Springer, 1990.

\bibitem{Timmes_1992}
F.~X. Timmes and S.~E. Woosley, \emph{The conductive propagation of nuclear
  flames. {I}---{D}egenerate {C + O} and {O + Ne + Mg} white dwarfs},
  \href{http://dx.doi.org/10.1086/171746}{\emph{Astrophys. J.} {\bf 396} (1992)
  649}.

\bibitem{Koester_1990}
D.~Koester and G.~Chanmugam, \emph{Physics of white dwarf stars},
  \href{http://dx.doi.org/10.1088/0034-4885/53/7/001}{\emph{Reports on Progress
  in Physics} {\bf 53} (1990) 837}.

\bibitem{Montero-Camacho:2019jte}
P.~Montero-Camacho, X.~Fang, G.~Vasquez, M.~Silva and C.~M. Hirata,
  \emph{{Revisiting constraints on asteroid-mass primordial black holes as dark
  matter candidates}},
  \href{http://dx.doi.org/10.1088/1475-7516/2019/08/031}{\emph{JCAP} {\bf 2019}
  (2019) 031} [\href{https://arxiv.org/abs/1906.05950}{{\tt
  arXiv:1906.05950}}].

\bibitem{Kippenhahn:2012zqe}
R.~Kippenhahn, A.~Weigert and A.~Weiss, \emph{Stellar Structure and Evolution}.
\newblock Springer, second~ed., 2012.

\bibitem{2017ASPC..509....3D}
P.~{Dufour}, S.~{Blouin}, S.~{Coutu}, M.~{Fortin-Archambault}, C.~{Thibeault},
  P.~{Bergeron} et~al., \emph{{The Montreal White Dwarf Database: A Tool for
  the Community}},  in \emph{20th European White Dwarf Workshop} (P.~E.
  {Tremblay}, B.~{Gaensicke} and T.~{Marsh}, eds.), vol.~509 of
  \emph{Astronomical Society of the Pacific Conference Series}, p.~3, Sheridan
  Books (Ann Arbor, Michigan), 2017.
\newblock [\href{https://arxiv.org/abs/1610.00986}{{\tt arXiv:1610.00986}}].

\bibitem{2016MNRAS.461.2100T}
P.~E. {Tremblay}, J.~{Cummings}, J.~S. {Kalirai}, B.~T. {G{\"a}nsicke},
  N.~{Gentile-Fusillo} and R.~{Raddi}, \emph{{The field white dwarf mass
  distribution}}, \href{http://dx.doi.org/10.1093/mnras/stw1447}{\emph{Mon.
  Not. Roy. Astron. Soc.} {\bf 461} (2016) 2100--2114}
  [\href{https://arxiv.org/abs/1606.05292}{{\tt arXiv:1606.05292}}].

\bibitem{Kepler:2006ns}
S.~O. Kepler, S.~J. Kleinman, A.~Nitta, D.~Koester, B.~G. Castanheira,
  O.~Giovannini et~al., \emph{{White Dwarf Mass Distribution in the SDSS}},
  \href{http://dx.doi.org/10.1111/j.1365-2966.2006.11388.x}{\emph{Mon. Not.
  Roy. Astron. Soc.} {\bf 375} (2007) 1315--1324}
  [\href{https://arxiv.org/abs/astro-ph/0612277}{{\tt
  arXiv:astro-ph/0612277}}].

\bibitem{polin2019nebular}
A.~Polin, P.~Nugent and D.~Kasen, \emph{Nebular models of sub-{C}handrasekhar
  mass type {Ia} supernovae: Clues to the origin of {Ca}-rich transients},
  \href{https://arxiv.org/abs/1910.12434}{{\tt arXiv:1910.12434}}.

\bibitem{Landau:1980egq}
L.~D. Landau and E.~M. Lifshitz, \emph{Course of Theoretical Physics, Volume 5:
  Statistical Physics, Part 1}.
\newblock Pergamon Press, 3rd~ed., 1980.

\bibitem{doi:10.1146/annurev.astro.38.1.191}
W.~Hillebrandt and J.~C. Niemeyer, \emph{Type {Ia} supernova explosion models},
  \href{http://dx.doi.org/10.1146/annurev.astro.38.1.191}{\emph{Annual Review
  of Astronomy and Astrophysics} {\bf 38} (2000) 191--230}.

\bibitem{Jones:2013dta}
D.~O. Jones et~al., \emph{{The Discovery of the Most Distant Known Type Ia
  Supernova at Redshift 1.914}},
  \href{http://dx.doi.org/10.1088/0004-637X/768/2/166}{\emph{Astrophys. J.}
  {\bf 768} (2013) 166} [\href{https://arxiv.org/abs/1304.0768}{{\tt
  arXiv:1304.0768}}].

\bibitem{Shapiro:2000abc}
S.~L. Shapiro and S.~A. Teukolsky, \emph{Black Holes, White Dwarfs, and Neutron
  Stars}.
\newblock Wiley-VCH, Weinheim, 2000.

\bibitem{Tolman:1939it}
R.~C. Tolman, \emph{Static solutions of {Einstein}'s field equations for
  spheres of fluid},
  \href{http://dx.doi.org/10.1103/PhysRev.55.364}{\emph{Phys. Rev.} {\bf 55}
  (1939) 364--373}.

\bibitem{Oppenheimer:1939cs}
J.~R. Oppenheimer and G.~M. Volkoff, \emph{On massive neutron cores},
  \href{http://dx.doi.org/10.1103/PhysRev.55.374}{\emph{Phys. Rev.} {\bf 55}
  (1939) 374--381}.

\bibitem{Salpeter:1961zz}
E.~E. Salpeter, \emph{{Energy and Pressure of a Zero-Temperature Plasma}},
  \href{http://dx.doi.org/10.1086/147194}{\emph{Astrophys. J.} {\bf 134} (1961)
  669--682}.

\bibitem{Kittel:1980abc}
C.~Kittel and H.~Kroemer, \emph{Thermal Physics}.
\newblock W.H. Freeman and Company, United States, 1980.

\bibitem{Potekhin:1999yv}
A.~Y. Potekhin, D.~A. Baiko, P.~Haensel and D.~G. Yakovlev, \emph{{Transport
  properties of degenerate electrons in neutron star envelopes and white dwarf
  cores}}, {\emph{Astron. Astrophys.} {\bf 346} (1999) 345}
  [\href{https://arxiv.org/abs/astro-ph/9903127}{{\tt
  arXiv:astro-ph/9903127}}].

\bibitem{Cox:1968abc}
J.~P. Cox and R.~T. Giuli, \emph{{Principles of Stellar Structure. Volume 1:
  Physical Principles}}.
\newblock Gordon and Breach, London, 1968.

\bibitem{Cox:1968def}
J.~P. Cox and R.~T. Giuli, \emph{{Principles of Stellar Structure. Volume 2:
  Applications to Stars}}.
\newblock Gordon and Breach, London, 1968.

\bibitem{Rybicki:2004abc}
G.~B. Rybicki and A.~P. Lightman, \emph{Radiative Processes in Astrophysics}.
\newblock Wiley-VCH, Weinheim, 2004.

\bibitem{Gasques:2005ar}
L.~R. Gasques, A.~V. Afanasjev, E.~F. Aguilera, M.~Beard, L.~C. Chamon, P.~Ring
  et~al., \emph{{Nuclear fusion in dense matter: Reaction rate and carbon
  burning}}, \href{http://dx.doi.org/10.1103/PhysRevC.72.025806}{\emph{Phys.
  Rev.} {\bf C72} (2005) 025806}
  [\href{https://arxiv.org/abs/astro-ph/0506386}{{\tt
  arXiv:astro-ph/0506386}}].

\bibitem{Yakovlev:2006fi}
D.~G. Yakovlev, L.~R. Gasques, M.~Beard, M.~Wiescher and A.~V. Afanasjev,
  \emph{{Fusion reactions in multicomponent dense matter}},
  \href{http://dx.doi.org/10.1103/PhysRevC.74.035803}{\emph{Phys. Rev.} {\bf
  C74} (2006) 035803} [\href{https://arxiv.org/abs/astro-ph/0608488}{{\tt
  arXiv:astro-ph/0608488}}].

\bibitem{WoosleyPrivate}
S.~E. Woosley. Private communication, 2019.

\bibitem{Griest:1989wd}
K.~Griest and M.~Kamionkowski, \emph{{Unitarity Limits on the Mass and Radius
  of Dark Matter Particles}},
  \href{http://dx.doi.org/10.1103/PhysRevLett.64.615}{\emph{Phys. Rev. Lett.}
  {\bf 64} (1990) 615}.

\bibitem{Tiburzi:2000rsq}
B.~C. Tiburzi and B.~R. Holstein, \emph{Bound states of a uniform spherical
  charge distributions--revisited!},
  \href{http://dx.doi.org/10.1119/1.19502}{\emph{Am. J. Phys.} {\bf 68} (2000)
  640}.

\bibitem{Foot:2010yz}
R.~Foot, \emph{{Do magnetic fields prevent mirror particles from entering the
  galactic disk?}},
  \href{http://dx.doi.org/10.1016/j.physletb.2011.04.012}{\emph{Phys. Lett.}
  {\bf B699} (2011) 230--232} [\href{https://arxiv.org/abs/1011.5078}{{\tt
  arXiv:1011.5078}}].

\bibitem{McDermott:2010pa}
S.~D. McDermott, H.-B. Yu and K.~M. Zurek, \emph{{Turning off the Lights: How
  Dark is Dark Matter?}},
  \href{http://dx.doi.org/10.1103/PhysRevD.83.063509}{\emph{Phys. Rev.} {\bf
  D83} (2011) 063509} [\href{https://arxiv.org/abs/1011.2907}{{\tt
  arXiv:1011.2907}}].

\bibitem{Clayton1983}
D.~Clayton, \emph{Principles of Stellar Evolution and Nucleosynthesis}.
\newblock University of Chicago Press, 1983.

\bibitem{Catalan:2008tr}
S.~Catalan, J.~Isern, E.~Garcia-Berro and I.~Ribas, \emph{{The initial-final
  mass relationship of white dwarfs revisited: effect on the luminosity
  function and mass distribution}},
  \href{http://dx.doi.org/10.1111/j.1365-2966.2008.13356.x}{\emph{Mon. Not.
  Roy. Astron. Soc.} {\bf 387} (2008) 1693}
  [\href{https://arxiv.org/abs/0804.3034}{{\tt arXiv:0804.3034}}].

\bibitem{Bahcall:2000nu}
J.~N. Bahcall, M.~H. Pinsonneault and S.~Basu, \emph{{Solar models: Current
  epoch and time dependences, neutrinos, and helioseismological properties}},
  \href{http://dx.doi.org/10.1086/321493}{\emph{Astrophys. J.} {\bf 555} (2001)
  990--1012} [\href{https://arxiv.org/abs/astro-ph/0010346}{{\tt
  arXiv:astro-ph/0010346}}].

\bibitem{Serenelli:2011py}
A.~M. Serenelli, W.~C. Haxton and C.~Pena-Garay, \emph{{Solar models with
  accretion. I. Application to the solar abundance problem}},
  \href{http://dx.doi.org/10.1088/0004-637X/743/1/24}{\emph{Astrophys. J.} {\bf
  743} (2011) 24} [\href{https://arxiv.org/abs/1104.1639}{{\tt
  arXiv:1104.1639}}].

\bibitem{Vinyoles:2016djt}
N.~Vinyoles, A.~M. Serenelli, F.~L. Villante, S.~Basu, J.~Bergstr\"om, M.~C.
  Gonzalez-Garcia et~al., \emph{{A new Generation of Standard Solar Models}},
  \href{http://dx.doi.org/10.3847/1538-4357/835/2/202}{\emph{Astrophys. J.}
  {\bf 835} (2017) 202} [\href{https://arxiv.org/abs/1611.09867}{{\tt
  arXiv:1611.09867}}].

\bibitem{Heger:2002by}
A.~Heger, C.~L. Fryer, S.~E. Woosley, N.~Langer and D.~H. Hartmann, \emph{{How
  massive single stars end their life}},
  \href{http://dx.doi.org/10.1086/375341}{\emph{Astrophys. J.} {\bf 591} (2003)
  288--300} [\href{https://arxiv.org/abs/astro-ph/0212469}{{\tt
  arXiv:astro-ph/0212469}}].

\bibitem{TOI1996}
R.~B. Firestone, \emph{Table of Isotopes}.
\newblock Wiley, 1st {CD}-{ROM}~ed., 1996.

\bibitem{Woosely:2003ng}
S.~E. Woosley, S.~Wunsch and M.~Kuhlen, \emph{{Carbon ignition in type Ia
  supernovae: an analytic model}},
  \href{http://dx.doi.org/10.1086/383530}{\emph{Astrophys. J.} {\bf 607} (2004)
  921} [\href{https://arxiv.org/abs/astro-ph/0307565}{{\tt
  arXiv:astro-ph/0307565}}].

\bibitem{Baym:1979zs}
G.~Baym and C.~Pethick, \emph{Physics of neutron stars},
  \href{http://dx.doi.org/10.1146/annurev.aa.17.090179.002215}{\emph{Annu. Rev.
  Astron. Astrophys.} {\bf 17} (1979) 415--443}.

\bibitem{Evans:2018bqy}
N.~W. Evans, C.~A.~J. O'Hare and C.~McCabe, \emph{{Refinement of the standard
  halo model for dark matter searches in light of the Gaia Sausage}},
  \href{http://dx.doi.org/10.1103/PhysRevD.99.023012}{\emph{Phys. Rev.} {\bf
  D99} (2019) 023012} [\href{https://arxiv.org/abs/1810.11468}{{\tt
  arXiv:1810.11468}}].

\bibitem{GarciaBerro:2007an}
E.~Garc\'ia-Berro, L.~G. Althaus, A.~H. C\'orsico and J.~Isern,
  \emph{{Gravitational settling of Ne-22 and white dwarf evolution}},
  \href{http://dx.doi.org/10.1086/527536}{\emph{Astrophys. J.} {\bf 677} (2008)
  473} [\href{https://arxiv.org/abs/0712.1212}{{\tt arXiv:0712.1212}}].

\bibitem{Renedo:2010vb}
I.~Renedo, L.~G. Althaus, M.~M. {Miller Bertolami}, A.~D. Romero, A.~H.
  C\'orsico, R.~D. Rohrmann et~al., \emph{{New cooling sequences for old white
  dwarfs}},
  \href{http://dx.doi.org/10.1088/0004-637X/717/1/183}{\emph{Astrophys. J.}
  {\bf 717} (2010) 183--195} [\href{https://arxiv.org/abs/1005.2170}{{\tt
  arXiv:1005.2170}}].

\bibitem{Kwok_1994}
S.~Kwok, \emph{Planetary nebulae: A modern view},
  \href{http://dx.doi.org/10.1086/133384}{\emph{Publications of the
  Astronomical Society of the Pacific} {\bf 106} (1994) 344}.

\bibitem{Unglaub:2008abc}
{K. Unglaub}, \emph{Mass-loss and diffusion in subdwarf b stars and hot white
  dwarfs: do weak winds exist?},
  \href{http://dx.doi.org/10.1051/0004-6361:20078019}{\emph{A\&A} {\bf 486}
  (2008) 923--940}.

\bibitem{2015SSRv..191..111F}
L.~{Ferrario}, D.~{de Martino} and B.~T. {G{\"a}nsicke}, \emph{{Magnetic White
  Dwarfs}}, \href{http://dx.doi.org/10.1007/s11214-015-0152-0}{\emph{Space Sci.
  Rev.} {\bf 191} (2015) 111--169}
  [\href{https://arxiv.org/abs/1504.08072}{{\tt arXiv:1504.08072}}].

\bibitem{1992A&A...257..534B}
E.~{Bravo}, J.~{Isern}, R.~{Canal} and J.~{Labay}, \emph{{On the contribution
  of Ne-22 to the synthesis of Fe-54 and Ni-58 in thermonuclear supernovae}},
  \href{http://adsabs.harvard.edu/abs/1992A%26A...257..534B}{\emph{Astron.
  Astrophys.} {\bf 257} (1992) 534--538}.

\bibitem{Bildsten:2001xb}
L.~Bildsten and D.~M. Hall, \emph{{Gravitational settling of 22ne in liquid
  white dwarf interiors}},
  \href{http://dx.doi.org/10.1086/319169}{\emph{Astrophys. J.} {\bf 549} (2001)
  L219} [\href{https://arxiv.org/abs/astro-ph/0101365}{{\tt
  arXiv:astro-ph/0101365}}].

\bibitem{Deloye:2002xb}
C.~J. Deloye and L.~Bildsten, \emph{{Gravitational settling of ${}^{22}$Ne in
  liquid white dwarf interiors-cooling and seismological effects}},
  \href{http://dx.doi.org/10.1086/343800}{\emph{Astrophys. J.} {\bf 580} (2002)
  1077--1090} [\href{https://arxiv.org/abs/astro-ph/0207623}{{\tt
  arXiv:astro-ph/0207623}}].

\bibitem{Burgers1969}
J.~Burgers, \emph{Flow Equations for Composite Gases}.
\newblock Academic Press, New York and London, 1969.

\bibitem{Olson:1975ts}
E.~Olson and M.~Bailyn, \emph{{Internal Structure of Multicomponent Static
  Spherical Gravitating Fluids}},
  \href{http://dx.doi.org/10.1103/PhysRevD.12.3030}{\emph{Phys. Rev.} {\bf D12}
  (1975) 3030--3036}.

\bibitem{1971IAUS...42...97V}
H.~M. {van Horn}, \emph{{Cooling of White Dwarfs}},  in \emph{White Dwarfs}
  (W.~J. {Luyten}, ed.),
  \href{https://ui.adsabs.harvard.edu/abs/1971IAUS...42...97V}{Proceedings of
  the IAU Symposium no. 42}, p.~97, Springer-Verlag (Dordrecht), 1971.

\bibitem{Kaplan:2018dqx}
D.~E. Kaplan and S.~Rajendran, \emph{{Firewalls in General Relativity}},
  \href{http://dx.doi.org/10.1103/PhysRevD.99.044033}{\emph{Phys. Rev.} {\bf
  D99} (2019) 044033} [\href{https://arxiv.org/abs/1812.00536}{{\tt
  arXiv:1812.00536}}].

\bibitem{Bondi_1952}
H.~Bondi, \emph{On spherically symmetrical accretion},
  \href{http://dx.doi.org/10.1093/mnras/112.2.195}{\emph{Monthly Notices of the
  Royal Astronomical Society} {\bf 112} (1952) 195--204}.

\bibitem{Eddington:1926abc}
A.~S. Eddington, \emph{The Internal Constitution of the Stars}.
\newblock Cambridge University Press, Cambridge, England, 1926.

\bibitem{WD1}
\url{http://www.montrealwhitedwarfdatabase.org/WDs/Gaia%20DR2%201116006982252923136/Gaia%20DR2%201116006982252923136.html}
  (accessed November 5, 2019).

\bibitem{WD6}
\url{http://www.montrealwhitedwarfdatabase.org/WDs/Gaia%20DR2%205037670392494744320/Gaia%20DR2%205037670392494744320.html}
  (accessed November 5, 2019).

\bibitem{WD3}
\url{http://www.montrealwhitedwarfdatabase.org/WDs/LP642-052/LP642-052.html}
  (accessed November 5, 2019).

\bibitem{WD7}
\url{http://www.montrealwhitedwarfdatabase.org/WDs/Gaia%20DR2%20502957442209211136/Gaia%20DR2%20502957442209211136.html}
  (accessed November 5, 2019).

\bibitem{WD2}
\url{http://www.montrealwhitedwarfdatabase.org/WDs/WD%202051-208/WD%202051-208.html}
  (accessed November 5, 2019).

\bibitem{WD4}
\url{http://www.montrealwhitedwarfdatabase.org/WDs/SDSS%20J090632.65+080715.9/SDSS%20J090632.65+080715.9.html}
  (accessed November 5, 2019).

\bibitem{WD5}
\url{http://www.montrealwhitedwarfdatabase.org/WDs/SDSS%20J220435.05+001242.9/SDSS%20J220435.05+001242.9.html}
  (accessed November 5, 2019).

\bibitem{Perets_2010}
H.~B. Perets, A.~Gal-Yam, P.~A. Mazzali, D.~Arnett, D.~Kagan, A.~V. Filippenko
  et~al., \emph{A faint type of supernova from a white dwarf with a helium-rich
  companion}, \href{http://dx.doi.org/10.1038/nature09056}{\emph{Nature} {\bf
  465} (2010) 322--325} [\href{https://arxiv.org/abs/0906.2003}{{\tt
  arXiv:0906.2003}}].

\bibitem{Kasliwal_2012}
M.~M. Kasliwal, S.~R. Kulkarni, A.~Gal-Yam, P.~E. Nugent, M.~Sullivan,
  L.~Bildsten et~al., \emph{Calcium-rich gap transients in the remote outskirts
  of galaxies},
  \href{http://dx.doi.org/10.1088/0004-637x/755/2/161}{\emph{Astrophys. J.}
  {\bf 755} (2012) 161} [\href{https://arxiv.org/abs/1111.6109}{{\tt
  arXiv:1111.6109}}].

\bibitem{Lunnan:2016ake}
R.~Lunnan et~al., \emph{{Two New Calcium-Rich Gap Transients in Group and
  Cluster Environments}},
  \href{http://dx.doi.org/10.3847/1538-4357/836/1/60}{\emph{Astrophys. J.} {\bf
  836} (2017) 60} [\href{https://arxiv.org/abs/1612.00454}{{\tt
  arXiv:1612.00454}}].

\bibitem{De:2018wpe}
K.~De et~al., \emph{{iPTF 16hgs: A double-peaked Ca-rich gap transient in a
  metal poor, star forming dwarf galaxy}},
  \href{http://dx.doi.org/10.3847/1538-4357/aadf8e}{\emph{Astrophys. J.} {\bf
  866} (2018) 72} [\href{https://arxiv.org/abs/1806.10623}{{\tt
  arXiv:1806.10623}}].

\bibitem{shen2019progenitors}
K.~J. Shen, E.~Quataert and R.~Pakmor, \emph{{The Progenitors of Calcium-Strong
  Transients}},
  \href{http://dx.doi.org/10.3847/1538-4357/ab5370}{\emph{Astrophys. J.} {\bf
  887} (2019) 180} [\href{https://arxiv.org/abs/1908.08056}{{\tt
  arXiv:1908.08056}}].

\bibitem{Frohmaier:2018fed}
C.~Frohmaier, M.~Sullivan, K.~Maguire and P.~E. Nugent, \emph{{The volumetric
  rate of calcium-rich transients in the local universe}},
  \href{http://dx.doi.org/10.3847/1538-4357/aabc0b}{\emph{Astrophys. J.} {\bf
  858} (2018) 50} [\href{https://arxiv.org/abs/1804.03103}{{\tt
  arXiv:1804.03103}}].

\bibitem{2011ApJ...734...38W}
S.~E. {Woosley} and D.~{Kasen}, \emph{{Sub-Chandrasekhar Mass Models for
  Supernovae}},
  \href{http://dx.doi.org/10.1088/0004-637X/734/1/38}{\emph{Astrophys. J.} {\bf
  734} (2011) 38} [\href{https://arxiv.org/abs/1010.5292}{{\tt
  arXiv:1010.5292}}].

\bibitem{KasenPrivateCommunication}
D.~Kasen. Private communication, 2017, 2019.

\bibitem{Fedderke:2013pbc}
M.~A. Fedderke, E.~W. Kolb, T.~Lin and L.-T. Wang, \emph{{Gamma-ray constraints
  on dark-matter annihilation to electroweak gauge and Higgs bosons}},
  \href{http://dx.doi.org/10.1088/1475-7516/2014/01/001}{\emph{JCAP} {\bf 1401}
  (2014) 001} [\href{https://arxiv.org/abs/1310.6047}{{\tt arXiv:1310.6047}}].

\bibitem{Krane:1988abc}
K.~S. Krane, \emph{Introductory Nuclear Physics}.
\newblock John Wiley \& Sons, Hoboken, NJ, 1988.

\bibitem{Shankar:1994abc}
R.~Shankar, \emph{Principles of Quantum Mechanics}.
\newblock Springer, second~ed., 1994.

\bibitem{1997A&A...323..415P}
A.~Y. {Potekhin}, G.~{Chabrier} and D.~G. {Yakovlev}, \emph{{Internal
  temperatures and cooling of neutron stars with accreted envelopes.}},
  {\emph{Astron. Astrophys.} {\bf 323} (1997) 415--428}
  [\href{https://arxiv.org/abs/astro-ph/9706148}{{\tt
  arXiv:astro-ph/9706148}}].

\bibitem{1990ApJS...74..291I}
N.~{Itoh}, K.~{Kojo} and M.~{Nakagawa}, \emph{{Relativistic free-free Gaunt
  factor of the dense high-temperature stellar plasma. II - Carbon and oxygen
  plasmas}}, \href{http://dx.doi.org/10.1086/191500}{\emph{Astrophys. J. Suppl.
  Ser.} {\bf 74} (1990) 291--314}.

\end{thebibliography}\endgroup
\end{document}